\documentclass[reprint,superscriptaddress,prb]{revtex4-1}

\usepackage{titlesec}%
\usepackage{subfigure}%
\usepackage{amsmath}%
\usepackage{amsfonts}%
\usepackage{amssymb}%
\usepackage{feynmp}%
\usepackage{graphicx}%
\usepackage{setspace}%
\usepackage{comment}%
\usepackage{dsfont}%
\usepackage{color}%
\usepackage{tikz}
\usetikzlibrary{arrows,calc}
\usepackage[pdftex,colorlinks=true,linkcolor=blue,citecolor=blue]{hyperref}

\newcommand{\tr}{\text{Tr}} 
\newcommand{\A}{\mathcal{A}} 
\newcommand{\F}{\mathcal{F}} 
\newcommand{\W}{\mathcal{W}} 
\newcommand{\C}{\mathcal{C}} 
\newcommand{\I}{\mathcal{I}} 
\newcommand{\w}{\wedge}
\newcommand{\ket}[1]{\left| #1 \right>} 
\newcommand{\bra}[1]{\left< #1 \right|} 
\newcommand{\braket}[2]{\left< #1 \vphantom{#2} \right| \left. #2 \vphantom{#1} \right>} 
\newcommand{\matrixel}[3]{\left<#1 \vphantom{#2} \vphantom{#3} \left| #2 \vphantom{#1} \vphantom{#3} \right| #3 \vphantom{#1} \vphantom{#2} \right>}
\newcommand{\rr}{d}

\numberwithin{equation}{section}

\begin{document}

\title{Electric Multipole Moments, Topological Multipole Moment Pumping, \\and Chiral Hinge States in Crystalline Insulators}

\author{Wladimir A. Benalcazar}
\affiliation{Department of Physics and Institute for Condensed Matter Theory, University of Illinois at Urbana-Champaign, IL 61801, USA}
\author{B. Andrei Bernevig}
\affiliation{Department of Physics, Princeton University, New Jersey 08544, USA}
\affiliation{Donostia International Physics Center, P. Manuel de Lardizabal 4, 20018 Donostia-San Sebasti\'an, Spain}\thanks{on sabbatical}
\affiliation{Laboratoire Pierre Aigrain, \'Ecole Normale Sup\'erieure-PSL Research University, CNRS, Universit\'e Pierre et Marie Curie-Sorbonne Universit\'es,
Universit\'e Paris Diderot-Sorbonne Paris Cit\'e, 24 rue Lhomond, 75231 Paris Cedex 05, France}\thanks{on sabbatical}
\affiliation{Sorbonne Universit\'es, UPMC Univ Paris 06, UMR 7589, LPTHE, F-75005, Paris, France}\thanks{on sabbatical}
\author{Taylor L. Hughes}
\email[To whom correspondence should be addressed at ]{hughest@illinois.edu}
\affiliation{Department of Physics and Institute for Condensed Matter Theory, University of Illinois at Urbana-Champaign, IL 61801, USA}

\date{\today}

\begin{abstract}
We extend the theory of dipole moments in crystalline insulators to higher multipole moments. As first formulated in Ref. \onlinecite{benalcazar2017quad}, we show that bulk quadrupole and octupole moments can be realized in crystalline insulators. In this paper, we expand in great detail the theory presented in Ref. \onlinecite{benalcazar2017quad}, and extend it to cover associated topological pumping phenomena, and a novel class of 3D insulator with chiral hinge states. We start by deriving the boundary properties of continuous classical dielectrics hosting only bulk dipole, quadrupole, or octupole moments. In quantum-mechanical crystalline insulators, these higher multipole bulk moments manifest themselves by the presence of boundary-localized moments of lower dimension, in exact correspondence with the electromagnetic theory of classical continuous dielectrics. In the presence of certain symmetries, these moments are quantized, and their boundary signatures are fractionalized. These multipole moments then correspond to new symmetry-protected topological phases. The topological structure of these phases is described by ``nested" Wilson loops, which we define. These Wilson loops reflect the bulk-boundary correspondence in a way that makes evident a hierarchical classification of the multipole moments. Just as a varying dipole generates charge pumping, a varying quadrupole generates dipole pumping, and a varying octupole generates quadrupole pumping. For non-trivial adiabatic cycles, the transport of these moments is quantized. An analysis of these interconnected phenomena leads to the conclusion that a new kind of Chern-type insulator exists, which has chiral, hinge-localized modes in 3D. We provide the minimal models for the quantized multipole moments, the non-trivial pumping processes and the hinge Chern insulator, and describe the topological invariants that protect them. 
\end{abstract}

\maketitle

\tableofcontents

\section{Introduction}

\begin{figure}
\centering
\includegraphics[width=\columnwidth]{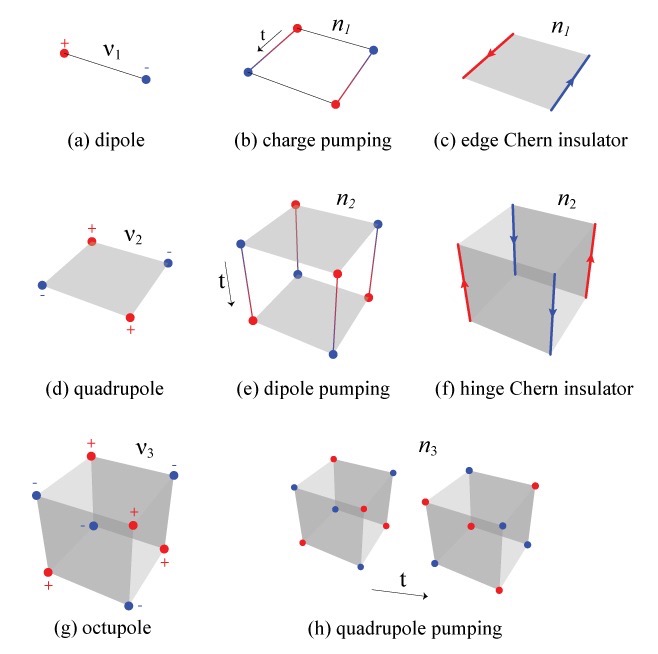}
\caption{(Color online) Multipole moments, associated multipole pumping processes, and derived topological insulators. (a,d,e) Dipole, quadrupole, and octupole insulators, respectively. Dots of different color represent corner-localized charges of opposite charge. When protected by symmetries, charges are quantized to either 0 or $\pm e/2$ as denoted by the $\mathbb{Z}_2$ topological indices $\nu_1$, $\nu_2$, and $\nu_3$, respectively. (b,e,h) Charge, dipole, and quadrupole pumping, respectively. Pumping over non-trivial closed cycles pumps quanta of charge, dipole, and quadrupole, described by Chern numbers $n_1$, $n_2$, and $n_3$, respectively. (c,f) Insulators with same topology as pumping processes: Chern insulator with chiral edge-localized modes (c), and Chern insulator with hinge-localized modes (f).}
\label{fig:summary}
\end{figure}

A successful theory describing the phenomenon of  bulk electric  polarization in crystalline insulators remained elusive for decades after the development of the band theory of crystals. The difficulty stemmed from the fact that
the macroscopic electric polarization of a periodic crystal cannot be unambiguously defined as the dipole of a unit cell~\cite{resta1992} and, therefore, the absolute macroscopic polarization of a crystal is ill defined. The recognition that only derivatives of the polarization are well-defined observables and correspond to experimental measurements\cite{resta1992} led to a resolution of this problem and to the formulation of what is now known as the modern theory of polarization~\cite{king-smith1993,vanderbilt1993,resta1993, resta1994,resta2007} in crystalline insulators. This theory is formulated in terms of Berry phases~\cite{berry1984, zak1989}, which account for the dipole moment densities in the bulk of a material, and it has its minimal realization in 1 dimension~\cite{SSH1979,kivelson1982} (1D). A bulk dipole moment manifests itself through the existence of boundary charges~\cite{vanderbilt1993} (Fig.~\ref{fig:summary}a). When the dipole moment densities vary over time, e.g., by an adiabatic evolution of an insulating Hamiltonian over time, electronic currents appear across the bulk of the material where the polarization is changing (Fig.~\ref{fig:summary}b)~\cite{king-smith1993, vanderbilt1993}. In particular, if adiabatic evolutions of the Hamiltonian are carried over closed cycles (i.e., those in which the initial and final Hamiltonians are the same), the electronic transport is quantized~\cite{thouless1983}. This quantization is given by a Chern number, and, mathematically, in systems with charge conservation, is closely related to the Hall conductance of a Chern insulator~\cite{thouless1982} (Fig.~\ref{fig:summary}c). 

A remarkable pattern develops in the topological objects describing these systems that follows a hierarchical mathematical structure as the dimensionality of space increases. For example, the expressions for the polarization $P_1$ \cite{zak1989,king-smith1993}, the hall conductance $\sigma_{xy}$ of a Chern insulator \cite{klitzing1980,thouless1982,haldane1988}, and the magneto-electric polarizability $P_3$ of a 3D time-reversal invariant or inversion symmetric topological insulator \cite{qi2008,essin2009,hughes2011inversion,turner2012}, are given by
\begin{align}
P_1&=-\frac{e}{2\pi}\int_{BZ} \tr \left[\A\right]
\label{eq:polarization}\\
\sigma_{xy}&=-\frac{e^2}{2\pi h}\int_{BZ} \tr\left[d \A+i \A\w \A\right]
\label{eq:HallConductance}\\
P_3&=-\frac{e^2}{4\pi h}\int_{BZ} \tr\left[\A \w d\A + \frac{2i}{3} \A \w \A \w \A \right],
\label{eq:MagnetoElectricPolarizability}
\end{align}
where $BZ$ is the Brillouin zone in one, two, and three spatial dimensions respectively, and $\A$ is the Berry connection, with components $[\A_{i}({\bf k})]^{mn}=-i\langle u^{m}_{\bf k}\vert\partial_{k_i}\vert u^{n}_{\bf k}\rangle$, where $\ket{u^{n}_{\bf k}}$ is the Bloch function of band $n$, and $m, n$ run only over occupied energy bands. 

This hierarchical mathematical structure positions the concept of charge polarization at the basis of the field of topological insulators and related phenomena. Fermionic SPTs with time-reversal, charge-conjugation, and/or chiral symmetries\cite{altland1997} in all spatial dimensions were categorized in a periodic classification table of topological insulators and superconductors\cite{schnyder2009,qi2008,kitaev2009}. Following this classification, many different groups have begun classifying SPTs protected by reflection\cite{teo2008,chiu2013,ryu2013,ueno2013,zhang2013,lau2016}, inversion\cite{fu2007,turner2010,hughes2011inversion,turner2012},  rotation~\cite{fu2011,fang2012,fang2013,teo2013, benalcazar2014}, non-symmorphic symmetries~\cite{liu2014, kobayashi2016, alexandradinata2016} and more~\cite{mong2010, jadaun2013,slager2013,morimoto2013,shiozaki2014, dong2016, chiu2016,liu2016,haruki2017, bradlyn2017,shiozaki2017}.

The mathematical topological invariants that characterize these phases are tied to quantized physical observables. For example, in 1D insulators in the presence of inversion symmetry, the polarization in Eq. \ref{eq:polarization} is quantized to either 0 or $e/2$ and is in exact correspondence with the Berry phase topological invariant~\cite{zak1989,qi2008,hughes2011inversion, turner2012}.  Recently, we showed the existence of quantized quadrupole and octupole moments, as well as quantized dipole currents, in crystalline insulators \cite{benalcazar2017quad}. The primary goal of this paper is to thoroughly develop the theory of quantized electromagnetic observables in topological crystalline insulators. In addition to the work presented in Ref. \onlinecite{benalcazar2017quad}, in this paper we discuss in more detail the observables of multipole moments and their relations, both in the classical continuum theory and in the quantum-mechanical crystalline theory and also discuss the extension of the theory of polarization to account for non-quantized higher multipole moments. To carry this out we systematically extend the theory of charge polarization in crystalline insulators by taking a different route than the extension suggested by the hierarchical mathematical structure evident in Eqs. \ref{eq:polarization}, \ref{eq:HallConductance}, and \ref{eq:MagnetoElectricPolarizability}, which deals primarily with polarizations. Our topological structure is also of hierarchical nature, but subtly involves the calculation of Berry phases of reduced sectors within the subspace of occupied energy bands. To find the relevant subspace we resolve the energy bands into spatially separated ``Wannier bands" through a Wilson-loop calculation, or equivalently, a diagonalization of a ground state projected position operator.  We call this structure `nested Wilson loops'. It goes one step beyond the previous developments in the understanding of topological insulator systems in terms of Berry phases \cite{yu2011,klich2011,alexandradinata2014,taherinejad2015,olsen2017}. At its core, this nested Wilson loop structure reflects the fact that even gapped edges of topological phases can signal a non-trivial bulk-boundary correspondence when the gapped edge Hamiltonian is topological itself and inherits such non-trivial topology from the bulk. 

This topological structure reveals that, in addition to bulk dipole moments, crystalline insulators can realize bulk quadrupole and octupole moments, as initially shown in Ref. \onlinecite{benalcazar2017quad} (Figs. \ref{fig:summary}d, g). In addition, this structure reveals other phenomena, detailed in this paper. When we allow for  the adiabatic deformation and evolution of Hamiltonians having non-zero quadrupole and octupole moments we find new types of quantized electronic transport and currents, extending what is already known in the case of the  adiabatic charge pumping (Fig. \ref{fig:summary}b)\cite{thouless1983}. In particular, the new types of adiabatic electronic currents are localized not in the bulk, but on edges or hinges of the material. They essentially amount to pumping a dipole or quadrupole across the bulk of the material, respectively (Figs. \ref{fig:summary}e, h). If the adiabatic process forms a closed cycle the transport is quantized, i.e., the amount of dipole or quadrupole being pumped is quantized. The first Chern number characterizes the 1D adiabatic pumping process; this process can be connected to a Chern insulator phase in one spatial dimension higher. The dipole pumping process in the 2D quadrupole system correspondingly predicts the existence of an associated 3D `hinge Chern insulator' having the same topological structure as a family of 2D quadrupole Hamiltonians forming an adiabatic evolution through a `non-trivial' cycle (i.e., a cycle that connects a quantized topological quadrupole insulator with a trivial insulator, while maintaining the energy gap open). This insulator has four hinge localized modes which are chiral and disperse in opposite directions at adjacent hinges, as shown schematically in Fig. \ref{fig:summary}f. In principle, the quadrupole pumping of the 3D octupole system would predict a 4D topological phase, though we will not discuss it any further here. 


Our focus throughout this paper is on tight-binding lattice models. A summary of the organization of this paper is detailed in the next subsection.  The  paper is self-contained, starting with a pedagogical description of the modern theory of polarization. Readers already familiar with the modern theory of electronic polarization, and the connection between Berry phase, Wannier functions, and Wilson loops can easily skip Section \ref{sec:Dipole_1D} after reading Section \ref{sec:ClassicalMultipoles}.

\subsection{Outline}
In Section \ref{sec:ClassicalMultipoles} we first define electric multipole moments within the classical electromagnetic theory, characterize their boundary signatures, and establish the criteria under which these moments are well defined. 

We then start the discussion of the dipole moment in crystalline insulators in 1 dimension (1D) in Section \ref{sec:Dipole_1D}, and in 2 dimensions (2D) in Section \ref{sec:Dipole_2D} and in Section \ref{sec:EdgePolarization}. Our formulation differs from the original one \cite{vanderbilt1993} in that, instead of referring to the relationship between electric current and change in electric polarization, we directly calculate the position of electrons in the crystal by means of diagonalizing the position operator projected into the subspace of occupied bands. This approach naturally connects the individual electronic positions with the polarization (i.e., dipole moment), as well as this polarization with the Berry phase accumulated by the subspace of occupied bands across the Brillouin zone of the crystal. Additionally, this approach provides us with eigenstates of well defined electronic position, which we then use to extend the formulation to higher multipole moments. 

In addition to this formulation, we discuss the symmetry constraints that quantize the dipole moments and present the case of the Su-Schrieffer-Hegger (SSH) model as a primitive model for the realization of the dipole symmetry-protected topological (SPT) phase. We further use extensions of this model that break the symmetries that protect the SPT phase and thus allow an adiabatic change in polarization and the appearance of currents. We will discuss the topological invariant that characterizes the quantization of charge transport in closed adiabatic cycles. 

In Section \ref{sec:Dipole_2D}, we extend the 1D treatment of the problem to 2D and introduce the concept of Wannier bands, which plays a crucial role in the description of higher multipole moments. We also characterize - in terms of Wannier bands - the topology of a Chern insulator and the Quantum Spin Hall insulator as examples, and make connections between the topology of a Chern insulator and the quantization of particle transport of Section \ref{sec:Dipole_1D}.

In Section \ref{sec:EdgePolarization} we describe the recently found phenomenon of edge-polarization \cite{vanderbilt2015} and its relation to corner charge. In particular, we use this as an example that allows discriminating corner charge arising from converging edge-localized dipole moments from the corner charge arising from higher multipole moments. 

We then describe the existence of the first higher multipole moment, the quadrupole moment, in Section \ref{sec:Quadrupole}. We first present the theory in terms of the diagonalization of position operators. Just as in the case of the dipole, the quadrupole moment is indicated by a topological quantity, which relates to the polarization of a Wannier band-resolved subspace within the subspace of occupied energy bands. From this formulation, we derive the conditions (i.e., the symmetries) under which the quadrupole moment quantizes to $\pm e/2$, realizing a quadrupole SPT. We then present a concrete minimal model with quadrupole moment. We describe the observables associated with it: the existence of edge polarization and corner charge, as well as the different symmetry-protected phases associated with this model and the nature of its phase transitions. We then break the symmetries that protect the SPT to cause adiabatic transport of charge, but in a pattern that amounts to a net pumping of dipole moment. This dipole moment transport can also be quantized in an analogous manner to the charge transport in a varying dipole. We describe the invariant associated with this quantization and the extension of this principle to the creation of unusual insulators, not described so far to the best of our knowledge, which present chiral hinge-localized dispersive modes due to its non-trivial topology. 

In Section \ref{sec:Octupole} we describe the existence of octupole moments. We describe the hierarchical topological structure that gives rise to higher multipole moments, as well as the minimal model that realizes a quantized octupole SPT. We also describe, by means of a concrete example, how the quantization of quadrupole transport can be realized. 

In Section \ref{sec:Discussion} we present a discussion that highlights and summarizes the main findings in this paper, and its implications in terms of future extensions of this work to other fermionic or bosonic systems, as well as a discussion on the anomalous nature of the boundaries of these multipole moment insulators. 


\section{The multipole expansion in the continuum electromagnetic theory}
\label{sec:ClassicalMultipoles}
Since the classical theory of multipole moments, even in the absence of a lattice, has various subtleties, we will spend time reviewing it in this Section. Our goal is to provide precise definitions for, and to extract the macroscopically observable signatures of,  the dipole, quadrupole, and octupole moments in insulators.

\subsection{Definitions}
\label{sec:MultipoleDefinitions}
\begin{figure}
\centering
\includegraphics[width=\columnwidth]{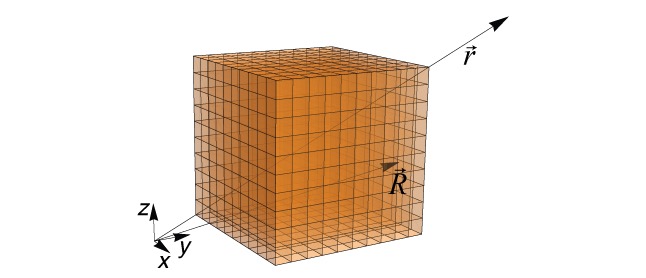}
\caption{(Color online) Macroscopic material divided in small voxels over which the multipole moment densities are calculated. Each voxel is labeled by its center point $\vec{R}$. $\vec{r}$ is the position (far) outside the material at which the potential is calculated.}
\label{fig:macroscopic_multipole}
\end{figure}

In this section we define multipole electric moments in macroscopic materials based on classical charge configurations in the absence of a lattice. We define a macroscopic material as one which can be divided into small volume elements (voxels), as shown in Fig.~\ref{fig:macroscopic_multipole}, over which multipole moment densities can be defined, and in such a way that these densities can be treated as continuous functions of the position at larger length scales. 
For a material divided into such voxels, the expression for the electric potential at position $\vec{r}$ due to a charge distribution over space is
\begin{align}
\phi(\vec{r})=\frac{1}{4\pi \epsilon} \sum_{\vec{R}} \int_{v(\vec{R})} d^3\vec{r'} \frac{\rho(\vec{r'}+\vec{R})}{|\vec{r}-\vec{R}-\vec{r'}|},
\label{eq:potential_macroscopic}
\end{align}
where $\rho(\vec{r})$ is the volume charge density, $\epsilon$ is the dielectric constant, $\vec{R}$ labels the voxel, and in the integral $r'$ runs through the volume $v(\vec{R})$ of voxel $\vec{R}$. Since the voxels are much smaller than the overall size of the material, we have that $|\vec{r'}| \ll |\vec{r}-\vec{R}|$ \textit{as long as $\vec{r}$ is outside of the material and sufficiently away from it}. Then, one can expand the potential \eqref{eq:potential_macroscopic} in powers of $1/|\vec{r}-\vec{R}|$ (see details in Appendix~\ref{sec:app_multipole_definitions}) to define the multipole moment densities 
 \begin{align}
 \rho(\vec{R})&=\frac{1}{v(\vec{R})} \int_{v(\vec{R})} d^3\vec{r'} \rho(\vec{r'}+\vec{R})\nonumber \\
p_i(\vec{R})&=\frac{1}{v(\vec{R})} \int_{v(\vec{R})} d^3\vec{r'} \rho(\vec{r'}+\vec{R}) r_i'\nonumber \\
q_{ij}(\vec{R})&=\frac{1}{v(\vec{R})} \int_{v(\vec{R})} d^3\vec{r'} \rho(\vec{r'}+\vec{R}) r_i' r_j'\nonumber \\
o_{ijk}(\vec{R})&=\frac{1}{v(\vec{R})} \int_{v(\vec{R})} d^3\vec{r'} \rho(\vec{r'}+\vec{R}) r_i' r_j' r_k'
\label{eq:multipole_moment_densities}
 \end{align}
which allow to write the terms in the expansion of the potential 
\begin{align}
\phi(\vec{r})= \sum_{l=0}^\infty \phi^l(\vec{r}),
\end{align}
as
\begin{align}
\phi^0(\vec{r})&=\frac{1}{4 \pi \epsilon} \int_{V}d^3\vec{R}  \left(\rho(\vec{R}) \frac{1}{|\vec{\rr}|} \right)\nonumber \\
\phi^1(\vec{r})&=\frac{1}{4 \pi \epsilon} \int_{V}d^3\vec{R} \left(p_i(\vec{R}) \frac{\rr_i}{|\vec{\rr}|^3}\right)\nonumber  \\
\phi^2(\vec{r})&=\frac{1}{4 \pi \epsilon} \int_{V}d^3\vec{R} \left(q_{ij}(\vec{R}) \frac{3 \rr_i \rr_j -|\vec{\rr}|^2 \delta_{ij}}{2|\vec{\rr}|^5}\right)\nonumber  \\
\phi^3(\vec{r})&=\frac{1}{4 \pi \epsilon} \int_{V}d^3\vec{R} \left(o_{ijk}(\vec{R}) \frac{5 \rr_i \rr_j \rr_k - 3 |\vec{\rr}|^2 \rr_k \delta_{ij}}{2|\vec{\rr}|^7}\right),
\label{eq:potential_multipole_moment_densities}
\end{align}
where $V$ is the total volume of the macroscopic material and $\vec{\rr} = \vec{r} - \vec{R}$. The potential $\phi^0(\vec{r})$ is due to the free `coarse-grained' charge density in Eq. \ref{eq:multipole_moment_densities}. In the limit of $v(\vec{R}) \to 0$, this coarse grained charge density is the original continuous charge density, and we recover the original expression \eqref{eq:potential_macroscopic}. In this case, all other multipole contributions identically vanish. 

\subsection{Dependence of the multipole moments on the choice of reference frame}
\label{sec:multipole_translation_invariance}
The multipole moments are in general defined with respect to a particular reference frame. For example, given a charge density per unit volume $\rho(\vec{r})$, consider the definition of the dipole moment
\begin{align}
P_i&=\int_v d^3\vec{r} \rho(\vec{r}) r_i.
\end{align}
If we shift the coordinate axes used in that definition such that our new positions are given by $r'_i = r_i + D_i$, and the charge density in this new reference frame is $\rho'(\vec{r'})=\rho(\vec{r})$, the dipole moment is now given by
\begin{align}
P'_i&=\int_v d^3\vec{r'} \rho'(\vec{r'}) r'_i\nonumber \\
&=\int_v d^3\vec{r} \rho(\vec{r}) (r_i+D_i)\nonumber \\
&=\int_v d^3\vec{r} \rho(\vec{r}) r_i+ D_i \int_v d^3\vec{r} \rho(\vec{r})\nonumber \\
&=P_i+ D_i Q
\end{align}
where $Q$ is the total charge. Notice, however, that the dipole moment is well defined for any reference frame if the total charge $Q$ vanishes. Similarly, a quadrupole moment transforms as
\begin{align}
Q'_{ij}=&\int_v d^3\vec{r'} \rho'(\vec{r'}) r'_i r'_j\nonumber \\
=&\int_v d^3\vec{r} \rho(\vec{r}) (r_i+D_i) (r_j+D_j)\nonumber \\
=&Q_{ij} +  P_i D_j+ D_i P_j +D_i D_j Q
\end{align}
which is not uniquely defined independent of the reference frame unless both the total charge and the dipole moments vanish. 
In general, for a multipole moment to be independent of the choice of reference frame, all of the lower moments must vanish.

\subsection{Boundary properties of multipole moments}
Now let us consider the macroscopic physical manifestations of the multipole moments. In all cases, we will consider the properties  that  appear at the boundaries of materials having non-vanishing multipole moments in their bulk. We consider each multipole density separately, assuming as indicated above, that all lower moments vanish.

\subsubsection{Dipole moment}
\label{sec:ClassicDipole}
\begin{figure}
\centering
\includegraphics[width=\columnwidth]{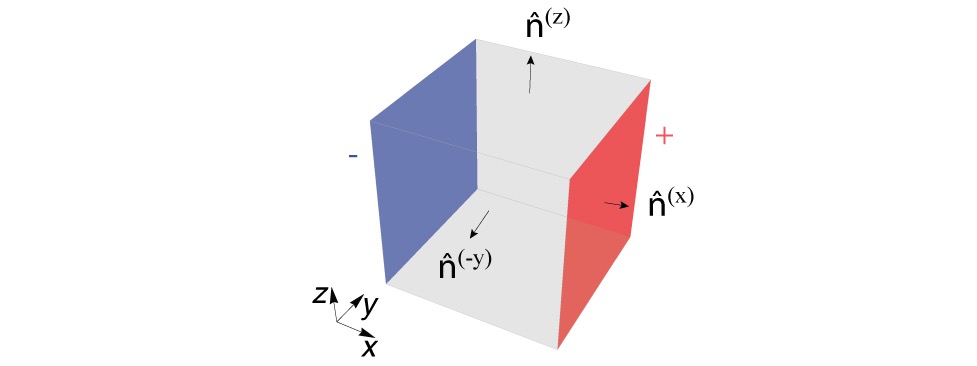}
\caption{(Color online)  Boundary charge in a material with uniform dipole moment per unit volume $p_x \neq 0$, $p_y=p_z=0$. Red (blue) color represents positive (negative) charge per unit area of magnitude $p_x$. The three arrows pointing out of the flat surfaces represent the unit vectors $\hat{n}^{(x)}, \hat{n}^{(-y)}$ and $\hat{n}^{(z)}$. The other three unit vectors are not shown for clarity of presentation.}
\label{fig:continuous_dipole}
\end{figure}

The potential due to a dipole moment density $p_i(\vec{R})$ is given by the second equation in \eqref{eq:potential_multipole_moment_densities}. As shown in Appendix \ref{sec:boundaries_EM_theory}, this potential can be recast in the form 
\begin{align}
\phi^1(\vec{r}) &= \frac{1}{4 \pi \epsilon} \oint_{\partial V} d^2\vec{R}  \left( n_i p_i \frac{1}{|\vec{r}-\vec{R}|} \right) \nonumber\\
&+ \frac{1}{4 \pi \epsilon} \int_V d^3\vec{R} \left(-\partial_i p_i \frac{1}{|\vec{r}-\vec{R}|} \right).
\end{align}
Since both terms scale as $1/|\vec{r}-\vec{R}|$, where $|\vec{r}-\vec{R}|$ is the distance from a point in the material to the observation point, we can define the charge densities
\begin{align}
\rho(\vec{R}) &= -\partial_i p_i(\vec{R})\nonumber \\
\sigma^{face}(\vec{R}) &= n_i p_i(\vec{R}). 
\label{eq:ChargeDensities_Dipole}
\end{align}
From now on, we will drop the label of the dependence of the variables on $\vec{R}$ for convenience. The first term is the volume charge density due to a divergence in the polarization, and the second is the areal charge density on the boundary of a polarized material. Hence, one manifestation of the dipole is a boundary charge as shown in Fig.~\ref{fig:continuous_dipole}. 

\subsubsection{Quadrupole moment}
\label{sec:ClassicQuadrupole}
\begin{figure}
\centering
\includegraphics[width=\columnwidth]{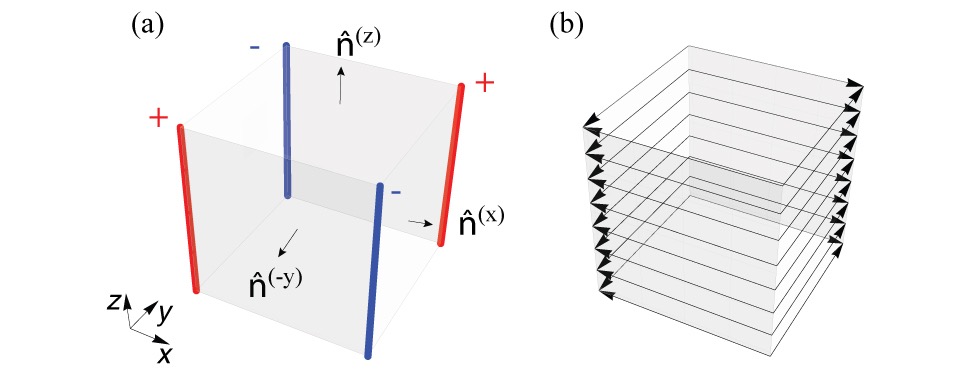}
\caption{(Color online)  Boundary properties of a cube with uniform quadrupole moment per unit volume $q_{xy}\neq 0$, $q_{yz}=q_{zx}=0$.  (a) Boundary charge. Red (blue) color represents positive (negative) charge densities per unit length of magnitude $q_{xy}$. (b) Boundary polarization. Arrows in represent boundary dipole moment per unit area of magnitude $q_{xy}.$ The unit vectors $\hat{n}^{(x)}, \hat{n}^{(-y)}$ and $\hat{n}^{(z)}$ are shown in (a) for reference.}
\label{fig:continuous_quadrupole}
\end{figure}

As shown in Appendix \ref{sec:boundaries_EM_theory}, the potential due to a quadrupole moment per unit volume $q_{ij}$ as listed in Eq. \ref{eq:potential_multipole_moment_densities} is equivalent to 
\begin{align}
\phi^2(\vec{r}) &=
\frac{1}{4 \pi \epsilon} \sum_{a,b} \int_{L_{ab}} d\vec{R} \left( \frac{1}{2} n^{(a)}_i n^{(b)}_j q_{ij}  \right) \frac{1}{\rr}\nonumber\\
&+\frac{1}{4 \pi \epsilon} \sum_a \int_{S_a} d^2\vec{R}  \left( - \partial_j n^{(a)}_i q_{ij}  \right) \frac{1}{\rr} \nonumber\\
&+\frac{1}{4 \pi \epsilon} \int_V d^3\vec{R} \left( \frac{1}{2}\partial_j \partial_i q_{ij}\right) \frac{1}{\rr}.
\label{eq:quad_potential}
\end{align}
This calculation was carried out for a system with a cubic geometry. $S_a$ represents the plane of surface normal to vector $\hat{n}^{(a)}$ and $L_{ab}$ represents the hinge at the intersection of surfaces with normal vectors $\hat{n}^{(a)}$ and $\hat{n}^{(b)}$. Since all the potentials scale as $1/\rr$, where $\vec{\rr}=\vec{r}-\vec{R}$ is the distance from the point in the material to the observation point, each expression in a parentheses can be interpreted as a charge density in its own right. Thus, we define the charge densities
\begin{align}
\lambda^{hinge\; a,b} &= \frac{1}{2} n^{(a)}_i n^{(b)}_j q_{ij}\nonumber \\
\sigma^{face\; a} &= - \partial_j \left( n^{(a)}_i  q_{ij} \right)\nonumber \\
\rho &= \frac{1}{2}\partial_j \partial_i q_{ij}.
\label{eq:ChargeDensitiesQuadrupole}
\end{align}
The first term is the charge density per length at the hinge $L_{ab}$  of the material. The second term is the area charge density at the boundary surface $S_{a}$ of the material due to a divergence in the quantity $n^{(a)}_i q_{ij}.$  Finally, the third term is the direct contribution of the quadrupole moment density to the volume charge density in the bulk of the material. For a cube with constant quadrupole moment $q_{xy}$ and open boundaries we illustrate the the charge density in Fig.~\ref{fig:continuous_quadrupole}a, as indicated by the expression for $\lambda^{hinge}$. Notice that the expression for the surface charge density $\sigma^{bound}$ could be written as
\begin{align}
\sigma^{face\; a}  &= - \partial_j p^{face\;a}_j,
\end{align} 
where  
\begin{align}
p^{face\; a}_j = n^{(a)}_i  q_{ij}
\label{eq:PolarizationDensityQuadrupole}
\end{align}
resembles the polarization for the volume charge density $\rho$ in Eq. \ref{eq:ChargeDensities_Dipole}. Thus, we interpret $p^{face\; a}_j$ as a bound dipole density (per unit area). This polarization exists on the surface perpendicular to $\hat{n}^{(a)}$ and runs parallel to that surface. An illustration of this polarization for a cube with constant quadrupole moment $q_{xy}$ is shown in Fig.~\ref{fig:continuous_quadrupole}b.

Notice, from \eqref{eq:ChargeDensitiesQuadrupole} and \eqref{eq:PolarizationDensityQuadrupole}, that the magnitudes of the hinge charge densities and the face dipole densities have the same magnitude as the quadrupole moment,
\begin{align}
|\lambda^{hinge} | = |p^{face}_j| = |q_{xy}|
\end{align}
since the implied sums over $i$ and $j$ in the first Eq. of \eqref{eq:ChargeDensitiesQuadrupole} cancel the factor $1/2$, because $q_{xy}=q_{yx}$.

\subsubsection{Octupole moment}
\label{sec:ClassicOctupole}
\begin{figure}%
\centering
\includegraphics[width=\columnwidth]{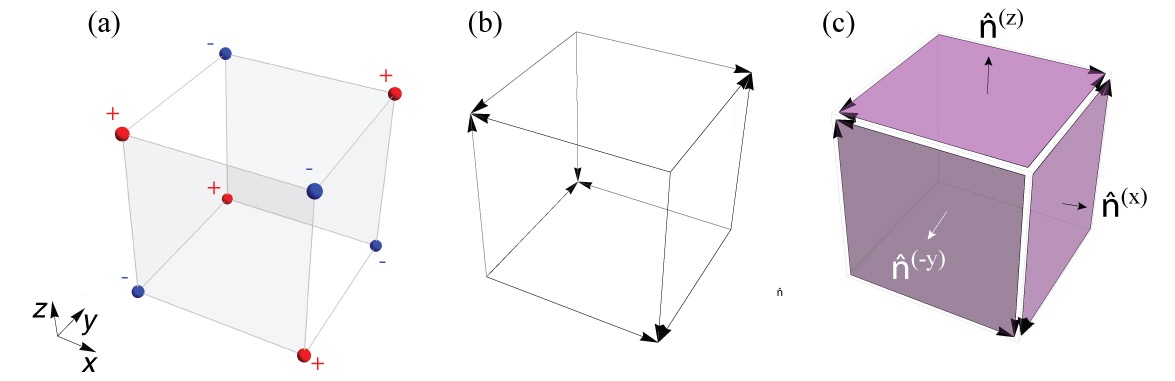}
\caption{(Color online)  Boundary properties of a cube with uniform octupole moment per unit volume $o_{xyz}.$. (a) Corner-localized charges. Red (blue) color represents positive (negative) charges with magnitude $o_{xyz}$. (b) Hinge-localized dipole moments per unit length of magnitude $o_{xyz}$. (c) Surface localized quadrupole moment densities. Purple squares represent quadrupole moments per unit area of magnitude $o_{xyz}$. The unit vectors $\hat{n}^{(x)}, \hat{n}^{(-y)}$ and $\hat{n}^{(z)}$ are shown in (c) for reference.}
\label{fig:continuous_octupole}
\end{figure}

Following a similar  procedure as that employed for the dipole and quadrupole moments, the potential due to an octupole moment per unit volume $o_{ijk}$ (Eq. \ref{eq:potential_multipole_moment_densities}) can be rewritten as 
\begin{align}
\phi^3(\vec{r}) &= \frac{1}{4 \pi \epsilon} \sum_{a,b,c} \frac{1}{6} n^{(a)}_i n^{(b)}_j n^{(c)}_k o_{ijk} \frac{1}{r}\nonumber\\
&+ \frac{1}{4 \pi \epsilon} \sum_{a,b} \int_{L_{ab}} d\vec{R} \left( -\frac{1}{2} n^{(a)}_i n^{(b)}_j \partial_k o_{ijk} \right) \frac{1}{\rr}\nonumber\\
&+ \frac{1}{4 \pi \epsilon} \sum_i \int_{S_a} d^2\vec{R} \left( \frac{1}{2} n^{(a)}_i \partial_j \partial_k o_{ijk} \right) \frac{1}{\rr}\nonumber\\
&+\frac{1}{4 \pi \epsilon} \int_V d^3\vec{R} \left(-\frac{1}{6} \partial_i \partial_j \partial_k o_{ijk} \right) \frac{1}{\rr},
\end{align}
from which we read off the various charge densities
\begin{align}
\delta^{corner\; a,b,c} &= \frac{1}{6} n^{(a)}_i n^{(b)}_j n^{(c)}_k o_{ijk}\nonumber\\
\lambda^{hinge\; a,b} &= -\frac{1}{2} n^{(a)}_i n^{(b)}_j \partial_k o_{ijk}\nonumber \\
\sigma^{face\; a} &= \frac{1}{2} n^{(a)}_i \partial_j \partial_k o_{ijk}\nonumber \\
\rho &= -\frac{1}{6} \partial_i \partial_j \partial_k o_{ijk}.
\label{eq:ChargeDensitiesOctupole}
\end{align}\noindent The new quantity $\delta^{corner\; a,b,c}$ represents localized charge bound at a corner where the three surfaces normal to $\hat{n}^{(a)}, \hat{n}^{(b)},$ and  $n^{(c)}$ intersect.
Comparing \eqref{eq:ChargeDensitiesOctupole} with the expressions for dipole and quadrupole moments we see that we can re-write the surface charge density per unit area, and the hinge charge density per unit length as
\begin{align}
\sigma^{face\;a} &= \frac{1}{2} \partial_j \partial_k q^{face\; a}_{jk}\nonumber \\
\lambda^{hinge\;a,b} &= -\partial_k  p^{hinge\;a,b}_k,
\end{align}
where 
\begin{align}
q^{face\; a}_{jk} &= n^{(a)}_i o_{ijk}\nonumber\\
p^{hinge\; a,b}_k &= \frac{1}{2} n^{(a)}_i n^{(b)}_j o_{ijk}
\label{eq:QuadrupoleAndPolarizationDensitiesOctupole}
\end{align}
are the quadrupole moment density per unit area on faces perpendicular to $\hat{n}^{(a)}$ and the polarization per unit length on hinges where surfaces normal to $\hat{n}^{(a)}$ and $\hat{n}^{(b)}$ intersect, respectively. These manifestations at the boundary are illustrated in Fig.~\ref{fig:continuous_octupole} for a cube with uniform octupole moment.

Notice, from \eqref{eq:ChargeDensitiesOctupole} and \eqref{eq:QuadrupoleAndPolarizationDensitiesOctupole}, that the magnitudes of the corner charge densities, the hinge dipole densities, and the face quadrupole densities have the same magnitude as the octupole moment,
\begin{align}
|\delta^{corner}|=|p^{hinge}_k| = |q^{face}_{jk}| = |o_{xyz}|
\end{align}
since the implied sums over $i$ and $j$ in the first Eq. of \eqref{eq:ChargeDensitiesOctupole} and second Eq. of \eqref{eq:QuadrupoleAndPolarizationDensitiesOctupole} cancel the prefactors of $1/6$ and $1/2$, respectively, because $o_{xyz}=o_{yzx}=o_{zxy}=o_{xzy}=o_{zyx}=o_{yxz}$.

\subsection{Bulk moments vs. boundary moments}
\label{sec:BulkVSBoundaryMoments}
\begin{figure}
\centering
\subfigure[]{
\begin{tikzpicture}[scale=.5]
 
		\coordinate (pp) at (1,1);
		\coordinate (pn) at (1,-1);
		\coordinate (np) at (-1,1);
		\coordinate (nn) at (-1,-1);
		
		\fill [pink] (1,1) circle(1);
		\fill [black] (0,1.1) circle(.1); \node[black, above] at (-0.2,1.1) {$\vec{r}_1$};
		\fill [black] (1.1,0) circle(.1); \node[black, right] at (1.1,-0.2) {$\vec{r}_2$};
		\fill [cyan!50!,opacity=0.5] (-2,1.1)--(1.1,1.1)--(1.1,-2)--(-2,-2);
		\node[black] at (-0.5,-0.5) {\begin{tabular}{c} $p=0$ \\ $q_{xy} = 0$\end{tabular}};
		\node[red] at (2.8,2.5) {$Q^{c}=p_1+p_2$};
		\draw [->,black,thick] (-2,1.1)--node[above left] {$\vec{p}_1$}(1,1.1) ;
		\draw [->,black,thick] (1.1,-2)--node[below right] {$\vec{p}_2$}(1.1,1) ;

\end{tikzpicture}
}\;\;
\subfigure[]{
\begin{tikzpicture}[scale=.5]
 
		\coordinate (pp) at (1,1);
		\coordinate (pn) at (1,-1);
		\coordinate (np) at (-1,1);
		\coordinate (nn) at (-1,-1);
		
		\fill [pink] (1,1) circle(1);
		\fill [cyan!50!,opacity=0.5] (-2,1.1)--(1.1,1.1)--(1.1,-2)--(-2,-2);
		\node[black] at (-0.5,-0.5) {\begin{tabular}{c} $p=0$ \\ $q_{xy} \neq 0$\end{tabular}};
		\node[red] at (2.2,2.5) {$Q^{c}=q_{xy}$};
		\draw [->,cyan,thick] (-2,1.1)--node[above left] {$|\vec{p}|=q_{xy}$}(1,1.1) ;
		\draw [->,cyan,thick] (1.1,-2)--node[below right] {$|\vec{p}|=q_{xy}$}(1.1,1) ;
		\fill [black] (0,1.1) circle(.1); \node[black, above] at (-0.2,1.1) {$\vec{r}_1$};
		\fill [black] (1.1,0) circle(.1); \node[black, right] at (1.1,-0.2) {$\vec{r}_2$};

\end{tikzpicture}
}
\caption{(Color online)  Corner charges due to (a) a pair of convergent dipoles and (b) a constant quadrupole. The most general case up to quadrupole expansion will have a superposition of both contributions.}
\label{fig:BulkVSBoundaryMoments}
\end{figure}
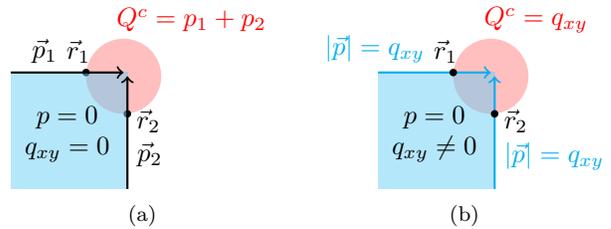

In this section we draw an important distinction between boundary observables arising from the presence of a bulk moments vs. boundary observables arising from ``free" moments of lower dimensionality attached to a boundary. The potential confusion is illustrated in Fig.~\ref{fig:BulkVSBoundaryMoments} where we consider a neutral, insulating material with no free charge in the bulk or boundary, so that all charge accumulation is induced by either dipole or quadrupole moments. In Fig.~\ref{fig:BulkVSBoundaryMoments}a there is charge accumulation where two boundary polarizations converge at a corner (in 2D) or a hinge (in 3D). These surface dipoles are meant to be a pure surface effect and not due to a bulk moment.  In Fig.~\ref{fig:BulkVSBoundaryMoments}b there are both surface polarizations and corner/hinge charge accumulation, but this time exclusively due to a quadrupole moment. The phenomenology in both cases are similar, so the natural question is how to distinguish the surface effect in Fig.~\ref{fig:BulkVSBoundaryMoments}a from the bulk effect in Fig.~\ref{fig:BulkVSBoundaryMoments}b.

Te be explicit, let us consider the 2D case. We first consider the existence of only boundary-localized dipole moments.
The contribution to charge density due to a dipole moment density $\vec{p}= \vec{p}(\vec{r})$ is
\begin{align}
\rho &= - \vec{\nabla} \cdot \vec{p}\nonumber\\
\sigma &= \vec{p} \cdot \vec{n}
\label{eq:DipoleVSQuad}
\end{align}
which is a restatement of \eqref{eq:ChargeDensities_Dipole}. The first term is the polarization-induced charge density per unit volume of the material, and $\vec{p} \cdot \vec{n}$ is the  charge density per unit area on a boundary surface with unit normal vector $\vec{n}$ induced by the bulk polarization $\vec{p}$. For the purpose of calculating the accumulated charge, let us consider an area $v$ which encloses the corner on which charge is accumulated, as shown by the red circle in Fig.~\ref{fig:BulkVSBoundaryMoments}a. To relate the induced charge in this volume to the polarization at its boundary we use the first Eq. in \eqref{eq:DipoleVSQuad}
\begin{align}
Q^{corner} &= \int_v \rho dv = \int_v \left(- \vec{\nabla} \cdot \vec{p} \right) dv \nonumber\\
&= -\oint_{\partial v} \vec{p} \cdot d\vec{s}\nonumber
\end{align} 
where in the second line we have applied Stokes' theorem, and where $\partial v$ is the boundary of area $v$.
We see from Fig. \ref{fig:BulkVSBoundaryMoments}a that the boundary dipoles $\vec{p}_1$ and $\vec{p}_2$ puncture the boundary of $v$. If we treat the polarizations as fully localized on the edge we can write 
\begin{align}
\vec{p}_1(\vec{r})=\hat{x} p_1 \delta(\vec{r}-\vec{r}_1)\nonumber\\
\vec{p}_2(\vec{r})=\hat{y} p_2 \delta(\vec{r}-\vec{r}_2),\nonumber
\end{align}
where $\vec{r}_1$ and $\vec{r}_2$ are shown in Fig.~\ref{fig:BulkVSBoundaryMoments}. Taking into account that the boundary $\partial v$ has normal vector $-\hat{x}$ at $\vec{r}_1$ and $-\hat{y}$ at $\vec{r}_2$, we have
\begin{align}
Q^{corner}=p_1 + p_2.
\label{eq:QcornerDipole}
\end{align}

In contrast, let us now consider the charge accumulation inside area $v$ due to a quadrupole moment $q_{xy},$ as shown in Fig.~\ref{fig:BulkVSBoundaryMoments}b. It follows from \eqref{eq:ChargeDensitiesQuadrupole} that, in this case, the induced charge is
\begin{align}
\rho = \frac{1}{2} \partial_j \partial_i q_{ij},
\end{align}
where summation is implied for repeated indices. The blue region has quadrupole density $q_{xy}=q_{yx} \neq 0$, and outside this region is vacuum. The total charge enclosed in the area $v$ (shown in red) is
\begin{align}
Q^{corner} &= \int_v \rho dv =  \int_v \left(\frac{1}{2} \partial_j \partial_i q_{ij} \right) dv\nonumber\\
&=  \frac{1}{2} \oint_{\partial v}  \left( \partial_i q_{ij} \right) n_j ds,
\end{align}
where in the second line we have applied Stokes' theorem. Here, $n_j$ is the $j^{th}$ component of the unit vector $\hat{n}$ normal to the boundary  $\delta v$. Since the quadrupole moment density is constant, there are only two places in $\partial v$ where the derivative $\partial_i q_{ij}$ does not vanish (see Fig.~\ref{fig:BulkVSBoundaryMoments}b):
(i) at $\vec{r}_1$ the unit vector normal to the boundary $\partial v,$ and pointing away from the area $v,$ is $\hat{n}=-\hat{x}$ and $\int_{-\epsilon}^{\epsilon} \partial_y q_{yx} dy= -q_{yx}$, and (ii) at $\vec{r}_2$ the unit vector normal to $\partial v$ pointing away from $v$ is $\hat{n}=-\hat{y}$, which leads to $\int_{-\epsilon}^{\epsilon} \partial_x q_{xy} dx = -q_{xy}$. Thus, the corner charge is,
\begin{align}
Q^{corner} &= \frac{1}{2} (q_{xy}+q_{yx}) = q_{xy}.
\label{eq:QcornerQuadrupole}
\end{align}

By comparing Eq. \ref{eq:QcornerDipole} with Eq. \ref{eq:QcornerQuadrupole}, we conclude that, in the case of only  boundary-localized ``free" dipole moments, the corner-localized charge is given by the sum of the converging boundary polarizations. However, in the case of a bulk quadrupole moment, the magnitude of the corner  charge matches the magnitude of the quadrupole moment. Since the boundary polarizations induced from a bulk quadrupole have the same magnitude as the quadrupole itself  (see Eq.~\ref{eq:PolarizationDensityQuadrupole}) adding up the two boundary polarizations in a similar way over-counts the corner charge. Heuristically the two boundary polarizations share the corner charge if arising from a bulk quadrupole moment, whereas they both contribute independently if arising from ``free" surface polarization. In summary, even though both cases in Fig.~\ref{fig:BulkVSBoundaryMoments} have edge-localized polarizations converging at a corner of the material, the resulting corner charge is not determined the same way from the boundary polarizations. For example, if we set $p_1=p_2=q_{xy}$ so that the magnitudes of the edge polarizations match in both cases, the case of converging edge polarizations Eq. \ref{eq:QcornerDipole} gives a corner charge $Q^{corner}=2q_{xy}$, while the case of a uniform quadrupole moment gives a corner charge $Q^{corner}=q_{xy}$. 

We now generalize the relations between bulk and boundary moments and their associated boundary charges. In 1D the difference between the total charge on the end of the system and the free charge (i.e., monopole moment) attached to the end is captured by the dipole moment
\begin{align}
Q^{end}-Q^{free}=p_x.
\end{align} In 2D the difference between the total corner charge and that coming from the total surface polarization contributions is determined by the bulk quadrupole moment
\begin{align}
Q^{corner}-p_{x}^{edge}-p_{y}^{edge}=-q_{xy}.
\end{align} Finally, in 3D, we can relate the octupole moment to the difference in the corner charge and the total surface quadrupole and total hinge polarization via
\begin{eqnarray}
&&Q^{corner}-\left( \sum_{i=x,y,z} p_{i}^{hinge}
+q^{face}_{xy}+q^{face}_{yz}+q^{face}_{xz}\right)\nonumber\\ &=&o_{xyz}.
\end{eqnarray} We have implicitly assumed in these three equations that the surfaces, hinges, and corners are all associated with positively oriented normal vectors. For simplicity we have also dropped $Q^{free}$ in the latter two equations: a free corner monopole has to be subtracted from the corner charge.


\subsection{Symmetries of the multipole moments}
\label{sec:classical_multipoles_symmetries}
Since we are primarily interested in cases where the multipole moments are quantized by symmetry, we need to consider their symmetry transformations. A full discussion of all the transformation properties of all of the components of every multipole moment can be done but would take us too far afield, so we only briefly comment on the simplest properties that provide useful physical intuition. 

We focus on systems with $d$-dimensional cubic-like symmetries, e.g., the crystal families of orthorhombic, tetragonal, and cubic materials. For a cubic point group, a non-zero, off-diagonal, $2^d$-pole configuration (e.g., $2^0\colon 1$ for charge, $ 2^1$ for dipole $p_x$, $2^2$ for quadrupole $q_{xy}$, and $2^3$ for octupole $o_{xyz}$) is only invariant under the $d$-dimensional ``tetrahedral" subgroup ($T(d)$) of the $d$-dimensional cubic symmetry group ($O(d)$). 
In 1D, $T(1)$ is just the identity operation. 
In 2D, $T(2)$ is the normal subgroup of the dihedral group $D_4$ (symmetries of the square) which contains the symmetries $\{1,C_4M_x, C_4M_y, C^2_4\}$, where $M_{x,y}$ is a reflection of only the $x,y$ coordinate respectively, and $C_4$ is the rotation by $\pi/2$. The quadrupole moment $q_{xy}$ is invariant under $T(2).$
In 3D, $o_{xyz}$ is invariant under the tetrahedral subgroup ($T(3) = T_d$) of the cubic group ($O(3) = O$).
 
Since the subgroup which leaves the $2^d$-pole invariant is a normal subgroup, we can consider the coset group, for example $O/T_d \equiv \mathbb{Z}_2.$ The trivial element of this coset represents all of the elements of $T(d)$, i.e., the ones that leave the multipole moment invariant. The non-trivial element represents the other transformations in $O$, all of which will cause the off-diagonal $2^d$-poles to switch sign. In 1D this is simple, as the full symmetry group is just $G = \{1, M_x\},$ and the polarization is invariant only under 1, so $G/1 = G \equiv \mathbb{Z}_2.$ In conclusion, under a symmetry in $G$ that projects onto the non-identity element of the $\mathbb{Z}_2$ factor group, the $2^d$-pole of a crystal insulator should be quantized. In addition, charge conjugation, $C$, quantizes the $2^d$-pole moment (note that each moment depends linearly on the charge). Under these symmetries the moment is odd, and is hence required to either vanish or be quantized to a non-trivial value allowed by the presence of the lattice.

Having defined the multipole moment densities in continuum electromagnetic theory, and having characterized their important observable properties, we now move to describe how they arise in crystalline insulators. We start with a review of dipole moments in 1D crystals, and sequentially advance our description towards bulk and edge dipole moments in 2D crystals, quadrupole moments in 2D crystals, and finally octupole moments in 3D crystals. Due to the dependence of the multipole moments on the origin of coordinates when lower multipole moments do not vanish, we assume in what follows that, \emph{for any multipole moment in question, all lower multipole moments vanish}.

\section{Bulk dipole moment in 1D crystals}
\label{sec:Dipole_1D}
Neutral one-dimensional crystals only allow for a dipole moment. In insulators, the electronic contribution to the polarization arises from the displacement of the electrons with respect to the ionic positive charges. In this section, to calculate the polarization, we  diagonalize the electronic projected position operator~\cite{kivelson1982,resta1998,yu2011, alexandradinata2014}, and construct the Wannier centers and Wannier functions~\cite{wannier1962,marzari1997}.  The polarization can then be easily extracted. In doing so, we will recover the result that the electronic polarization is given by the Berry phase accumulated by the parallel transport of the subspace of occupied bands across the Brillouin zone (BZ). The electronic polarization can have a topological nature in the presence of certain symmetries\cite{zak1989, king-smith1993}.

\subsection{Preliminary considerations}
\label{sec:PrelimConsiderations}
\begin{figure}%
\centering
\includegraphics[width=\columnwidth]{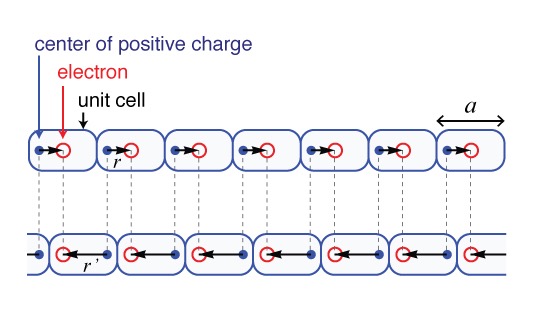}
\caption{(Color online)  Ambiguity in the definition of the electronic positions. Two 1D lattices with one atomic site (blue dots) and one electron (red circles) per unit cell. Although the two physical configurations for the two 1D lattices are the same, the electronic positions $r$ and $r'=r-a$ differ by a lattice constant due to the difference in the definitions of their unit cells.}
\label{fig:ambiguity_electronic_positions}
\end{figure}

Let us first consider insulators with discrete translation symmetry, but simply composed of point charges. As seen in section \ref{sec:multipole_translation_invariance}, the polarization is well defined only if it has zero net charge.  Discrete translation symmetry implies that it is sufficient to characterize the polarization by considering a single unit cell. Thus, given a definition of a unit cell, and a coordinate frame fixed within it, the dipole moment density per unit length is given by
\begin{align}
\lambda = \frac{1}{a} \left(\sum_{\alpha=1}^{N_{nuclei}} q_\alpha R_\alpha + \sum_{\alpha=1}^{N_{elec}} -e r_\alpha \right),
\end{align}
where $R_\alpha$ are the positions of the positive charges (i.e. the atomic nuclei), $r_\alpha$ are the electronic positions, and $a$ is the lattice constant (from now on, we will set $a=1$ for simplicity, unless otherwise specified). We are free to reposition the coordinate frame so that its origin is at the center of charge of the atomic nuclei, i.e., at  
\begin{align}
R_c = \frac{1}{Q_{nuclei}}\sum_{\alpha=1}^{N_{nuclei}} q_\alpha R_\alpha,
\end{align}
where $Q_{nuclei}=\sum_{\alpha=1}^{N_{nuclei}} q_\alpha $ is the total positive charge within the unit cell. This choice of coordinate frame cancels out the contribution to the polarization density due to positive charges.  Although the coordinate frame is now fixed, there is still an ambiguity in the definition of the unit cell, as illustrated in Fig.~\ref{fig:ambiguity_electronic_positions}, where the same lattice  charge configuration is shown with two definitions of the unit cell. 
In both cases the locations of both ionic centers (blue dots) and electrons (red circles) are the same, but the electronic positions relative to the ionic charges in the same cell (black arrows), $r$ and $r'$, differ by a lattice constant, i.e., $r' = r-a$. This difference has  no physical meaning, and thus the ambiguity is removed by making the identification 
\begin{align}
r_\alpha \equiv r_\alpha \;\;\; \mbox{mod }a,
\label{eq:mod_a}
\end{align}
where $a$ is the lattice constant. 

With this important subtlety in mind,  we now describe the quantum mechanical theory of electronic polarization in crystals developed by King-Smidth, Vanderbilt, and Resta \cite{king-smith1993,vanderbilt1993,resta2007}. This theory characterizes the bulk dipole moment, and is commonly know as the modern theory of polarization. At the core, the approach is as follows: since the electronic wavefunctions are distributed over the material, we calculate their positions by solving for the eigenvalues of the periodic position operator $\hat{x}$ projected into the subspace of occupied bands \cite{yu2011,alexandradinata2014}. These eigenvalues, or \emph{Wannier centers} \cite{wannier1962}, will then map the quantum mechanical problem into the classical problem of point charges \cite{vanderbilt1993}. Notably, we find that the eigenfunctions associated to these centers are useful in the formulation of higher multipole moments, as we will see for the case of quadrupole (Section \ref{sec:Quadrupole}) and octupole (Section \ref{sec:Octupole}) moments.

\subsection{The large Wilson loop, Wannier centers and Wannier functions}
\label{sec:PxP_1D}
The position operator for the electrons in a crystal with $N$ unit cells and $N_{orb}$ orbitals per unit cell is~\cite{resta1998}
\begin{align}
\hat{x}=\sum_{R,\alpha} c^\dagger_{R,\alpha}\ket{0} e^{-i \Delta_k (R+r_\alpha)}  \bra{0} c_{R,\alpha},
\end{align}
where $\alpha \in 1\ldots N_{orb}$ labels the orbital, $R \in 1 \ldots N$ labels the unit cell, $r_\alpha$ is the position of orbital $\alpha$ relative to the center of positive charge within the unit cell or, more generally, relative to the (fixed) origin of system of coordinates (see Section \ref{sec:PrelimConsiderations}), and $\Delta_k=2\pi/N$ (remember we have set $a=1$). Consider the discrete Fourier transform
\begin{align}
c_{R,\alpha} &= \frac{1}{\sqrt{N}}\sum_k e^{-i k(R+r_\alpha)}c_{k,\alpha}\nonumber \\
c_{k,\alpha} &= \frac{1}{\sqrt{N}}\sum_R e^{i k(R+r_\alpha)}c_{R,\alpha},
\end{align}
where $k \in \Delta_k\cdot(0, 1, \ldots N-1)$. We impose the boundary conditions
\begin{align}
c_{R+N, \alpha}=c_{R,\alpha} \rightarrow c_{k+G,\alpha} = e^{i Gr_\alpha}c_{k,\alpha},
\label{eq:app_creation_operator_boundary_conditions}
\end{align}
where $G$ is a reciprocal lattice vector (the phase $e^{i Gr_\alpha}$ is generally $e^{i {\bf G} \cdot {\bf r_\alpha}}$, and can be positive or negative depending on the choice of origin).  In this new basis, we can alternatively write the position operator as
\begin{align}
\hat{x}&=\sum_{k,\alpha} c^\dagger_{k+\Delta_{k},\alpha} \ket{0} \bra{0} c_{k,\alpha},
\end{align}
as well as the second quantized Hamiltonian
\begin{align}
H=\sum_k c^\dagger_{k,\alpha} [h_k]^{\alpha,\beta} c_{k,\beta},
\label{eq:app_Hamiltonian}
\end{align}
where summation is implied over repeated orbital indices. Due to the periodicity \eqref{eq:app_creation_operator_boundary_conditions}, the Hamiltonian $h_k$ obeys
\begin{align}
h_{k+G} = V^{-1}(G) h_k V(G),
\label{eq:app_hamiltonian_boundary_conditions}
\end{align}
where
\begin{align}
[V(G)]^{\alpha,\beta}=e^{-iGr_\alpha}\delta_{\alpha,\beta}.
\label{eq:Vmatrix}
\end{align}
We diagonalize this Hamiltonian as
\begin{align}
[h_k]^{\alpha,\beta} = \sum_n [u^n_k]^\alpha \epsilon_{n,k} [u^{*n}_k]^\beta,
\end{align}
where $[u^n_k]^\alpha$ is the $\alpha$-th component of the eigenstate $\ket{u^n_k}$. To enforce the periodicity \eqref{eq:app_hamiltonian_boundary_conditions}, we impose the periodic gauge
\begin{align}
[u^n_{k+G}]^\alpha = [V^{-1}(G)]^{\alpha,\beta} [u^n_k]^\beta.
\label{eq:UPeriodicGauge}
\end{align}
This diagonalization allows us to write Eq.~\ref{eq:app_Hamiltonian} as
\begin{align}
H = \sum_{n,k} \gamma^\dagger_{n,k} \epsilon_{n,k} \gamma_{n,k},
\end{align}
where 
\begin{align}
\gamma_{n,k} = \sum_\alpha [u^{*n}_k]^\alpha c_{k,\alpha}
\label{eq:GammaNK}
\end{align}
is periodic in the BZ, as it obeys 
\begin{align}
\gamma_{n,k} =\gamma_{n,k+G}.
\end{align}

As we are interested in insulators at zero temperature, we will focus on the occupied electron bands. We hence build the projection operator into occupied energy bands
\begin{align}
P^{occ}=\sum_{n=1}^{N_{occ}} \sum_k \gamma^\dagger_{n,k}\ket{0} \bra{0}\gamma_{n,k},
\end{align}
where $N_{occ}$ is the number of occupied energy bands. From now on we assume that summations over bands include only occupied energy bands.
We now proceed to diagonalize the position operator projected into the subspace of occupied bands\cite{resta1998}
\begin{align}
P^{occ} \hat{x} P^{occ} &= \sum_{n,k}\sum_{n',k'} \gamma^\dagger_{n,k}\ket{0} \bra{0}\gamma_{n',k'} \times \nonumber\\
&\left( \sum_{q,\alpha}  \bra{0}\gamma_{n,k} c^\dagger_{q+\Delta_k,\alpha}\ket{0} \bra{0}c_{q,\alpha} \gamma^\dagger_{n',k'}\ket{0} \right).
\end{align}
From \ref{eq:GammaNK} we have
$\bra{0} \gamma_{n,k}c^\dagger_{q,\alpha}\ket{0} = [u^{*n}_k]^\alpha \delta_{k,q}$, so the projected position operator reduces to
\begin{align}
P^{occ} \hat{x} P^{occ} &= \sum_{m,n=1}^{N_{occ}}\sum_{k} \gamma^\dagger_{m,k+\Delta_k}\ket{0} \braket{u^m_{k+\Delta_k}}{u^n_k} \bra{0}\gamma_{n,k}
\label{eq:PxP_1D}
\end{align}
where we have adopted the notation $\braket{u^m_q}{u^n_k}=\sum_\alpha [u^{*m}_q]^\alpha [u^n_k]^\alpha$ ($\braket{u^m_k}{u^n_q} \neq \delta_{m,n} \delta_{k,q}$ in general. They only obey $\braket{u^m_k}{u^n_k} = \delta_{m,n}$).

The matrix $G_k$ with components $[G_k]^{mn}=\braket{u^m_{k+\Delta_k}}{u^n_k}$ is not unitary due to the discretization of $k$. However, it is unitary in the thermodynamic limit, as seen in Appendix~\ref{sec:app_WilsonLine_ThermodynamicLimit}. To render it unitary for finite $N$, consider the singular value decomposition\cite{souza2001}
\begin{align}
G = U D V^\dagger,
\end{align}
where $D$ is a diagonal matrix. The failure of $G$ to be unitary is manifest in the fact that the (real valued) singular values along the diagonal of $D$ are less than 1. Therefore, we define, at each $k$, 
\begin{align}
F = U V^\dagger
\end{align}
which is unitary. We refer to $F_k$ as the Wilson line element at $k$. In the thermodynamic limit, $N \rightarrow \infty$, we have that $[F_k]^{mn}=[G_k]^{mn}$.
To diagonalize the projected position operator, let us write the eigenvalue problem:
\begin{align}
(P^{occ}\hat{x}P^{occ}) \ket{\Psi^{j}} = E^j \ket{\Psi^{j}},
\label{eq:PxP_eigenproblem}
\end{align}
which, in the basis $\gamma_{n,k} \ket{0}$, adopts the following form
\begin{align}
\left( \begin{array}{ccccc}
0 & 0 & 0 & \ldots & F_{k_N}\\
F_{k_1} & 0 & 0 & \ldots & 0\\
0 & F_{k_2} & 0 & \ldots & 0\\
\vdots & \vdots & \vdots & \ddots & \vdots\\
0 & 0 & 0 & \ldots & 0\\
\end{array} \right)
\left( \begin{array}{c}
\nu_{k_1}\\
\nu_{k_2}\\
\nu_{k_3}\\
\vdots\\
\nu_{k_N}\\
\end{array} \right)^j=
E^j
\left( \begin{array}{c}
\nu_{k_1}\\
\nu_{k_2}\\
\nu_{k_3}\\
\vdots\\
\nu_{k_N}\\
\end{array} \right)^j,
\label{eq:PxP_diagonalization}
\end{align}
where $k_1=0$, $k_2 = \Delta_k$, $\dots$, $k_N= \Delta_k (N-1)$, and $j \in 1\ldots N_{occ}$.
Here we have replaced $G_k$ in Eq.~\ref{eq:PxP_eigenproblem} by $F_k$ to restore the unitary character of the Wilson line elements. By repeated application of the equations above, one can obtain the relation
\begin{align}
\W_{k_f \leftarrow k_i} \ket{\nu^j_{k_i}} = (E^j)^{(k_f-k_i)/\Delta_k} \ket{\nu^j_{k_f}},
\label{eq:app_parallel_transport_1}
\end{align}
where we are adopting the bra-ket notation $\ket{\nu^j_{k_l}}$ for the vector formed by the collection of values $[\nu^j_{k_l}]^n$, for $n \in 1 \ldots N_{occ}$. We define the discrete Wilson line as
\begin{align}
\W_{k_f \leftarrow k_i} =F_{k_f-\Delta_k} F_{k_f-2\Delta_k} \ldots F_{k_i+\Delta_k} F_{k_i}
\end{align}
For a large Wilson loop, i.e. a Wilson line that goes across the entire Brillouin zone (from now on, by Wilson loop we refer exclusively to large Wilson loops), Eq.~\ref{eq:app_parallel_transport_1} results in the eigenvalue problem
\begin{align}
\W_{k+2\pi \leftarrow k} \ket{\nu^j_k} = (E^j)^N \ket{\nu^j_k}.
\end{align}
where the subscript $k$ labels the starting point, or \textit{base point}, of the Wilson loop. While the Wilson-loop eigenstates depend on the base point, its eigenvalues do not. Furthermore, since the Wilson loop is unitary, its eigenvalues are simply phases
\begin{align}
(E^j)^N = e^{i 2\pi \nu^j}
\end{align}
which has $N$ solutions
\begin{align}
E^{j,R} &= e^{i 2\pi \nu^j/N + i 2\pi R /N}\nonumber\\
&=e^{i \Delta_k (\nu^j + R)}
\end{align}
for $R \in 0 \ldots N-1$. The phases $\nu^j$ are the Wannier centers. They correspond to the positions of the electrons relative to the center of the unit cells. The eigenfunctions of the Wilson loop at different base points are related to each other (up to a $U(1)$ gauge, which we now fix to be the identity) by the parallel transport equation
\begin{align}
\ket{\nu^j_{k_f}} = e^{-i(k_f-k_i) \nu^j} \W_{k_f \leftarrow k_i} \ket{\nu^j_{k_i}},
\label{eq:app_parallel_transport_2}
\end{align}
which is a restatement of Eq.~\ref{eq:app_parallel_transport_1}. Since $j \in 1 \ldots N_{occ}$ and $R \in 0 \ldots N-1$, there are as many projected position operator eigenstates and eigenvalues as there are states in the occupied bands. Given the normalized Wilson-loop eigenstates, the eigenstates of the projected position operator, which now reads
\begin{align}
(P^{occ}\hat{x}P^{occ}) \ket{\Psi^j_R} = e^{i \Delta_k (\nu^j + R)}\ket{\Psi^j_R}
\end{align}
are
\begin{align}
\ket{\Psi^j_R} = \frac{1}{\sqrt{N}} \sum_{n=1}^{N_{occ}}\sum_k \left[ \nu^j_k \right]^n e^{-i k R} \gamma^\dagger_{nk}\ket{0},
\label{eq:WannierFunctions_1D}
\end{align}
where $\left[ \nu^j_k \right]^n$ is the $n^\text{th}$ component of the $j^\text{th}$ Wilson-loop eigenstate $\ket{\nu^j_k}$. 
This form of the solution follows directly from \eqref{eq:PxP_diagonalization}.
We call these functions the \textit{Wannier functions} (WF). Here, $j \in 1 \ldots N_{occ}$ labels the WF and $R \in 0\ldots N-1$ identifies the unit cell to which they are associated. These states obey
\begin{align}
\braket{\Psi^i_{R_1}}{\Psi^j_{R_2}} = \delta_{i,j} \delta_{R_1,R_2},
\end{align}
i.e., they form an orthonormal basis of the subspace of occupied bands of the Hamiltonian. 
Before using these results to calculate the polarization, let us comment on the gauge freedom of the Wannier functions. If $\ket{\nu^j_{k_0}}$ is the eigenstate of $\W_{k_0+2\pi \leftarrow k_0}$, then so is $e^{i\phi_0}\ket{\nu^j_{k_0}}$. Naively, one could assign different phases $e^{i\phi_k}$ to each of the $\ket{\nu^j_k}$ in the expansion of \eqref{eq:WannierFunctions_1D}. However, this is not allowed, because the phases of the Wilson-loop eigenstates at subsequent crystal momenta $k$ are fixed to $e^{i\phi_0}$ by the parallel transport relation $\eqref{eq:app_parallel_transport_2}$  --which is our gauge-fixing condition. Thus, the Wannier functions \eqref{eq:WannierFunctions_1D} inherit only an overall phase factor $e^{i\phi_0}$, as expected.

\subsection{Polarization}
The prescription detailed above for the diagonalization of $P^{occ}\hat{x}P^{occ}$ reveals that the expected value of the electronic positions relative to the center of positive charge within the unit cell is given by the  Wannier centers, which are encoded in the phases of the Wilson-loop eigenvalues, i.e., in
\begin{align}
\W_{k+2\pi \leftarrow k} \ket{\nu^j_k}= e^{i 2\pi \nu^j} \ket{\nu^j_k}.
\end{align}
For $j=1\ldots N_{occ}$, the Wannier centers are the collection of values $\{\nu^j\}$. There are $N_{occ}$ Wannier centers associated to each unit cell, and there are $N_{occ}$ electrons per cell in the ground state. The electronic contribution to the dipole moment, measured as the electron charge times the displacement of the electrons from the center of the unit cell is proportional to
\begin{align}
p=\sum_j \nu^j.
\label{eq:polarization_WannierCenters}
\end{align}
In the expression above we have set the electron charge $e=1$ for convenience in the reminder of the paper, unless otherwise noted. 
The expression \eqref{eq:polarization_WannierCenters} is true for any unit cell due to translation invariance, and thus it is a bulk property of the crystal.
Since the Wannier centers are the phases of the eigenvalues of the Wilson loop, we can alternatively write the polarization as
\begin{align}
p=-\frac{i}{2\pi} \mbox{Log } \mbox{Det} \left[ \W_{k+2\pi \leftarrow k}\right].
\label{eq:polarization_Det}
\end{align}
Furthermore, in the thermodynamic limit (see Appendix \ref{sec:app_WilsonLine_ThermodynamicLimit}), if we write the Wilson loop in terms of the Berry connection 
\begin{align}
[\A_k]^{mn}= -i \bra{u^m_k}\partial_k \ket{u^n_k},
\label{eq:Berry_connection}
\end{align}
we have
\begin{align}
p&=-\frac{i}{2\pi} \text{Log } \text{Det} \left[ e^{-i \int_k^{k+2\pi} \A_k dk}\right]\nonumber\\
&=-\frac{1}{2\pi} \int_k^{k+2\pi} \tr [\A_k] dk\;\;\;\mbox{mod 1},
\label{eq:app_polarization_continuum_1d}
\end{align}
which is the well known expression for the polarization in the modern theory of polarization \cite{king-smith1993,vanderbilt1993,resta2007}. The electronic polarization is proportional to the Berry phase that the subspace of occupied bands $P^{occ}_{\bf k}= \ket{u^n_{\bf k}}\bra{u^n_{\bf k}}$ accumulates as it is parallel-transported around the BZ.

\subsubsection{Polarization and gauge freedom}
If the  electrons are `reassigned' to new unit cells, the polarization with the new assignment changes by an integer (see Fig.~\ref{fig:ambiguity_electronic_positions}). Mathematically, this is evident in \eqref{eq:polarization_WannierCenters} from the fact that the Wannier centers $\nu^j$, defined as the $\log$ of a $U(1)$ phase, are also defined mod 1. In the expression \eqref{eq:app_polarization_continuum_1d}, it is not obvious \emph{a priori} how this ambiguity appears. However, this expression for the polarization is not gauge invariant in the following sense. One is free to choose a different ``gauge" for the functions $\ket{u^n_k}$,
\begin{align}
\left.\ket{u'^{m}_k} = \sum_n [U_k]^{mn} \ket{u^n_k}\right. .
\end{align} The Slater determinant that forms the many-body insulating wavefunction is left invariant by this transformation.  The gauge transformation leads to a changed connection
\begin{align}
\A'_k= U^\dagger_k \A_k U_k - i  U^\dagger_k \partial_{k} U_k.
\label{eq:app_new_connection}
\end{align}
This new adiabatic connection gives a polarization
\begin{align}
p'&=p+\frac{i}{2\pi}\int_k^{k+2\pi} dk \tr \left[U^\dagger_k \partial_{k} U_k \right]\nonumber\\
&=p+\frac{i}{2\pi}\int_k^{k+2\pi} dk \tr \left[ \partial_{k} \ln U_k \right]\nonumber\\
&=p+\frac{i}{2\pi} \tr \left[ \ln U_k\right] \bigg|_k^{k+2\pi} \nonumber\\
&=p+\frac{i}{2\pi} \ln \left[ \det U_k\right] \bigg|_k^{k+2\pi} \nonumber\\
&=p+\frac{i}{2\pi} \sum_i \left[ i\phi_i(k+2\pi) - i \phi_i(k)\right] \nonumber\\
&=p+n
\end{align}
where $n$ is an integer. In the second to last line, $\{\phi_i(k)\}$ are the phases of the eigenvalues of $U_k$. The fact that $U_k$ is periodic in $k$ implies that the phases of its eigenvalues can differ at most by a multiple of $2\pi$ between $k$ and $k+2\pi$. Thus, we see that different gauge choices may vary the polarization, but only by integers. 

In what follows we will use the Wilson loop formulation of the polarization instead of the expression \eqref{eq:app_polarization_continuum_1d} written in terms of the gauge-dependent Berry connection. We will later see that the formulation in terms of Wilson loops has a key additional advantage: the \emph{Wilson-loop eigenfunctions} give us access to the Wannier functions \eqref{eq:WannierFunctions_1D}, which in turn allow as to generalize the concept of a quantized dipole moment, as discussed in the next subsection, to quantized higher multipole moments.

\subsection{Symmetry protection and quantization}
\label{sec:DipoleQuantization}
The polarization can be restricted to specific values in the presence of symmetries. For example, a two-band inversion-symmetric insulator at half filling has only one electron per unit cell. Thus, the electron center of charge has to be located at either the atomic center or halfway  between centers, as any other position of the electron violates inversion symmetry. We say that in this case the polarization is `quantized' to be either 0, for electrons at atomic sites, or 1/2, for electrons in between atomic sites. In what follows, we show how symmetries impose constraints on the allowed values of the Wannier centers and consequently on the polarization. For that purpose, we refer to the relations for Wilson loops\cite{alexandradinata2014} that are detailed in Appendix \ref{sec:WilsonLoopsSymmetry}. We first define the notation for Wilson loops. We denote a Wilson loop with base point $k,$ and with parallel transport towards increasing values of momentum until reaching $k+2\pi$ as
\begin{align}
\W_{x,k} \equiv F_{k+N \Delta_k} F_{k+(N-1)\Delta_k} \ldots F_{k+\Delta_k} F_{k},
\label{eq:WilsonLoopDefinitions1}
\end{align}
where $F_k$ is the unitary matrix resulting from the singular value decomposition of $G_k$, which has components $[G_k]^{mn}=\braket{u^m_{k+\Delta_k}}{u^n_k}$ (see Section \ref{sec:PxP_1D}). Similarly, denote the Wilson loop with base point $k$ that advances the parallel transport towards decreasing values of momentum until reaching $k-2\pi$ as
\begin{align}
\W_{-x,k} \equiv F_{k-N \Delta_k} F_{k-(N-1)\Delta_k} \ldots F_{k-\Delta_k} F_{k}.
\label{eq:WilsonLoopDefinitions2}
\end{align}
These Wilson loops obey
\begin{align}
\W_{-x,k} = \W^\dagger_{x,k}
\end{align}
as shown in Appendix \ref{sec:WilsonLoopsSymmetry}. We now show the quantization of the polarization in 1D crystals due to inversion and chiral symmetries.

\subsubsection{Inversion symmetry}
\label{sec:Dipole_under_Inversion}
A crystal with inversion symmetry obeys
\begin{align}
\hat{\I} h_k \hat{\I}^{-1} = h_{-k},
\label{eq:HamiltonianUnderInversion_1d}
\end{align}
where $\I$ is the unitary ($\I^{-1} = \I^\dagger$) inversion operator. As shown in Appendix \ref{sec:WilsonLoopsSymmetry}, in the presence of inversion the Wilson loops obey
\begin{align}
B_{\I,k} \W_{x,k} B^\dagger_{\I,k} &\stackrel{\I}{=} \W^\dagger_{x,-k}
\label{eq:WilsonLoopUnderInversion_1D}
\end{align}
where $B^{mn}_{\I,k}=\matrixel{u^m_k}{\I}{u^n_{-k}}$ is the unitary `sewing' matrix that connects the states at $\ket{u^m_k}$ and $\ket{u^m_{-k}}$ having equal energies (see Appendix \ref{sec:WilsonLoopsSymmetry} for details). Since the Wilson-loop eigenvalues are independent of the base point, Eq. \ref{eq:WilsonLoopUnderInversion_1D} implies that the set of Wilson-loop eigenvalues has to be equal to its complex conjugate, which implies, for the set of Wannier centers,
\begin{align}
\{ \nu_j \} \stackrel{\I}{=} \{ -\nu_j \} \quad \mbox{mod 1}.
\label{WannierConstraintInversion_1D}
\end{align}
This forces the Wannier centers to be either $0$, $1/2$, or to come in complex conjugate pairs $\{\nu, -\nu\}$. Physically, inversion implies that the electrons have to either be: (i) centered at an atomic site ($\nu = 0$), (ii) in between sites ($\nu=1/2$), or (iii) to come in pairs arranged on opposite sides of each atomic center and equally distant from it ($\{\nu$, $-\nu\}$). In the first and third cases, the polarization is 0, while in the second case it is $1/2$. Hence, in general, we have that
\begin{align}
p \stackrel{\I}{=} - p\;\;\;\mbox{mod 1}.
\end{align}
That is, under inversion,
\begin{align}
p \stackrel{\I}{=}0 \mbox{ or } 1/2.
\label{eq:DipoleQuantizationInversion}
\end{align}
This quantization under inversion symmetry allows for an alternative way of calculating the Wannier centers. From \eqref{eq:HamiltonianUnderInversion_1d} it follows that at the inversion-symmetric momenta $k_*=0,\pi$ we have
\begin{align}
[\hat{\I}, h_{k_*}]=0.
\end{align}
Thus, the eigenstates of the Hamiltonian at $k_*$ can be chosen to be simultaneous eigenstates of the inversion operator
\begin{align}
\hat{\I} \ket{u_{k_*}} = \I(k_*) \ket{u_{k_*}},
\end{align}
where $\I(k_*)$ are the inversion eigenvalues at momenta $k_*=0,\pi$.
The inversion eigenvalues can then be used as labels for the inversion representation at $k^*$ that the occupied bands take. If the representation is the same at $k=0$ and $k=\pi$, the topology is trivial, and the polarization is zero. However, if the representations at these two points of the BZ differ, we have a non-trivial topology associated with a non-zero polarization \cite{turner2010,hughes2011inversion,turner2012}. We can encode these relations in the expression
\begin{align}
e^{i 2\pi p}=\I(0) \I^*(\pi),
\label{eq:polarization1D_InversionEigenvalues}
\end{align}
where the asterisk stands for complex-conjugation. 
A formal and complete derivation of the relation between Wilson-loop eigenvalues and inversion eigenvalues was first shown in Ref. \onlinecite{alexandradinata2014}. The relations between inversion and Wilson-loop eigenvalues that we will use are shown in Tables \ref{tab:map_I_W_1bands} and \ref{tab:map_I_W_2bands}.

\begin{table}[t]
\begin{center}
\begin{tabular}{ccc}
 $\hat{\I}$ eigenval. \;\;&  $\hat{\I}$ eigenval. \;\;& $\W$ eigenval. \\
 at $k=0$\;\; & at $k=\pi$\;\; &\\
\hline
 $+$&$+$ & $+1$ \\
 $+$&$-$ & $-1$ \\
\end{tabular}
\end{center}
\caption{Relation between inversion and Wilson-loop eigenvalues for an insulator with one occupied band. $\hat{\I}$ is the inversion operator. $\W$ is the Wilson loop. The signs $\pm$ represent $\pm1$  if $\I^2=+1$ or $\pm i$ if $\I^2=-1$.}
\label{tab:map_I_W_1bands}
\end{table}

\begin{table}[t]
\begin{center}
\begin{tabular}{ccc}
 $\hat{\I}$ eigenval. \;\;&  $\hat{\I}$ eigenval. \;\;& $\W$ eigenval. \\
 at $k=0$\;\; & at $k=\pi$\;\; &\\\hline
 $(++)$&$(++)$ & $(+1,+1)$ \\
 $(++)$&$(+-)$ & $(+1,-1)$ \\
 $(++)$&$(--)$ & $(-1,-1)$ \\
 $(+-)$&$(+-)$ & $(c.c.)$
\end{tabular}
\end{center}
\caption{Relation between inversion and Wilson-loop eigenvalues for an insulator with two occupied bands. $\hat{\I}$ is the inversion operator. $\W$ is the Wilson loop. The signs $\pm$ represent $\pm1$ if $\I^2=+1$ or $\pm i$ if $\I^2=-1$. $c.c.$ stands for complex conjugate pair of values of magnitude 1.}
\label{tab:map_I_W_2bands}
\end{table}

\subsubsection{Chiral symmetry}
Although less evident, chiral (sublattice) symmetry also quantizes the polarization. Chiral symmetry implies that the Bloch Hamiltonian obeys
\begin{align}
\Pi h_k \Pi^{-1} = -h_k,
\label{eq:HamiltonianUnderChiral_1d}
\end{align}
where $\Pi$ is the unitary ($\Pi^{-1} = \Pi^\dagger$) chiral operator. Under this symmetry, the Wilson loop obeys
\begin{align}
B_{\Pi,k} \W^{occ}_k B^\dagger_{\Pi,k} &\stackrel{\Pi}{=} \W^{unocc}_k.
\label{eq:WilsonLoopUnderChiral_1D}
\end{align}
Here, $\W^{occ}_k$ ($\W^{unocc}_k$) is the Wilson loop at base point $k$ over occupied (unoccupied) bands, and $B^{mn}_{\Pi, k}=\matrixel{u^m_k}{\Pi}{u^n_k}$ is a sewing matrix that connects states $\ket{u^m_k}$ and $\ket{u^n_k}$ having opposite energies, that is, such that $\epsilon_{m,k} = - \epsilon_{n,k}$. Eq. \ref{eq:WilsonLoopUnderChiral_1D} implies that the Wannier centers from the occupied $\nu^j$ bands equal those calculated from the unoccupied bands $\eta^j$,
\begin{align}
\{ \nu_j \} \stackrel{chiral}{=} \{ \eta_j \} \quad \mbox{mod 1}
\end{align}
and thus,
\begin{align}
p^{occ} \stackrel{chiral}{=} p^{unocc}.
\label{eq:polarization_chiral_1}
\end{align}
It is important to recall that to have strict chiral symmetry as we assume here, the number of occupied bands in a gapped system will be equal to the number of unoccupied bands. To conclude our argument, an additional consideration is necessary: The Hilbert space over all bands (occupied \emph{and} unoccupied) is topologically trivial. Thus, the polarization that results from both the occupied and unoccupied bands is necessarily also trivial, i.e., 
\begin{align}
p^{occ}+p^{unocc}=0\;\;\;\mbox{mod 1},
\end{align}
which leads to
\begin{align}
p^{occ} \stackrel{chiral}{=} -p^{unocc}\;\;\;\mbox{mod 1}.
\label{eq:polarization_chiral_2}
\end{align}
From \eqref{eq:polarization_chiral_1} and \eqref{eq:polarization_chiral_2} we conclude that
\begin{align}
p \stackrel{chiral}{=} 0 \mbox{ or } 1/2,
\end{align}
i.e., the polarization is quantized in the presence of chiral (sublattice) symmetry. 

In what follows, we discuss the features of a system with non-zero polarization by studying the minimal model that realizes the dipole phase. In general a bulk polarization per unit length of $p$ manifests itself at the boundary in the existence of bound surface charges of magnitude $p$, in exact correspondence to the classical electromagnetic theory [cf. Eq.~\ref{eq:ChargeDensities_Dipole}]. Consequently, the topological dipole phase exhibits quantized, fractional boundary charge of $\pm e/2$, which can be protected, e.g.,  by inversion or chiral symmetries. Additionally, we give a concrete example of adiabatic current being pumped in this model\cite{ricemele1982,atala2013,wang2013,lu2016}.

\subsection{Minimal model with quantized polarization in 1D}

A minimal model for an insulator with bulk polarization in one dimension is the Su-Schrieffer-Hegger model \cite{SSH1979}, which describes a chain with alternating strong and weak bonds between atoms, as in polyacetylene \cite{SSH1979}. A tight-binding schematic of this structure is shown in Fig.~\ref{fig:dipole_lattice}a,b. Its Hamiltonian is
\begin{align}
H^{SSH}=\sum_R \left(\gamma c^{\dagger}_{R,1} c_{R,2}+\lambda c^{\dagger}_{R,2} c_{R+1,1}+c.c.\right),
\label{eq:SSH_RealSpace}
\end{align}
where $\gamma$ and $\lambda$ are hopping terms within and between unit cells respectively. Its corresponding Bloch Hamiltonian in momentum space is
\begin{align}
h^{SSH}(k) = \left( \begin{array}{cc}
0 & \gamma + \lambda e^{-ik}\\
\gamma + \lambda e^{ik} & 0 
\end{array}\right),
\label{eq:SSHBlochHamiltonian}
\end{align}
where the basis of the matrix follows the numbering in Fig.  \ref{fig:dipole_lattice}a. More compactly, we will write this, and the Hamiltonians to come, in terms of the Pauli matrices $\sigma_i$, for $i=1,2,3$:
\begin{align}
h^{SSH}(k) = [\gamma + \lambda \cos(k)] \sigma_1 + \lambda \sin(k) \sigma_2.
\label{eq:SSH_Hamiltonian}
\end{align}
The SSH model has energies
\begin{align}
\epsilon(k) = \pm \sqrt{(\lambda^2 + 2\lambda \gamma \cos(k) + \gamma^2 )}.
\end{align}
The model is  gapped unless $|\gamma| = |\lambda|$. Thus, at half filling, the SSH model is an insulator, unless $\gamma = \lambda$ ($\gamma = - \lambda$) where the bands touch at the $k=\pi$ ($k=0$) points of the BZ and the system is metallic.

\subsubsection{Symmetries}
\begin{figure}[t]%
\centering
\includegraphics[width=\columnwidth]{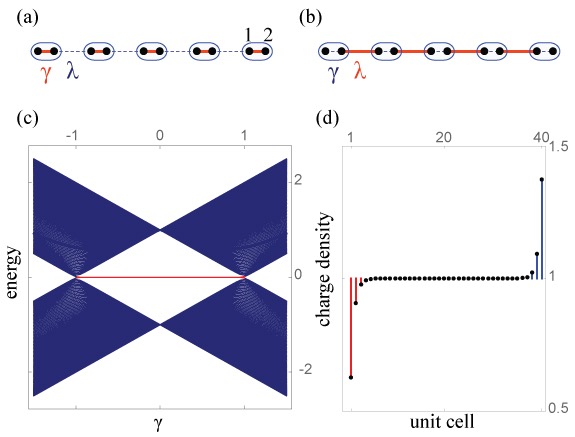}
\caption{(Color online)  Su-Schrieffer-Hegger model with Hamiltonian \eqref{eq:SSH_RealSpace}. (a) Trivial phase ($|\gamma|> |\lambda|$). (b) Topological dipole phase ($|\gamma|<|\lambda|$). (c) Energy spectrum for a chain with open boundaries as a function of $\gamma$ when $\lambda = 1$. Red energies correspond to two degenerate edge-localized states. (d) Electron density in the topological dipole phase ($\lambda=1$, $\gamma=0.5$). The total charge at the edges is $\pm e/2$ relative to background.}
\label{fig:dipole_lattice}
\end{figure}

The Hamiltonian \ref{eq:SSHBlochHamiltonian} has inversion symmetry $\I h(k) \I^\dagger = h(-k)$, with $\I = \sigma_1,$ and chiral symmetry $\Pi h(k) \Pi^\dagger = -h(k)$ with $\Pi = \sigma_3$. Thus, this model has quantized polarization: $p=0$ for $|\gamma|>|\lambda|$ and $p=1/2$ for  $|\gamma|<|\lambda|$. At $|\gamma|=|\lambda|$, the energy gap closes. This crossing is necessary to change the insulating phase from one with $p=0$ to $p=1/2,$ or vice versa. Thus, the polarization is an index that labels two distinct phases, the `trivial' $p=0$ phase and the `non-trivial' or `dipole' phase $p=1/2$. This is the simplest example of a \emph{symmetry protected topological} (SPT) phase, because the two phases are clearly distinguished \emph{only} in the presence of the symmetries that quantize the dipole moment. However, both the trivial and the ``non-trivial" state are described in terms of localized Wannier states -- therefore a more appropriate term for the ``non-trivial" state is an obstructed atomic limit  \cite{bradlyn2017}. An illustration of these two phases and the transition point is shown in Fig.~\ref{fig:dipole_lattice}c, where the spectrum of the open-boundary Hamiltonian is parametrically plotted as a function of $\gamma$, for a fixed value of $\lambda=1$.  

\subsubsection{Quantization of the boundary charge}
In an SSH crystal with open boundaries, one consequence of the quantization of the bulk polarization to $e/2$ in the dipole phase is the appearance of $\pm e/2$ charge at its edges.  This accumulation is due to the existence two, degenerate and edge-localized modes. In the presence of chiral symmetry, the edge mode energies are pinned to zero, and the edge modes are eigenstates of the chiral operator. In the absence of chiral symmetry, the zero energy protection of the edge modes is lost; chiral-breaking terms lift the energies of the edge modes away from 0, but they will remain degenerate (resulting in a twofold degenerate ground-state at half-filling) as long as inversion is preserved in the system with open boundaries.

To determine a fixed sign for the polarization one must weakly break the degeneracy of the edge modes. For $N$ unit cells, half filling implies that there are $N$ electrons, $N-1$ of which fill bulk states. The extra electron thus will fill one of the edge states, but if they are degenerate, the electron cannot pick which state to fill. Splitting the degeneracy infinitesimally is enough to decide which end mode is filled, thus choosing the `sign' of the dipole. In the SSH model, the symmetry breaking can be achieved by adding the term $ \delta \sigma_3$ to \eqref{eq:SSH_Hamiltonian} for an infinitesimal value of delta $\delta$. Notice that $\sigma_3$ breaks both chiral and inversion symmetries, as required.

\subsection{Charge pumping}
\label{sec:DipolePumping}
In this section we describe the pumping of electronic charge in insulators by means of adiabatic deformations of the Hamiltonian. Originally conceived by Thouless \cite{thouless1983} as a method to extract current out of an insulator, this mechanism also has a well established connection with the quantum anomalous Hall effect~\cite{qi2008}. We exploit an analogous connection in Section \ref{sec:HingeInsulator} to construct an insulator with chiral hinge states that has the same topology as a 2D quadrupolar pumping cycle. In what follows, we describe two concrete examples of charge pumping. We start with a pedagogical example that allows us to closely follow the motion of the Wannier centers during the adiabatic evolution. However, this model requires a piecewise continuous parametrization. Therefore we also describe a pumping with a fully continuous parametrization - although it is less obvious pictorially.  

\begin{figure}[t]%
\centering
\includegraphics[width=\columnwidth]{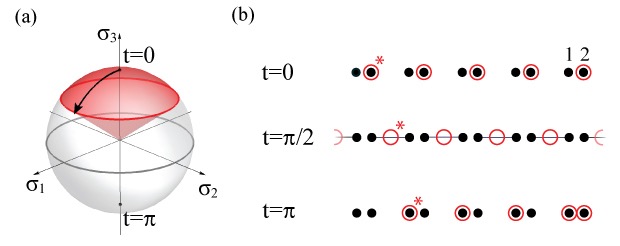}
\caption{(Color online)  Adiabatic pumping of an electron in the SSH model (Eq. \ref{eq:SSH_pumping}) parametrized by \eqref{eq:DipolePumping_spherical} during the first half of the cycle (the second half of the cycle generates no transport). (a) As the adiabatic parameter $t$ advances, the Berry phase of the occupied band across the BZ increases proportional to the solid angle enclosed by the ground-state projector on the Bloch sphere. (b) Electronic positions at $t=0,\pi/2,\pi$ that illustrate how the positions advance proportionally to the Berry phase illustrated in (a).}
\label{fig:dipole_pumping}
\end{figure}

The pedagogical example uses the SSH model as follows. Consider the SSH Hamiltonian \eqref{eq:SSH_Hamiltonian} with additional on-site energies $\delta \sigma_3$, which breaks the chiral and inversion symmetries,
\begin{align}
h^{SSH}_\delta = \left[\gamma+\lambda \cos(k)\right] \sigma_1 + \lambda \sin(k) \sigma_2 + \delta \sigma_3.
\label{eq:SSH_pumping}
\end{align}
We modify the parameters $\lambda$, $\gamma$, and $\delta$ adiabatically:
\begin{align}
(\delta, \lambda, \gamma)=\left\{
\begin{array}{ll}
(\cos(t),\sin(t),0)& 0< t \leq \pi\\
(\cos(t),0,|\sin(t)|) & \pi < t \leq 2\pi
\label{eq:DipolePumping_spherical}
\end{array}
\right.
\end{align}
where $t$ is the adiabatic parameter. This parametrization represents an evolution of a family of Hamiltonians through a closed cycle that returns to the original configuration when $t=2\pi p$ for integer $p.$ In the process, however, an electron is transferred from the left to the right at each unit cell. The first half of the cycle is illustrated in Fig.~\ref{fig:dipole_pumping}. 
At $t=0$ the the Hamiltonian is $+\sigma_3$, i.e., it is in the atomic limit, and at half filling the basis sites `2' are occupied. In Fig.~\ref{fig:dipole_pumping}a, this corresponds to the north pole of the Bloch sphere. As time progresses, the hopping amplitude $\lambda$ increases while keeping $\gamma = 0$, which results in the wave functions progressively leaking into sites `1' of the neighboring unit cells to their right. At $t=\pi/2$ the occupancy of the sites is uniform, with Wannier center in between unit cells. This is the dipole phase, with $p=1/2$. Then, for  $\pi/2 < t< \pi$ the hopping amplitude decreases while the on-site potentials reverse sign. Thus, the eigenstates increasingly occupy states `1'. At $t=\pi$, the Hamiltonian is $-\sigma_3$, and only sites `1' are occupied. In Fig.~\ref{fig:dipole_pumping}a, this corresponds to the south pole of the Bloch sphere. During this first half of the cycle, electrons have crossed one unit cell to the right. Topologically, the entire Bloch sphere of the Hamiltonian in Eq. \ref{eq:SSH_pumping} has been swept, which is characterized by a Chern number $n=1$. The second half of the cycle does not cause transport, as it switches the electron occupancy from sites `1' back to sites `2' on the \emph{same} unit cell (since $\lambda=0$). At $t= 3 \pi/2$, the system is again inversion-symmetric, and the occupancy is uniform, with Wannier center in the middle of the unit cell. This is the $p=0$ phase. Hence we can think of the above interpolation as an cycle between the $p=1/2$ and the $p=0$ phases, during which, as explicitly shown, an electron has been moved from one side of the chain to the other. 

\begin{figure}[t]%
\centering
\includegraphics[width=\columnwidth]{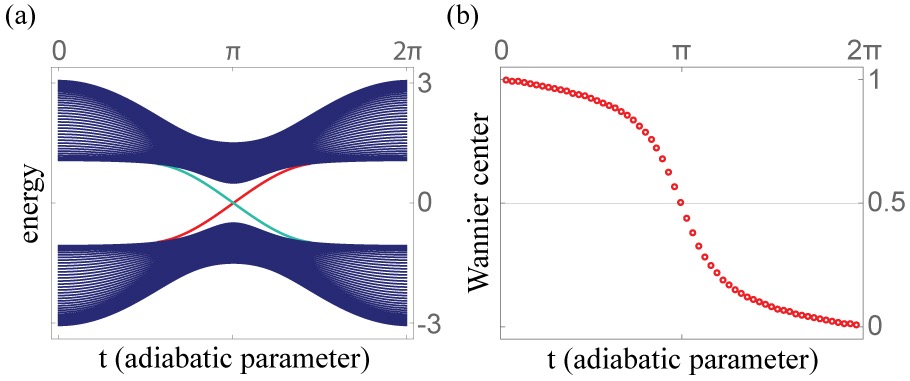}
\caption{(Color online)  Adiabatic pumping of an electron with the adiabatic family of Hamiltonians \eqref{eq:dipole_pumping_toroidal}. (a) Energies when boundaries are open as function of the adiabatic parameter $t$. (b) Wannier centers in the unit cell as function of the adiabatic parameter $t$. In both plots $\gamma = 0.5$, $m=1$.}
\label{fig:dipole_pumping_toroidal}
\end{figure}

Although the pumping method described above has an intuitive pictorial representation, we also want to generate electronic adiabatic pumping with a fully continuous parametrization. If we can find such a representation then we can use it to generate a lattice model in one spatial dimension higher which will be topologically equivalent to a quantum anomalous Hall (Chern) insulator. This is carried out by reinterpreting the adiabatic parameter as an additional momentum quantum number for a 2D system\cite{qi2008}. One way to realize this is by the family of Hamiltonians
\begin{align}
h(k,t) =& [\gamma + \cos(k)] \sigma_1 + \sin(k) \sigma_2 \nonumber\\
&+ m \sin(t) \sigma_3 + [1+ m \cos(t)] \sigma_2,
\label{eq:dipole_pumping_toroidal}
\end{align}
where $t$ is the adiabatic parameter. 
Fig.~\ref{fig:dipole_pumping_toroidal} shows the energy bands and the Wannier centers as a function of the adiabatic parameter. Eq. \ref{eq:dipole_pumping_toroidal} encloses a monopole of Berry flux as it sweeps out a torus $T^2$ instead of  a sphere $S^2$ as in \eqref{eq:DipolePumping_spherical}. In Fig.~\ref{fig:dipole_pumping_toroidal}, $\gamma =0.5$ and $m=1$. Thus, at $t=0$ the system is in the trivial phase, while at $t=\pi$ the lattice is in the SSH dipole phase. Correspondingly, we see that at $t=0$ there are no zero-energy modes when boundaries are open (Fig.~\ref{fig:dipole_pumping_toroidal}a), and the Wannier center (and consequently its polarization) is at a value of zero (Fig.~\ref{fig:dipole_pumping_toroidal}b). At $t=\pi$, there are two states with zero energy when we have open boundaries, and the Wannier center is at $\nu=1/2$. Finally at $t=2\pi$ the system has returned back to its initial state in the atomic limit after the charge has moved by one unit cell.

\section{Bulk dipole moment in 2D crystals}
\label{sec:Dipole_2D}
We now investigate the existence of dipole moments in 2D crystals. Without loss of generality, we calculate the position operator along $x$ projected into the occupied bands\cite{resta1998}
\begin{align}
P^{occ} \hat{x} P^{occ} &= \sum_{k} \gamma^\dagger_{m,(k_x+\Delta_{k_x}, k_y)}\ket{0} \bra{0}\gamma_{n,(k_x,k_y)}\times \nonumber \\
&\braket{u^n_{(k_x+\Delta_{k_x},k_y)}}{u^m_{(k_x,k_y)}}
\end{align}
which is similar to Eq.~\ref{eq:PxP_1D}, but with the extra quantum number $k_y$. Importantly, notice that the operator is diagonal in $k_y$. Thus, all the findings in Section \ref{sec:Dipole_1D} follow through in this case too, but with the extra label $k_y$. In particular, for a large Wilson loop $\W_{x, {\bf k}}$, which  has ${\bf k}=(k_x,k_y)$ as its base point and runs along increasing values of $k_x$, as the obvious extension to 2D of definition \eqref{eq:WilsonLoopDefinitions1}, we have
\begin{align}
\W_{x, {\bf k}} \ket{\nu^j_{x,{\bf k}}} = (E^j)^N \ket{\nu^j_{x,{\bf k}}},
\label{eq:WilsonLoopEigenproblem2D}
\end{align}
where 
\begin{align}
(E^j)^N = e^{i 2\pi \nu^j_x(k_y)}
\end{align}
are its eigenvalues and $\ket{\nu^j_{x,{\bf k}}}$ its eigenstates.
The WFs along $x$ are then
\begin{align}
\ket{\Psi^j_{R_x,k_y}} = \frac{1}{\sqrt{N_x}} \sum_{n=1}^{N_{occ}}\sum_{k_x} \gamma^\dagger_{n,\bf k}\ket{0} \left[ \nu^j_{x,{\bf k}} \right]^n e^{-i k_x R_x},
\label{eq:app_MLWF_2D}
\end{align}
where ${\bf k}=(k_x,k_y)$ is the crystal momentum, with $k_{x,y} = n_{x,y} \Delta_{k_{x,y}}$, for $n_{x,y} \in 0,1,\ldots, N_{x,y}-1$ and $\Delta_{k_{x,y}}=2\pi/N_{x,y}$. 
These functions obey
\begin{align}
\braket{\Psi^j_{R_x,k_y}}{\Psi^{j'}_{R'_x,k'_y}} = \delta_{j,j'} \delta_{R_x,R'_x} \delta_{k_y,k'_y},
\end{align}
i.e., they form an orthonormal basis of the subspace of occupied energy bands of the Hamiltonian. 
For the Wilson-loop eigenstates $\ket{\nu^j_{x,{\bf k}}}$, the subscript $x$ specifies the direction of its Wilson loop, and ${\bf k}$ specifies its base point, so, for example, Eq. \ref{eq:WilsonLoopEigenproblem2D} is explicitly written as
\begin{align}
\W_{(k_x+2\pi,k_y) \leftarrow (k_x,k_y)} \ket{\nu^j_{x,(k_x,k_y)}} = (E^j)^N \ket{\nu^j_{x,(k_x,k_y)}}.
\end{align}
Although the phases $\nu^j_x(k_y)$ of the eigenvalues of the Wilson loop $\W_{x,\bf k}$ do not depend on $k_x$, in general they do depend on $k_y$. 
Thus, the polarization for one-dimensional crystals translates into polarization as a function of $k_y$ in its 2D counterpart, that is
\begin{align}
p_x(k_y) &=\sum_{j=1}^{N_{occ}} \nu^j_x(k_y) = -\frac{i}{2\pi} \text{Log Det}[\W_{x,\bf k}],
\end{align}
which, in the thermodynamic limit becomes
\begin{align}
p_x(k_y) &=-\frac{1}{2\pi}\int_{0}^{2\pi}\tr[\A_{x,\bf k}] dk_x,
\label{eq:PxKy}
\end{align}
where ${\bf k}=(k_x,k_y)$ and $[\A_{x,\bf k}]^{mn}= -i \bra{u^m_{\bf k}} \partial_{k_x} \ket{u^n_{\bf k}}$ is the non-Abelian Berry connection (where $m,n$ run over occupied energy bands). The total polarization along $x$ is
\begin{align}
p_x = \frac{1}{N_y} \sum_{k_y}p_x(k_y).
\label{eq:Polarization_2D_discrete}
\end{align}
In the thermodynamic limit, $\frac{1}{N_y} \sum_{k_y} \to \frac{1}{2\pi}\int dk_y$, the polarization in 2D crystals is
\begin{align}
p_x &=-\frac{1}{(2\pi)^2} \int_{BZ} \tr [\A_{x,\bf k}] d^2\bf k.
\label{eq:Polarization_2D}
\end{align}
Here $BZ$ is the 2D Brillouin zone. 
The 2D polarization is thus given by the vector ${\bf p}=(p_x,p_y)$, where each component $p_i$ is calculated using \eqref{eq:Polarization_2D} with $[\A_{i,\bf k}]^{mn}= -i \bra{u^m_{\bf k}} \partial_{k_i} \ket{u^n_{\bf k}}$, for $i=x,y$.

\subsection{Symmetry protection and quantization}
As in 1D, the polarization in 2D can have the values 0 or 1/2 (in appropriate units) under the presence of certain symmetries. In this section we consider the symmetries that protect the quantization of the polarization in 2D. The conclusions detailed below follow from the  symmetry transformations of the Wilson loops derived in Appendix \ref{sec:WilsonLoopsSymmetry}.

\subsubsection{Reflection symmetries}
In the presence of reflection symmetries $M_x: x \rightarrow -x$ and $M_y: y \rightarrow -y$, the Bloch Hamiltonian obeys 
\begin{align}
\hat{M}_x h_{(k_x,k_y)} \hat{M}^{-1}_x &= h_{(-k_x,k_y)},\nonumber\\
\hat{M}_y h_{(k_x,k_y)} \hat{M}^{-1}_y &= h_{(k_x,-k_y)},
\label{eq:HamiltonianUnderReflection_2d}
\end{align}
respectively. The polarization along $x$ as a function of $k_y$ \eqref{eq:PxKy} under these symmetries obeys
\begin{align}
p_x(k_y) \stackrel{M_x}{=} -p_x(k_y)\nonumber\\
p_x(k_y) \stackrel{M_y}{=} p_x(-k_y),
\end{align}
and similarly for $p_y(k_x)$. These relations imply that, under $M_x$, 
\begin{align}
p_x(k_y) \stackrel{M_x}{=}0 \mbox{ or } 1/2
\label{eq:pxkyUnderReflection}
\end{align}
(and similarly for $p_y(k_x)$ under $M_y$). This quantized value can be easily computed by comparing the reflection representations at the reflection-invariant lines in the BZ. Concretely, from \eqref{eq:HamiltonianUnderReflection_2d} it follows that
\begin{align}
[\hat{M}_x,h_{(k_{*x},k_y)}] = 0
\end{align}
for $k_{*x}=0,\pi$ and for $k_y \in [-\pi,\pi)$. Thus, following with the rationale in Section \ref{sec:Dipole_under_Inversion} for the case of inversion in 1D, the polarization $p_x(k_y)$ under reflection symmetry $M_x$ can be found by calculating
\begin{align}
e^{i 2\pi p_x(k_y)}=m_x(0,k_y) m^*_x(\pi,k_y).
\label{eq:pxky_ReflectionEigenvalues}
\end{align}
where $m_x(k_{*x},k_y)$ are the reflection eigenvalues at the reflection invariant lines of the BZ and the superscript asterisk stands for complex-conjugation (in the case of double-groups for which reflection symmetries have complex eigenvalues).
Now, the polarization at fixed $k_y$ can be thought of as the polarization of a 1D Bloch Hamiltonian $h(k_x,k_y)$, for $k_x \in [-\pi,\pi)$. It follows from \eqref{eq:pxkyUnderReflection} that, under reflection $M_x$, this 1D Hamiltonian has quantized polarization. Since this polarization is a topological index, a change in this index $p_x(k_y)$ across $k_y$ is only possible if the Hamiltonian $h(k_x,k_y)$ closes the gap at certain values of $k_y$. Thus, for Hamiltonians that are gapped in energy for all $k_x,k_y \in [-\pi,\pi)$, their polarizations $p_x(k_y)$ are not only quantized, but also continuous across $k_y \in [-\pi,\pi)$. This implies that the overall polarization is also quantized,
\begin{align}
p_i \stackrel{M_i}{=}0 \mbox{ or } 1/2
\label{eq:pxUnderReflection}
\end{align}
for $i=x,y$. 

\subsubsection{Inversion symmetry}
\label{sec:p2DInversion}
Under inversion symmetry,
\begin{align}
\I h_{\bf k} \I^{-1}_x = h_{-{\bf k}},
\label{eq:HamiltonianUnderInversion_2d}
\end{align}
we have the relation
\begin{align}
p_x(k_y) \stackrel{\I}{=} -p_x(-k_y)\;\;\mbox{mod 1}.
\label{eq:pxkyUnderInversion}
\end{align}
This implies that the polarization \eqref{eq:Polarization_2D_discrete} obeys
\begin{align}
p_x \stackrel{\I}{=} -p_x\;\;\mbox{mod 1},
\end{align}
i.e., under inversion symmetry the polarization is quantized:
\begin{align}
p_x \stackrel{\I}{=} 0 \mbox{ or } 1/2.
\label{eq:pxUnderInversion}
\end{align}
However, the restriction \eqref{eq:pxkyUnderInversion} does not quantize the polarization at each $k_y$, as in \eqref{eq:pxkyUnderReflection} for reflection symmetries. This allows for $p_x(k_y)$ to acquire any value $[0,1]$, except at the inversion symmetric momenta $k_{*y}=0,\pi$, where we have
\begin{align}
p_x(k_{*y}) \stackrel{\I}{=}0 \mbox{ or } 1/2.
\end{align}

\begin{figure} [t]
\centering
\begin{tikzpicture}[scale=1.2]
 
 
 		\draw [->,black] (-1.5,0)--(1.5,0) node[below] {$k_x$};
	  	\draw [->,black] (0,-1.5)--(0,1.5) node[right] {$k_y$};		
		
		 \draw [red, line width=.7mm,dashed] (-1,.0)--(1,0);
	 	 \draw [red, line width=.7mm,dashed] (-1,1)--(1,1);
		 \draw [red, line width=.7mm,dashed] (-1,-1)--(1,-1);
		 \draw [blue, thick] (0,-1)--(0,1);
	 	 \draw [blue,thick] (-1,-1)--(-1,1);
	 	 \draw [blue,thick] (1,-1)--(1,1);
	  
		\fill [black] (0,0) circle (2pt) node[below left] {${\bf \Gamma}$};
		\fill [black] (1,0) circle (2pt) node[below left] {${\bf X}$};
		\fill [black] (0,1) circle (2pt) node[below left] {${\bf Y}$};
		\fill [black] (1,1) circle (2pt) node[below left] {${\bf M}$};
		
\end{tikzpicture}
\caption{(Color online) Brillouin zone in 2D and its reflection invariant points and lines of the Brillouin zone. In the presence of reflection symmetries $M_{x,y}$ and inversion $\I$, the Hamiltonian at the solid blue (dashed red) lines commutes with $\hat{M}_x$ ($\hat{M}_y$). At the points ${\bf \Gamma}=(0,0)$, ${\bf X}=(\pi,0)$, ${\bf Y}=(\pi,0)$, and ${\bf M}=(\pi,\pi)$ the Hamiltonian also commutes with $\hat{\I}$.}
\label{fig:2DBZ}
\end{figure}
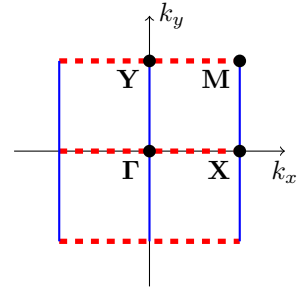

The two values, $p_x(0)$ and $p_x(\pi)$, are topological indices that are related to the parity of the Chern number (defined in Eq.~\ref{eq:ChernNumber})\cite{hughes2011inversion,turner2012}
\begin{align}
e^{i\pi n} = e^{i 2\pi p_x(0)}e^{i 2\pi p_x(\pi)}.
\end{align}
This relationship between the parity of the Chern number and the polarizations $p_x(0)$, $p_x(\pi)$ will become apparent in the discussion to follow in Section \ref{sec:WannierBands_AllModels}. 
Using \eqref{eq:polarization1D_InversionEigenvalues}, this expression reduces to
\begin{align}
e^{i\pi n} = \I({\bf \Gamma})\I({\bf X})\I({\bf Y})\I({\bf M}),
\end{align}
where the momenta $\bf \Gamma$, $\bf X$, $\bf Y$, and $\bf M$ are shown in Fig.~\ref{fig:2DBZ}.  (Note that the inversion eigenvalues are real for both single and double groups, and hence complex conjugation is not necessary).

The polarization of an insulator with a non-zero Chern number is a subtle matter and requires special care because of the partial occupation of the chiral edge states\cite{coh2009}. We will only consider the polarization of insulators with vanishing Chern number. When the Chern number is zero, the polarization can be determined from the inversion eigenvalues of the occupied bands via\cite{hughes2011inversion,turner2012}
\begin{align}
p_x \stackrel{\I}{=}\begin{cases}
0 & \mbox{if }\I({\bf \Gamma})\I({\bf X})=+1 \mbox{ and } \I({\bf Y})\I({\bf M})=+1\\
1/2 & \mbox{if }\I({\bf \Gamma})\I({\bf X})=-1 \mbox{ and } \I({\bf Y})\I({\bf M})=-1
\end{cases}.
\label{eq:pxInversionEigenvalues}
\end{align}
For a system with vanishing Chern number, the polarization $p_x(k_y)$ for $k_y =0$ and $k_y=\pi$  are identical. Hence we only need to compare either $\I({\bf \Gamma})$ and $\I({\bf X})$ together or $\I({\bf Y})$ and $\I({\bf M})$ together to determine $p_x.$ Similarly, $p_y$ can be inferred by
\begin{align}
p_y \stackrel{\I}{=}\begin{cases}
0 & \mbox{if }\I({\bf \Gamma})\I({\bf Y})=+1 \mbox{ and } \I({\bf X})\I({\bf M})=+1\\
1/2 & \mbox{if }\I({\bf \Gamma})\I({\bf Y})=-1 \mbox{ and } \I({\bf X})\I({\bf M})=-1
\end{cases}.
\label{eq:pyInversionEigenvalues}
\end{align}

\begin{figure}
\centering
\subfigure[]{
\includegraphics[width=0.35\columnwidth]{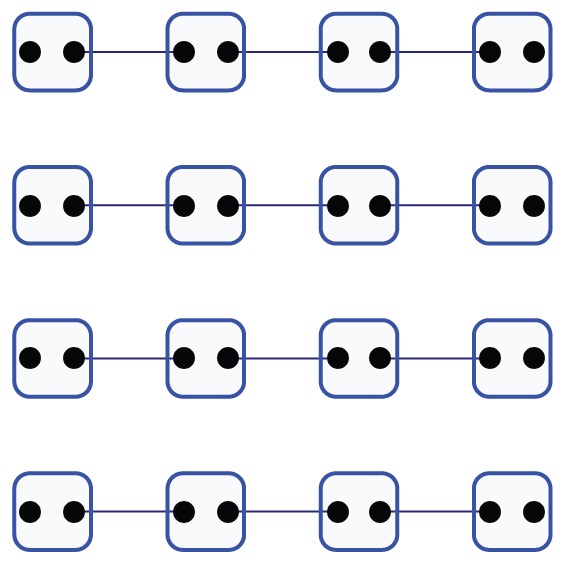}
}\;\;
\subfigure[]{
\begin{tikzpicture}[scale=1]
 
 		\fill [cyan!20!] (1,1)--(-1,1)--(-1,-1)--(1,-1)--(1,1);
 
 		\draw [->,black] (-1.5,0)--(1.5,0) node[below] {$k_x$};
	  	\draw [->,black] (0,-1.5)--(0,1.5) node[left] {$k_y$};		
		
		 \draw [black] (-.1,0)--(1,0);
	 	 \draw [black] (-1,1)--(1,1);
		 \draw [black] (-1,-1)--(1,-1);
		 \draw [black] (0,-1)--(0,1);
	 	 \draw [black] (-1,-1)--(-1,1);
	 	 \draw [black] (1,-1)--(1,1);
	  
		\fill [red] (0,0) circle (1.5pt) node[above right] {${\bf -}$};
		\fill [red] (1,0) circle (1.5pt) node[above right] {${\bf +}$};
		\fill [red] (0,1) circle (1.5pt) node[above right] {${\bf -}$};
		\fill [red] (1,1) circle (1.5pt) node[above right] {${\bf +}$};
		
\end{tikzpicture}
}
\caption{(Color online)  A weak topological insulator formed by stacking 1D insulators in the topological dipole phase. (a) Lattice. (b) Inversion eigenvalues at the high symmetry points $\bf \Gamma$, $\bf X$, $\bf Y$, and $\bf M$.}
\label{fig:2BandWeakTI}
\end{figure}
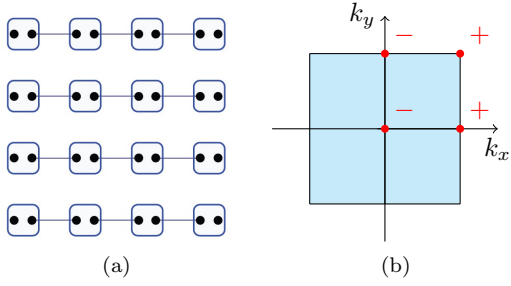

A simple realization of an insulator with polarization $(p_x,p_y)=(1/2,0)$ is shown in Fig.~\ref{fig:2BandWeakTI}a, which consists of a series of 1D SSH chains in the topological dipole phase oriented along $x$ and stacked along $y$. It has inversion eigenvalues as shown in Fig.~\ref{fig:2BandWeakTI}b. Such a stacked insulator is called a \emph{weak topological insulator} (weak TI)\cite{asahi2012,teo2013,benalcazar2014,hughes2014}, because, although it is a 2D system, its non-trivial topology is essentially one dimensional. Thus, they can be realized by stacking layers of 1D topological insulators. In this particular case the ground state of the system can be described by localized Wannier states in both the trivial and non-trivial phases, and hence we could again identify it with an obstructed atomic limit \cite{bradlyn2017}.  In general, the polarization of a weak TI is described by an index~\cite{asahi2012,teo2013,benalcazar2014}
\begin{align}
{\bf G} = p_x {\bf b_y} + p_y {\bf b_x},
\label{eq:WeakIndexVector}
\end{align}
where $p_x$ and $p_y$ are the polarizations \eqref{eq:Polarization_2D}, which can be determined by \eqref{eq:pxInversionEigenvalues} and \eqref{eq:pyInversionEigenvalues}, and $\bf b_x$, $\bf b_y$ are unit reciprocal lattice vectors of the crystal.

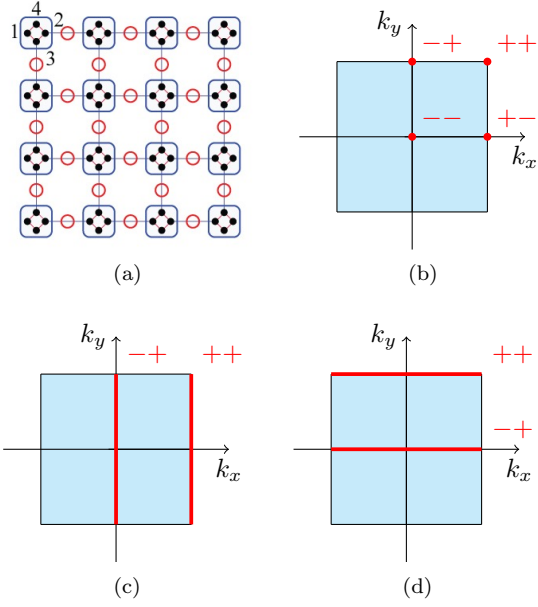
\begin{figure}
\centering
\subfigure[]{
\includegraphics[width=0.4\columnwidth]{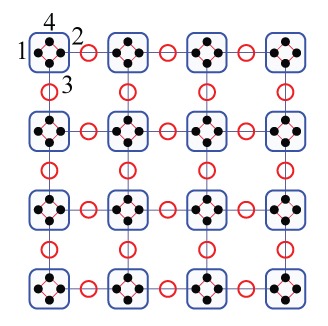}
}\;\;
\subfigure[]{
\begin{tikzpicture}[scale=1]
 
 		\fill [cyan!20!] (1,1)--(-1,1)--(-1,-1)--(1,-1)--(1,1);
 
 		\draw [->,black] (-1.5,0)--(1.5,0) node[below] {$k_x$};
	  	\draw [->,black] (0,-1.5)--(0,1.5) node[left] {$k_y$};		
		
		 \draw [black] (-.1,0)--(1,0);
	 	 \draw [black] (-1,1)--(1,1);
		 \draw [black] (-1,-1)--(1,-1);
		 \draw [black] (0,-1)--(0,1);
	 	 \draw [black] (-1,-1)--(-1,1);
	 	 \draw [black] (1,-1)--(1,1);
	  
		\fill [red] (0,0) circle (1.5pt) node[above right] {${\bf --}$};
		\fill [red] (1,0) circle (1.5pt) node[above right] {${\bf +-}$};
		\fill [red] (0,1) circle (1.5pt) node[above right] {${\bf -+}$};
		\fill [red] (1,1) circle (1.5pt) node[above right] {${\bf ++}$};
		
\end{tikzpicture}
}\\
\subfigure[]{
\begin{tikzpicture}[scale=1]
 
 		\fill [cyan!20!] (1,1)--(-1,1)--(-1,-1)--(1,-1)--(1,1);
 
 		\draw [->,black] (-1.5,0)--(1.5,0) node[below] {$k_x$};
	  	\draw [->,black] (0,-1.5)--(0,1.5) node[left] {$k_y$};		
		
		 \draw [black] (-.1,0)--(1,0);
	 	 \draw [black] (-1,1)--(1,1);
		 \draw [black] (-1,-1)--(1,-1);
		 \draw [black] (0,-1)--(0,1);
	 	 \draw [black] (-1,-1)--(-1,1);
	 	 \draw [black] (1,-1)--(1,1);
	  
		\draw [-,red,line width=.5mm] (0,-1)--(0,1) node[above right] {${\bf -+}$};
		\draw [-,red,line width=.5mm] (1,-1)--(1,1) node[above right] {${\bf ++}$};

\end{tikzpicture}
}\;\;
\subfigure[]{
\begin{tikzpicture}[scale=1]
 
 		\fill [cyan!20!] (1,1)--(-1,1)--(-1,-1)--(1,-1)--(1,1);
 
 		\draw [->,black] (-1.5,0)--(1.5,0) node[below] {$k_x$};
	  	\draw [->,black] (0,-1.5)--(0,1.5) node[left] {$k_y$};		
		
		 \draw [black] (-.1,0)--(1,0);
	 	 \draw [black] (-1,1)--(1,1);
		 \draw [black] (-1,-1)--(1,-1);
		 \draw [black] (0,-1)--(0,1);
	 	 \draw [black] (-1,-1)--(-1,1);
	 	 \draw [black] (1,-1)--(1,1);
	  
		\draw [-,red,line width=.5mm] (-1,0)--(1,0) node[above right] {${\bf -+}$};
		\draw [-,red,line width=.5mm] (-1,1)--(1,1) node[above right] {${\bf ++}$};

\end{tikzpicture}
}
\caption{(Color online)  Insulator with Bloch Hamiltonian $h^1({\bf k})$ as in the first Eq. of \eqref{eq:Hamiltonians12}. Periodic boundaries are imposed, so that top and bottom edges, as well as left and right edges, are identified. (a) Lattice. Hopping terms (black lines) have strength 1. Couplings within unit cells (red lines) have strength $\gamma \ll 1$. Red circles indicate 2D Wannier centers. (b) Inversion eigenvalues at the high symmetry points $\bf \Gamma$, $\bf X$, $\bf Y$, and $\bf M$. (c) $M_x$ eigenvalues along the $(0,k_y)$ and $(\pi,k_y)$ invariant lines. (d) $M_y$ eigenvalues along the $(k_x,0)$ and $(k_x,\pi)$ invariant lines.}
\label{fig:h1}
\end{figure}

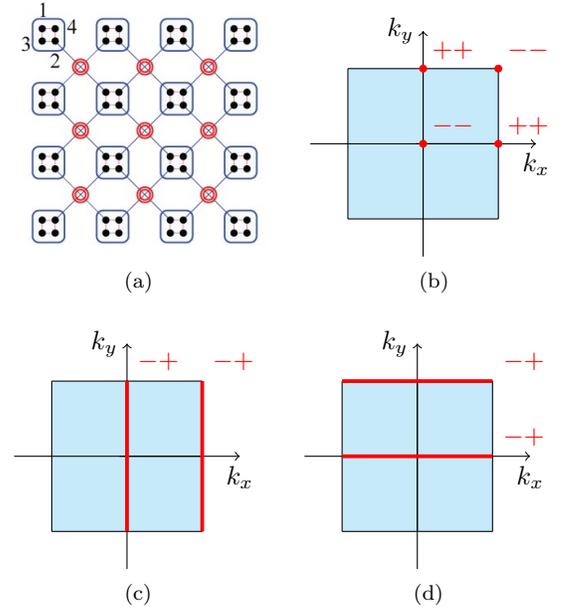
\begin{figure}
\centering
\subfigure[]{
\includegraphics[width=0.4\columnwidth]{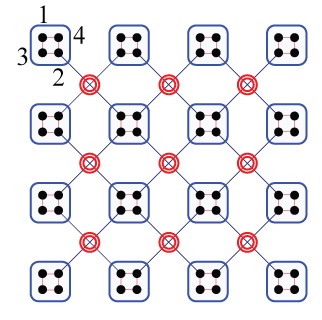}
}\;\;
\subfigure[]{
\begin{tikzpicture}[scale=1]
 
 		\fill [cyan!20!] (1,1)--(-1,1)--(-1,-1)--(1,-1)--(1,1);
 
 		\draw [->,black] (-1.5,0)--(1.5,0) node[below] {$k_x$};
	  	\draw [->,black] (0,-1.5)--(0,1.5) node[left] {$k_y$};		
		
		 \draw [black] (-.1,0)--(1,0);
	 	 \draw [black] (-1,1)--(1,1);
		 \draw [black] (-1,-1)--(1,-1);
		 \draw [black] (0,-1)--(0,1);
	 	 \draw [black] (-1,-1)--(-1,1);
	 	 \draw [black] (1,-1)--(1,1);
	  
		\fill [red] (0,0) circle (1.5pt) node[above right] {${\bf --}$};
		\fill [red] (1,0) circle (1.5pt) node[above right] {${\bf ++}$};
		\fill [red] (0,1) circle (1.5pt) node[above right] {${\bf ++}$};
		\fill [red] (1,1) circle (1.5pt) node[above right] {${\bf --}$};
		
\end{tikzpicture}
}\\
\subfigure[]{
\begin{tikzpicture}[scale=1]
 
 		\fill [cyan!20!] (1,1)--(-1,1)--(-1,-1)--(1,-1)--(1,1);
 
 		\draw [->,black] (-1.5,0)--(1.5,0) node[below] {$k_x$};
	  	\draw [->,black] (0,-1.5)--(0,1.5) node[left] {$k_y$};		
		
		 \draw [black] (-.1,0)--(1,0);
	 	 \draw [black] (-1,1)--(1,1);
		 \draw [black] (-1,-1)--(1,-1);
		 \draw [black] (0,-1)--(0,1);
	 	 \draw [black] (-1,-1)--(-1,1);
	 	 \draw [black] (1,-1)--(1,1);
	  
		\draw [-,red,line width=.5mm] (0,-1)--(0,1) node[above right] {${\bf -+}$};
		\draw [-,red,line width=.5mm] (1,-1)--(1,1) node[above right] {${\bf -+}$};

\end{tikzpicture}
}\;\;
\subfigure[]{
\begin{tikzpicture}[scale=1]
 
 		\fill [cyan!20!] (1,1)--(-1,1)--(-1,-1)--(1,-1)--(1,1);
 
 		\draw [->,black] (-1.5,0)--(1.5,0) node[below] {$k_x$};
	  	\draw [->,black] (0,-1.5)--(0,1.5) node[left] {$k_y$};		
		
		 \draw [black] (-.1,0)--(1,0);
	 	 \draw [black] (-1,1)--(1,1);
		 \draw [black] (-1,-1)--(1,-1);
		 \draw [black] (0,-1)--(0,1);
	 	 \draw [black] (-1,-1)--(-1,1);
	 	 \draw [black] (1,-1)--(1,1);
	  
		\draw [-,red,line width=.5mm] (-1,0)--(1,0) node[above right] {${\bf -+}$};
		\draw [-,red,line width=.5mm] (-1,1)--(1,1) node[above right] {${\bf -+}$};

\end{tikzpicture}
}
\caption{(Color online)  Insulator with Bloch Hamiltonian $h^2({\bf k})$ as in the second Eq. of \eqref{eq:Hamiltonians12}. Periodic boundaries are imposed, so that top and bottom edges, as well as left and right edges, are identified. (a) Lattice. Hopping terms (black lines) have strength 1. Couplings within unit cells (red lines) have strength $\gamma \ll 1$. Red circles indicate 2D Wannier centers. (b) Inversion eigenvalues at the high symmetry points $\bf \Gamma$, $\bf X$, $\bf Y$, and $\bf M$. (c) $M_x$ eigenvalues along $(0,k_y)$ and $(\pi,k_y)$. (d) $M_y$ eigenvalues along $(k_x,0)$ and $(k_x,\pi)$.}
\label{fig:h2}
\end{figure}

In the case of insulators with multiple occupied bands, the single inversion eigenvalue $\I({\bf k_{*}})$ in \eqref{eq:pxInversionEigenvalues} and \eqref{eq:pyInversionEigenvalues}, for ${\bf k_{*}}={\bf \Gamma}$, $\bf X$, $\bf Y$, and $\bf M$, is replaced by the multiplication of the inversion eigenvalues of all the occupied bands at ${\bf k}_*$. For example, consider the two insulators with Bloch Hamiltonians
\begin{align}
h^{1}({\bf k}) &= \left[\cos(k_x) \tau_x + \sin(k_x) \tau_y \right] \nonumber\\
&\oplus \left[\cos(k_y) \tau_x + \sin(k_y) \tau_y \right]\nonumber\\
&+\gamma \tau_x \otimes (\tau_0+\tau_x),\nonumber \\~\nonumber\\
h^{2}({\bf k}) &= \left[\cos(k_x-k_y) \tau_x + \sin(k_x-k_y) \tau_y \right] \nonumber\\
&\oplus \left[\cos(k_x+k_y) \tau_x + \sin(k_x+k_y) \tau_y \right]\nonumber\\
&+\gamma \tau_x \otimes (\tau_0+\tau_x), 
\label{eq:Hamiltonians12}
\end{align}
which have lattices, inversion and reflection eigenvalues as shown in Fig.~\ref{fig:h1} and Fig.~\ref{fig:h2}, respectively. Their weak indices are ${\bf G}^1=(1/2,1/2)$ and ${\bf G}^2=(0,0)$, respectively. In the case of $h^1({\bf k})$, the two Wannier centers of the two occupied bands are $(\nu_x,\nu_y)=(0,1/2)$ and $(\nu_x,\nu_y)=(1/2,0)$, as indicated by the red circles in Fig.~\ref{fig:h1}a. This leads to a non-trivial polarization along both directions. In the case of $h^2({\bf k})$, on the other hand, the two Wannier centers have the same value $(\nu_x,\nu_y)=(1/2,1/2)$, leading to trivial polarization when combined. Notice, from the inversion eigenvalues shown in Fig.~\ref{fig:h2}b, that although $h^2({\bf k})$ has trivial polarization, it is not a trivial atomic limit insulator (a trivial insulator has all inversion eigenvalues at all high symmetry points equal), it is an obstructed atomic limit where the Wannier centers are located away from the atom positions \cite{bradlyn2017}. We also point out that the inversion and reflection eigenvalues in Fig.~\ref{fig:h1} and \ref{fig:h2} are compatible with the relations shown in Table \ref{tab:map_I_W_2bands}.

A more comprehensive classification of topological crystalline insulators takes into account the full structure of the inversion eigenvalues of the occupied bands, or, more generally, the point group corepresentations on the subspace of occupied bands, to construct \emph{crystalline topological invariants}~\cite{teo2008,chiu2013,ryu2013,ueno2013,zhang2013,lau2016,fu2007,turner2010,hughes2011inversion,turner2012,fu2011,fang2012,fang2013,teo2013, benalcazar2014,liu2014, kobayashi2016, alexandradinata2016, mong2010, jadaun2013,slager2013,morimoto2013,shiozaki2014, dong2016, chiu2016, liu2016,haruki2017, bradlyn2017,shiozaki2017}. Such a classification, however, is outside of the scope of this paper. 

\subsection{Wilson loops and Wannier bands}
\label{sec:WannierBands_AllModels}
\begin{figure}[t]%
\centering
\includegraphics[width=\columnwidth]{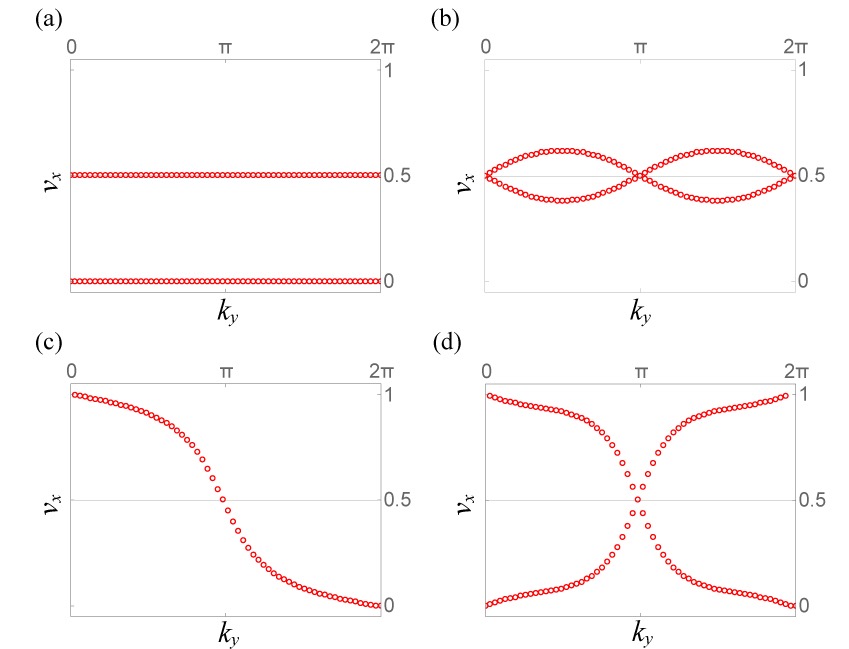}
\caption{(Color online)  Wannier bands in (a) weak topological insulator, (b) trivial insulator, (c) Chern insulator, and (d) QSH insulator. The Wannier band in (b) is twofold degenerate. (a,b) have Hamiltonians \eqref{eq:Hamiltonians12}, respectively. (c,d) have Hamiltonians \eqref{eq:Hamiltonians_AllModels} with $m=1$ and $m=3$, respectively.}
\label{fig:WannierBands_AllModels}
\end{figure}

We now introduce the concept of \emph{Wannier bands} as the set of Wannier centers along $x$ as a function of $k_y$, $\nu_x(k_y)$, or, vice versa, as the set of Wannier centers along $y$ as a function of $k_x$, $\nu_y(k_x)$. Unless otherwise specified, we will use the generic term \emph{Wannier bands} to refer to $\nu_x(k_y)$. The Wannier bands have associated \emph{hybrid Wannier functions} \eqref{eq:app_MLWF_2D} that are localized along $x$ but are Bloch-like along $y$. Although the term `hybrid Wannier function' is rather general, as more than one definition exist to refer to partially-localized states \cite{resta2001, vanderbilt2014, vanderbilt2015, troyer2016}, here we refer exclusively to the eigenstates of the projected position operator along one direction as a function of the perpendicular crystal momentum, as in Eq. \ref{eq:app_MLWF_2D}. They will be useful in the formulation of higher multipole moments in Section \ref{sec:Quadrupole}. Fig.~\ref{fig:WannierBands_AllModels} shows the Wannier bands for (a) the insulator $h^1(\bf k)$ and (b) the insulator $h^2(\bf k)$, as well as for (c) a Chern insulator, and (d) a Quantum Spin Hall (QSH) insulator. These last two insulators have corresponding Hamiltonians
\begin{align}
h^{Chern}({\bf k}) &= \sin(k_x) \tau_x + \sin(k_y) \tau_y \nonumber\\
&+ \left[m+\cos(k_x)+\cos(k_y)\right] \tau_z,\nonumber \\ \nonumber\\
h^{QSH}({\bf k}) &= \sin(k_x) (\Gamma_{zx}+\Gamma_{xx})\nonumber \\ 
&+ \sin(k_y) (\Gamma_{yx}+\Gamma_{0y}) \nonumber\\
&+ \left[2-m-\cos(k_x)-\cos(k_y)\right] \Gamma_{0z},
\label{eq:Hamiltonians_AllModels}
\end{align}
where $\Gamma_{ij} = \sigma_i \otimes \tau_j$, and $\sigma_i$ ($\tau_i$) are Pauli matrices corresponding to the spin (orbital) degrees of freedom. We consider these models at half filling. At this filling all of them are insulators. 

The models $h^1({\bf k})$ and $h^2({\bf k})$, described in Section \ref{sec:p2DInversion}, admit the construction of 2D Wannier centers, because their projected position operators $P^{occ} \hat{x} P^{occ}$ and $P^{occ} \hat{y} P^{occ}$ commute. Although this property is not true in general, even for some trivial insulators, these two models serve the illustrative purpose of mapping the electronic wave functions to classical point-charges in 2D \cite{vanderbilt1993}. Indeed, the electron Wannier centers in these two models can be essentially located by inspection. Specifically, with vanishing couplings within unit cells ($\gamma=0$), reflection symmetry then implies that, at half-filling (with 2 electrons per unit cell), the electron positions have to be as shown with red circles in Figs. \ref{fig:h1}a and \ref{fig:h2}a. Their Wannier bands are compatible with these electronic positions. Having two occupied bands, these insulators have two Wannier bands each. For $h^1(\bf k)$, the $\nu_x(k_y)$ bands are $\nu^1_x(k_y)=0$, $\nu^2_x(k_y)=1/2$. These values are fixed by reflection symmetry (Fig.~\ref{fig:h1}c), and are compatible with the electronic positions in Fig.~\ref{fig:h1}a. For $h^2(\bf k)$, we have $\nu^1_x(k_y)$ and $\nu^2_x(k_y)$ coming in opposite pairs. This is allowed by its reflection eigenvalues (Fig.~\ref{fig:h2}c). Notice, however, that the inversion eigenvalues in this model (Fig.~\ref{fig:h2}b) impose certain degeneracies in the Wannier values, $\nu^1_x(0)=\nu^2_x(0)=1/2$ and  $\nu^1_x(\pi)=\nu^2_x(\pi)=1/2$. Thus, the two electronic positions have to be degenerate at a value of $1/2$, as shown in the pictorial representation of Fig.~\ref{fig:h2}a. When Wannier bands $\nu_y(k_x)$ are calculated in these two models, we obtain the similar plots.

The case of the Chern insulator and the QSH insulator are not as straightforward to interpret because in these systems $P^{occ} \hat{x} P^{occ}$ and $P^{occ} \hat{y} P^{occ}$ do not commute. Thus, it is not possible to map the electronic wave functions to `point-like' charges as in the case of the insulators in \eqref{eq:Hamiltonians12}. In the case of the Chern insulator, the Wannier band $\nu_x(k_y)$ winds around 1 time across the 1D BZ $k_y \in [0,2\pi)$. In general, a Chern insulator will wind around $n$ times, where
\begin{align}
n = \frac{1}{2 \pi}\int_{BZ} d^2{\bf k} \tr[\F({\bf k})]
\label{eq:ChernNumber}
\end{align}
is the Chern number of the Chern Insulator. Here $\F({\bf k})=\partial_{k_x}\A_{y,\bf k}-\partial_{k_y}\A_{x, \bf k}+i [\A_{x,\bf k},\A_{y,\bf k}]$ is the Berry curvature. To see how the Chern number encodes this winding of the Wannier bands, let us take the simple case in which $[\A_{x,\bf k},\A_{y,\bf k}]=0$. Furthermore, let us make the gauge choice $\partial_{k_x}\A_{y,\bf k}=0$. Then we have
\begin{align}
n &= \frac{1}{2 \pi}\int_{BZ} d^2{\bf k} \tr[\F({\bf k})]\nonumber\\
& = \frac{1}{2 \pi}\int_{BZ} d^2{\bf k} \left(-\partial_{k_y}\tr[\A_{x, \bf k}]\right)\nonumber\\
& = \int_0^{2\pi} dk_y \partial_{k_y} \left(-\frac{1}{2 \pi}\int_0^{2\pi} dk_x \tr[\A_{x, \bf k}]\right)\nonumber\\
& = \int_0^{2\pi} dk_y \partial_{k_y} p_x(k_y).
\label{eq:ChernNumber_WindingPolarization}
\end{align}
Notice the resemblance of the Wannier bands in Fig.~\ref{fig:WannierBands_AllModels}c with those as a function of the adiabatic parameter $t$ in Fig.~\ref{fig:dipole_pumping_toroidal}b. Indeed, both insulators are systems with topology indexed by $n=1$; if in Eq. \ref{eq:dipole_pumping_toroidal} we make the change $t \rightarrow k_y$, the system becomes a Chern insulator. The reverse procedure, termed \emph{dimensional reduction}, is one method for a hierarchical classification of topological insulators \cite{qi2008}. The dimensional reduction `connects' the 2D Chern insulator with the adiabatic pumping of charge by means of a changing bulk dipole moment in 1D. 

In general, this type of dimensional hierarchy \emph{mathematically} connects topological insulators of different dimensions, having the dipole moment as its starting point in 1D. However, this connection does not provide a natural physical generalization of the 1D dipole moment to higher multipole moments. In Section \ref{sec:Quadrupole} we show that, in order to generate a classification that generalizes the 1D dipole moment to higher multipole moments in higher dimensions, the notion of Wannier bands is crucial. 


\subsubsection{Symmetry constraints on Wannier bands}
The Wannier bands, being related to the position of electrons in the lattice, are constrained in the presence of symmetries. In Appendix \ref{sec:WilsonLoopsSymmetry}, we show that the constraints due to time reversal (TR), chiral ($\Pi$), and charge conjugation ($CC$) symmetries are
\begin{align}
\left\{ \nu_x^i(k_y) \right\} &\stackrel{TR}{=} \left\{ \nu_x^i(-k_y) \right\}\nonumber\\
\left\{ \nu^i_x(k_y) \right\} &\stackrel{\Pi}{=} \left\{ \eta^i_x(k_y)\right\}\nonumber\\
\left\{ \nu^i_x(k_y) \right\} &\stackrel{CC}{=} \left\{ \eta^i_x(-k_y) \right\}
\label{eq:WannierBands_under_symm1}
\end{align}
mod 1. In the last two relations, the values $\{ \eta^i_x(k_y) \}$ are Wannier bands calculated over \emph{unoccupied} energy bands. The Chern insulator with Hamiltonian as in the first Eq. of \eqref{eq:Hamiltonians_AllModels} breaks TR symmetry, because its Wannier bands (Fig.~\ref{fig:WannierBands_AllModels}c) are not symmetric with respect to $k_y=0$, as required by the first Eq. in \eqref{eq:WannierBands_under_symm1}. In contrast, the QSH insulator with Hamiltonian as in the second Eq. of \eqref{eq:Hamiltonians_AllModels} shows Wannier bands compatible with TR symmetry (Fig.~\ref{fig:WannierBands_AllModels}d)\cite{yu2011,prodan2011,soluyanov2012}. Indeed, the QSH insulator has non-trivial topology protected by TR symmetry due to Kramers degeneracy. This protection is also manifest in the degeneracy of the Wannier values (see Appendix \ref{sec:WilsonLoopsSymmetry}).

Additionally, the constraints due to the reflection ($M_x, M_y$), inversion ($\I$) and $C_4$ symmetries are
\begin{align}
\left\{\nu_x^i(k_y) \right\} &\stackrel{M_x}{=} \left\{ -\nu_x^i(k_y) \right\}\nonumber\\
\left\{ \nu_x^i(k_y) \right\} &\stackrel{M_y}{=} \left\{ \nu_x^i(-k_y) \right\}\nonumber\\
\left\{\nu_x^i(k_y)\right\} &\stackrel{\I}{=}\left\{- \nu_x^i(-k_y)\right\}\nonumber\\
\left\{ \nu^i_x(k_y) \right\} &\stackrel{C_4}{=} \left\{ \nu^i_y(k_x=-k_y)\right\}\nonumber\\
\left\{ \nu^i_y(k_x) \right\} &\stackrel{C_4}{=} \left\{ -\nu^i_x(k_y=k_x)\right\}
\label{eq:WannierBands_under_symm2}
\end{align}
mod 1 (see Appendix \ref{sec:WilsonLoopsSymmetry}). Recall that in 2D inversion $\mathcal{I}$ and $C_2$ transform the coordinates the same way, hence the constraints on the Wannier bands due to $C_2$ are the same as those generated by $\mathcal{I}$ in 2D.  Now, notice in particular that in the presence of the reflection symmetry $M_x$, the first relation implies that the Wannier bands are either flat bands locked to $0$ or $1/2,$ or can disperse, but must occur in  pairs $\{-\nu_x(k_y),\nu_x(k_y)\}$. Since in gapped systems the values of $\nu_x^i(k_y)$ cannot change abruptly from $0$ to $1/2$ across different values of $k_y \in [-\pi,\pi)$, $M_x$ reflection implies that the polarization is either $p_x=0$ or $1/2$. This is the case in the insulators $h^1({\bf k})$ and $h^2({\bf k})$ with Hamiltonians \eqref{eq:Hamiltonians12}, having Wannier bands as in Fig.~\ref{fig:WannierBands_AllModels}a,b. Notice that these descriptions are compatible with the constraints on the polarization in Eq.~\ref{eq:pxkyUnderReflection} and Eq.~\ref{eq:pxUnderReflection}. Indeed, for spinless insulators in 2D, the constraints due to $M_x$ (which for spinless fermions has real eigenvalues $\pm 1$) on $\nu_x(k_y)$ at each $k_y$ are the same as the constraints due to $\I$ in 1D (see Eq~\ref{WannierConstraintInversion_1D}). Thus, Table~\ref{tab:map_I_W_2bands} in 1D is extended to Table~\ref{tab:EigenvaluesRelations} in 2D. These relations between inversion, reflection, and Wilson loop eigenvalues can be verified in the insulators $h^1({\bf k})$ and $h^2({\bf k})$, defined in \eqref{eq:Hamiltonians12}. Each of  these two insulators have both inversion and reflection symmetries with eigenvalues as shown in Fig.~\ref{fig:h1} and Fig.~\ref{fig:h2}, and with Wannier values as shown in Fig.~\ref{fig:WannierBands_AllModels}a,b.
\begin{table}
\begin{center}
\begin{tabular}{l l l}
$\hat{Q}$ eigenval.\;\;& $\hat{Q}$ eigenval.\;\;& Eigenval. of  \\
 at ${\bf k_*}$ \;\;& at ${\bf k^*+G/2}$\;\; & $\W_{{{\bf k}^*}+{\bf G}
\leftarrow \bf k^*}$\\
\hline\\
$(++)$ & $(++)$ & $(1,1)$\\
$(++)$ & $(+-)$ & $(1,-1)$\\
$(++)$ & $(--)$ & $(-1,-1)$\\
$(+-)$ & $(+-)$ & $(\lambda,\lambda^*)$
\end{tabular}
\end{center}
\caption{Relation between eigenvalues of $\hat{Q}= \hat{\I}$, $\hat{M}_x$ or $\hat{M}_y$ and Wilson loops. $\pm$ are the eigenvalues of reflection or inversion operators at high-symmetry momenta $\bf k^*$ and ${{\bf k}^*}+\bf G/2$ over the subspace of two occupied energy bands. The corresponding Wilson loop eigenvalues are for the Wilson loop in the direction of the reciprocal lattice vector $\bf G$\cite{alexandradinata2014}. The signs $\pm$ represent $\pm1$ if $\hat{Q}^2=+1$ or $\pm i$ if $\hat{Q}^2=-1$.}
\label{tab:EigenvaluesRelations}
\end{table}

\subsubsection{Wannier bands and the edge Hamiltonian}
\label{sec:WannierBandsEdgeHamiltonian}
Being unitary, we can express the Wilson loop as the exponential of a Hermitian matrix,
\begin{align}
\W_{\C,\bf k} \equiv e^{i H_{\W_\C}(\bf k)}.
\end{align}
We refer to $H_{\W_\C}(\bf k)$ as the \textit{Wannier Hamiltonian}. Notice that in the definition above, the argument $\bf k$ of the edge Hamiltonian is the base point of the Wilson loop. The eigenvalues of $H_{\W_\C}({\bf k})$ are precisely the Wannier bands, $\{2\pi \nu_x(k_y)\}$ or $\{2\pi \nu_y(k_x)\}$, which only depend on the coordinate of $\bf k$ normal to $\C$, e.g., in two-dimensions, the eigenvalues depend on $k_y$ for $\C$ along $k_x$ and vice versa. 

The Wannier Hamiltonian $H_{\W_\C}(\bf k)$ has been shown to be adiabatically connected with the Hamiltonian at the edge perpendicular to $\C$ \cite{klich2011}. We remark here that the map is not an exact identification, but rather, a map that preserves the topological properties of the Hamiltonian at the edge. The Wannier bands, being the spectrum of $H_{\W_\C}(\bf k)$, are adiabatically connected with the energy spectrum of the edge. Indeed, we see from Fig.~\ref{fig:WannierBands_AllModels} that this interpretation correctly describes the edge properties of the systems in Eq. \eqref{eq:Hamiltonians_AllModels}. For example, we recognize the standard edge state patterns for the Chern insulator and the QSH insulator, while the weak topological insulator has a flat band of edge states as expected for an ideal system with vanishing correlation length. 

Let us now mention some useful relations obeyed by the Wannier Hamiltonian. If we denote with  $-\C$ the contour $\C$ but in reverse order, it follows that
\begin{align}
\W_{-\C,\bf k} = \W^\dagger_{\C,\bf k} = e^{-i H_{\W_\C}(\bf k)},
\label{eq:WannierHamiltonian}
\end{align}
thus, we make the identification
\begin{align}
H_{\W_{-\C}}({\bf k}) = -H_{\W_{\C}}({\bf k}).
\end{align}
The transformations of Wilson loops under the symmetries studied here are derived in detail in Appendix \ref{sec:WilsonLoopsSymmetry}. 
Insulators with a lattice symmetry obey
\begin{align}
g_{\bf k} h_{\bf k} g_{\bf k}^\dagger = h_{D_g{\bf k}},
\label{eq:app_Hamiltonian_under_symmetry}
\end{align}
where $g_{\bf k}$ is the unitary operator
\begin{align}
g_{\bf k} = e^{-i(D_g {\bf k})\cdot {\bf \delta}}U_g.
\end{align}
$U_g$ is an $N_{orb} \times N_{orb}$ matrix that acts on the internal degrees of freedom of the unit cell, and $D_g$ is an operator in momentum space sending ${\bf k} \rightarrow D_g {\bf k}$. In real space, on the other hand, we have ${\bf r} \rightarrow D_g{\bf r} + {\bf \delta}$, where $\bf \delta=0$ in the case of symmorphic symmetries, or takes a fractional value (in unit-cell units) in the case of non-symmorphic symmetries.

Using the definition of the Wannier Hamiltonian \eqref{eq:WannierHamiltonian}, we can rewrite the expression for the transformation of Wilson loops in Appendix \ref{sec:WilsonLoopsSymmetry} into the form
\begin{align}
B_{g,\bf k} H_{\W_\C}({\bf k}) B^\dagger_{g,\bf k} = H_{\W_{D_g \C}}(D_g \bf k),
\end{align}
where
\begin{align}
B^{mn}_{g,\bf k}=\matrixel{u^m_{D_g{\bf k}}}{g_{\bf k}}{u^n_{\bf k}}
\end{align}
is the unitary sewing matrix that connects states at ${\bf k}$ with those at $D_g{\bf k}$ which have the same energy. 

Hence, we can interpret the usual sewing matrix $B_{g,\bf k}$ for the bulk Hamiltonian as a symmetry operator of the edge/Wannier Hamiltonian. In particular, we have
\begin{align}
B_{M_x,\bf k} H_{\W_x}({\bf k}) B^\dagger_{M_x,\bf k} &= -H_{\W_x}(M_x \bf k)\nonumber \\
B_{M_y,\bf k} H_{\W_x}({\bf k}) B^\dagger_{M_y,\bf k} &= H_{\W_x}(M_y \bf k)\nonumber \\
B_{\I,\bf k} H_{\W_x}({\bf k}) B^\dagger_{\I,\bf k} &= -H_{\W_x}(-\bf k).
\label{eq:app_Wannier_Hamiltonian_symmetries}
\end{align}

\section{Edge dipole moments in 2D crystals}
\label{sec:EdgePolarization}
\begin{figure}[t]%
\centering
\includegraphics[width=\columnwidth]{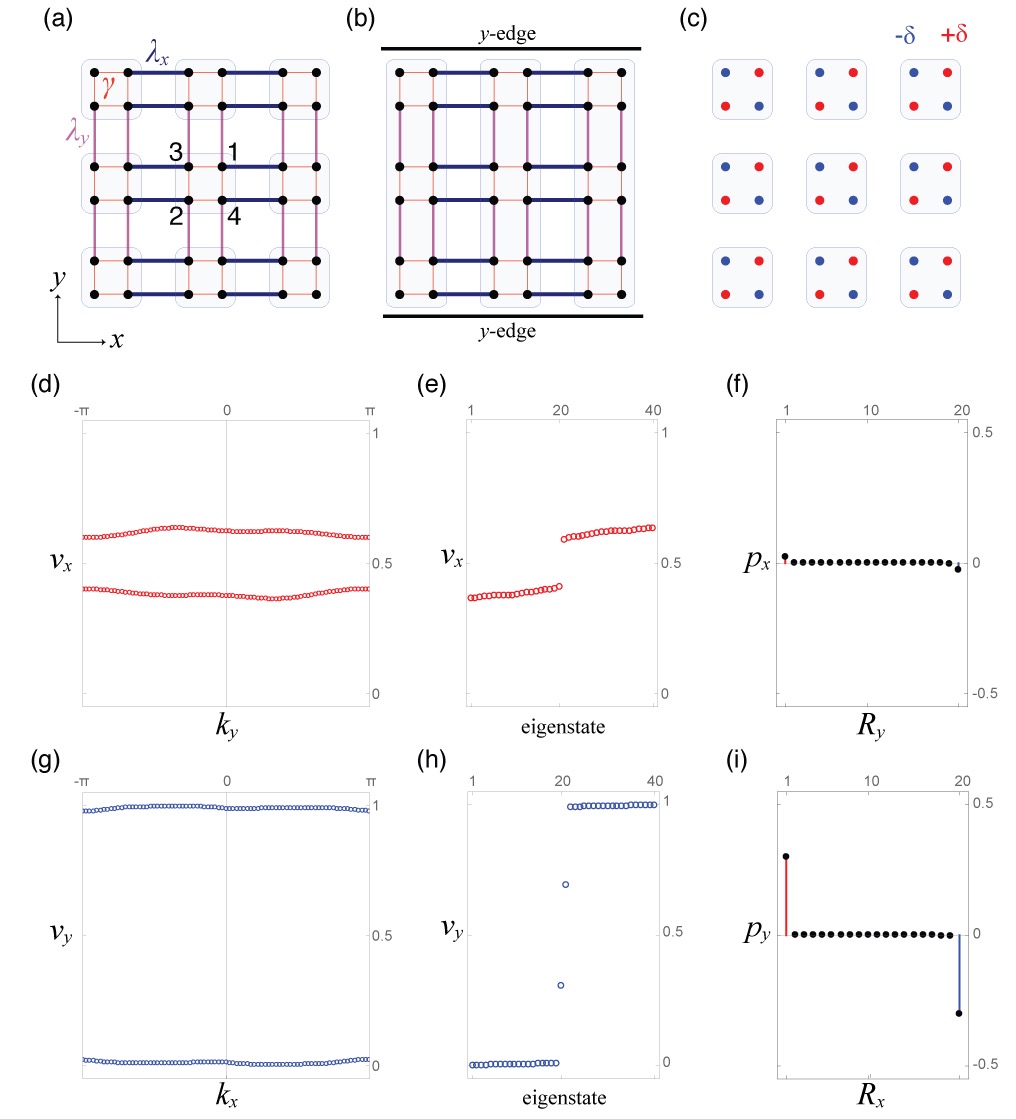}
\caption{(Color online)  Edge polarization in insulator with Hamiltonian \eqref{eq:hEdgePolarizationHamiltonian} (inversion-symmetric). (a) Lattice with periodic boundaries. (b) Lattice with periodic boundaries along $x$ and open along $y$. Long vertical rectangles are redefined, effective unit cells in the pseudo-1D lattice construction. (c) On-site potential on the lattice. Red (blue) sites represent the on-site energies of $+\delta$ ($-\delta$) that break $M_x$ and $M_y$ symmetry but preserve inversion. (d,g) Wannier bands $\nu_x(k_y)$ and $\nu_y(k_x)$ for the configuration in (a). (e) Wannier values $\nu_x$ and (f) polarization $p_x(R_y)$ for the configuration in (b) which has an open boundary. (h) Wannier values $\nu_y$ and (i) polarization $p_y(R_x)$ for a configuration as in (b) but with boundaries open along $x$ and closed along $y$. In all plots we set $\lambda_x=1$, $\lambda_y=0.5$, $\gamma=0.1$, $\delta=0.2$, and the strength of the small perturbation to break chiral and time-reversal symmetries (see text) to 0.15. We verify that this perturbation does not close the energy gap.}
\label{fig:h2EdgePolarization}
\end{figure}

Before discussing the bulk quadrupole moment in 2D insulators, we take the intermediate step of studying 
2D crystalline insulators which 
may give rise to \emph{edge-localized} polarizations \cite{vanderbilt2015}. In particular, we describe the procedure to calculate the position-dependent polarization in an insulator, and then we show in an example how the edge polarization arises.
We start by considering a 2D crystal with $N_x \times N_y$ sites. For calculating the polarization along $x$ as a function of position along $y$, we choose  the insulator to have periodic boundary conditions along $x$ and open boundary conditions along $y$. In this configuration there is no crystal momenta $k_y$, and we can treat this crystal as a wide, pseudo-1D lattice by absorbing the labels $R_y \in 1 \ldots N_y$ into the internal degrees of freedom. We are essentially forming a redefined unit cell that extends along the entire length of the crystal in the $y$-direction. This is shown schematically in Fig.~\ref{fig:h2EdgePolarization}b. Thus, the formulation in Section \ref{sec:PxP_1D} follows through in this case, with the redefinition:
\begin{align}
c_{k,\alpha} &\rightarrow c_{k_x,R_y,\alpha}
\end{align}
which allows us to write the second-quantized Hamiltonian as
\begin{align}
H = \sum_{k_x} c^\dagger_{k_x,R_y,\alpha} [h_{k_x}]^{R_y,\alpha,R'_y,\beta} c_{k_x,R'_y,\beta},
\label{eq:Pseudo1DHamiltonian}
\end{align}
for $\alpha, \beta \in 1\ldots N_{orb},$ and $R_y,R'_y \in 1 \ldots N_y$. In the above redefinitions, notice that, since the boundaries remain closed along $x$,  $k_x$ is still a good quantum number. We diagonalize this Bloch Hamiltonian as
\begin{align}
[h_{k_x}]^{R_y,\alpha,R'_y,\beta} = \sum_n [u^n_{k_x}]^{R_y,\alpha} \epsilon_{n,k_x} [u^{*n}_{k_x}]^{R'_y\beta},
\label{eq:Pseudo1DBlochHamiltonian}
\end{align}
where $n \in 1 \ldots N_{orb} \times N_y$.
So, if the 2D Bloch Hamiltonian with periodic boundary conditions along $x$ and $y$, $h_{(k_x,k_y)}$, has $N_{occ}$ occupied bands, its associated pseudo-1D Bloch Hamiltonian $h_{k_x}$ in \eqref{eq:Pseudo1DBlochHamiltonian} has $N_{occ} \times R_y$ occupied bands. We can diagonalize the Hamiltonian \eqref{eq:Pseudo1DHamiltonian} as
\begin{align}
H = \sum_{n,{k_x}} \gamma^\dagger_{n,{k_x}} \epsilon_{n,k_x} \gamma_{n,{k_x}},
\end{align}
where
\begin{align}
\gamma_{n,k_x} = \sum_{R_y,\alpha} [u^{*n}_{k_x}]^{R_y,\alpha} c_{k_x,R_y,\alpha}.
\label{eq:pseudo1Danihilation_operator}
\end{align}
Following Section \ref{sec:PxP_1D}, the matrices
\begin{align}
[G_{k_x}]^{mn} \equiv \sum_{R_y,\alpha}[u^{*m}_{k_x+\Delta k_x}]^{R_y,\alpha} [u^n_{k_x}]^{R_y,\alpha},
\end{align}
are used in the construction of the Wilson line elements $[F_{k_x}]^{mn}$ and subsequently the Wilson loops $[\W_{k_x+2\pi \leftarrow k_x}]^{mn}$, where $m,n \in 1 \dots N_{occ} \times N_y$. Notice that the size of these Wilson-loop matrices is $N_y$-times larger than the size of Wilson-loop matrices when both boundaries are closed in the crystal.

The hybrid Wannier functions have the same form as in \eqref{eq:WannierFunctions_1D}:
\begin{align}
\ket{\Psi^j_{R_x}} = \frac{1}{\sqrt{N_x}} \sum_{n=1}^{N_{occ} \times N_y}\sum_{k_x} \left[ \nu^j_{k_x} \right]^n e^{-i k_x R_x} \gamma^\dagger_{n,k_x}\ket{0},
\label{eq:WannierFunctions_pseudo1D}
\end{align}
for $j \in 1 \ldots N_{occ} \times N_y$, $R_x \in 1\ldots N_x$, and where $\left[ \nu^j_{k_x} \right]^n$ is the $n^{th}$ component of the $j^{th}$ Wilson-loop eigenstate $\ket{\nu^j_{k_x}}$, and $\gamma^\dagger_{n,k_x}$ is given in \eqref{eq:pseudo1Danihilation_operator}.
In order to spatially resolve the $x$-component of the polarization along the $y$ direction, we calculate the probability density of the hybrid Wannier functions \eqref{eq:WannierFunctions_pseudo1D},
\begin{align}
\rho^{j,R_x}(R_y) &= \sum_{R'_x,\alpha} \braket{\Psi^j_{R_x}}{\phi^{R_y, \alpha}_{R'_x}}\braket{\phi^{R_y, \alpha}_{R'_x}}{\Psi^j_{R_x}}\nonumber\\
&=\frac{1}{N_x} \sum_{k_x, \alpha} \left| [u^n_{k_x}]^{R_y, \alpha}[\nu^j_{k_x}]^n\right|^2
\end{align}
(in the first equation above no sums are implied over repeated indices). Notice that there is no dependence on the unit cell $R_x$ --as expected since the density is translationally invariant in the $x$ direction. Thus, we can write $\rho^{j,R_x}$ simply as $\rho^j$. This probability density then allows us to resolve the hybrid Wannier functions \eqref{eq:WannierFunctions_pseudo1D} along the $y$-direction. In particular, it will let us determine whether any of these functions are localized at the (open) edges at $R_y=0,N_y$. This probability density also allows us to calculate the $x$-component of the polarization via
\begin{align}
p_x(R_y) = \sum_j \rho^j(R_y) \nu_x^j 
\label{eq:y_resolved_x_polarization}
\end{align}
which is resolved at each site $R_y$. 

\begin{figure}[t]%
\centering
\includegraphics[width=\columnwidth]{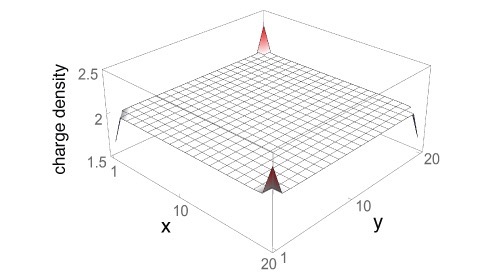}
\caption{(Color online)  Electronic charge density in an insulator with Hamiltonian \eqref{eq:hEdgePolarizationHamiltonian} with full open boundaries. There are boundary charges at  the four corners.}
\label{fig:h2CornerCharge}
\end{figure}

We now illustrate the existence of edge polarization with an example. Consider the insulator with Bloch Hamiltonian
\begin{align}
h(\bf k) &= \left( \begin{matrix}
\delta \tau_0 & q(\bf k)\nonumber\\
q^\dagger(\bf k) & -\delta \tau_0
\end{matrix} \right), \\ q(\bf k)&=\left(\begin{matrix}
\gamma + \lambda_x e^{i k_x} & \gamma + \lambda_y e^{i k_y}\\
\gamma + \lambda_y e^{-i k_y} & \gamma + \lambda_x e^{-i k_x}
 \end{matrix}\right),
\label{eq:hEdgePolarizationHamiltonian}
\end{align}\noindent where $\tau_0$ is the $2\times 2$ identity matrix, and $\tau_{a},$ $a=1 , 2, 3$ are Pauli matrices.
A tight-binding representation of this model is shown in Fig.~\ref{fig:h2EdgePolarization}a. $\gamma$ is the strength of the coupling within unit cells, represented by red lines in Fig.~\ref{fig:h2EdgePolarization}a, and $\lambda_{x,y}$ are the strengths of horizontal and vertical hoppings between nearest neighbor cells. 
$\delta$ is the amplitude of an on-site potential (Fig.~\ref{fig:h2EdgePolarization}c) that breaks reflection symmetry along $x$ and $y,$ but maintains inversion symmetry.  When $\delta=0$, this model has reflection and inversion symmetries, with operators $\hat{M}_x=\tau_x \otimes \tau_0$, $\hat{M}_y=\tau_x \otimes \tau_x$, and $\mathcal{I}=\tau_0 \otimes \tau_x$. 

This insulator also has fine-tuned chiral and time-reversal symmetries. However, since we are only interested in protection due to spatial symmetries, we add a small perturbation to \eqref{eq:hEdgePolarizationHamiltonian} in our numerics of the form:
\begin{align}
h_{pert}({\bf k}) &= EE \cos(k_x) + OE \sin(k_x)\nonumber\\
&+EE \cos(k_y) + EO \sin(k_y),\nonumber
\end{align} 
where $EE$, $OE$, and $EO$ are $4\times 4$ random matrices that obey
\begin{align}
[EE,\hat{M}_x]&=0,\;\; [EE,\hat{M}_y]=0,\nonumber\\
\{OE,\hat{M}_x\}&=0,\;\; [OE,\hat{M}_y]=0,\nonumber\\
[EO,\hat{M}_x]&=0,\;\; \{EO,\hat{M}_y\}=0,
\end{align}
and with entries in the range $[0,1]$. These nearest-neighbor perturbations break the chiral and time-reversal symmetries, while preserving the reflection symmetries along both $x$ and $y$, as well as inversion symmetry. These perturbations are added to ensure that the interesting features do not rely on these fine-tuned symmetries.

\begin{figure}[t]%
\centering
\subfigure[]{
\begin{tikzpicture}[scale=1]
 
 		\fill [cyan!20!] (1,1)--(-1,1)--(-1,-1)--(1,-1)--(1,1);
 
 		\draw [->,black] (-1.5,0)--(1.5,0) node[below] {$k_x$};
	  	\draw [->,black] (0,-1.5)--(0,1.5) node[left] {$k_y$};		
		
		 \draw [black] (-.1,0)--(1,0);
	 	 \draw [black] (-1,1)--(1,1);
		 \draw [black] (-1,-1)--(1,-1);
		 \draw [black] (0,-1)--(0,1);
	 	 \draw [black] (-1,-1)--(-1,1);
	 	 \draw [black] (1,-1)--(1,1);
	  
		\draw [-,red,line width=.5mm] (0,-1)--(0,1) node[above right] {${\bf --}$};
		\draw [-,red,line width=.5mm] (1,-1)--(1,1) node[above right] {${\bf ++}$};

\end{tikzpicture}
}\;\;
\subfigure[]{
\begin{tikzpicture}[scale=1]
 
 		\fill [cyan!20!] (1,1)--(-1,1)--(-1,-1)--(1,-1)--(1,1);
 
 		\draw [->,black] (-1.5,0)--(1.5,0) node[below] {$k_x$};
	  	\draw [->,black] (0,-1.5)--(0,1.5) node[left] {$k_y$};		
		
		 \draw [black] (-.1,0)--(1,0);
	 	 \draw [black] (-1,1)--(1,1);
		 \draw [black] (-1,-1)--(1,-1);
		 \draw [black] (0,-1)--(0,1);
	 	 \draw [black] (-1,-1)--(-1,1);
	 	 \draw [black] (1,-1)--(1,1);
	  
		\draw [-,red,line width=.5mm] (-1,0)--(1,0) node[above right] {${\bf -+}$};
		\draw [-,red,line width=.5mm] (-1,1)--(1,1) node[above right] {${\bf -+}$};

\end{tikzpicture}
}\\
\subfigure[]{
\includegraphics[width=\columnwidth]{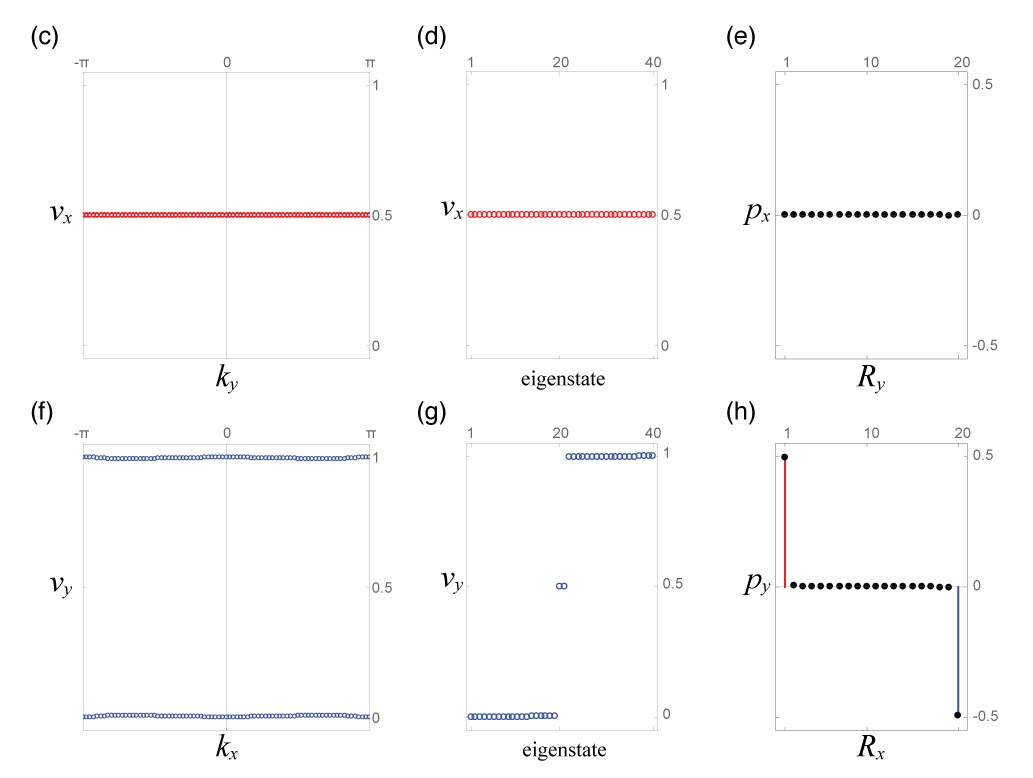}
}
\caption{(Color online) Insulator with Bloch Hamiltonian \eqref{eq:hEdgePolarizationHamiltonian} and $\delta=0$. Here, we set $\lambda_x>\lambda_y$. (a) $M_x$ eigenvalues along the $(0,k_y)$ and $(\pi,k_y)$. (b) $M_y$ eigenvalues along the $(k_x,0)$ and $(k_x,\pi)$. (c) Wannier bands $\nu_x(k_y)$. (d) Wannier values $\nu_x$ when boundaries are open along $y$. (e) $p_x(R_y)$ for configuration as in (d). (f) Wannier bands $\nu_y(k_x)$. (g) Wannier values $\nu_y$ when boundaries are open along $x$. Values at $\nu_x=0.5$ have edge localized eigenstates. (h) $p_y(R_x)$ for configuration as in (g). The parameters used here are as in Fig.~\ref{fig:h2EdgePolarization} but with $\delta=0$, except for (e) and (h), for which $\delta=10^{-4}$.}
\label{fig:h2EdgePolarizationQuantized}
\end{figure}
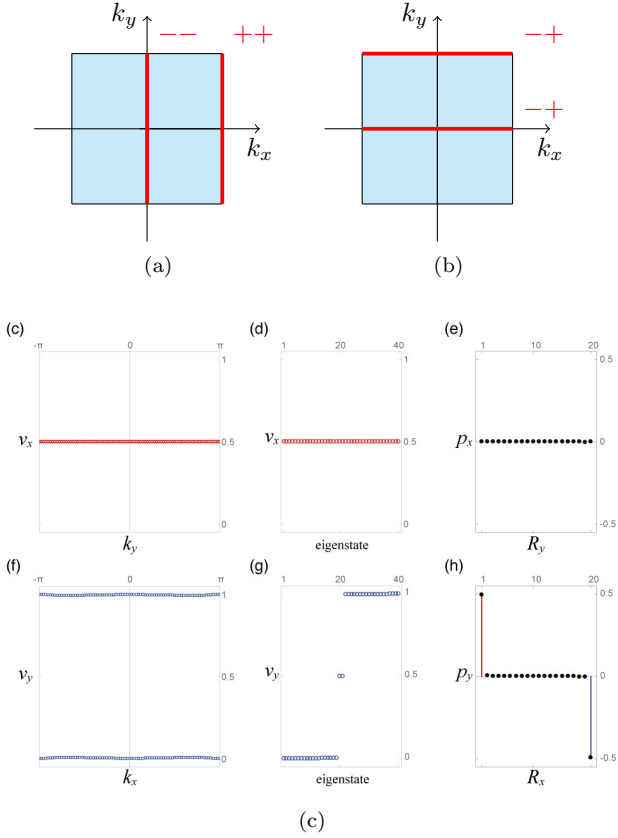

We first consider the general case of generating non-quantized edge polarizations by breaking reflection symmetries (Fig.~\ref{fig:h2EdgePolarization} and Fig.~\ref{fig:h2CornerCharge}), and later on discuss the special case in which these edge polarizations are quantized by restoring reflection symmetries (Fig. \ref{fig:h2EdgePolarizationQuantized}). In both cases, however, preserving inversion symmetry is necessary in order to have an overall vanishing bulk polarization. In particular, in order to have well defined edge polarizations, we require that the edges do not accumulate charge and are neutral, hence, the bulk of the insulator should not be polarized.

For the general case of non-quantized edge polarizations, we consider $\lambda_x > \lambda_y$ and $\gamma \ll |\lambda_x-\lambda_y|$. Under these conditions, the crystal is an insulator at half filling. The  inversion eigenvalues of the occupied bands come in $\pm1$ pairs at all symmetry points. Therefore, there is no protection of degeneracies in the Wannier bands, as we can see from the plots of $\nu_x(k_y)$ and $\nu_y(k_x)$ shown in Fig.~\ref{fig:h2EdgePolarization}d,g. If all the boundaries of the system are closed, the crystal has uniform, vanishing bulk polarization protected by inversion symmetry. If instead we open the boundaries along $y$, as in Fig.~\ref{fig:h2EdgePolarization}b, we can use the formulation from earlier in this section to treat this crystal as a pseudo-1D insulator, with redefined unit cells as shown by the long vertical rectangles in Fig.~\ref{fig:h2EdgePolarization}b. The Wannier values $\nu^j_x$, for $j \in 1 \ldots 40$ ($N_{occ}=2$, $N_y=20$), obtained from this calculation are shown in Fig.~\ref{fig:h2EdgePolarization}e. Using these values and their associated hybrid-Wannier functions, we calculate the polarization $p_x(R_y)$ using \eqref{eq:y_resolved_x_polarization}. This is shown in Fig.~\ref{fig:h2EdgePolarization}f. Remarkably, although the polarization vanishes in the bulk, there is \emph{edge-localized polarization} parallel to the edge. 
If, instead of opening the boundaries along $y$, we choose to open them in $x$, we observe the Wannier values $\nu^j_y$, for $j \in 1 \ldots 40$, and polarization $p_y(R_x)$ shown in Fig.~\ref{fig:h2EdgePolarization}h,i, respectively. We find a vanishing polarization in the bulk,  but non-zero edge-localized polarizations tangent to the edge. For our choice of $\lambda_x> \lambda_y$  the values of $|p_y|$ localized at the $R_x=0,N_x$ edges are larger than the values $|p_x|$ localized at $R_y=0,N_y$.

To complete the picture, we ask what happens if the edge polarization is terminated at a corner. Fig.~\ref{fig:h2CornerCharge} shows the charge density in this insulator \eqref{eq:hEdgePolarizationHamiltonian} with both boundaries open. We see that, relative to the background charge density of $2e$ per unit cell, there are corner-localized charges $Q^{corner}$. These charges and the edge polarizations obey 
$Q^{corner}= p^{edge\;x} + p^{edge\;y}$, as expected for insulators with vanishing bulk dipole and quadrupole moments (Section \ref{sec:BulkVSBoundaryMoments}). As such, this polarization is purely a surface effect and not generated by a bulk quadrupole moment.

Now let us consider the case with reflection symmetry. As is typical for these types of calculations we still must break the reflection symmetries infinitesimally by setting $0<\delta \ll \gamma, \lambda_{x,y}$. This infinitesimal, non-zero $\delta$ perturbation breaks reflection symmetries softly and is necessary to choose an unambiguous sign for the quantized edge polarizations. In a lattice with full open boundary conditions, this perturbation serves to  split the degeneracy of the four corner-localized mid-gap modes to determine how they are filled at half-filling. This then chooses the signs of the corner charges in a way consistent with the choice of edge polarizations.

We find that for  $\lambda_x>\lambda_y$ we  have $Q^{corner}=p^{edge\;y}=1/2$ and $p^{edge\;x}=0$, while for  $\lambda_y>\lambda_x$ we find $Q^{corner}=p^{edge\;x}=1/2$ and $p^{edge\;y}=0.$  Let us focus on the case $\lambda_x>\lambda_y$. By setting $\delta=0$, the reflection eigenvalues for this insulator are indicated in Fig.~\ref{fig:h2EdgePolarizationQuantized}a,b. Based on the analysis of reflection eigenvalues summarized in Table~\ref{tab:EigenvaluesRelations}, we conclude that $M_x$ fixes the Wannier bands to $\nu^{1,2}_x(k_y)=1/2$, as in Fig~\ref{fig:h2EdgePolarizationQuantized}c, while $M_y$ does not restrict the values of $\nu^{1,2}_y(k_x)$ to either 0 or $1/2.$ Instead, they only have to obey $\nu^{1}_y(k_x)=-\nu^{2}_y(k_x)$, as in Fig.~\ref{fig:h2EdgePolarizationQuantized}f. If we now open the boundaries along $y$ and calculate $\nu^j_x$, we obtain degenerate values $\nu^j_x=1/2$ (Fig.~\ref{fig:h2EdgePolarizationQuantized}d), which result in $p_x(R_y)=0$ (Fig.~\ref{fig:h2EdgePolarizationQuantized}e). If we instead open the boundaries along $x$ and calculate $\nu^j_y$ we obtain the gapped bands which have corresponding Wannier eigenstates that have weight primarily in the bulk of the sample. Interestingly, in addition to the gapped bulk Wanner states, we find a pair of Wannier values pinned at $1/2$ that have Wannier eigenstates localized at the edges $R_x=0$ and $R_x=N_x$ (Fig.~\ref{fig:h2EdgePolarizationQuantized}g). It is this pair of states that results in the edge polarization of $\pm 1/2$, as shown in Fig.~\ref{fig:h2EdgePolarizationQuantized}h (a small value of $\delta=10^{-4}$ was chosen for plots e and h to break Wannier degeneracies).

In contrast to this phenomenology, we will see in Section \ref{sec:Quadrupole} that insulators with quadrupole moments also have edge-localized polarizations and corner-localized charges, but, unlike in the present case, these boundary properties obey $|Q^{corner}|= |p^{edge\;x}| = |p^{edge\;y}|$, as required for a quadrupole (see Section \ref{sec:BulkVSBoundaryMoments}).

\section{Bulk quadrupole moment in 2D crystals}
\label{sec:Quadrupole}
Any quadrupole insulator should have three basic properties: (i) its bulk dipole moment must vanish, otherwise the quadrupole moment is ill defined (see Section \ref{sec:multipole_translation_invariance}); (ii) the insulator must have at least two occupied bands, since a crystal with one occupied band can only generate dipole moments (a quadrupole is made from two separated dipoles); and (iii) it should have edges that are insulators themselves, as only insulating edges can host edge-localized polarization (hence, edges should not host gapless states and thus the bulk must have Chern number $n=0$). 

From the classical analysis of Section \ref{sec:ClassicQuadrupole} we concluded that the boundary signatures of an ideal 3D insulator with only bulk quadrupole moment density are the existence of charge density per unit length at hinges $\lambda^{hinge\;a,b}=\frac{1}{2}n^{(a)}_i n^{(b)}_j q_{ij}$ and polarization density per unit area at faces $p^{face\;a}=n^{(a)}_i q_{ij}$, where in these two expressions summation is implied over repeated indices. In 2D, these expressions reduce to corner charges and edge polarization density per unit length, respectively:
\begin{align}
Q^{corner \; a, b}&=\frac{1}{2}n^{(a)}_{i}n^{(b)}_{j}q_{ij}\nonumber\\ 
p_{j}^{edge\; a}&=n^{(a)}_{i}q_{ij}.
\label{eq:QuadSignatures}
\end{align} 
In the expressions above, $q_{ij}$ is the quadrupole moment per unit area, with $q_{xy} = q_{yx} \neq 0$, $q_{xx} = q_{yy}=0$.

The insulator with Hamiltonian \eqref{eq:hEdgePolarizationHamiltonian} meets all the three basic requirements: it is an insulator with two electrons per unit cell, zero bulk polarization and no Chern number. Furthermore, it does have corner charges when both boundaries are open. However, it fails to meet the relations \eqref{eq:QuadSignatures}; its edge polarizations are not of the same magnitude as its corner charge. Instead, they obey $Q^{corner}=p^{edge\;x}+p^{edge\;y}$ (see Section \ref{sec:EdgePolarization}) and can be accounted for by the theory of polarization up to dipole moments (see Section \ref{sec:BulkVSBoundaryMoments}).

In this section, we describe a model realization of a symmetry-protected quadrupole insulator-- an insulator with vanishing dipole moment and fractional, quantized quadrupole moment-- that manifests through the predicted the boundary signatures of Eq.~\ref{eq:QuadSignatures}. 
This model has two occupied bands, and a vanishing Chern number. Crucially, \emph{its pair of Wannier bands are gapped, and each Wannier band can have an associated Berry phase}. Physically, this corresponds to a bulk configuration in which two parallel dipoles aligned along one direction are separated along its perpendicular direction. We first describe the formalism of Wannier band topology for 2D insulators, and then describe the observables of a quadrupole insulator. We will then explore the quantization of dipole pumping resulting from non-trivial adiabatic cycles that connect the quadrupole and trivial phases, and end with a description of an insulator with hinge localized chiral modes in 3D which exhibits the same topology as the dipole pumping process.

\subsection{Wannier-sector polarization}
\label{sec:TopologyWannierBands}
\begin{figure}[t]%
\centering
\includegraphics[width=\columnwidth]{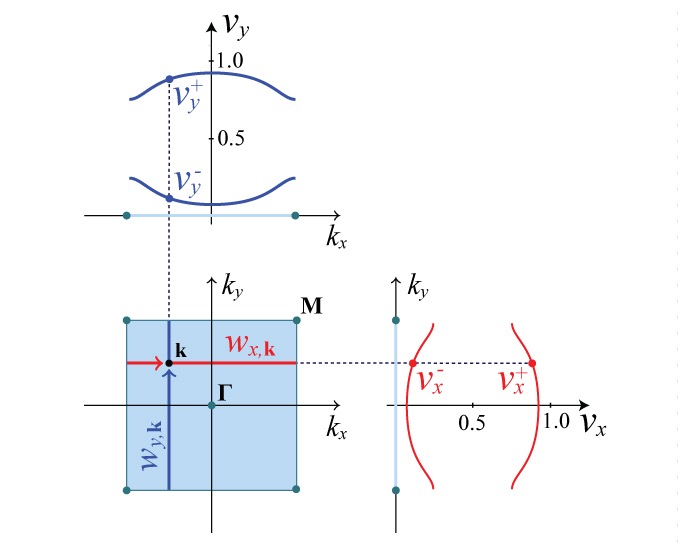}%
\caption{(Color online)  Gapped Wannier bands $\nu^\pm_x(k_y)$ (red lines to the right of the BZ) and $\nu^\pm_y(k_x)$ (blue lines above of the BZ) of the quadrupole insulator with Hamiltonian \eqref{eq:QuadHamiltonian}.}
\label{fig:WannierBands}
\end{figure}

In this section we study the topology of Wannier bands $\nu^j_x(k_y)$  that are gapped across the entire 1D BZ $k_y \in (-\pi,\pi]$, a minimal example of which is shown in Fig.~\ref{fig:WannierBands}. Due to the gap in the Wannier spectrum around $\nu_x=1/2$ we can broadly define two Wannier sectors
\begin{align}
\nu^-_x &= \{\nu^j_x(k_y), \mbox{ s.t. } \nu^j_x(k_y) \mbox{ is below the Wannier gap} \}\nonumber \\
\nu^+_x &= \{\nu^j_x(k_y), \mbox{ s.t. } \nu^j_x(k_y) \mbox{ is above the Wannier gap} \}.\nonumber
\end{align}
Since the Wannier bands are defined mod 1, we adopt the convention of defining the Wannier sectors $\nu^-_x \in [0,1/2)$ and  $\nu^+_x \in [1/2,1)$.

We then choose those above or below the gap and form the projector
\begin{align}
P_{\nu_x}&=\sum_{j=1}^{N_W}\sum_{R_x, k_y} \ket{\Psi^j_{R_x,k_y}}\bra{\Psi^j_{R_x,k_y}}\nonumber\\
&=\sum_{j=1}^{N_W} \sum_{n,m=1}^{N_{occ}} \sum_{\bf k} \gamma^\dagger_{n,\bf k}\ket{0}[\nu^j_{x,\bf k}]^n [\nu^{*j}_{x,\bf k}]^m \bra{0} \gamma_{m,\bf k},
\end{align}
where $\sum_j^{N_W}$ is a summation over all Wannier bands in the sector $\nu_x$, for $\nu_x = \nu^+_x$ or $\nu^-_x$. $N_W$ is the number of Wannier bands in sector $\nu_x,$ $R_x \in 0 \ldots N_x-1$ labels the unit cells, $k_y=\Delta_{k_y} n_y$, for $\Delta_{k_y}=2\pi / N_y,$ and $n_y \in 0,1, \ldots, N_y-1$ is the crystal momentum along $y$. 

We are interested in studying the topological properties of the subspace spanned by $P_{\nu_x}$ in the reduced quasi-1D BZ $k_y \in (-\pi,\pi]$. As we will see in Section \ref{sec:WannierBandsEdgeHamiltonian}, the topology of the Wannier sectors is related to the topology of the physical edge Hamiltonian. As such, it will provide a bulk measure of the edge topology. In particular, we want to diagonalize the position operator,
\begin{align}
\hat{y} &= \sum_{{\bf R}, \alpha} c^\dagger_{{\bf R},\alpha} \ket{0}e^{-i \Delta_{k_y} (R_y+r_{\alpha,y})} \bra{0} c_{{\bf R},\alpha}\nonumber \\
&=\sum_{k_x,k_y,\alpha} c^\dagger_{k_x,k_y+\Delta_{k_y},\alpha} \ket{0} \bra{0} c_{k_x,k_y,\alpha},
\end{align}
projected into the Wannier sector $\nu_x$,
\begin{align}
P_{\nu_x} \hat{y} P_{\nu_x} =&\sum_{j,j'=1}^{N_W}\sum_{\bf k} \sum_{n,m,n',m'=1}^{N_{occ}} \gamma^\dagger_{n,{\bf k}+{\bf \Delta_{k_y}}}\ket{0}  \bra{0}\gamma_{n',\bf k}\times\nonumber\\
&\left([\nu^j_{x,{\bf k}+{\bf \Delta_{k_y}}}]^n [\nu^{*j}_{x,{\bf k}+{\bf \Delta_{k_y}}}]^m \times \right. \nonumber\\
&\left. \braket{u^m_{{\bf k}+{\bf \Delta_{k_y}}}}{u^{m'}_{\bf k}}[\nu^{j'}_{x,{\bf k}}]^{m'}[\nu^{j'*}_{x,{\bf k}}]^{n'} \right).
\end{align}

To simplify the notation let us define the \textit{Wannier band basis}
\begin{align}
\ket{w^j_{x,\bf k}} = \sum_{n=1}^{N_{occ}}\ket{u^n_{\bf k}} [\nu^j_{x,\bf k}]^n
\label{eq:WannierBasis}
\end{align}
for $j \in 1\ldots N_W$. This basis obeys,
\begin{align}
\braket{w^j_{x,\bf k}}{w^{j'}_{x, \bf k}} = \delta_{j,j'}.
\end{align}
However, in general $\braket{w^j_{x,\bf k}}{w^{j'}_{x, \bf q}} \neq \delta_{j,j'} \delta_{\bf k, \bf q}$.
The projected position operator then reduces to
\begin{align}
P_{\nu_x} \hat{y} P_{\nu_x} =&\sum_{j,j'=1}^{N_W}\sum_{\bf k} \sum_{n,n'=1}^{N_{occ}} \gamma^\dagger_{n,{\bf k}+{\bf \Delta_{k_y}}}\ket{0}  \bra{0}\gamma_{n',\bf k}\times\nonumber\\
&\left([\nu^j_{x,{\bf k}+{\bf \Delta_{k_y}}}]^n \braket{w^j_{x,{\bf k}+{\bf \Delta_{k_y}}}}{w^{j'}_{x,\bf k}}[\nu^{j'*}_{x,{\bf k}}]^{n'} \right),
\end{align}
Notice that the operator is diagonal in $k_x$. Explicitly,
\begin{align}
P_{\nu_x} \hat{y} P_{\nu_x} =&\sum_{k_x,k_y} \sum_{n,n'=1}^{N_{occ}} \gamma^\dagger_{n,(k_x,k_y+ \Delta_{k_y})}\ket{0} \times \nonumber\\
&[F^{\nu_x}_{y,(k_x,k_y)}]^{n,n'} \bra{0}\gamma_{n',(k_x,k_y)},
\end{align}
where
\begin{align}
 [F^{\nu_x}_{y,(k_x,k_y)}]^{n,n'}&=\sum_{j,j'=1}^{N_W}[\nu^j_{x,(k_x,k_y+\Delta_{k_y})}]^n \times \nonumber\\
&\braket{w^j_{x,(k_x,k_y+\Delta_{k_y})}}{w^{j'}_{x,(k_x,k_y)}}[\nu^{j'*}_{x,{(k_x,k_y)}}]^{n'}.
\end{align}
To diagonalize $P_{\nu_x} \hat{y} P_{\nu_x} $, we calculate the Wilson loop along $y$
\begin{align}
[{\W}^{\nu_x}_{y, \bf k}]^{n,n'} &= F^{\nu_x}_{y,{\bf k}+N_y {\bf \Delta_{k_y}}} \ldots F^{\nu_x}_{y,{\bf k}+ {\bf \Delta_{k_y}}} F^{\nu_x}_{y,{\bf k}}\nonumber\\
&= [\nu^j_{x,{\bf k}+N_y{\bf \Delta_{k_y}}}]^n [\tilde\W^{\nu_x}_{y,\bf k}]^{j,j'}[\nu^{j'*}_{x,{\bf k}}]^{n'}\nonumber\\
&= [\nu^j_{x,\bf k}]^n [\tilde\W^{\nu_x}_{y,\bf k}]^{j,j'}[\nu^{j'*}_{x,{\bf k}}]^{n'},
\end{align}
for $n,n' \in 1\ldots N_{occ}$, and $j,j' \in 1\ldots N_W$. $\tilde\W^{\nu_x}_{y,\bf k}$ is the Wilson loop along $y$ over the Wannier sector $\nu_x$ performed over the Wannier band basis,
\begin{align}
[\tilde\W^{\nu_x}_{y,\bf k}]^{j,j'} =& \braket{w^j_{x,{\bf k}+N_y{\bf \Delta_{k_y}}}}{w^{r}_{x,{\bf k}+(N_y-1) {\bf \Delta_{k_y}}}} \times \nonumber\\
&\bra{w^{r}_{x,{\bf k}+(N_y-1) {\bf \Delta_{k_y}}}} \ldots \nonumber\\
&\dots \ket{w^{s}_{x,{\bf k}+{\bf \Delta_{k_y}}}}\braket{w^s_{x,{\bf k}+{\bf \Delta_{k_y}}}}{w^{j'}_{x,\bf k}}.
\label{eq:WilsonLoop_WannierBasis}
\end{align}
In the expression above, summation is implied over repeated indices $r,\ldots,s \in 1 \ldots N_W$ over all Wannier bands in the Wannier sector $\nu_x$. 

Since $N_W<N_{occ}$, this Wilson loop \eqref{eq:WilsonLoop_WannierBasis} is calculated over a subspace \textit{within} the subspace of occupied energy bands. In general, we will indicate an operator written \emph{in a Wannier band basis with a tilde}, while no tilde indicates that it is written in the basis of energy bands. Since we have used $\nu_x$ as the label for the Wannier bands along $x$, we will use the labels $\nu^{\nu_x}_y$ for the eigenvalues and eigenvectors for the Wilson-loop along $y$ carried out for the Wannier band sector $\nu_x$.  This Wilson loop diagonalizes as
\begin{align}
\tilde\W^{\nu_x}_{y,\bf k}  \ket{\nu^{\nu_x,j}_{y,\bf k}} = e^{i 2\pi \nu^{\nu_x,j}_y(k_x)} \ket{\nu^{\nu_x,j}_{y,\bf k}}
\end{align}
for $j \in 1 \ldots N_W$.
The polarization over the Wannier sector $\nu_x$ at $k_x$ is then given by the sum of the $N_W$ phases $\nu^{\nu_x}_y(k_x)$,
\begin{align}
p^{\nu_x}_y(k_x) = \sum_{j=1}^{N_{\nu_x}} \nu^{\nu_x,j}_y(k_x) \;\;\mbox{mod 1}.
\end{align}
This can be written as
\begin{align}
p^{\nu_x}_y(k_x) = -\frac{i}{2\pi} \text{Log Det}[\tilde\W^{\nu_x}_{y,\bf k}].
\end{align}
The total polarization of the Wannier bands $\nu_x$ is
\begin{align}
p^{\nu_x}_y = \frac{1}{N_x} \sum_{k_x} p^{\nu_x}_y(k_x).
\label{eq:Wannier_sector_polarization}
\end{align}
In the thermodynamic limit it becomes
\begin{align}
p^{\nu_x}_y &=-\frac{1}{(2\pi)^2} \int_{BZ} \tr \left[\tilde\A^{\nu_x}_{y,\bf k}\right] d^2\bf k
\label{eq:PolarizationWannierSector}
\end{align}
where $\tilde\A^{\nu_x}_{y,\bf k}$ is the Berry connection of Wannier bands $\nu_x$ having components 
\begin{align}
[\tilde\A^{\nu_x}_{y,\bf k}]^{j,j'} = -i \bra{w^j_{x,\bf k}} \partial_{k_y} \ket{w^{j'}_{x,\bf k}},
\label{eq:WannierBerryConnection}
\end{align}
where $j,j' \in 1 \ldots N_W$ run over the Wannier bands in Wannier sector $\nu_x$.

The Wannier-sector polarization has a physical significance. In the bulk of the material, a Wannier gap for the Wilson loop along $x$ implies the existence of a spatial separation between electrons along $x$. For example, for the Wannier bands of Fig. \ref{fig:WannierBands}, electrons in the sector $\nu^-_x$ are on the left side of the unit cell, and those in sector $\nu^+_x$ are on its right side. Thus, a non-zero polarization in the $y$-direction of such a Wannier sector represents a translation, up or down, of the electrons of that Wannier sector. 

\begin{figure}[t]%
\centering
\includegraphics[width=\columnwidth]{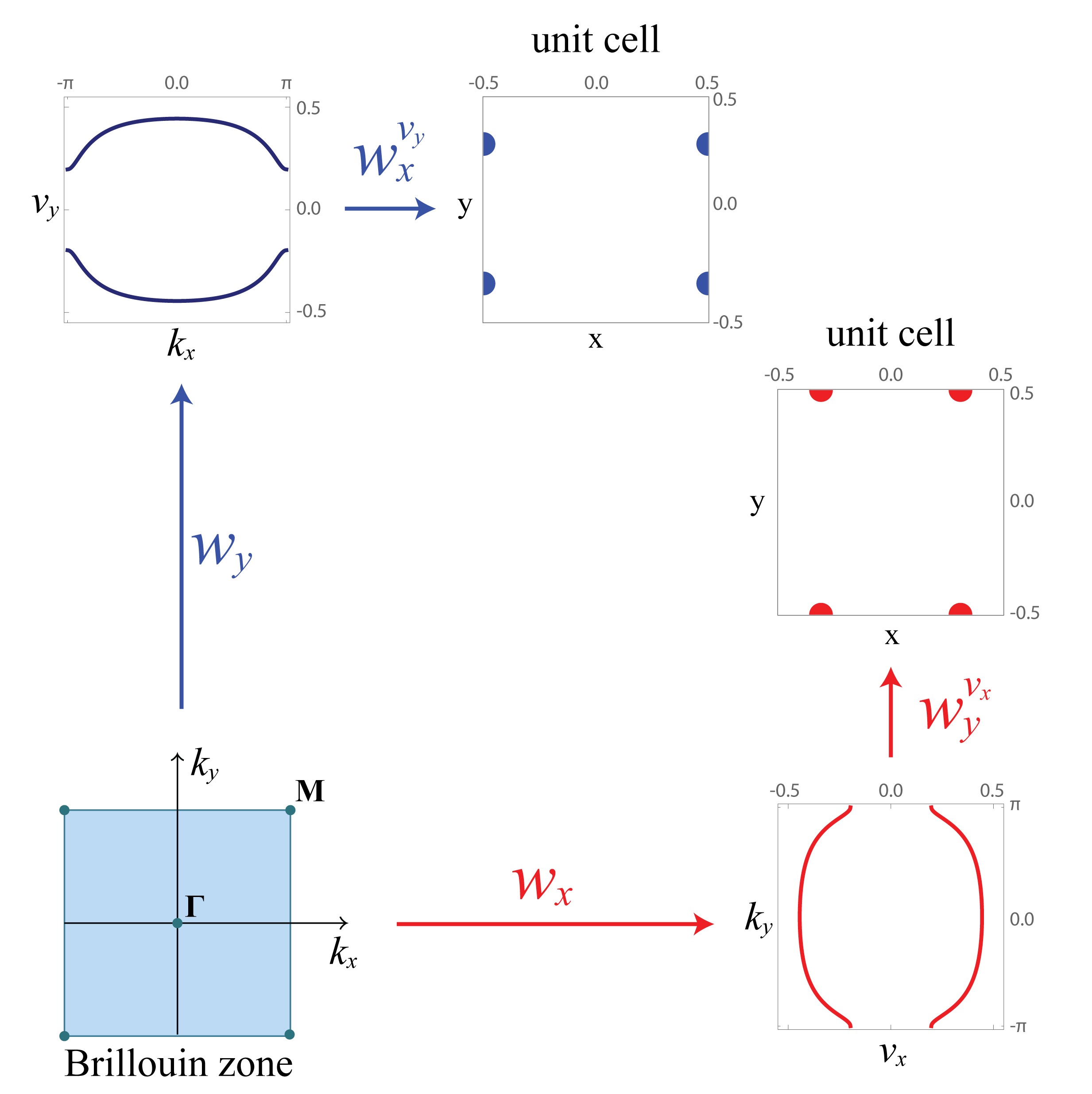}
\caption{(Color online)  Maximally localized Wannier centers of electrons within the unit cell of a quadrupole insulator in the non trivial phase (see Hamiltonian \eqref{eq:QuadHamiltonian} for the minimal quadrupole insulator) upon calculation of the Wilson loop $\W_{y,{\bf k}}$ followed by $\W^{\nu_y}_{x,{\bf k}}$ (blue path) or $\W_{x,{\bf k}}$ followed by $\W^{\nu_x}_{y,{\bf k}}$ (red path).The overall localizations do not coincide for these two paths because $[P^{occ}\hat{x}P^{occ},P^{occ}\hat{y}P^{occ}] \neq 0$. The quadrupole insulator has two electrons per unit cell.}
\label{fig:Quad_BulkWannierPolarization}
\end{figure}

We can see a simple example of this for the insulator $h^1({\bf k})$ in \eqref{eq:Hamiltonians12} from Section \ref{sec:WannierBands_AllModels}. Since its Wannier bands are flat, we can assign the values $(\nu^-_x,\nu^+_x)=(0,1/2)$ (see Fig.~\ref{fig:WannierBands_AllModels}a). The polarization along $y$ of each of these Wannier bands gives the centers $(\nu^{\nu^-_x}_y,\nu^{\nu^+_x}_y)=(1/2,0)$. Thus, the electron at the center of the unit cell along $x$ is inbetween unit cells along $y$, and vice versa (see Fig.~\ref{fig:h1}a). For this model the position operators along $x$ and $y$ projected into the full subspace of occupied bands actually commute, and therefore this system has a full 2D Wannier representation, as shown in Fig.~\ref{fig:h1}a. The concept illustrated here, however, is also valid in cases that do not admit a 2D Wannier representation, i.e., when the projected position operators along different directions do not commute, as long as the Wannier bands are gapped. This is what happens in a model that realizes a topological quadrupole insulator. This is schematically represented in Fig.~\ref{fig:Quad_BulkWannierPolarization}. The solid half-circles represent the maximally localized Wannier centers of the electrons in a minimal quadrupole insulator, to be described in detail in Section \ref{sec:QuadMinimalModel}. 

Consider Fig.~\ref{fig:Quad_BulkWannierPolarization}. If we first diagonalize the Wilson loop along $x$ at each $k_y$, we find two Wannier bands, $\pm \nu_x(k_y)$, separated by a gap. Since these Wannier bands depend on the value of $k_y$, this is not an extremely sharp resolution of the position of the electronic charge along $x$. However, the existence of the Wannier gap (see Fig.~\ref{fig:WannierBands}) implies that there are two clouds of electrons, one corresponding to each Wannier sector: one to the right of the center of the unit cell and another one to its left (the center of the unit cell is at the center of each panel). Thus, we can assign an average $x$-coordinate to each cloud by calculating the average values of the Wannier bands, i.e., $\nu^{1,2}_x = 1/N_y \sum_{k_y} \nu^{1,2}_x(k_y)$. Furthermore, we can resolve the position of \emph{each} of these two electronic clouds along $y$ by calculating the nested Wilson loop \eqref{eq:WilsonLoop_WannierBasis}. This yields $p^{\pm \nu_x}_y=1/2$ in the non trivial quadrupole phase, as in Fig.~\ref{fig:Quad_BulkWannierPolarization}, or $p^{\pm \nu_x}_y=0$ in the trivial phase (not shown). In Fig.~\ref{fig:Quad_BulkWannierPolarization} we also show the localization if we first resolve the electronic positions along $y$ and then along $x$. The electronic localizations thus depend on the order in which the positions along $x$ and $y$ are resolved. This is a testament to the fact that the projected position operators along $x$ and $y$ do not commute.

Notice that, in the two paths for electron localization represented in Fig. \ref{fig:Quad_BulkWannierPolarization}, the overall bulk polarization vanishes, which is a requirement for a well-defined quadrupole moment (see Section \ref{sec:multipole_translation_invariance}). Since the overall polarization in the bulk has contributions from both Wannier sectors, it follows that
\begin{align}
p^{\nu^+_x}_y+p^{\nu^-_x}_y = 0 \;\;\mbox{mod 1}
\end{align}
for a well-defined quadrupole moment. This can be enforced by inversion symmetry, as shown in Appendix~\ref{sec:WilsonLoopsSymmetry}.

\subsection{Symmetry protection and quantization of the Wannier-sector polarization}
\label{sec:WannierPolarization_quantization}
Under reflections  $M_x$, $M_y$, and inversion $\mathcal{I}$ the Wannier-sector polarizations obey
\begin{align}
p^{\nu^+_x}_y &\stackrel{M_x}{=} p^{\nu^-_x}_y \nonumber\\
p^{\nu^\pm_x}_y &\stackrel{M_y}{=} -p^{\nu^\pm_x}_y\;\; \implies p^{\nu^\pm_x}_y\stackrel{M_y}{=}0 \mbox{ or } 1/2 \nonumber\\
p^{\nu^+_x}_y &\stackrel{\mathcal{I}}{=} -p^{\nu^-_x}_y
\label{eq:WannierPolarization_SymmetryConstraints}
\end{align}
mod 1. These relations are derived in Appendix \ref{sec:WilsonLoopsSymmetry}. The relations for $p^{\nu^\pm_y}_x$ are the same as the above with the exchange of labels $x \leftrightarrow y$. We also note that in 2D the constraints generated by $C_2$ symmetry are the same as those generated by inversion $\mathcal{I}$ symmetry.

In the expressions \eqref{eq:WannierPolarization_SymmetryConstraints}, $M_y$ directly quantizes $p^{\nu^\pm_x}_y$. $M_x$, on the other hand, also requires $\I$ to quantize $p^{\nu^\pm_x}_y$ (see first and third relations in Eq. \ref{eq:WannierPolarization_SymmetryConstraints}). Since in spinless systems the existence of both reflection symmetries implies the existence of inversion symmetry, the Wannier-sector polarizations $p^{\nu^\pm_x}_y$ and $p^{\nu^\pm_y}_x$ (calculated from $H_{\W_{x}}$ and $H_{\W_{y}}$ respectively) of a system with both reflection symmetries take quantized values
\begin{align}
p^{\nu^\pm_x}_y, p^{\nu^\pm_y}_x \stackrel{M_x,M_y}{=} 0 \mbox{ or } 1/2.
\label{eq:WannierPolarization_quantization}
\end{align}
The quantization due to reflection symmetries can be used to compute the Wannier-sector polarization in a simpler way. The Wannier band basis  \eqref{eq:WannierBasis} obeys
\begin{align}
\hat{M}_y \ket{w^\pm_{x,(k_x,k_y)}}=\alpha^{\pm}_{M_y}(k_x,k_y)\ket{w^\pm_{x,(k_x,-k_y)}} 
\label{eq:Wannier_basis_reflection}
\end{align}
with a $U(1)$ phase $\alpha^{\pm}_{M_y}(k_x,k_y)$ (see Appendix \ref{sec:WilsonLoopsSymmetry}). In particular, at the reflection-invariant momenta, $k_{*y}=0 \text{ and }\pi$,  $\alpha^{\pm}_{M_y}(k_x, k_{*y})$ are the eigenvalues of the reflection representation of $\ket{w^\pm_{x,\bf k}}$ at $(k_x,k_{*y}).$ These can take the values $\pm 1$  $(\pm i)$ for spinless (spinfull) fermions. If the representation is the same (different) at $k_{*y} = 0$ and $k_{*y} = \pi$, the Wannier-sector polarization is trivial (non-trivial) \cite{hughes2011inversion}. Thus, in reflection-symmetric insulators the Wannier-sector polarization can then be computed by
\begin{align}
\mbox{exp}\left\{i2\pi p^{\nu^\pm_x}_y\right\} = \alpha^{\pm}_{M_y}(k_x,0) \alpha^{\pm\ast}_{M_y}(k_x,\pi),
\label{eq:WannierPolarization_reflection}
\end{align}
where the superscript asterisk stands for complex-conjugation. The Wannier-sector polarization then takes the quantized values 
\begin{align}
p^{\nu^\pm_x}_y \stackrel{M_y}{=} \begin{cases}
0 & \mbox{if trivial}\\
1/2 & \mbox{if non-trivial}
\end{cases}.\nonumber
\end{align}
%

\subsection{Conditions for the existence of  a Wannier gap}
\label{sec:GappedWannierBands_conditions}
In the previous subsection we saw that reflection symmetries quantize the Wannier-sector polarization. This Wannier-sector polarization is well defined only if the Wannier bands are gapped. In this section we show that, in order to have gapped Wannier bands in the presence of reflection symmetries, $M_x$ and $M_y$, the reflection operators must not commute. We will prove this by showing that if the reflection operators commute, we necessarily have gapless Wannier bands\cite{benalcazar2017quad}, i.e., 
\begin{align}
[\hat{M}_x,\hat{M}_y] = 0 \implies \mbox{Gapless Wannier bands}.
\end{align}
Two natural ways to have noncommuting reflection symmetries are to have a model with magnetic flux so that reflection is only preserved up to a gauge transformation (see Appendix~\ref{sec:app_symmetries_up_to_gauge}), or to have spin-1/2 degrees of freedom. For our simple quadrupole model below we chose the former interpretation for the simplicity of the description. 

For a crystal with $N_{occ}$ occupied energy bands and reflection and inversion symmetries $M_x$, $M_y$, and $\I$, such that $[\hat{M}_x,\hat{M}_y]=0$, various cases need to be considered\cite{benalcazar2017quad}:
\begin{itemize}
\item $N_{occ}=2$: In this case, the Wannier bands $\nu_x(k_y)$ are necessarily gapless at $k_y=0,\pi$.
\item $N_{occ} = 4$: In this case, the Wannier bands $\nu_x(k_y)$ can be generically gapped, with each Wannier sector being two-dimensional. The two eigenvalues of the nested Wilson loop over the Wannier sector $\nu_x$, for either $\nu_x = \nu^+_x$ or $\nu^-_x$ can be shown to come in pairs $(\nu^{\nu_x}_y(k_x),-\nu^{\nu_x}_y(k_x))$ at $k_x=0,\pi$. This implies that $p^{\nu_x}_y = 0$.
\item $N_{occ} = 4n$: In this case, we have a generalized version of the $N_{occ}=4$ case.
\item $N_{occ} = 4n + 2$: In this case, the Wannier bands split as in the $4n$ case plus a leftover set of two bands, and the Wannier spectrum is gapless.
\item $N_{occ}$ is odd: In this case, the Wannier bands split as in one of the even band cases above plus one extra band. The extra band is always gapless, as its value is necessarily 0 or 1/2 since it does not have partner (see Section \ref{sec:Dipole_under_Inversion}).
\end{itemize}
Here we will elaborate on the first case. The other cases for generic numbers of bands are detailed in Appendix~\ref{sec:GappedWannierBands_conditions2}.
Consider a spinless crystal with reflection symmetries $M_x$ and $M_y$. For a tight binding Hamiltonian $h(k_x,k_y)$, these symmetries are expressed by Eq.~\ref{eq:HamiltonianUnderReflection_2d}.
Such a system also has the inversion symmetry expressed by Eq.~\ref{eq:HamiltonianUnderInversion_2d}. The reflection operators $\hat{M}_{x,y}$ and the inversion operator $\hat{\I}$ are related by
\begin{align}
\hat{\I} = \hat{M}_y \hat{M}_x.
\end{align}\noindent We should be careful at this point to note that in some 2D systems this definition of inversion is problematic since the operator $\hat{\I}$ should obey $\hat{\I}^2=1.$ If we have $\hat{M}_y \hat{M}_x=- \hat{M}_y \hat{M}_x$, as we will encounter later, then by this definition $\hat{\I}^2=-1$ and so we should more precisely identify $\hat{M}_y \hat{M}_x$ as a $C_2$ rotation operator in 2D instead.

The special high-symmetry points and lines in the Brillouin zone, at which a given operator $\hat{Q}$ (we consider $\hat{Q}=\hat{M}_{x}$, $\hat{M}_{y}$ and $\hat{\I}$) commutes with the Hamiltonian,
\begin{align}
[h_{\bf k^*},\hat{Q}] = 0.
\end{align}
are shown in Fig.~\ref{fig:2DBZ} in blue for $M_x$, red for $M_y$, and black dots for all $M_x$, $M_y$, and $\I$. We are interested in the conditions under which we have gapped Wannier bands along both $x$ and $y$. 
This can be inferred from the $\hat{Q}$ eigenvalues at the high-symmetry points \cite{alexandradinata2014}, as shown in Table~\ref{tab:EigenvaluesRelations}.

There we see that, in order to have gapped Wannier bands, the Wilson loop eigenvalues must come in complex conjugate pairs $(\lambda, \lambda^*)$ not pinned at $1$ or $-1.$ Hence, we require that, along each of the blue and red lines of Fig.~\ref{fig:2DBZ}, the reflection eigenvalues come in pairs $(+-)$. To these two conditions (one along blue lines for $\hat{M}_x$ eigenvalues and another one along red lines for $\hat{M}_y$ eigenvalues) we need to add the third requirement that the inversion eigenvalues must also come in pairs $(+-)$ at the high-symmetry points ${\bf k}_* =  \bf \Gamma, X, Y, M$ of the BZ. If the condition is not also met for the inversion eigenvalues, the complex conjugate pairs are still forced to be a pair of $1$ or pair of $-1$ at $k_y=0,\pi$ for Wilson loops along $k_x$, and at $k_x=0,\pi$ for Wilson loops along $k_y$. We will now see that this third condition is not possible to meet simultaneously with the first two conditions if the reflection operators commute. If we have
\begin{align}
[\hat{M}_x,\hat{M}_y]=0,
\end{align}
it is possible to simultaneously label the energy states at the high-symmetry points by their reflection eigenvalues, e.g., $\ket{m_x,m_y}$, where
\begin{align}
\hat{M}_x \ket{m_x,m_y} &= m_x \ket{m_x,m_y}\nonumber \\
\hat{M}_y \ket{m_x,m_y} &= m_y \ket{m_x,m_y},
\label{eq:ReflectionEigenproblem}
\end{align}
for reflection eigenvalues $m_{x,y}=\pm$. Since the inversion operator is
$\hat{\I} = \hat{M}_x \hat{M}_y =\hat{M}_y \hat{M}_x$, the inversion eigenvalues are
\begin{align}
\hat{\I} \ket{m_x,m_y} = m_x m_y \ket{m_x,m_y}.
\end{align}
The two combinations of states that have $\hat{M}_x$ and $\hat{M}_y$ eigenvalues of $(+-)$ are listed in Table~\ref{tab:app_states_pm_mx_eigenvalues}. 
\begin{table}
\begin{center}
\begin{tabular}{c c c c}
states  &$M_x$ eigenval. &  $M_y$ eigenval. &  $\I$ eigenval.\\
\hline
$\left(\ket{++},\ket{--}\right)$ & $(+-)$ & $(+-)$ & $(-1)^s(+1,+1)$\\
$\left(\ket{+-},\ket{-+}\right)$ & $(+-)$ & $(-+)$ & $(-1)^s(-1,-1)$\\
\end{tabular}
\end{center}
\caption{States with eigenvalues $(+-)$ for $M_x$ and $M_y$ and their $\I$ eigenvalues at the high-symmetry points. The signs $\pm$ represent $\pm1$ if $\hat{M}^2_{x,y}=+1$ or $\pm i$ if $\hat{M}^2_{x,y}=-1$. Also, $s=0$ ($s=1$) if $\hat{M}^2_{x,y}=+1$ ($\hat{M}^2_{x,y}=-1$).}
\label{tab:app_states_pm_mx_eigenvalues}
\end{table}
However, those two options do not meet the third condition of having $\hat{\I}$ eigenvalues $(+-)$ at the high-symmetry points. Instead, the $\hat{\I}$ eigenvalues at the high-symmetry points are always either $(++)$ or $(--)$. These inversion eigenvalues imply that the Wilson loop eigenvalues come in complex conjugate pairs $\lambda,\lambda^*$, with $\lambda = \lambda^* = +1 \text{ or }-1$. Thus, along the high-symmetry lines (blue and red lines of Fig.~\ref{fig:2DBZ}) the Wilson loops have eigenvalues
\begin{align}
(\lambda,\lambda^*) \rightarrow (1,1) \mbox{ or } (-1,-1),
\end{align}
i.e., at those lines the Wannier bands close the gap. Conversely, if instead of imposing the conditions of having $(+-)$ for both $\hat{M}_x$ and $\hat{M}_y$ eigenvalues, we started by  first fixing $(+-)$ for $\hat{\I}$ eigenvalues, at most only one of the reflection eigenvalues will be $(+-)$. The other one will necessarily have either $(++)$ or $(--)$. An example of this case is insulator \eqref{eq:hEdgePolarizationHamiltonian} with $\delta=0$, with reflection eigenvalues shown in Fig. \ref{fig:h2EdgePolarizationQuantized}a,b.

\subsection{Simple model with topological quadrupole moment}
\label{sec:QuadMinimalModel}

\begin{figure}[t]%
\centering
\includegraphics[width=\columnwidth]{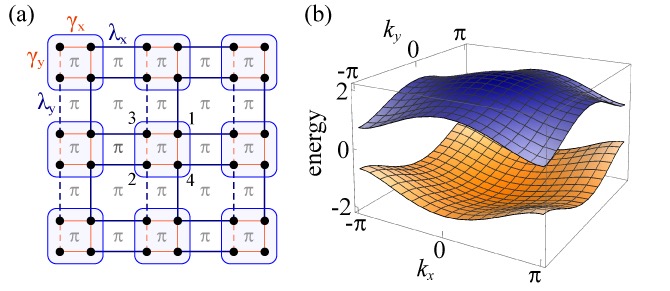}
\caption{(Color online)  Lattice (a) and energy spectrum (b) of the minimal model with quadrupole moment density having the Bloch Hamiltonian Eq. \ref{eq:QuadHamiltonian}. In (a), dashed lines have a negative sign to account for a flux of $\pi$ threading each plaquette. In (b) each energy band is twofold degenerate.}
\label{fig:quad_lattice}
\end{figure}

We now focus on the detailed description of a model for an insulator with a quadrupole moment. The minimal model is a 2D crystal with two occupied bands. For simplicity we choose a microscopic representation consisting of four spinless fermion orbitals with Hamiltonian
\begin{align}
H^q &= \sum_{\bf R} \left[
\gamma_x \left(c^\dagger_{{\bf R},1}c_{{\bf R},3}+ c^\dagger_{{\bf R},2}c_{{\bf R},4} + h.c. \right)\right. \nonumber\\
&+\gamma_y \left(c^\dagger_{{\bf R},1}c_{{\bf R},4}- c^\dagger_{{\bf R},2}c_{{\bf R},3}+ h.c. \right) \nonumber\\
&+\lambda_x \left(c^\dagger_{{\bf R},1}c_{{\bf R+\hat{x}},3}+ c^\dagger_{{\bf R},4}c_{{\bf R+\hat{x}},2}+ h.c. \right) \nonumber\\
&\left.+\lambda_y \left(c^\dagger_{{\bf R},1}c_{{\bf R+\hat{y}},4}- c^\dagger_{{\bf R},3}c_{{\bf R+\hat{y}},2}+ h.c. \right) \right],
\label{eq:QuadRealSpaceHamiltonian}
\end{align}
where $c^\dagger_{{\bf R},i}$ is the creation operator for degree of freedom $i$ in unit cell $\bf R$, for $i=1,2,3,4$ as shown in Fig.~\ref{fig:quad_lattice}a. $\gamma_{x,y}$ are hopping matrix elements within a unit cell. Here ${\bf \hat{x}}=(1,0)$ and ${\bf \hat{y}}=(0,1)$, so that $\lambda_{x,y}$ represent the amplitudes of hopping to nearest neighbor unit cells along $x,y$ respectively. The negative signs, represented by the dashed lines in Fig.~\ref{fig:quad_lattice}a, are a gauge choice for the  $\pi$-flux threaded through each plaquette (including within the unit cell itself). The corresponding Bloch Hamiltonian is
\begin{align}
h^q({\bf k})&=\left[\gamma_x + \lambda_x  \cos(k_x)\right] \Gamma_4 + \lambda_x \sin(k_x) \Gamma_3\nonumber\\
&+ \left[\gamma_y + \lambda_y \cos(k_y)\right] \Gamma_2 + \lambda_y \sin(k_y) \Gamma_1,
\label{eq:QuadHamiltonian}
\end{align}
where $\Gamma_0= \tau_3 \otimes \tau_0$, $\Gamma_k=-\tau_2 \otimes \tau_k$, $\Gamma_4=\tau_1 \otimes\tau_0$ for $k=1,2,3$, where $\tau_{1,2,3}$ are Pauli matrices, and $\tau_0$ is the $2 \times 2$ identity matrix. The energy bands are
\begin{align}
\epsilon({\bf k})=\pm \sqrt{\epsilon^2_x(k_x)+ \epsilon^2_y(k_y)}
\label{eq:quad_energy}
\end{align}
where $\epsilon_i(k_i)=\sqrt{\gamma_i^2 + 2 \gamma_i \lambda_i \cos(k_i) + \lambda_i^2}$ for $i=x,y$. Each of the upper and lower energy bands is twofold degenerate. 

This Hamiltonian is gapped across the entire bulk Brillouin zone (BZ) unless $\vert\gamma_x/\lambda_x\vert=1$ \emph{and} $\vert \gamma_y/\lambda_y\vert=\pm1$. A plot of the energy spectrum in the 2D BZ is shown in Fig.~\ref{fig:quad_lattice}b for $\gamma_x/\lambda_x=\gamma_y/\lambda_y=0.5$.
We consider this system at half filling, so that only the lowest 2 bands are occupied. This Hamiltonian has vanishing polarization, and zero Chern number for the entire range of parameters for which it is gapped. Thus, it meets the preliminary requirements of an insulator with quadrupole moment density (outlined at the beginning of Section \ref{sec:Quadrupole}). The projected position operators along $x$ and $y$ do not commute at half filling, and the Hamiltonian has a pair of gapped Wannier bands, as shown in Fig.~\ref{fig:WannierBands}. 

In the present form, this Hamiltonian has symmetries that quantize the Wannier-sector polarizations $p^{\nu^\pm_x}_y, p^{\nu^\pm_y}_x = 0$ or $1/2$, which we describe in the following subsection. Associated to this quantization is the existence of sharply quantized corner charges and edge polarizations, in agreement with \ref{eq:QuadSignatures}. Upon breaking the symmetries that quantize the quadrupole moment, a generalized version of this model can generate values of quadrupole moment satisfying $q_{xy} \in (0,1]$. As an extension, we will see that when coupling this system to an adiabatically varying parameter, a quantum of dipole can be pumped in a way analogous to the quantum of charge pumped in the case of a cyclically varying bulk dipole moment (Section \ref{sec:DipolePumping}).

\subsubsection{Symmetries}
\label{sec:quad_symmetries}

The quadrupole moment $q_{xy}$ in 2D is even under the group $T(2)$, which contains the operations $\{1,C_4M_x, C_4M_y, C^2_4\}$ (see  Section ~\ref{sec:classical_multipoles_symmetries}), where $M_{x,y}$ are reflections, and $C_4$ is the rotation by $\pi/2$ around the $z$-axis. This implies that none of the symmetries $\{C_4M_x, C_4M_y, C^2_4\}$ quantize the quadrupole moment $q_{xy}$ in crystalline insulators. On the other hand, the reflection operations $M_{x,y}$ and $C_4$,  transform $q_{xy}$ to $-q_{xy}$. Hence, crystalline insulators \emph{with vanishing bulk dipole moment} having any of $\{M_x, M_y, C_4\}$ will have a well-defined, quantized quadrupole moment, though most insulators may simply just have a vanishing moment.

The quadrupole model with Bloch Hamiltonian \eqref{eq:QuadHamiltonian}  has the reflection symmetries of \eqref{eq:HamiltonianUnderReflection_2d}
with operators
\begin{align}
\hat{M}_x = i \tau_1 \otimes \tau_3,\quad
\hat{M}_y = i \tau_1 \otimes \tau_1,
\label{eq:QuadSymmetryOperators}
\end{align}
as well as $C_2$ symmetry
\begin{align}
\hat{r}_2h^q({\bf k}) \hat{r}^\dagger_2 = h^q(-{\bf k})
\end{align}
with the $C_2$ rotation operator
\begin{align}
\hat{r}_2=\hat{M}_x \hat{M}_y&= -i\tau_0 \otimes \tau_2.
\end{align}
Notice that $C_2$ symmetry for this model resembles the inversion symmetry \eqref{eq:HamiltonianUnderInversion_2d}.
The reflection and $C_2$ operators obey $\hat{M}_{x,y}^2=-1$ and $\hat{r}_2^2=-1$. The point group of $h^q({\bf k})$ in \eqref{eq:QuadHamiltonian} is thus the quaternion group
\begin{align}
Q=\left\langle \bar{e},\hat{M}_x,\hat{M}_y, \hat{r}_2\left|\begin{array}{*{20}c}\bar{e}^2=1,\\ \hat{M}^2_x = \hat{M}^2_y = \hat{r}_2^2 = \hat{M}_x\hat{M}_y\hat{r}_2=\bar{e}\end{array}\right.\right\rangle
\label{eq:quad_quaternion_group}
\end{align}
with $\bar{e}=-1$. The quaternion group is of order 8, with elements  $\{\pm1,\pm\hat{M}_{x,y},\pm\hat{r}_2\}$.
The three operators \eqref{eq:QuadSymmetryOperators} each have eigenvalues $\{-i,-i,+i,+i\}$. 
Due to the $\pi$-flux threading each plaquette, the reflection operators do not commute, instead, they obey
\begin{align}
[\hat{M}_x,\hat{M}_y]&= -2i \tau_0 \otimes \tau_2\nonumber\\
\{\hat{M}_x,\hat{M}_y\}&=0.
\label{eq:QuadReflectionSymmetriesCommutations}
\end{align} 
The energy band degeneracy is protected at the high symmetry points of the BZ by the non-commutation of the reflection operators $\hat{M}_x$, $\hat{M}_y$ (see Appendix \ref{sec:EnergyDegeneracyProtection}). 
Thus, it is not possible to lift the twofold degeneracy of the energy bands at those points while preserving both reflection symmetries. Indeed, at each of the high-symmetry points of the BZ, the subspace of occupied bands lies in the two-dimensional representation of the quaternion group. 

Since $C_2$ transforms the Bloch Hamiltonian $h^q({\bf k})$ the same way as $\I$ does, $C_2$ symmetry quantizes the bulk dipole moment in $h^q({\bf k})$ to $\bf p=0$, as required for an insulator with well-defined quadrupole moment. $M_x$ or $M_y$ then quantize the quadrupole moment of $h^q({\bf k})$ to either $q_{xy}=0$ or $1/2$. The three symmetries in $h^q({\bf k})$, $M_x$, $M_y,$ and $C_2$ are simultaneously present due to the fact that the existence of two of them implies the existence of the third one. 

Alternatively, $C_4$ also quantizes $q_{xy}$. If we set $\gamma_x = \gamma_y$ and $\lambda_x = \lambda_y$, $h^q(\bf k)$ has also $C_4$ symmetry,
\begin{align}
\hat{r}_4 h^q({\bf k}) \hat{r}_4^\dagger = h^q(R_4{\bf k}), \quad 
\hat{r}_4=\left(
\begin{array}{cc}
0 & \tau_0\\
-i \tau_2 & 0
\end{array} \right),
\label{eq:QuadC4RotationSymmetries}
\end{align}
where $R_4$ is the rotation by $\pi/2$ of the crystal momentum, i.e., $R_4(k_x,k_y)=(k_y,-k_x)$. The $C_4$ rotation operator obeys $\hat{r}_4^2=\hat{r}_2$ and $\hat{r}_4^4=-1$ (the minus sign is due to the flux per unit cell) and has eigenvalues $\{e^{\pm i \pi/4},e^{\pm i 3\pi/4}\}$.

Finally, $h^q({\bf k})$, as written in Eq. \ref{eq:QuadHamiltonian}, lies in class BDI, i.e.,  it has time-reversal, chiral  and charge conjugation symmetries
\begin{align}
\Theta h^q({\bf k}) \Theta^{-1} &= h^q(-{\bf k}), \quad \Theta=K \nonumber\\
\Pi h^q({\bf k}) \Pi^{-1} &= -h^q({\bf k}), \quad \Pi = \Gamma^0 \nonumber\\
C h^q({\bf k}) C^{-1} &= - h^q(-{\bf k}), \quad C = \Gamma^0 K.
\end{align} However, these symmetries are not necessary for quantization of the quadrupole moment. In fact, we show in Appendix \ref{sec:app_QuadBreakingSymmetriesSimulation} that we can break all of these symmetries and still preserve the quantization of the quadrupole observables as long as the reflection symmetries are preserved.

\subsubsection{Boundary signatures of the quadrupole phase}
\label{sec:quad_boundary_signatures}
\begin{figure}[t]%
\centering
\includegraphics[width=\columnwidth]{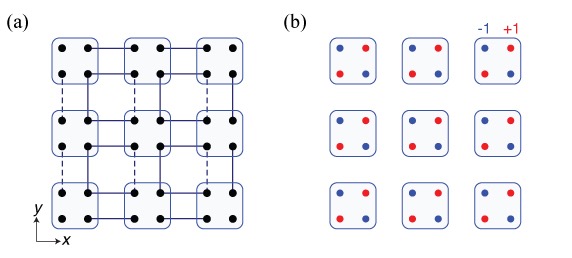}
\caption{(Color online)  (a) Schematic of quadrupole model in the limit $\gamma_x=\gamma_y=0$. (b) On-site perturbation that breaks reflection symmetries in $x$ and $y$ while preserving $C_2$ symmetry.}
\label{fig:quad_terms}
\end{figure}

Eq.~\ref{eq:QuadSignatures} gives the physical signatures of the quadrupole phase. With open boundaries, edge-localized polarizations exist, which can generate observable charge or currents as indicated by
\begin{align}
Q^{edge\; a} &= -\partial_j p_{j}^{edge\; a} \nonumber\\
J_{j}^{edge\; a} &= \partial_t p_{j}^{edge\; a}.
\label{eq:QuadEdgeSignatures}
\end{align} 
When two perpendicular boundaries are open, the edge polarizations along the boundaries generate a quadrupole pattern (see Fig.~\ref{fig:continuous_quadrupole}), and the corner hosts charges having the same magnitude as the edge polarizations. 
To illustrate these symmetry-protected signatures it will be convenient to use the Hamiltonian \eqref{eq:QuadHamiltonian} in the limit $\gamma_x=\gamma_y=0$, as shown in Fig.~\ref{fig:quad_terms}a. In this limit it is straightforward to identify the localized 1D boundary TIs associated with the edge polarization by eye, as well as the degenerate, mid-gap modes responsible for the corner charges. 


Before we proceed we point out a common subtlety in the calculation of electric moments. In the symmetry-protected topological phases with quantized quadrupole moment, the observables are not unambiguously defined, e.g., a value of $q_{xy}=1/2$ is equivalent to a value of $-1/2$, and similarly for $p^{edge}$ and $Q^{corner}.$ This occurs when $h^q({\bf k})$ has the quantizing symmetries that transform $q_{xy}$ to $-q_{xy}$.
In order to unambiguously calculate the observables of the quadrupole insulator in a many-body ground state, we infinitesimally break all the symmetries that quantize the quadrupole so it will evaluate to a number close to, but not equal to, either $1/2$ or $-1/2$. For that purpose, we consider the Hamiltonian
\begin{align}
h^q_\delta({\bf k}) = h^q({\bf k})+ \delta \Gamma_0,
\label{eq:QuadHamiltonian_with_delta}
\end{align}
where $h^q({\bf k})$ is the pristine, reflection-symmetric Hamiltonian in Eq. \ref{eq:QuadHamiltonian} with quantized boundary signatures, and $\Gamma_0 = \tau_3 \otimes \tau_0$ represents an on-site potential with the pattern shown in Fig.~\ref{fig:quad_terms}b. This potential obeys $[\Gamma^0,\hat{M}_x] \neq 0$, $[\Gamma^0,\hat{M}_y] \neq 0$ and $[\Gamma^0,\hat{r}_2]=0$, i.e., it breaks the quantizing reflection symmetries of the quadrupole moment (infinitesimally for $\delta \ll \gamma_{x,y}, \lambda_{x,y}$), but, crucially, retains $C_2$ symmetry, which maintains a vanishing, quantized value of the bulk dipole moment. In this Section we will keep $\delta \ll \gamma_{x,y}, \lambda_{x,y}$.

\begin{figure}[t]%
\centering
\includegraphics[width=\columnwidth]{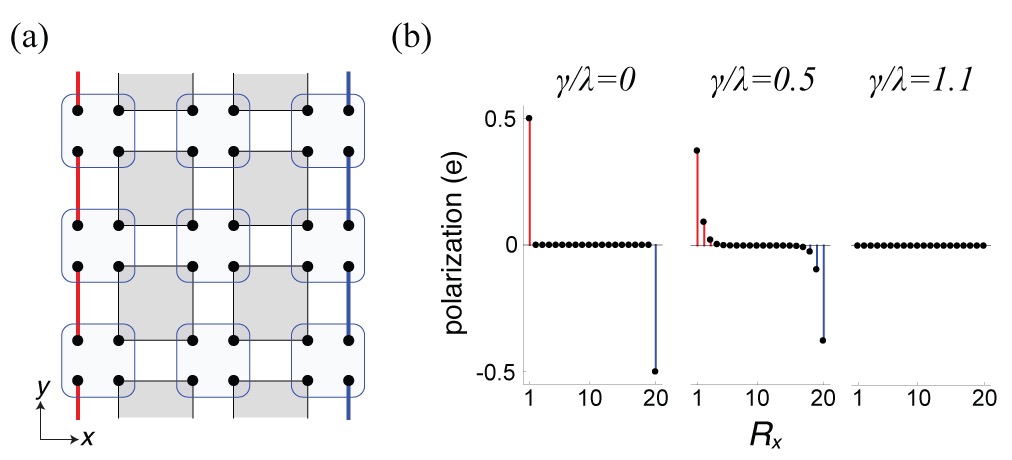}
\caption{(Color online)  Edge polarization in the quadrupole insulator. (a) Schematic of quadrupole model in the limit $\gamma_x=\gamma_y=0$ with open boundaries along $x$ and closed along $y$. Gray squares are bulk plaquettes over which the polarization is zero. Red and blue lines represent edge-localized 1D quantized dipole moments.  (b) Polarization along $y$ as a function of $x$ for $\gamma/\lambda=0,0.5,1.1$ when an infinitesimal on-site perturbation as in Fig.~\ref{fig:quad_terms}b is added.}
\label{fig:quad_polarization}
\end{figure}

\emph{Edge polarization:}
A direct consequence of the non-trivial bulk topology of the quadrupole phase is the existence of edge polarization. This polarization is tangent (i.e., parallel), to the edge. Consider first the quadrupole insulator in the  limit $\gamma_x=\gamma_y=0$ of Eq. \ref{eq:QuadHamiltonian_with_delta}, with $\delta \ll \lambda=\lambda_x=\lambda_y$, and having open boundaries along $x$ but closed along $y$, as shown in Fig.~\ref{fig:quad_polarization}a. In the bulk, the electrons are connected via hopping on the square plaquettes, and form hybridized orbitals localized on the squares in the zero-correlation length limit (shaded squares in Fig.~\ref{fig:quad_polarization}a). The overall electronic displacement in these plaquettes is zero (see Appendix~\ref{sec:ReflectionEigenvalues_WannierValues}). At the edges, however, electrons are only connected vertically as in the 1D symmetry-protected dipole phase of the SSH model (compare red and blue edges of Fig.~\ref{fig:quad_polarization}a with Fig.~\ref{fig:dipole_lattice}b), and thus form hybridized orbitals localized on dimers. The small value of $\delta$ breaks the reflection symmetries and infinitesimally displaces the electrons away from $1/2$ to `choose a sign' for the edge polarizations, as shown in the first plot of Fig.~\ref{fig:quad_polarization}b, where we plot the polarization along $y$ resolved in space along $x$, $p_y(R_x)$;  this is calculated using the prescription in Section \ref{sec:EdgePolarization} that results in Eq.~\ref{eq:y_resolved_x_polarization}. 
If we turn on  $\gamma_x$ and $\gamma_y$, the edge polarization remains quantized to 1/2 (although it exponentially penetrates into the bulk), as long as $|\gamma_{x}/\lambda_{x}|<1$ and $|\gamma_{y}/\lambda_{y}|<1$, as shown in the second plot of Fig.~\ref{fig:quad_polarization}b. If, on the other hand, $|\gamma_{x}/\lambda_{x}|>1$ or $|\gamma_{y}/\lambda_{y}|>1$, the edge polarization drops to zero, as seen in the third plot of Fig.~\ref{fig:quad_polarization}b. 

\emph{Corner charge:}
In Section \ref{sec:Dipole_1D} we saw that a 1D bulk dipole moment per unit length $q$ is associated with edge-localized charges $\pm q$. This leads to the conclusion that, in the quadrupole insulator with full open boundaries, the edge-localized dipole moments will accumulate corner charge. Thus, if edge dipole moments per unit length of $q$ exist, we would expect a corner charge $\pm 2q$. However, the corner charge in the quadrupole insulator $h^q({\bf k})$ has equal magnitude to the edge-polarization, i.e., $q$, following \eqref{eq:QuadSignatures}. Hence, since the contributions from edge dipole moments alone over-count the corner charge, there has to be an additional direct contribution from the bulk to the corner charge.

\begin{figure}[t]%
\centering
\includegraphics[width=\columnwidth]{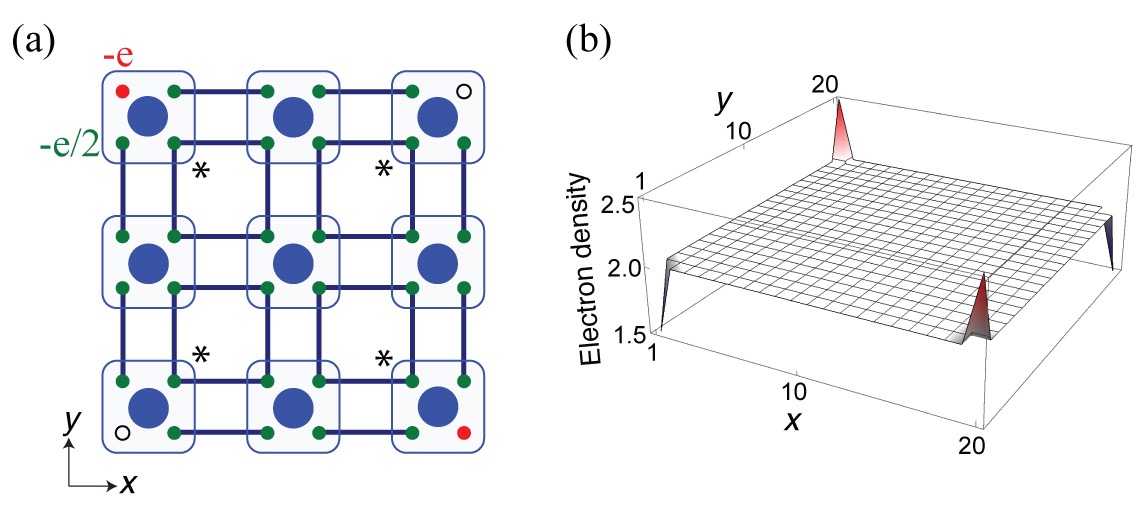}
\caption{(Color online)  Corner charge in the quadrupole insulator. (a) Schematic of charge in the limit $\gamma_x=\gamma_y=0$ when an infinitesimal perturbation as in Fig.  \ref{fig:quad_terms}b is included. The lines connecting sites represent localized hybrid electron orbitals in the many-body ground state at half filling. This is an exact representation of the ground state in the zero-correlation length limit. Each blue, green, and red circle represents charges of $+2e$ (ionic), $-e/2$, and $-e$, respectively. White circles do not have charge. (b) Simulation of the charge density for $\lambda_x=\lambda_y=1$, $\gamma_x=\gamma_y=10^{-3}$, and $\delta = 10^{-3}$. Sites in the square plaquette orbitals marked with a $\ast$ closest to them represent half-charge contributions to the corner charge from the bulk orbitals. }
\label{fig:quad_charge}
\end{figure}

The different contributions to the corner charge can be clearly illustrated in the limit $\gamma_x=\gamma_y=0$, as shown in Fig.~\ref{fig:quad_charge}a. The large blue circles represent an ionic charge of $+2e$ per unit cell, which is constant across unit cells. Each unit cell has four electronic degrees of freedom, and thus, at half filling, each unit cell contributes 2 electrons. The sites connected by lines represent localized hybrid orbitals of the occupied electrons in the many-body ground state. In the bulk there are two localized square orbitals on each inter-cell plaquette, and the electrons in these orbitals have equal weight on each site of the plaquette. On the edges there are localized inter-cell dimer orbitals, one per dimer, where the electrons have equal weight on each site of the dimer.  In this limit where  $\gamma_x=\gamma_y=0$, each of the green sites in the bulk has an electronic charge of $-e/2$ coming from the two square-localized orbitals, each contributing $1/4$ of an electron. Similarly, each of the green sites on the edge SSH chains has an electronic charge of $-e/2$. Finally, there are two red and two white circles at the corners. Each of these degrees of freedom are decoupled from the rest in this limit, and therefore have exactly zero energy. These are the corner modes associated with the fractional corner charge in the topological quadrupole phase. Out of the four mid-gap corner modes, two should be filled at half-filling. A small value of $\delta \ll \lambda_{x,y}$ in \eqref{eq:QuadHamiltonian_with_delta} breaks this degeneracy in a manner where $C_2$ symmetry is preserved, and unambiguously specifies which modes are to be filled. When $\delta>0$, each of the red circles are occupied since they are at lower energy than the white circles. Thus each red circle contributes $-e$ to the charge at its corner unit cell. The white circles, on the other hand, remain unoccupied and do not contribute to the electronic corner charge. Notice that in the bulk and the edges, the positive atomic charge cancels the electronic charge. In the corner unit cells, however, there is a total charge of $\pm e/2$. Just as in the case of the edge polarization, the corner charge persists as long as $|\gamma_{x,y}/\lambda_{x,y}|<1,$ and drops to zero otherwise. An example of the distribution of electronic charge density for $\gamma_{x,y} \neq 0$ is shown in Fig.~\ref{fig:quad_charge}b.

The Hamiltonian \eqref{eq:QuadHamiltonian} in the limit $\gamma_{x,y}=0$, shown in Fig.~\ref{fig:quad_charge}a, illustrates two important characteristics of the quadrupole: (i) the fractionalization of the corner charge does not come from the edge polarizations alone, i.e., the contributions due to the non-trivial polarizations give an overall integer contribution to the corner charge. The fractionalization of the corner charge comes from the bulk charge density, and in this simple limit it comes from the corners of the plaquette orbitals that are closest to the corners (circles marked with a $\ast$ in Fig.~\ref{fig:quad_charge}a). (ii) Despite the existence of two topological edge dipole moments in the non-trivial phase, there is one zero-energy mode per corner. This is contrary to the conventional notion that a domain between two SSH chains, both of which are in the topological phase, should not trap a stable mid-gap mode. 
The apparent paradox is resolved because the protected topological corner mode is a simultaneous eigenstate of \emph{both} edge Hamiltonians along the $x$ and $y$-edges. This is evident in the pictorial representation of Fig.~\ref{fig:quad_charge}a, but can be confirmed in a more general setting by an explicit calculation of the corner mode eigenstate, as shown in Appendix \ref{sec:DomainWall}. \emph{Indeed, the corner states are not traditional 1D domain wall states, and represent a new mechanism to generate such modes on the boundary of a 2D system}. 

In order to understand how the boundary polarization arises in the topological quadrupole phase, it is useful to study the topology of the Wannier bands, and how this topological structure manifests at boundaries. We focus on this in the following three Sections. 

\subsubsection{Topological classes of the Wannier bands}
\label{sec:QuadTopologicalPhases}
Under $M_x$, $M_y$ and $C_2$, the Wannier-sector polarizations obey the relations in Eq. \eqref{eq:WannierPolarization_SymmetryConstraints}. In the quadrupole insulator $h^q({\bf k})$, these relations imply that: (i) all the Wannier-sector polarizations, $p^{\nu^\pm_y}_x$ and $p^{\nu^\pm_x}_y$, are quantized (see Eq.~\ref{eq:WannierPolarization_quantization}), and (ii) out of these four polarizations two are redundant due to $C_2$ symmetry. Specifically, we can re-write the third expression in \eqref{eq:WannierPolarization_SymmetryConstraints} as
\begin{align}
p^{\nu^+_y}_x+p^{\nu^-_y}_x & \stackrel{C_2}{=}0\;\;\;\mbox{mod 1} \nonumber\\
p^{\nu^+_x}_y+p^{\nu^-_x}_y & \stackrel{C_2}{=}0\;\;\;\mbox{mod 1},
\label{eq:WannierPolarizationUnderInversion}
\end{align}
which is  the statement that the total dipole moment vanishes, as is needed for a well defined quadrupole moment (see Section \ref{sec:multipole_translation_invariance}),
\begin{align}
{\bf p}=(p_x,p_y)= {\bf 0},
\end{align}
where $p_x = p^{\nu^+_y}_x+p^{\nu^-_y}_x$ and $p_y = p^{\nu^+_x}_y+p^{\nu^-_x}_y$. 

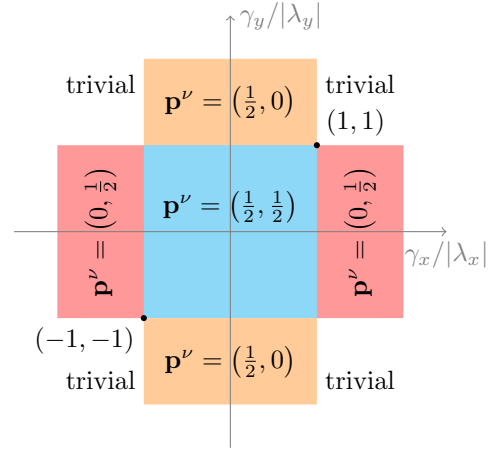
\begin{figure}
	\centering
	\begin{tikzpicture}[scale=1.15]
	
	\coordinate (pp) at (1,1);
	\coordinate (pn) at (1,-1);
	\coordinate (np) at (-1,1);
	\coordinate (nn) at (-1,-1);
	
	\fill [cyan!40!] (pp)--(pn)--(nn)--(np);
	\fill [orange!40!] (pp)--(1,2)--(-1,2)--(np);
	\fill [orange!40!] (pn)--(1,-2)--(-1,-2)--(nn);
	\fill [red!40!] (pp)--(2,1)--(2,-1)--(pn);
	\fill [red!40!] (np)--(-2,1)--(-2,-1)--(nn);
	
	\draw [->,gray] (-2.5,0)--(2.5,0) node[below] {$\gamma_x/|\lambda_x|$};
	\draw [->,gray] (0,-2.5)--(0,2.5) node[right] {$\gamma_y/|\lambda_y|$};
	
	\fill [black] (pp) circle (1pt) node[above right] {$(1,1)$};
	\fill [black] (nn) circle (1pt) node[below left] {$(-1,-1)$};
	
	\node[above] at (1.5,1.5) {trivial};
	\node[above] at (-1.5,1.5) {trivial};
	\node[below] at (1.5,-1.5) {trivial};
	\node[below] at (-1.5,-1.5) {trivial};		
	\node[align = center] at (0,1.5) {${\bf p}^{\nu}=\left(\frac{1}{2},0\right)$};
	\node[align = center] at (0,-1.5) { ${\bf p}^{\nu}=\left(\frac{1}{2},0\right)$};
	\node[align = center, rotate=90] at (1.5,0) { ${\bf p}^{\nu}=\left(0,\frac{1}{2}\right)$};
	\node[align = center, rotate=90] at (-1.5,0) { ${\bf p}^{\nu}=\left(0,\frac{1}{2}\right)$};
	\node[align = center, above] at (0,0) { ${\bf p}^{\nu}=\left(\frac{1}{2},\frac{1}{2}\right)$};
		
	\end{tikzpicture}
	\caption{(Color online)  Diagram of topological classes for the Wannier bands of the insulator $h^q({\bf k})$ with Bloch Hamiltonian Eq. \ref{eq:QuadHamiltonian}. The indices ${\bf p}^\nu$ are defined in Eq.~\ref{eq:WannierBandIndex}. The trivial class has ${\bf p}^\nu=(0,0)$.}
	\label{fig:quad_PhaseDiagram}
\end{figure}

Due to the relations \eqref{eq:WannierPolarizationUnderInversion}, only two independent Wannier-sector polarizations are necessary to specify the topological class of the Wannier bands, and we can define the index
\begin{align}
{\bf p}^{\nu} \equiv (p^{\nu^-_y}_x,p^{\nu^-_x}_y).
\label{eq:WannierBandIndex}
\end{align}
Under $M_x$, $M_y$ and $C_2$, the classification of the Wannier band topology in $h^q({\bf k})$ is $\mathbb{Z}_2 \times \mathbb{Z}_2$. A diagram of these classes is shown in Fig.~\ref{fig:quad_PhaseDiagram} as a function of the ratios $\gamma_x/|\lambda_x|$ and $\gamma_y/|\lambda_y|$.  The central square of the diagram in the ranges $\gamma_x/|\lambda_x| \in [-1,1]$ and $\gamma_y/|\lambda_y| \in [-1,1]$ is the region in parameter space having ${\bf p}^{\nu}=(1/2,1/2)$. Additionally, there are two regions in parameter space with ${\bf p}^{\nu}=(0,1/2),$ and two more with ${\bf p}^{\nu}=(1/2,0)$, as well as four regions in the trivial topological class ${\bf p}^{\nu}=(0,0)$.

\begin{figure}[t]%
	\centering
	\includegraphics[width=\columnwidth]{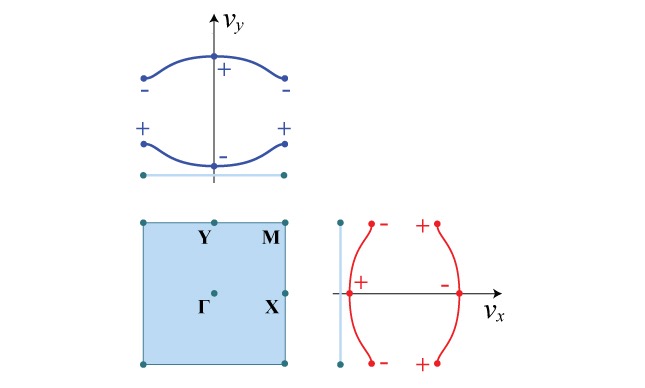}%
	\caption{(Color online)  Reflection eigenvalues $\alpha^\pm_{M_y}(k_x,k_{*y})$ (red signs) and $\alpha^\pm_{M_x}(k_{*x},k_y)$ (blue signs) of the Wannier bands of the occupied energy bands in the topological quadrupole phase. Here $k_{*x,y}=0,\pi$. For the unoccupied bands, all signs are inverted.}
	\label{fig:WannierBands_ReflectionEigenvalues}
\end{figure}

In the presence of reflection symmetries, ${\bf p}^\nu$ can be determined by the reflection representation of the Wannier bands at the high-symmetry lines 
\begin{align}
\hat{M}_y \ket{w^\pm_{x,(k_x,k_{*y})}} & =\alpha^{\pm}_{M_y}(k_x,k_{*y})\ket{w^\pm_{x,(k_x,k_{*y})}} \nonumber\\
\hat{M}_x \ket{w^\pm_{y,(k_{*x},k_y)}} & =\alpha^{\pm}_{M_x}(k_{*x},k_y)\ket{w^\pm_{y,(k_{*x},k_y)}} ,
\label{eq:WannierBands_reflection}
\end{align}
for $k_{*x,y}=0,\pi$ (see Section \ref{sec:WannierPolarization_quantization}). In the ${\bf p}^\nu=(1/2,1/2)$ class, the values of $\alpha^\pm_{M_x}(k_{*x},k_y)$ and $\alpha^\pm_{M_y}(k_x,k_{*y})$ are as shown in Fig.~\ref{fig:WannierBands_ReflectionEigenvalues}. For each of the topological classes of the Wannier bands shown in  Fig.~\ref{fig:quad_PhaseDiagram}, the corresponding $\alpha$ values are shown in Table \ref{tab:quad_phases_ReflectionEigenvalues}. Using these values we can evaluate the Wannier-sector polarizations according to Eq. \ref{eq:WannierPolarization_reflection}. In $h^q({\bf k})$, these lead to the Wannier-sector polarization (for the lower Wannier bands) shown in the phase diagram in Fig.~\ref{fig:quad_PhaseDiagram}.

\begin{table}
	\begin{center}
		\begin{tabular}{lcccc}
			&${\bf p}^\nu=(1/2,1/2)$\;\; & ${\bf p}^\nu=(1/2, 0)$ \;\; & ${\bf p}^\nu=(0, 1/2)$\\
			\hline \\
			$\alpha_{M_x}^-(0,k_y)$&$-$ & $-$ & $\pm$\\
			$\alpha_{M_x}^-(\pi,k_y)$&$+$ & $+$ & $\pm$\\
			\\
			$\alpha_{M_y}^-(k_x,0)$&$+$ & $\pm$ & $+$\\
			$\alpha_{M_y}^-(k_x,\pi)$&$-$ & $\pm$ & $-$\\
			\end{tabular}
	\end{center}
	\caption{Reflection eigenvalues of lower Wannier bands in the different topological classes of the Wannier bands of the Hamiltonian in Eq. \ref{eq:QuadHamiltonian}. The upper (lower) values in the ${\bf p}^\nu=(1/2,0)$ phase correspond to the upper (lower) blocks  in the phase diagram. The upper (lower) values in the ${\bf p}^\nu=(0,1/2)$ phase correspond to the blocks to the left (right) in the phase diagram in Fig.~\ref{fig:quad_PhaseDiagram}.}
	\label{tab:quad_phases_ReflectionEigenvalues}
\end{table}

Since calculating the $\alpha$ values requires finding the Wannier basis $\ket{w^\pm_{x, \bf k}}$ or $\ket{w^\pm_{y,\bf k}}$, and finding these bases requires calculating non-Abelian Wilson loops, it would be more convenient to have an easier alternative for determining the $\alpha$ values. Eq. \ref{eq:app_Wilson_loop_under_My} in Appendix \ref{sec:WilsonLoopsSymmetry} implies that the Wilson loop $\W_{x,\bf k}$, under $M_y$, obeys 
\begin{align}
[B_{M_y,(k_x,k_{*y})},\W_{x,(k_x,k_{*y})}]=0
\end{align} 
at $k_{*y}=0,\pi$. Thus, both the Wilson loop and the sewing matrix $B_{M_y,(k_x,k_{*y})}$, which encodes the reflection representation at the reflection invariant lines $(k_x,k_{*y})$ for $k_x \in [-\pi,\pi)$, can be simultaneously diagonalized and hence they have common eigenstates. Thus, at the reflection invariant lines in momentum space shown in Fig.~\ref{fig:2DBZ}, it is possible to label the subspace of occupied bands by their respective reflection eigenvalues. The subspaces along each of these lines can be divided into two sectors: one labeled by reflection eigenvalue $+i$, and another one labeled by a reflection eigenvalue $-i$. We can then calculate Abelian Wilson loops in each of these sectors separately. This will directly tell us the Wannier values associated with each reflection representation. A detailed calculation of this is shown in Appendix \ref{sec:ReflectionEigenvalues_WannierValues} in the limit $\gamma_x=\gamma_y=0$. 
Upon obtaining the values of $\alpha$ and their corresponding reflection eigenvalues, the Wannier-sector polarization can be calculated using Eq.  \ref{eq:WannierPolarization_reflection}. In some cases this method, which relies on resolving states according to their symmetry eigenvalues, will be easier to apply than the full non-Abelian formulation.

\subsubsection{Transitions between the topological classes of Wannier bands}
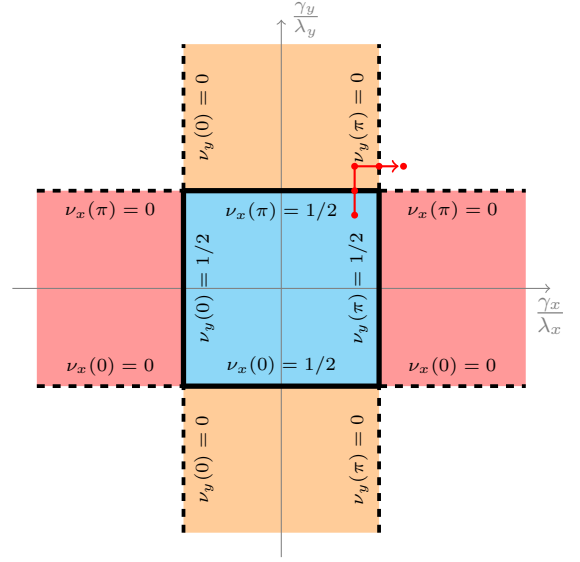
\begin{figure}
	\centering
	\begin{tikzpicture}[scale=1.3]
	
	\coordinate (pp) at (1,1);
	\coordinate (pn) at (1,-1);
	\coordinate (np) at (-1,1);
	\coordinate (nn) at (-1,-1);
	
	\fill [cyan!40] (pp)--(pn)--(nn)--(np);
	\fill [orange!40!] (pp)--(1,2.5)--(-1,2.5)--(np);
	\fill [orange!40!] (pn)--(1,-2.5)--(-1,-2.5)--(nn);
	\fill [red!40!] (pp)--(2.5,1)--(2.5,-1)--(pn);
	\fill [red!40!] (np)--(-2.5,1)--(-2.5,-1)--(nn);
	
	\draw [->,gray] (-2.75,0)--(2.75,0) node[below] {\large $\frac{\gamma_x}{\lambda_x}$};
	\draw [->,gray] (0,-2.75)--(0,2.75) node[right] {$\large \frac{\gamma_y}{\lambda_y}$};
	
	
	\draw[black,line width=.7mm] (np)--node[below]{\scriptsize $\nu_x(\pi)=1/2$} (pp);
	\draw[black,line width=.5mm,dashed] (-2.5,1)--node[below]{\scriptsize $\nu_x(\pi)=0$} (np);
	\draw[black,line width=.5mm,dashed] (pp)--node[below]{\scriptsize $\nu_x(\pi)=0$} (2.5,1);
	
	\draw[black,line width=.7mm] (nn)--node[above]{\scriptsize $\nu_x(0)=1/2$}(pn);
	\draw[black,line width=.5mm,dashed] (-2.5,-1)--node[above]{\scriptsize $\nu_x(0)=0$}(nn);
	\draw[black,line width=.5mm,dashed] (pn)--node[above]{\scriptsize $\nu_x(0)=0$}(2.5,-1);
	
	\draw[black,line width=.7mm] (nn) --node[below, rotate=90]{\scriptsize $\nu_y(0)=1/2$} (np);
	\draw[black,line width=.5mm,dashed] (-1,-2.5) --node[below, rotate=90]{\scriptsize $\nu_y(0)=0$} (nn);
	\draw[black,line width=.5mm,dashed] (np) --node[below, rotate=90]{\scriptsize $\nu_y(0)=0$} (-1,2.5);
	
	\draw[black,line width=.7mm] (pn) --node[above, rotate=90]{\scriptsize $\nu_y(\pi)=1/2$} (pp);
	\draw[black,line width=.5mm,dashed] (1,-2.5) --node[above, rotate=90]{\scriptsize $\nu_y(\pi)=0$} (pn);
	\draw[black,line width=.5mm,dashed] (pp) --node[above, rotate=90]{\scriptsize $\nu_y(\pi)=0$} (1,2.5);
	
	\draw[->,thick, red](0.75,0.75)--(0.75,1.25)--(1.2,1.25);
	\fill [red] (0.75,0.75) circle (1pt);
	\fill [red] (0.75,1) circle (1pt);
	\fill [red] (0.75,1.25) circle (1pt);
	\fill [red] (1,1.25) circle (1pt);
	\fill [red] (1.25,1.25) circle (1pt);
	
	\end{tikzpicture}
	\caption{(Color online) Diagram of Wannier band transitions in model \eqref{eq:QuadHamiltonian}.  At transitions there is Wannier gap closing at either $\nu_x(k_y=0,\pi)=0$ or $\nu_y(k_x=0,\pi)=0$ (dashed lines) or at either $\nu_x(k_y=0,\pi)=1/2$ or $\nu_y(k_x=0,\pi)=1/2$ (solid thick lines). }
	\label{fig:quad_phase_transitions}
\end{figure}

\begin{figure}%
	\centering
	\includegraphics[width=1\columnwidth]{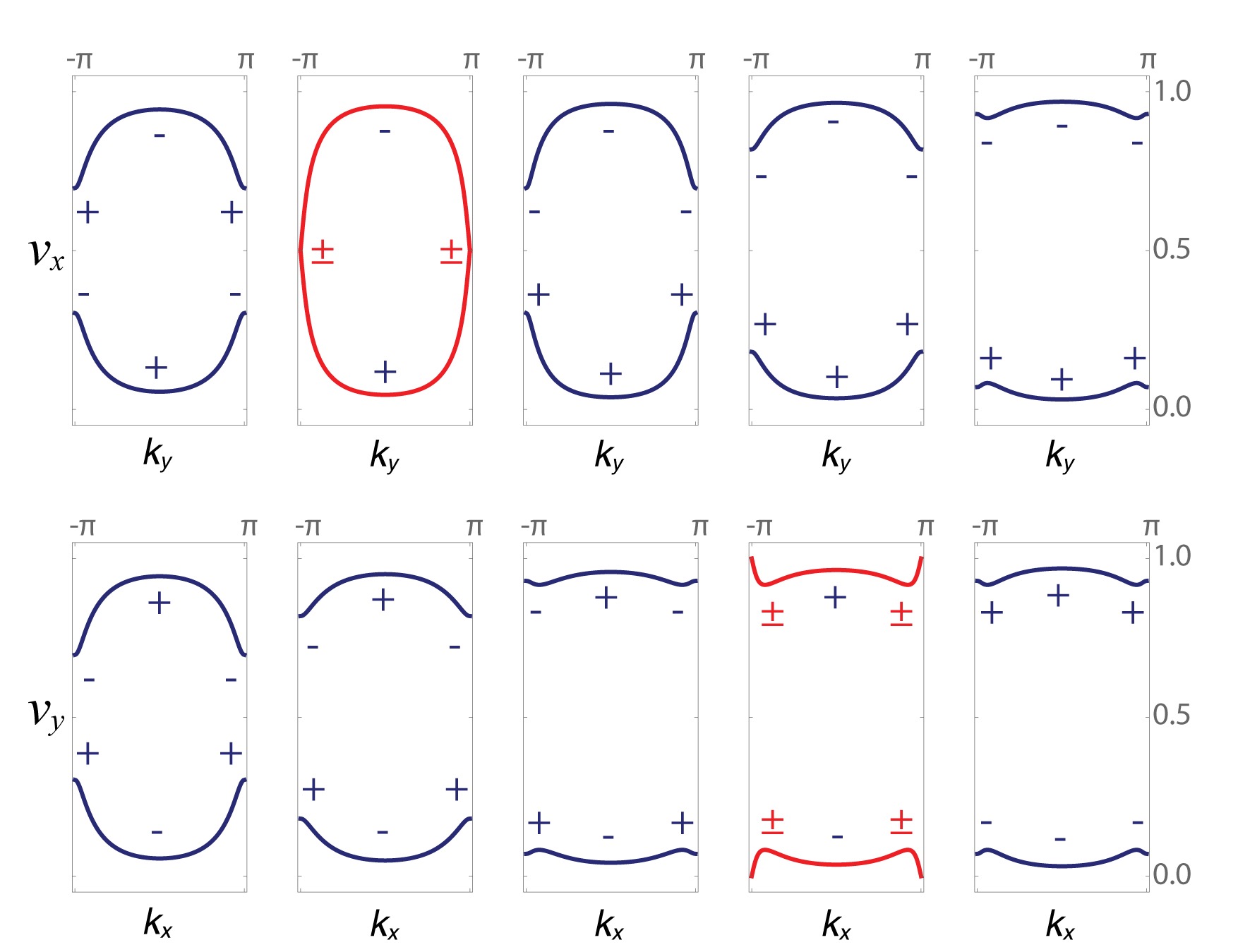}%
	\caption{(Color online)  Wannier bands $\nu_x(k_y)$ (first line) and $\nu_y(k_x)$ (second line). The parameters used are, from left to right $(\gamma_x/\lambda_x, \gamma_y/\lambda_y) = (0.75, 0.75) \rightarrow (0.75, 1) \rightarrow (0.75, 1.25) \rightarrow (1, 1.25) \rightarrow (1.25, 1.25)$, as in the red lines of Fig.~\ref{fig:quad_phase_transitions}.}
	\label{fig:IndirectPhaseTransition_WannierBands}
\end{figure}

At the transitions between topological classes indicated in Fig.~\ref{fig:quad_PhaseDiagram}, the Wannier gap closes. This is analogous to the closing of the energy gap in phase transitions between distinct symmetry-protected topological phases. Fig.~\ref{fig:quad_phase_transitions} shows the momentum and Wannier value locations at which the Wannier gap closes at all the topological class transitions in Fig.~\ref{fig:quad_PhaseDiagram}. With reflection symmetries $M_x$ and $M_y$, the Wannier gap can close at two Wannier values, $\nu=0$ or $1/2$. 
Consider, for example, the path in parameter space shown by the red line in Fig.~\ref{fig:quad_phase_transitions} that starts in the ${\bf p}^\nu=(1/2,1/2)$ class and ends in the trivial class ${\bf p}^\nu=(0,0)$ via the intermediate class ${\bf p}^{\nu}=(1/2,0)$. In Fig.~\ref{fig:IndirectPhaseTransition_WannierBands} we plot the two Wannier bands for each of the five Hamiltonians corresponding to the red dots in Fig.~\ref{fig:quad_phase_transitions}. In the ${\bf p}^\nu=(1/2,1/2)$ class, both Wannier bands are gapped and non-trivial. At the first transition point, the Wannier bands $\nu_x(k_y)$ become gapless at $k_y=\pi$ as they become twofold degenerate at $\nu_x(k_y=\pi)=1/2$. The bands $\nu_y(k_x)$, on the other hand, remain gapped at all $k_x \in (-\pi,\pi]$. On the other side of this transition, in the ${\bf p}^{\nu}=(1/2,0)$ class, the Wannier bands $\nu_x(k_y)$ become gapped again, but this time they have trivial topology (i.e., $p^{\nu^-_x}_y=0$). 

As the ${\bf p}^{\nu}=(1/2,0)$ class approaches the transition into the trivial class ${\bf p}^\nu=(0,0)$, another Wannier gap closing event occurs. This time, however, it is the $\nu_y(k_x)$ bands that close the gap at the $k_x=\pi$ point. They acquire the twofold degenerate value of $\nu_y(k_x=\pi)=0$. On the other side of the transition, in the trivial class, both Wannier bands are gapped and have trivial topology. At transitions from the ${\bf p}^\nu=(1/2,1/2)$ class to the trivial class ${\bf p}^\nu=(0, 0)$ other than the one indicated by the red line in Fig.~\ref{fig:IndirectPhaseTransition_WannierBands}, transitions can occur by closing the Wannier gaps of $\nu_y$ ($\nu_x$) at $k_x = 0$ or $\pi$ ($k_y=0$ or $\pi$), as indicated in Fig.~\ref{fig:quad_phase_transitions}. In all cases, however, the Wannier gap always closes at the value $\nu=1/2$ as it leaves the ${\bf p}^\nu=(1/2,1/2)$ class to either the ${\bf p}^{\nu}=(0,1/2)$ or ${\bf p}^{\nu}=(1/2,0)$ classes, and then at the value $\nu=0$ from these classes to the trivial ${\bf p}^{\nu}=(0,0)$ class. This is not accidental. 

\subsubsection{Bulk-boundary correspondence for Wannier bands and edge polarization}
\label{sec:EdgeTopology}

We saw that transitions between different topological classes of Wannier bands close the Wannier gap. Therefore, at a physical boundary between insulators having different Wannier classes, the Wannier gap is also expected to close. We denote this property as a bulk-boundary correspondence for Wannier bands. Consider, for example, the quadrupole insulator \eqref{eq:QuadRealSpaceHamiltonian} with closed boundaries along $x$ and open along $y$. By redefining the unit cell of this 2D crystal to get an effective 1D crystal with a unit cell of $N_{orb}\times N_y$ sites, we can obtain a Bloch Hamiltonian $h^q(k_x)$, with only one crystal momentum $k_x$. We write $h^q(k_x)$ to differentiate it from the Bloch Hamiltonian $h^q({\bf k})=h^q(k_x,k_y)$, which has full periodic boundaries. While $h^q({\bf k})$ has Wilson loops $\W_{x,{\bf k}}$ with Wannier bands $\nu^\pm_x(k_y)$, $h^q(k_x)$ has Wilson loops $\W_{x, k_x}$ with Wannier values $\nu^j_x$, for $j \in 1\ldots 2 N_y$ (at half filling), as defined in Section \ref{sec:EdgePolarization}. 

The bulk-boundary correspondence for Wannier bands then implies that, if the Wannier bands $\nu^\pm_x(k_y)$ of $h^q(\bf k)$ have non-trivial topology, i.e., if $p^{\nu^\pm_x}_y =1/2$, there will be $y$-edge-localized eigenstates of the Wilson loop $\W_{x, k_x}$ of $h^q(k_x)$ with eigenvalue $\nu_x=0$ or $1/2$ (as protected by $M_x$). We denote these as Wannier edge states. Hence, the Wannier values of the insulator with open boundaries along $y$ are gapless, and the gapless states are localized at the edges. If, on the other hand, $p^{\nu^\pm_x}_y =0$, then there are no edge-localized eigenstates of $\W_{x, k_x}$; i.e., no Wannier edge states. For the former case in which $p^{\nu^\pm_x}_y =1/2,$ the topological modes, localized at $R_y=0$ and $R_y=N_y$, are extended along $x$. Their Wannier value, which is either $\nu_x=0$ or $1/2,$  indicates their dipole moment along $x$. These modes are thus responsible for the edge-localized tangential polarization. Whether the topological Wannier edge modes have $\nu_x=0$ or $1/2$ (the only two allowed values under $M_x$) is determined by the value of $p^{\nu^\pm_y}_x$. 

The connection between the bulk property $p^{\nu^\pm_x}_y =1/2$ and the existence of Wannier edge states can be seen as follows. The bulk Wannier bands $\nu^\pm_x(k_y)$, being gapped, allow us two define two maximally-localized Wannier centers along $x$: one arising from the $\nu_x^+$ Wannier sector, which is localized to the right (horizontally) of the center of the unit cell, and another arising from the $\nu_x^-$ Wannier sector, which is localized to the left (horizontally) of the center of the unit cell. $p^{\nu^\pm_x}_y =1/2$ tells us that each of these Wannier centers is displaced by half of a unit cell along $y,$ hence giving rise to Wannier edge states when the system's boundaries are opened. When $p^{\nu^\pm_x}_y$ is exactly quantized, which Wannier center of the ${\nu^\pm_x}$ Wannier Bands is displaced up or down is ambiguous in the bulk of the insulator. However, when the boundaries are open, $C_2$ symmetry guarantees that there will be one Wannier edge mode on each of the lower and upper surfaces (as opposed to, say, two edge modes on the upper surface and none on the lower). 

\begin{figure*}[t]%
\centering
\includegraphics[width=\textwidth]{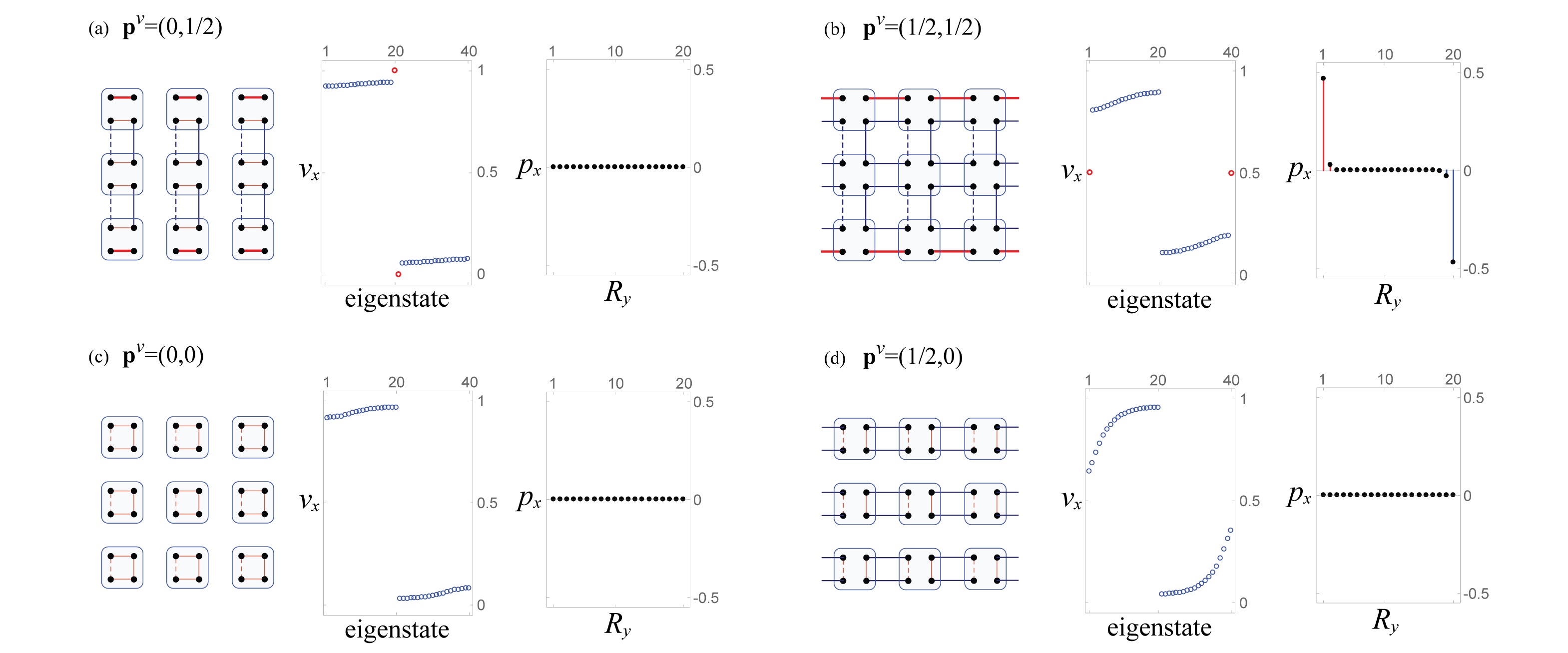}%
\caption{(Color online) Representative tight-binding Hamiltonians, Wannier values $\{\nu^j_x\}$, for $j \in 1 \ldots 2N_y$, and polarization $p_x(R_y)$ for all topological classes ${\bf p}^\nu$ of $h^q(\bf k)$. The tight-binding Hamiltonians have closed boundaries along $x$ and open along $y$, and are drawn with $\gamma_{x,y},\lambda_{x,y}=0$ whenever possible to ease the visualization of the edge states and tangential polarizations. The Wannier values $\{\nu^j_x\}$ and polarizations $p_x(R_y)$ are calculated in the same topological class as the Hamiltonians on their left, but with parameters $\lambda_x=\lambda_y=1$ in all four cases and $(\gamma_x,\gamma_y)=(1.25,0.25)$ in (a), $(\gamma_x,\gamma_y)=(0.25,0.25)$ in (b), $(\gamma_x,\gamma_y)=(1.25,1.25)$ in (c), and $(\gamma_x,\gamma_y)=(0.25,1.25)$ in (d). (a) and (b) have pairs of Wannier edge states (red thick circles) with degenerate values $\nu_x=0$ and $\nu_x=1/2$, respectively. (c) and (d) do not have Wannier edge states. Only (b) has a non-trivial edge polarization and quantized quadrupole moment. A value of $\delta=10^{-3}$ was set in the calculation of $p_x(R_y)$ for all cases to choose the sign of the quadrupole.}
\label{fig:quad_bulk-boundary}
\end{figure*}

The properties described above can be visualized in Fig.~\ref{fig:quad_bulk-boundary}, which shows the simplest tight-binding Hamiltonians (by setting $\gamma_{x,y}, \lambda_{x,y} = 0$ whenever possible) in all of the four Wannier classes of $h^q({\bf k})$. Next to each of the tight-binding lattices we show the Wannier values $\{\nu^j_x\}$, for $j \in 1\ldots2N_y$ (i.e., the eigenvalues of $\W_{x,k_x}$) for $h^q(k_x)$ in the same Wannier-band topological class, as well as their resulting polarizations $p_x(R_y)$.

In the cases in which $p^{\nu^\pm_x}_y =1/2$ (Fig.~\ref{fig:quad_bulk-boundary}a,b), $2N_y-2$ states are spread over the bulk, while two--the  topological Wannier edge states--are localized on the edges; one at $R_y=0$ and the other one at $R_y=N_y$. These two Wannier edge states have dipole moments along $x$ equal to the value of $p^{\nu^\pm_y}_x$. For $p^{\nu^\pm_y}_x=0$  (Fig.~\ref{fig:quad_bulk-boundary}a) the edge states have zero dipole moment; note by inspection of the tight-binding lattice that the edge states are each an SSH chain in the trivial phase. For $p^{\nu^\pm_y}_x=1/2$  (Fig.~\ref{fig:quad_bulk-boundary}b)  the edge states have non-trivial, half integer dipole moment; note by inspection of the tight-binding lattice that these edge states are each an SSH chain in the non-trivial phase.

When $p^{\nu^\pm_x}_y =0$, on the other hand, there are no Wannier edge states in the $W_{x,k_x}$ spectrum of $h^q(k_x)$. This is independent of the value of $p^{\nu^\pm_y}_x$ (Fig.~\ref{fig:quad_bulk-boundary}c,d).
Thus, whether the Wanner bands $\nu^\pm_y(k_x)$ are trivial or topological (Fig.~\ref{fig:quad_bulk-boundary}c,d, respectively), the absence of Wannier edge states automatically leads to a vanishing dipole moment along $x$ at both $y$-edges.

From the analysis above it follows that, of all the four classes in $h^q(\bf k)$, only the class ${\bf p}^\nu=(1/2,1/2)$ exhibits non-trivial edge polarization. Correspondingly, only this class has corner-localized charges of $1/2$ when the boundaries along both $x$ and $y$ are open. Hence, only the ${\bf p}^\nu=(1/2,1/2)$ class has non-trivial quantized quadrupole moment $q_{xy}=1/2$, while all the other three classes have trivial quantized quadrupole moment $q_{xy}=0$.



\subsubsection{Topological phases in the quadrupole insulator}
Now that we have identified which Wannier topological classes (in the presence of $M_x$ and $M_y$) have non-trivial quadrupole moments, we look into the symmetry-protected topological quadrupole phases. The quadrupole insulator has a quantized phase protected not only by $M_x$ and $M_y$, but also by $C_4$ symmetry. That is, we could choose either the combination of $M_x$ and $M_y$ or $C_4$ to protect the quadrupole moment. We analyze these two types of protection separately. 

\paragraph{Reflection symmetric phases}

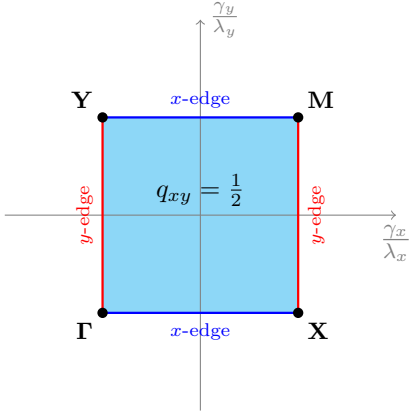
\begin{figure}[t]
	\centering
	\begin{tikzpicture}[scale=1.3]
	
	\coordinate (pp) at (1,1);
	\coordinate (pn) at (1,-1);
	\coordinate (np) at (-1,1);
	\coordinate (nn) at (-1,-1);
	
	\fill [cyan!40] (pp)--(pn)--(nn)--(np);
	\node[align = center, above] at (0,0) {$q_{xy}=\frac{1}{2}$};
	
	\draw [->,gray] (-2,0)--(2,0) node[below] {\large $\frac{\gamma_x}{\lambda_x}$};
	\draw [->,gray] (0,-2)--(0,2) node[right] {$\large \frac{\gamma_y}{\lambda_y}$};
	
	\draw[blue,line width=.3mm] (np)--node[above]{\scriptsize $x$-edge} (pp);	
	\draw[blue,line width=.3mm] (nn)--node[below]{\scriptsize $x$-edge}(pn);
	\draw[red,line width=.3mm] (nn) --node[above, rotate=90]{\scriptsize $y$-edge} (np);
	\draw[red,line width=.3mm] (pn) --node[below, rotate=90]{\scriptsize $y$-edge} (pp);

	\fill [black] (pp) circle (1.5pt) node[above right] {\small ${\bf M}$};
	\fill [black] (pn) circle (1.5pt) node[below right] {\small ${\bf X}$};
	\fill [black] (np) circle (1.5pt) node[above left] {\small ${\bf Y}$};
	\fill [black] (nn) circle (1.5pt) node[below left] {\small ${\bf \Gamma}$};	
	
	\end{tikzpicture}
	\caption{(Color online) Phase diagram of the quadrupole insulator $h^q({\bf k})$ with Hamiltonian \eqref{eq:QuadHamiltonian}.  Transitions close the bulk energy gap when $C_4$ symmetry is preserved at the indicated points of the BZ. Transitions close the edge energy gap when $M_x$ and $M_y$ reflections are preserved at the indicated edges. }
	\label{fig:QuadPhases}
\end{figure}

A diagram of the topological quadrupole phases of the insulator $h^q(\bf k)$ is shown in Fig.~\ref{fig:QuadPhases} as a function of the ratios $\gamma_x/|\lambda_x|$ and $\gamma_y/|\lambda_y|$.  The central square of the diagram in the ranges $\gamma_x/|\lambda_x| \in (-1,1)$ and $\gamma_y/|\lambda_y| \in (-1,1)$ has a quantized quadrupole moment $q_{xy}=1/2$, as it has the boundary signatures of Eq.~\ref{eq:QuadSignatures}. Outside of this region, there is a trivial quantized quadrupole $q_{xy}=0$.

\begin{figure}[t]%
	\centering
	\includegraphics[width=\columnwidth]{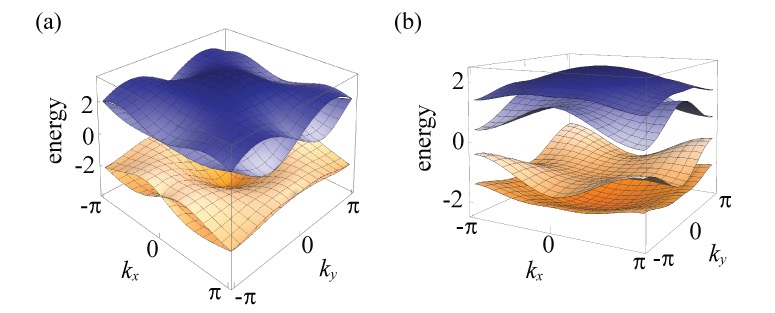}%
	\caption{(Color online) Energy spectrum for quadrupole models that preserve (a) only $M_x$ and $M_y$ symmetries (Hamiltonian described in Appendix~\ref{sec:app_QuadBreakingSymmetriesSimulation}), and (b) only $C_4$ symmetry (Hamiltonian described in Appendix~\ref{sec:QuadC4}). Notice that in both cases, $C_2$ is also preserved.}
	\label{fig:quad_spectra_quaternion_c4}
\end{figure}

Following the paradigm for the topological characterization of crystalline symmetry-protected topological phases \cite{fu2007,fu2011,hughes2011inversion,turner2012,fang2012,teo2013, benalcazar2014,bradlyn2017}, we look at the symmetry group representations that the subspace of occupied bands take at the high-symmetry points of the BZ. The point group of the quadrupole insulator $h^q({\bf k})$ is the quaternion group \eqref{eq:quad_quaternion_group}, which has the character table shown in Appendix \ref{sec:CharacterQuaternion}. This group has 4 one-dimensional representations and 1 two-dimensional representation. The points of the BZ invariant under this group are $\bf k_*=\Gamma, X, Y, M$. At all of these points, there is a twofold degeneracy in the spectrum protected by the non-commutation of the $\hat{M}_x$ and $\hat{M}_y$ operators (see Appendix~\ref{sec:EnergyDegeneracyProtection}). Symmetry-allowed perturbations can be added to lift most of the twofold degeneracy of the bulk energy bands of $h^q({\bf k})$ (given in  Eq.~\ref{eq:quad_energy}), however the degeneracy will persist at all $\bf k_*$ points of the BZ (Fig.~\ref{fig:quad_spectra_quaternion_c4}a). 

Since the occupied bands at each of the $\bf k_*$ points lies in a two-dimensional representation of the point group, it is expected, from this point of view, that the energy bands are twofold degenerate in energy at the $\bf k_*$ points. However, since the group admits only one two-dimensional representation, one cannot construct the typical bulk crystalline topological invariants (the representations are the same at each $\bf k_*$), and hence the topological structure is `hidden' from the point of view of the energy bands. Instead, the topology in the presence of $M_x$ and $M_y$ is captured by the topological classes of the Wannier bands. From the character table in Appendix~\ref{sec:CharacterQuaternion} it follows that the reflection and $C_2$ eigenvalues of the occupied energy bands at each $\bf k_*$ all come in  $(+i,-i)$ pairs. Indeed, these values are necessary to have gapped Wannier bands, $\nu^\pm_x(k_y)$ and $\nu^\pm_y(k_x)$, as shown in Fig.~\ref{fig:WannierBands} (pairs of reflection or $C_2$ eigenvalues other than $(+i,-i)$, inevitably lead to at least one pair of Wannier bands being gapless, see Section~\ref{sec:GappedWannierBands_conditions}). The Wannier bands can belong to different topological classes, as discussed in Section~\ref{sec:QuadTopologicalPhases}, and are identified by a pair of Wannier-sector polarizations as in Eq. \eqref{eq:WannierBandIndex}. Since the $q_{xy}=1/2$ phase coincides with the Wannier band topology having ${\bf p}^\nu=(1/2,1/2)$, we can construct the index for the reflection symmetry-protected quadrupole phase as
\begin{align}
q_{xy} \stackrel{M_x,M_y}{=} p^{\nu^-_x}_y p^{\nu^-_y}_x + p^{\nu^+_x}_y p^{\nu^+_y}_x
\label{eq:QuadIndexWannierPolarization}
\end{align}
which takes values
\begin{align}
q_{xy} \stackrel{M_x,M_y}{=} \begin{cases}
0 & \mbox{if trivial}\\
1/2 & \mbox{if non-trivial}
\end{cases}.
\label{eq:QuadIndexWannierPolarizationCases}
\end{align}
The expression \eqref{eq:QuadIndexWannierPolarization} resembles the classical expression for a quadrupole; it is the multiplication of two `coordinates' (one measures the displacement along $x$ and the other one along $y$) added over the charges (two electrons in this case). Due to the constraint \eqref{eq:WannierPolarizationUnderInversion}, the index simplifies to
\begin{align}
q_{xy} \stackrel{M_x,M_y}{=} 2 p^{\nu^-_x}_y p^{\nu^-_y}_x.
\label{eq:QuadIndexWannierPolarization2}
\end{align}

Accordingly, both the edge polarizations and the corner charges of the insulator in this non-trivial SPT phase are quantized. Appendix \ref{sec:app_QuadBreakingSymmetriesSimulation} shows simulations that break all symmetries except the quantizing reflection symmetries $M_x$ and $M_y$ to verify the quantization of the corner charge and edge polarization, as long as the symmetry-breaking perturbations do not close the Wannier gaps.  The protection due to Wannier band topology--instead of bulk energy band topology--is a new mechanism of topological protection. This protection implies that, at symmetry-preserving boundaries between a topological phase and the vacuum, edges are not required to be gapless, i.e., it allows for the possibility of gapped edges. These edges, however, are topological themselves. The Wannier band protection mechanism then leads to the existence of gapped, symmetry-preserving edges which are topological. This idea can be extended far beyond this example so that we can generate bulk topological phases with many types of gapped, surface SPTs.  In some sense these phases represent a simpler version of the gapped, symmetry preserving surfaces of 3D strong topological phases which must instead be topologically ordered\cite{vishwanath2013physics,bonderson2013,wang2013gapped,chen2014symmetry,metlitski2015symmetry}. 

\paragraph{$C_4$ symmetric phases}

In the presence of $C_4$ symmetry, the quadrupole minimal model $h^q({\bf k})$, with Hamiltonian \eqref{eq:QuadHamiltonian}, is in either the trivial phase, $q_{xy}=0$, or the topological phase, $q_{xy}=1/2$. Unlike the case in which symmetries $M_x$ and $M_y$ protect the quadrupole moment, under $C_4$ symmetry the quadrupole moment is protected by the topology of the \emph{bulk energy bands}. Accordingly, a topological index can be built by comparing the rotation representations of the subspace of occupied energy bands at the $C_4$-symmetric points of the BZ, ${\bf k}_{\star}={\bf \Gamma}$ and ${\bf M}$ \cite{teo2013, benalcazar2014}. Since $C_4$ symmetry only has one-dimensional representations, it does not protect degeneracies in the energy bands of $h^q({\bf k})$. An example of this lack of protection is shown in Fig.~\ref{fig:quad_spectra_quaternion_c4}b, where we show the energy bands for a Hamiltonian based on $h^q(\bf k)$ that has flux other than $\pi$ at each plaquette. This modification in the Hamiltonian breaks $M_x$ and $M_y$ but preserves $C_4$, as detailed in Appendix~\ref{sec:QuadC4}. 

In order to define the topological index, consider $h^q(\bf k_\star)$, with $\gamma_x=\gamma_y=\gamma$ and $\lambda_x=\lambda_y=\lambda$, which obeys
\begin{align}
[\hat{r}_4, h^q({\bf k_{\star}})]=0.
\end{align}
From this it follows that the eigenstates of $h^q(\bf k_\star)$ are also eigenstates of the rotation operator. Thus, the occupied states at ${\bf k}_{*}$ obey
\begin{align}
\hat{r}_4 \ket{u^n_{{\bf k}_{\star}}} = r^n_4({\bf k}_{\star}) \ket{u^n_{{\bf k}_{\star}}},
\end{align}
where $n=1,2$ labels the occupied states, and $r^n_4({\bf k}_{*})$ is the rotation eigenvalue of the $n^{th}$ occupied state at the $C_4$-invariant momentum ${\bf k}_{\star}$. 

\begin{figure}
	\subfigure[]{
		\begin{tikzpicture}[scale=.6]
		\draw[<->,gray](-1.5,0)--(1.5,0) node[below]{\scriptsize $k_x$};
		\draw[<->,gray](0,-1.5)--(0,1.5) node[left]{\scriptsize $k_y$};
		\coordinate (p1) at (1,1);
		\coordinate (p2) at (1,-1);
		\coordinate (p3) at (-1,-1);
		\coordinate (p4) at (-1,1);
		\fill[blue!10!](1,1)--(1,0)--(0,0)--(0,1)--(1,1);
		\draw[](p1)--(p2)--(p3)--(p4)--(p1);
		\fill [blue] (0,0) circle (2.5pt) node[below left] {\scriptsize ${\bf \Gamma}$};
		\fill [red] (p1) circle (2.5pt) node[below left] {\scriptsize ${\bf M}$};
		\fill [white] (-2.2,0) circle (1.5pt);
		\fill [white] (2.2,0) circle (1.5pt);
		\fill [white] (0,-2.2) circle (1.5pt);
		\fill [white] (0,2.2) circle (1.5pt);
		\end{tikzpicture}}
	\subfigure[]{
		\begin{tikzpicture}[scale=.6]
		\draw[<->,gray](-1.5,0)--(1.5,0) node[right]{\scriptsize Re};
		\draw[<->,gray](0,-1.5)--(0,1.5) node[above]{\scriptsize Im};
		\draw [gray] (0,0) circle (1);
		\fill [red] (0.707,0.707) circle (3pt) node[above right] {\scriptsize $r^+_4({\bf M})$};
		\fill [blue] (-0.707,0.707) circle (3pt) node[above left] {\scriptsize $r^-_4({\bf \Gamma})$};
		\fill [blue] (-0.707,-0.707) circle (3pt) node[below left] {\scriptsize $r^+_4({\bf \Gamma})$};
		\fill [red] (0.707,-0.707) circle (3pt) node[below right] {\scriptsize $r^-_4({\bf M})$};
		\fill [white] (0,-2.2) circle (1pt);
		\fill [white] (-2.2,0) circle (1.5pt);
		\fill [white] (2.2,0) circle (1.5pt);
		\fill [white] (0,-2.2) circle (1.5pt);
		\fill [white] (0,2.2) circle (1.5pt);
		\end{tikzpicture}}\\
	\subfigure[]{
		\begin{tikzpicture}[scale=.6]
		\draw[<->,gray](-1.5,0)--(1.5,0) node[right]{\scriptsize Re};
		\draw[<->,gray](0,-1.5)--(0,1.5) node[above]{\scriptsize Im};
		\draw [gray] (0,0) circle (1);
		\fill [magenta] (-0.707,0.707) circle (3pt); 
		\node[align = left] at (-1.407,1) {\scriptsize \color{red} $r^-_4({\bf M})$ \\ \scriptsize \color{blue} $r^-_4({\bf \Gamma})$};
		\fill [magenta] (-0.707,-0.707) circle (3pt);
		\node[align = left] at (-1.407,-1){\scriptsize \color{blue} $r^+_4({\bf \Gamma})$ \\ \scriptsize \color{red} $r^+_4({\bf M})$};
		\fill [white] (0,-2) circle (1pt);
		\fill [white] (-2.2,0) circle (1.5pt);
		\fill [white] (2.2,0) circle (1.5pt);
		\end{tikzpicture}}
	\subfigure[]{
		\begin{tikzpicture}[scale=.6]
		\draw[<->,gray](-1.5,0)--(1.5,0) node[right]{\scriptsize Re};
		\draw[<->,gray](0,-1.5)--(0,1.5) node[above]{\scriptsize Im};
		\draw [gray] (0,0) circle (1);
		\fill [magenta] (0.707,0.707) circle (3pt);
		\node[align = right] at (1.4,1) {\scriptsize \color{red} $r^+_4({\bf M})$ \\ \scriptsize \color{blue} $r^+_4({\bf \Gamma})$};
		\fill [magenta] (0.707,-0.707) circle (3pt);
		\node[align = right] at (1.4,-1){\scriptsize \color{blue} $r^-_4({\bf \Gamma})$ \\ \scriptsize \color{red} $r^-_4({\bf M})$};
		\fill [white] (0,-2) circle (1pt);
		\fill [white] (-2.2,0) circle (1.5pt);
		\fill [white] (2.2,0) circle (1.5pt);
		\end{tikzpicture}}
	\caption{(Color online) Rotation representations for the occupied bands of the  quadrupole model \eqref{eq:QuadHamiltonian} in the presence of $C_4$ symmetry \eqref{eq:QuadC4RotationSymmetries}. (a) BZ and its $C_4$-invariant momenta ${\bf k}_{*} = {\bf \Gamma}, {\bf M}$ (b-d) Let $\lambda_x=\lambda_y=1$, $\gamma_x = \gamma_y = \gamma$.  $C_4$ rotation eigenvalues at ${\bf k}_{*}$ for (b) quadrupole phase, $|\gamma| < 1$, (c) trivial phase with $\gamma>1$, and (d) trivial phase with $\gamma<-1$.}
	\label{fig:quadRotationEigenvalues}
\end{figure}
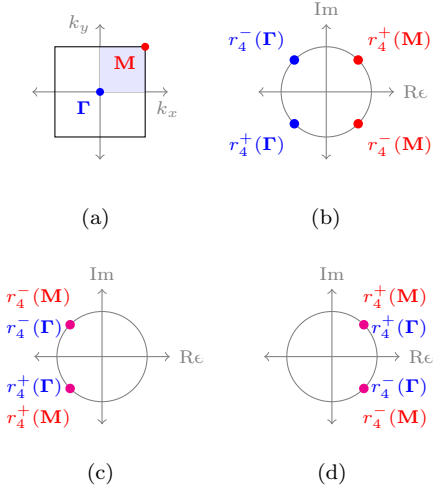

To build the index, we first recall that the eigenvalues of the $C_2$ operator, $\hat{r}_2=\hat{r}_4^2$, for the occupied bands at the $\bf k_\star$ are $\pm i$  (due to the $\pi$-flux per unit cell, the $C_2$ operator squares to $-1$ even for spinless systems). Thus, the $C_4$ eigenvalues in the occupied energy bands come in pairs $r^+_4(\bf k_{\star})$, $r^-_4(\bf k_{\star})$, such that, 
\begin{align}
\left(r^+_4(\bf k_{\star}) \right)^2 &= +i\nonumber\\
\left(r^-_4(\bf k_{\star}) \right)^2 &= -i.
\end{align}
We can then take either the $r^+_4(\bf k_{\star})$ values or the $r^-_4(\bf k_{\star})$ values to construct the index 
\begin{align}
e^{i 2\pi q_{xy}} \stackrel{C_4}{=} r^+_4({\bf M}) r^{+*}_4({\bf \Gamma}) = r^-_4({\bf M}) r^{-*}_4({\bf \Gamma}),
\label{eq:QuadIndexRotationEigenvalues}
\end{align}
which takes the values of $e^{i 2\pi q_{xy}}=\pm 1$ in the trivial or topological quadrupole phases, respectively. This corresponds to quantized values of the quadrupole of
\begin{align}
q_{xy} \stackrel{C_4}{=}\begin{cases}
0 & \mbox{if trivial}\\
1/2 & \mbox{if non-trivial}
\end{cases}.
\label{eq:QuadIndexRotationEigenvaluesCases}
\end{align}
This index is independent of the choice of $C_4$ rotation operator, provided that the same operator is used at both $\bf \Gamma$ and $\bf M$. For example, for the choice of $\hat{r}_4$ of \eqref{eq:QuadC4RotationSymmetries}, which obeys $\hat{r}^4_4=-\tau_0 \otimes \tau_0$, its eigenvalues are $e^{\pm i \pi/4}$, $e^{\pm i 3\pi/4}$, and the rotation eigenvalues  in the trivial and topological quadrupole phases of \eqref{eq:QuadHamiltonian} under $C_4$ symmetry are schematically shown in Fig.~\ref{fig:quadRotationEigenvalues}. 

The bulk energy gap of $h^q({\bf k})$ closes at the $C_4$-preserving transitions. In such transitions, the bulk energy gap closes at $\bf \Gamma$, $\bf X$, $\bf Y$, or $\bf M$, as indicated by the dots in Fig.~\ref{fig:QuadPhases}. At these transitions, the rotation eigenvalues of the occupied energy bands change from the configuration in Fig.~\ref{fig:quadRotationEigenvalues}b to those in either (c) or (d). 

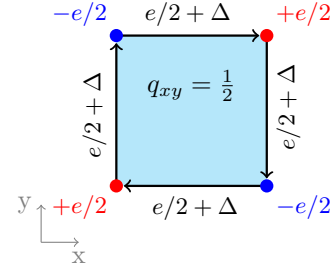
\begin{figure}
	\centering
	\begin{tikzpicture}[scale=1]
	
	\coordinate (pp) at (1,1);
	\coordinate (pn) at (1,-1);
	\coordinate (np) at (-1,1);
	\coordinate (nn) at (-1,-1);
	
	\fill [cyan!50!,opacity=0.5] (pp)--(pn)--(nn)--(np);
	\node[align = center, above] at (0,0) {$q_{xy}=\frac{1}{2}$};
	
	\draw [->,gray] (-2,-1.75)--(-1.5,-1.75) node[below] {x};
	\draw [->,gray] (-2,-1.75)--(-2,-1.25) node[left] {y};
	
	\draw[->,black,line width=.3mm] (np)--node[above]{\small $e/2+\Delta$} (.9,1);	
	\draw[->,black,line width=.3mm] (pn)--node[below]{\small $e/2+\Delta$}(-.9,-1);
	\draw[->,black,line width=.3mm] (nn) --node[above, rotate=90]{\small $e/2+\Delta$} (-1,.9);
	\draw[->,black,line width=.3mm] (pp) --node[below, rotate=90]{\small $e/2+\Delta$} (1,-.9);

	\fill [red] (pp) circle (2.5pt) node[above right] {\small $+e/2$};
	\fill [blue] (pn) circle (2.5pt) node[below right] {\small $-e/2$};
	\fill [blue] (np) circle (2.5pt) node[above left] {\small $-e/2$};
	\fill [red] (nn) circle (2.5pt) node[below left] {\small $+e/2$};	
	
	\end{tikzpicture}
	\caption{(Color online) Schematic of a $C_4$-symmetric insulator that breaks $M_x$ and $M_y$ in the topological phase. It has quantized corner charges $\pm e/2$ but not quantized edge polarizations $e/2+\Delta$, in a $C_4$ symmetric pattern.}
	\label{fig:QuadC4EdgesAndCorners}
\end{figure}

We finally point out that, since $C_4$ symmetry does not protect the Wannier-sector polarizations, the quantization of the edge polarizations is not guaranteed in the presence of $C_4$ symmetry. For example, if $C_4$-symmetric perturbations having hopping terms between nearest neighbor unit cells are added, the observables of the Hamiltonian can be modified as schematically shown in Fig.~\ref{fig:QuadC4EdgesAndCorners} (we note that if only fluxes other than $\pi$ are put on each plaquette to break the reflection symmetries of $h^q({\bf k})$ then the edge polarizations remain quantized). Even though the edge polarizations are not quantized, (i) the corner charge remains quantized, and (ii) the relation between edge polarizations and corner charge still implies the existence of a \emph{quantized} quadrupole moment, on top of which edge dipoles of magnitude $\Delta$ are overlapped in a $C_4$ symmetric pattern (see Section~\ref{sec:BulkVSBoundaryMoments}).

\subsubsection{Phase transitions in the quadrupole insulator}
The closing of either the energy gap or the Wannier gap is a property dictated by the bulk band parameters (for the latter we note that calculations of both sets of hybrid Wannier bands requires the bulk wavefunctions). In this section, we describe how the phase transitions in $h^q({\bf k})$ manifest at the boundaries. In the following description, we set $\lambda_x=\lambda_y=1$ for simplicity.

\begin{figure}[t!]%
	\centering
	\includegraphics[width=\columnwidth]{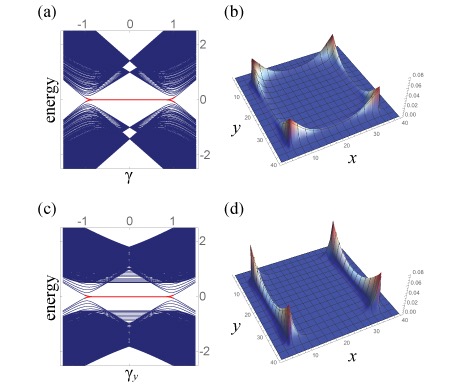}
	\caption{(Color online)  Two types of quadrupole phase transitions for Hamiltonian \eqref{eq:QuadHamiltonian} with full open boundaries. For (a,b) we have a $C_4$, $M_x$, and $M_y$ symmetric Hamiltonian. (a) Energy bands as a function of $\gamma=\gamma_x=\gamma_y$. (b) Probability density function of the zero-energy modes as the system approaches the transition with $C_4$, $M_x$, $M_y$ symmetric Hamiltonian, having $(\gamma_x,\gamma_y)=(0.75,0.75)$. (c,d) have $M_x$, and $M_y$ but not $C_4$ (c) energy bands as a function of $\gamma_y$ while fixing $\gamma_x=0.5$. (d) Probability density function of the zero-energy modes as the system approaches the transition for a Hamiltonian having $(\gamma_x,\gamma_y)=(0.5,0.75)$. In the simulations, there are $40 \times 40$ unit cells. For the purpose of illustration, unit cells in the range $N_{x,y} \in [5,34]$ are in the topological quadrupole phase with $(\gamma_x,\gamma_y)$ as indicated. Unit cells outside of $N_{x,y} \in [5,34]$ are in the trivial phase with $(\gamma_x,\gamma_y)$=(2,2). All unit cells have $\lambda_x=\lambda_y=1$.}
	\label{fig:quad_PhaseTransition1}
\end{figure}

We start with the $C_4$, $M_x,$ and $M_y$-symmetric transition with full open boundaries. The energy bands for this system as a function of the parameter $\gamma=\gamma_x=\gamma_y$ are shown in Fig.~\ref{fig:quad_PhaseTransition1}a. In the topological phase, the red lines denote the \emph{corner-localized, four-fold degenerate} modes, which are characteristic of the topological quadrupole phase, as seen in Fig. \ref{fig:quad_PhaseTransition1}b. During the transition, the \emph{bulk} energy gap closes and the corner localized states hybridize and fuse into the bulk and are no longer present in the trivial phase.

If we now drive the transition by varying $\gamma_y$ while keeping $|\gamma_x|<1$, as in Fig~\ref{fig:quad_PhaseTransition1}c, only the energy gap of the edge parallel to $y$ closes. Consequently, as the transition is approached, the four corner modes hybridize in pairs along the edge parallel to $y$, as seen in Fig. \ref{fig:quad_PhaseTransition1}d. Recall that this phase transition is associated with a closing of the Wannier gap at $\nu_x(k_y=\pi)=1/2$ when the system has periodic boundary conditions (second column, first row, of Fig.~\ref{fig:IndirectPhaseTransition_WannierBands}). Hence, we conclude that a gap closing of the Wannier bands results in an energy gap closing in the 1D \emph{$x$-edge} Hamiltonian\footnote{Although a bulk transition implies closing the edge energy gap, the contrary is not true, i.e., a closing of the edge energy gap as a result of a phase transition \emph{of the edge} does not imply a transition in the bulk; in this later case, the edge transition just causes the non-trivial edge at the boundary of the quadrupole to recede in order to exclude the outermost trivial edge that underwent the transition.}. This relation between the bulk property of Wannier gap closing and the energy gap closing of the edge can be inferred from the adiabatic mapping connecting the Wannier bands to the Hamiltonian of the edge as detailed in Section~\ref{sec:WannierBandsEdgeHamiltonian}. 

\begin{figure}[t!]%
	\centering
	\includegraphics[width=\columnwidth]{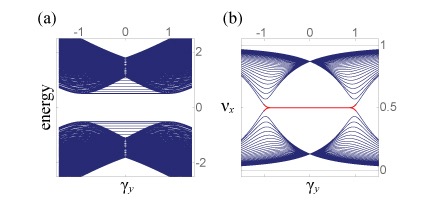}
	\caption{(Color online)  (a) Energy bands and (b) Wannier bands $\nu^j_x$, for $j \in 1\ldots 2N_y$,  with closed boundaries along $x$ and open along $y$ as a function of $\gamma_y$ while fixing $\gamma_x=0.5$. In all plots $\lambda_x=\lambda_y=1$. Red lines indicate the 2-fold degenerate states with polarization 1/2 localized at the open edges.}
	\label{fig:quad_PhaseTransition2}
\end{figure}

Indeed, one can verify that the energy gap closing occurs along the $x$-edge by repeating the calculation in (c), but for a quadrupole insulator with periodic boundaries along $x$. This is shown in Fig. \ref{fig:quad_PhaseTransition2}a. In contrast to what happens in Fig~\ref{fig:quad_PhaseTransition1}c, the energy bands in  Fig. \ref{fig:quad_PhaseTransition2}a do not close the gap. In this setup, in which boundaries along $x$ ($y$) are closed (open), this transition can be visualized by plotting the Wannier bands $\nu^j_x$, for $j \in 1 \ldots 2N_y$, as a function of the parameter $\gamma_y$. This is shown in Fig.~\ref{fig:quad_PhaseTransition2}b. The red line in this figure indicates the \emph{twofold Wannier-degenerate} states that are localized \emph{at the two opposite $y$-edges}, having Wannier value of $1/2$. These states are characteristic of the quadrupole phase, and are responsible for the quantized edge polarizations (see Section~\ref{sec:EdgeTopology}). Analogous to the corner-localized modes in the energy plots, the edge-localized modes in the Wannier plots hybridize as the Wannier gap closes, and fuse into the bulk outside of the quadrupole phase. Physically, this plot illustrates that (even in the absence of corners) the edge polarizations are clear signatures that persist only as long as the bulk is in the quadrupole topological phase. The mapping described in Section~\ref{sec:WannierBandsEdgeHamiltonian} that adiabatically maps the Wannier bands of Fig.~\ref{fig:quad_PhaseTransition2}b to the edge energies in Fig.~\ref{fig:quad_PhaseTransition1}c is consistent with this phenomenology.

In the phase diagram in Fig. \ref{fig:QuadPhases} for the quadrupole topological phases of $h^q({\bf k})$, the blue and red lines indicate the edges at which the energy bands close for $M_x$, $M_y$-preserving phase transitions, and the black dots indicate the points of the BZ at which the bulk energy bands close for $C_4$-preserving phase transitions. 

Let us make some final notes about the multi-critical nature of the bulk phase transition. We see that we only find a bulk phase transition in our phase diagram when $C_4$ symmetry is preserved. However, our phase diagram is implicitly assuming that both mirror symmetries are preserved since every point in the phase diagram has mirror symmetry by design. Hence, if we have both mirror symmetries \emph{and} $C_4$ symmetry we naturally have a bulk critical point where a transition occurs via a double Dirac point in momentum space. If we remove $C_4$ symmetry but preserve mirror, then we have already seen in detail that we will not generically have a bulk critical point separating the quadrupole phase from the completely trivial phase. Additionally, there is one more option we have not discussed which is to preserve $C_4,$ but break \emph{both} mirrors. We need to break both mirrors because their product is proportional to the $C_2$ rotation operator, and hence must be preserved. This implies that both mirrors are either preserved or both broken. In this scenario there will still generically be a bulk gap closing when transitioning out of the quadrupole phase. However, the direct transition to a trivial insulator will be replaced by a two-step process with an intermediate phase separating the quadrupole insulator from the trivial insulator. As one tunes a single parameter, the quadrupole phase will first transition to a Chern insulator with a bulk gap closing at a single Dirac point. Then as the parameter is further tuned, the Chern insulator will transition to the trivial phase through a second single Dirac point. Thus, breaking mirrors will split the direct quadrupole-to-trivial transition into two single Dirac cone transitions with an intermediate Chern insulator phase. The Chern insulator phase is not compatible with mirror symmetry, and hence does not appear in the phase diagram if mirror symmetry is preserved. 

%

\subsection{Dipole pumping}
\label{sec:QuadPumping}

We now break the symmetries that protect the topological quadrupole phase by adding perturbations to the Hamiltonian \eqref{eq:QuadHamiltonian}. As a result, the quadrupole observables lose their quantization. We will see in particular that a new type of electronic pumping occurs, that of a dipole current. 

\begin{figure}[t]%
\centering
\includegraphics[width=\columnwidth]{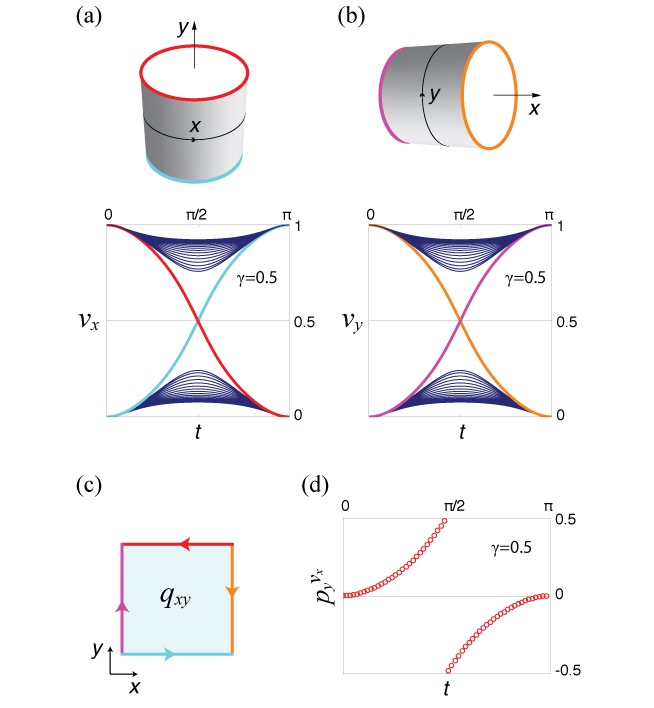}
\caption{(Color online)  Adiabatic pumping \eqref{eq:QuadPumpingSpherical} for the first part of the cycle $0<t \leq \pi$. (a,b) Wannier bands as function of adiabatic parameter $t$ when boundaries along $y$ (a) or $x$ (b) are open. Wannier bands crossing the gap have eigenstates localized at the edges indicated in their respective cylinders. (c) Overall pattern of the edge-localized charge pumping. (d) Wannier-sector polarization as a function of adiabatic parameter $t$.}
\label{fig:pumping_WannierBands}
\end{figure}

Breaking the symmetries that quantize the quadrupole can occur in the following scenarios:
\begin{itemize}
\item Perturbation breaks $M_i$ and $C_2$ symmetries but keeps $M_j$. This quantizes $p^{\pm \nu_i}_j$ and the bulk dipole moment $p_j$ but does not quantize $p^{\pm \nu_j}_i$ nor the bulk dipole moment $p_i$. Here $i,j=x,y$ and $i \neq j$.
\item Perturbation breaks $M_x$ and $M_y$ symmetries but keeps $C_2$ symmetries. This does not quantize $p^{\pm \nu_y}_x$ nor $p^{\pm \nu_x}_y$, but keeps the total bulk dipole moment quantized, e.g., $\bf p = 0$.
\end{itemize}
We concentrate on the second scenario, because we are interested in pumping arising exclusively from the bulk quadrupole moment and not from dipole moment contributions. A perturbation that breaks both reflection symmetries while preserving $C_2$ is the one we have used in Eq. \ref{eq:QuadHamiltonian_with_delta} to choose the `sign' of the quadrupole. The simplest and most illustrative pumping process consists of the adiabatic evolution of the insulator in Eq.~\ref{eq:QuadHamiltonian_with_delta} parametrized by $t$ according to 
\begin{align}
(\delta, \lambda, \gamma)=\left\{
\begin{array}{ll}
(\cos(t),\sin(t),0)& 0< t \leq \pi\\
(\cos(t),0,|\sin(t)|) & \pi < t \leq 2\pi
\label{eq:QuadPumpingSpherical}
\end{array}
\right.,
\end{align}
where for simplicity we have chosen $\lambda_x=\lambda_y=\lambda$ and $\gamma_x=\gamma_y=\gamma$.
During this process, the twofold degenerate energies in the bulk remain gapped for all times $t$: $\epsilon({\bf k},t)= \pm \sqrt{1+\sin^2(t)}$.
Similarly, the energy gaps of the edges also remain gapped, which is crucial for adiabaticity. Fig.~\ref{fig:pumping_WannierBands}a,b shows the transport corresponding to the adiabatic evolution \eqref{eq:QuadPumpingSpherical} during the first half of the cycle. In Fig.~\ref{fig:pumping_WannierBands}a (Fig.~\ref{fig:pumping_WannierBands}b)  boundaries are closed along $x$ ($y$) and open along $y$ ($x$), as schematically indicated by the cylinders. The plots track the Wannier values during the first half of the cycle during which all the inter-unit cell transport takes place. The dark blue lines correspond to Wannier eigenstates that extend over the bulk of the material. Each of the red, cyan, orange, and purple lines, on the other hand, correspond to one Wannier center whose wave function localizes at an edge of the material, as indicated by the corresponding lines on the cylinders.
At $t=0$ the system is in the trivial Hamiltonian $\Gamma_0$. This is a momentum-independent Hamiltonian which represents an insulator in the atomic limit, and therefore all its Wannier values are zero. As the system is adiabatically deformed, the on-site perturbation becomes smaller and the hopping amplitudes increase. Two Wannier sectors appear, as well as Wannier eigenstates localized at the edges. At $t=\pi/2$ the reflection symmetries are restored, and we encounter the topological quadrupole phase of model \eqref{eq:QuadHamiltonian}, which has Wannier-sector polarization $p^{\nu^\pm_i}_j=1/2$ and consequent edge-localized polarizations of $1/2$. The evolution continues, with hopping terms fading and on-site terms increasing magnitude, but with the opposite sign as in the range $t \in [0,\pi/2)$. As we approach the end of the first half of the cycle, $t\rightarrow \pi$, the system approaches the trivial phase, and $p^{\nu^\pm_i}_j \rightarrow 1=0$ mod 1. While the bulk Wannier states show no net transport after the half cycle, the edge-localized Wannier states show net transport of one electron from right to left by one unit cell on the upper boundary, and one from left to right by one unit cell on the lower boundary in Fig.~\ref{fig:pumping_WannierBands}a. Fig.~\ref{fig:pumping_WannierBands}b shows a similar pattern. The combined overall pattern of transport, however, is not that of a circulating current. Instead, it is consistent with the a quadrupole pattern where the bulk dipole remains fixed to zero, as shown in Fig.~\ref{fig:pumping_WannierBands}c. During the second half of the cycle, $\pi < t \leq 2\pi$, the Hamiltonian remains in the the trivial atomic limit phase, and thus causes no electronic transport.

This adiabatic evolution is associated with a Berry flux and Chern number of the Wannier bands as calculated via
\begin{align}
\Delta q_{xy}&= \int_0^{2\pi} d\tau \partial_{\tau} p^{\nu^\pm_i}_j(\tau)=1
\label{eq:quad_pumping_invariant}
\end{align}
for $i,j=x,y$ and $i \neq j$.
In essence, this pumping process results in an \emph{edge-localized Thouless pumping process}, associated with the changing edge polarizations due to the adiabatic change of the bulk quadrupole moment. During a full cycle there is quantized transport captured by the winding of the Wannier-sector polarizations \eqref{eq:quad_pumping_invariant}, as shown numerically in Fig.~\ref{fig:pumping_WannierBands}d (the Wannier-sector polarization winds completely in the range $t \in [0,\pi)$ because the Berry flux for $t \in [\pi,2\pi)$ is zero, since the Hamiltonian remains in the atomic-limit phase during this second half of the cycle). 

\begin{figure}[t]%
\centering
\includegraphics[width=\columnwidth]{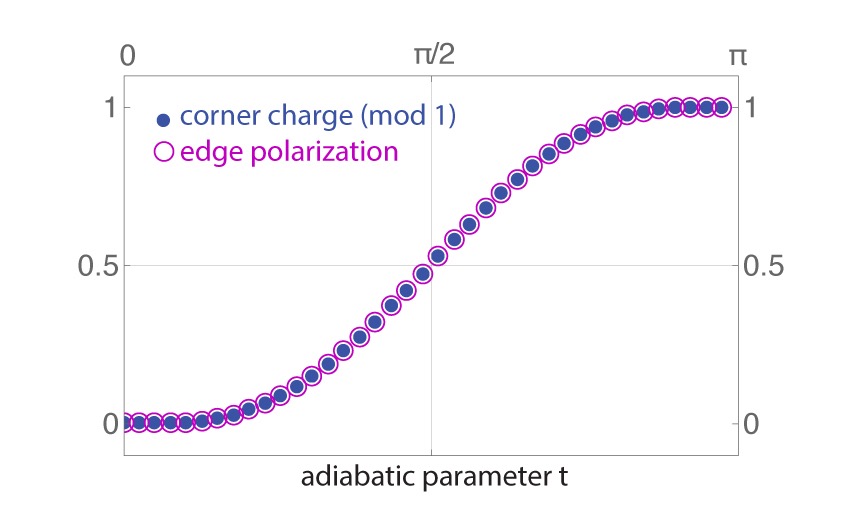}
\caption{(Color online)  Corner charge (blue solid dots) and edge polarization as function of pumping parameter $t$ for the parametrization of \eqref{eq:QuadPumpingSpherical} during the first part of the cycle $0<t \leq \pi$.}
\label{fig:quad_pumping_observables}
\end{figure}

The pumping process detailed above provides us with a family of Hamiltonians which, we claim, have quadrupole moments that range from 0 to 1. To confirm this, we track the corner charge and the edge polarizations during the entire cycle. The prescription for the calculation of the edge polarization is shown in Section \ref{sec:EdgePolarization}. Fig.~\ref{fig:quad_pumping_observables} shows a plot of the instantaneous corner charge as well as the instantaneous edge polarization (we find that the magnitudes of the edge polarizations along all edges are equal). The edge polarizations are calculated by integrating the contributions to tangential polarization up to the middle of the crystal, i.e., as
\begin{align}
p^{edge\;-y} &= \sum_{R_y=1}^{\frac{N_y}{2}} p_x(R_y)\nonumber\\
p^{edge\;+y} &= \sum_{R_y=\frac{N_y}{2}+1}^{N_y} p_x(R_y)
\end{align}
where $p_x(R_y)$ is given by Eq.~\ref{eq:y_resolved_x_polarization}, and similarly for $p^{edge\; \pm x}$. The corner charges are calculated by integrating the charge density over a quadrant of the crystal, i.e., as
\begin{align}
Q^{corner \;-x,\;-y} = \sum_{R_x=1}^{\frac{N_x}{2}} \sum_{R_y=1}^{\frac{N_y}{2}} \rho(\bf R),
\end{align}
where $\rho(\bf R)$ is is the charge density over occupied states. A similar calculation follows for the other three corner charges in $Q^{corner \; \pm x,\; \pm y}$.
The edge polarizations and the corner charges are the same, in agreement with \eqref{eq:QuadSignatures},  over the entire range of the pumping process. The bulk polarization remains zero during this process since $C_2$ symmetry is always preserved. 

\begin{figure}[t]%
\centering
\includegraphics[width=\columnwidth]{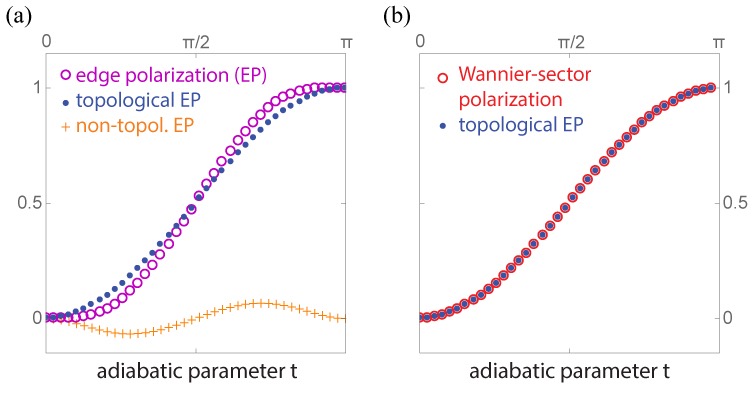}
\caption{(Color online)  Contributions to the edge polarization during the adiabatic pumping \eqref{eq:QuadPumpingSpherical} during the first part of the cycle $0<t \leq \pi$. (a) Total edge polarization (purple circles), topological contribution (blue solid dots), and non-topological contribution (orange + signs). (b) Topological contribution to the edge polarization (solid blue dots) and Wannier-sector polarization (red circles).}
\label{fig:quad_pumping_comparison }
\end{figure}

When calculating the overall edge polarization, following the prescription of Section~\ref{sec:EdgePolarization}, there is a subtlety. We find that there are two contributions that can be differentiated: one contribution is captured by the edge-localized eigenstates of the Wilson loop with open boundaries in one direction, which we call `topological'; the other, `non-topological' contribution, comes from eigenstates of the Wilson loop distributed over the bulk. These separate contributions are shown in Fig.~\ref{fig:quad_pumping_comparison }a. Numerically, the topological contribution is easily discriminated because its Wannier value is situated within the Wannier gap for $t \in (0,\pi)$ (see Fig.~\ref{fig:pumping_WannierBands}a,b). At $t=0,\pi$, on the other hand, all Wannier values vanish. We find that the Wannier-sector polarization \eqref{eq:PolarizationWannierSector} reflects the values of the topological term, as shown in Fig.~\ref{fig:quad_pumping_comparison }b, but does not capture the non-topological contribution. Hence, the Wannier-sector polarizations, i.e., the polarizations of the effective edge Hamiltonian should be treated as a symmetry protected topological invariant and not as a quantitative measure of the exact edge polarization when the symmetries are relaxed. We conjecture that this is because the effective edge Hamiltonian is only adiabatically connected to the physical edge Hamiltonian\cite{klich2011}, thus only topological properties are necessarily preserved. Importantly, the total edge polarizations and corner charges are all quantized to $0$ or $1/2$ in the trivial or topological symmetry-protected quadrupole phases, respectively, and the non-topological contribution to edge polarization vanishes in the presence of the quantizing symmetries. Hence, although the Wannier-sector polarization does not describe the precise value of the edge polarization and corner charge  when there is a bulk contribution to the edge polarization, it does correctly describe the topological properties of the quadrupole. Besides providing the correct quantized values of corner charge and edge polarization in the SPT phases, the quantization of dipole pumping is also correctly given by Eq. \ref{eq:quad_pumping_invariant}.

\begin{figure}[t]%
\centering
\includegraphics[width=\columnwidth]{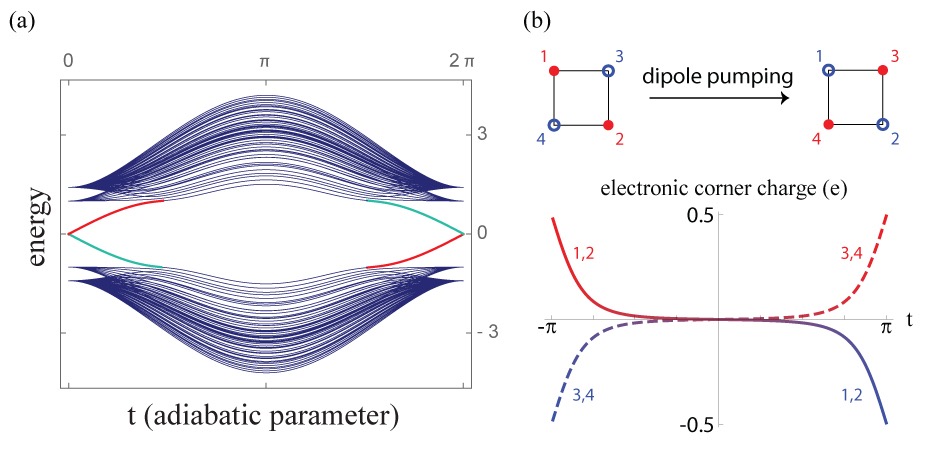}
\caption{(Color online) Adiabatic pumping \eqref{eq:quad_pumping_torus} with open boundaries in \textit{both} directions. $t$ is the pumping adiabatic parameter. (a) Energy spectrum. Green (red) lines are twofold degenerate and have corresponding modes that localize at opposite corners. (b) Corner charge during pumping. Open blue (solid red) circles represent a corner charge of $+e/2$ ($-e/2$) at the beginning and end points of the pumping process. Charge inversion amounts to pumping a quantum of dipole moment.}
\label{fig:QuadPumpingOpenBoundaries}
\end{figure}

\emph{Opening the boundaries:}
Let us now elaborate on the pumping process from the point of view of the corner charges. The pattern of electronic current shown in Fig.~\ref{fig:pumping_WannierBands}c suggests that charge flows from one pair of opposite corners to the other pair. A direct calculation of the energy bands for this type of adiabatic pumping when boundaries \emph{along both $x$ and $y$} are open should reflect this pattern. Since it will become useful, let us do this by using an alternative parametrization of the pumping process which varies continuously over the entire adiabatic cycle (as opposed to pumping \eqref{eq:QuadPumpingSpherical}, which is continuous piecewise). It is given by
\begin{align}
h^q_{pump}({\bf k},t)&= (-m \cos(t)+1) (\Gamma^4+\Gamma^2) - m \sin(t) \Gamma^0\nonumber\\
&+\cos(k_x) \Gamma^4 + \sin(k_x) \Gamma^3\nonumber\\
&+ \cos(k_y) \Gamma^2 + \sin(k_y) \Gamma^1.
\label{eq:quad_pumping_torus}
\end{align}
The pumping process \eqref{eq:quad_pumping_torus} maintains $C_2$ symmetry, with $\hat{r}_2=-i \tau_0 \otimes \tau_2$, at all values of the adiabatic parameter $t \in [0,2\pi)$, which locks the polarization to zero.
For $0<m<2$ ($-2<m<0$) the insulator is in the quadrupole (trivial) phase at $t=0$ and in the trivial (quadrupole) phase at $t=\pi$, while for $|m|>2$ there is no dipole pumping, as the insulator is in the trivial phase at both $t=0,\pi$.  Fig.~\ref{fig:QuadPumpingOpenBoundaries} shows the adiabatic evolution of this Hamiltonian with $m=1$ and open boundaries along both directions. In  Fig.~\ref{fig:QuadPumpingOpenBoundaries}a the bulk energies (marked in dark blue) are gapped. The energies that cross the bulk energy gap (marked in red and green) are each twofold degenerate (i.e., there are a total of four gap-crossing states), and correspond to the corner-localized states. Each pair of twofold degenerate states localize at opposite corners. At half filling, the result of pumping is to change the values of the charges at the corners by $e$, as seen in Fig.~\ref{fig:QuadPumpingOpenBoundaries}b, so that the final quadrupole is equivalent to the original one upon a rotation by $90^\circ$. In Fig.~\ref{fig:QuadPumpingOpenBoundaries}b, we start at a time $-\pi+\epsilon$ and finish at $\pi-\epsilon$, for $\epsilon \ll1$, so that we clearly define the initial sign of the quadrupole by slightly deviating away from the perfectly symmetric SPT quadrupole phase. 

Although the pumping \eqref{eq:QuadPumpingSpherical} also reflects the characteristics we just described, the convenience of the parametrization \eqref{eq:quad_pumping_torus} will become evident when we make a connection between dipole pumping processes and a new type of topological insulator in one higher spatial dimension in the next subsection \ref{sec:HingeInsulator}. 

\subsection{Topological Insulator with hinge-localized chiral modes}
\label{sec:HingeInsulator}
In Section \ref{sec:DipolePumping} we saw that adiabatic charge pumping in 1D insulators by means of a changing dipole moment is characterized by the winding of the Wannier eigenvalues as a function of the adiabatic parameter. This winding is equivalent to a Chern number in the mixed momentum-adiabatic parameter space. If we rename the adiabatic parameter $t$ in the model with a torus parameterization to a new momentum variable, e.g., $t \rightarrow k_y,$ the resulting 2D model is a Chern insulator characterized by the usual 2D Chern number over the BZ. An analogous connection exists in the case of the quantization of adiabatic dipole pumping by means of a changing quadrupole moment. 

\begin{figure}[t]%
\centering
\includegraphics[width=\columnwidth]{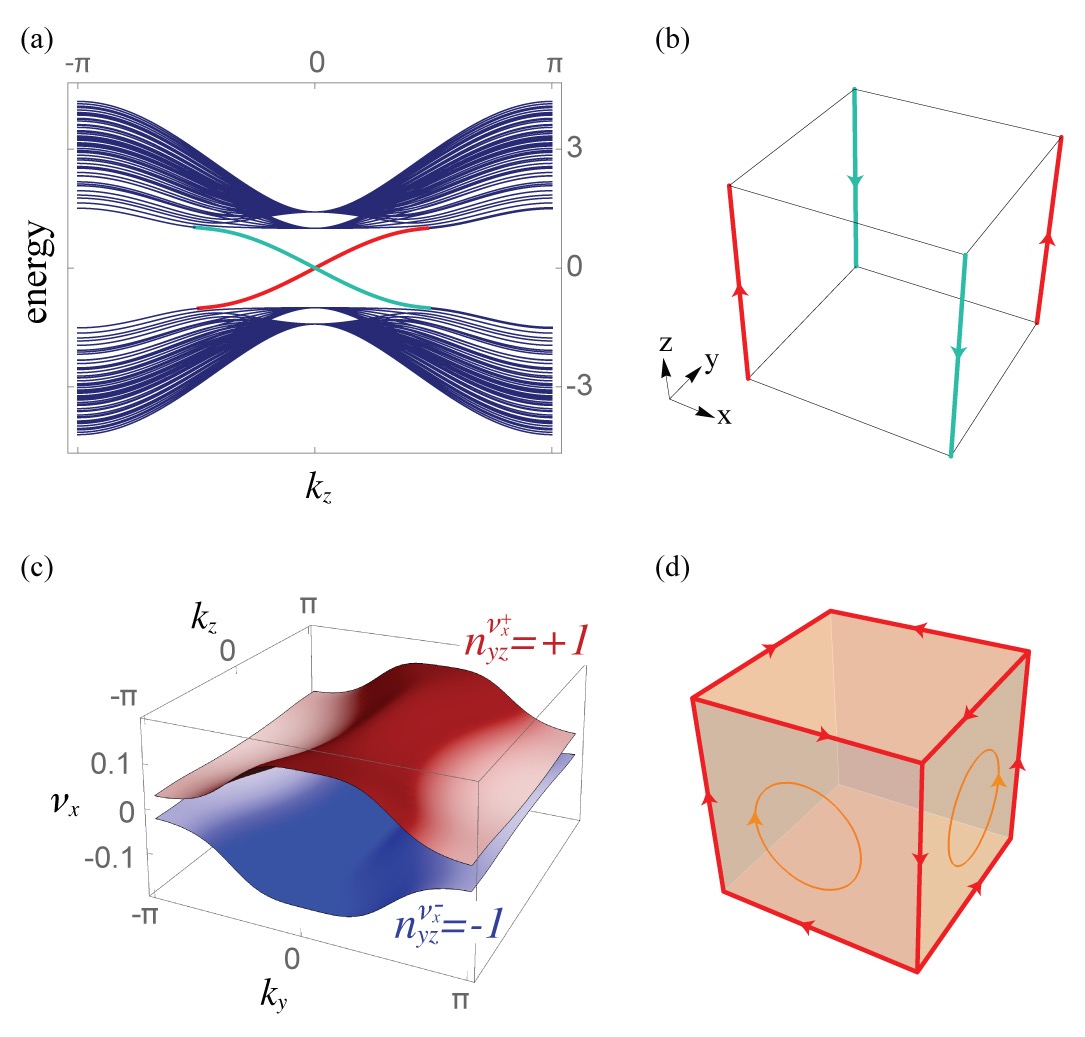}
\caption{(Color online) A crystalline insulator with chiral, edge-localized modes that disperse in equal directions at opposed corners and opposite directions in adjacent ones. This insulator is in the same topological class as the pumping \eqref{eq:quad_pumping_torus}. Both can be identified via the map $t \rightarrow k_z$. (a) Energy dispersion for a system with open boundaries along $x$ and $y$ but closed boundaries along $z$. (b) Hinge localized modes. Arrows indicate direction of dispersion in the presence of an electric field along $z$. (c) Wannier bands, each having a non-zero Chern number defined in \eqref{eq:ChernWannierSector}. (d) Illustration of the compatibility relationship between Chern invariants \eqref{eq:ChernWannierSectorConstraints}. Circles indicate direction of chiral currents compatible with the hinge currents of (b).}
\label{fig:HingeChernInsulator}
\end{figure}

If we substitute $t \rightarrow k_z$ in Eq. \eqref{eq:quad_pumping_torus} the resulting model is the Hamiltonian of a 3D insulator with a winding quadrupole invariant along $k_z$. Fig.~\ref{fig:HingeChernInsulator}a shows the dispersion of this insulator when boundaries are open along \emph{both} $x$ and $y,$ but closed along $z$. Notice that this is in essence the same plot as that in Fig. \ref{fig:QuadPumpingOpenBoundaries}a. The interpretation, however, is different. The corner-localized modes during the adiabatic pumping now map to edge localized modes that are \emph{chiral} and carry current in a quadrupolar fashion when an electric field along $z$ is applied. A schematic of this insulator is shown in Fig.~\ref{eq:quad_pumping_torus}b. These hinge-localized modes are protected by the Wannier-band Chern number
\begin{align}
n^{\nu_x}_{yz} &=\frac{1}{(2\pi)^2} \int_{BZ} \tr \left[\tilde\F^{\nu_x}_{yz,\bf k}\right] d^3\bf k,
\label{eq:ChernWannierSector}
\end{align}
where 
\begin{align}
\tilde\F^{\nu_i}_{jk,{\bf k}} = \partial_j \tilde\A^{\nu_i}_{k,\bf k} - \partial_k \tilde\A^{\nu_i}_{j,\bf k}+i[\tilde\A^{\nu_i}_{j,\bf k},\tilde\A^{\nu_i}_{k,\bf k}],
\end{align}
for $i,j,k=x,y,z$ and $i\neq j \neq k$, is the Berry curvature over the Wannier bands $\nu_i,$ and $\tilde\A^{\nu_i}_{j,\bf k}$ is the Berry connection of the $\nu_i$ Wannier sector, defined in \eqref{eq:WannierBerryConnection}. A plot of these Wannier bands is shown in Fig. \ref{fig:HingeChernInsulator}c. They are gapped and each of them carry a Chern number (instead of just a Berry phase like the 2D quadrupole model). Notice that we always have
\begin{align}
n^{\nu^-_i}_{jk} = -n^{\nu^+_i}_{jk}.
\end{align}

From this analysis we conclude that 3D insulators have additional anisotropic topological indices that signal the presence of chiral, hinge-localized states parallel to $x, y$ or $z$. For example, in the insulator of Fig.~\ref{fig:HingeChernInsulator}, we have
\begin{align}
n^{\nu^+_x}_{yz} &= - n^{\nu^+_y}_{zx} = 1\nonumber\\
n^{\nu^+_z}_{xy} &= 0.
\label{eq:ChernWannierSectorConstraints}
\end{align}
In general, this type of cyclic relationship is kept. Thus, unlike the weak indices for polarization \eqref{eq:WeakIndexVector}, which are each independent of each other, the Chern numbers $n^{\nu_i}_{jk}$ defined in \eqref{eq:ChernWannierSector} are related by similar constraints to \eqref{eq:ChernWannierSectorConstraints}, as otherwise the hinge-localized modes would give incompatible hinge current flows. See Fig. \ref{fig:HingeChernInsulator}d for an illustration of the compatibility conditions. While the lateral surfaces have chiral currents described by the first Eq. in \ref{eq:ChernWannierSectorConstraints}, the upper and lower surfaces have currents in a quadrupole pattern.

\section{Bulk octupole moment in 3D crystals}
\label{sec:Octupole}
The natural extension of the quadrupole moment in 2D is the octupole moment in 3D.
In this section we discuss in detail the calculation of the quantized octupole moment and describe a simple model that realizes it. We discuss both the SPT phase with quantized boundary signatures, and an adiabatic pumping process. In particular, for the latter we will see that an adiabatic cycle can pump a quantum of quadrupole moment. 

\subsection{Simple model with quantized octupole moment in 3D}
\begin{figure}[t]%
\centering
\includegraphics[width=\columnwidth]{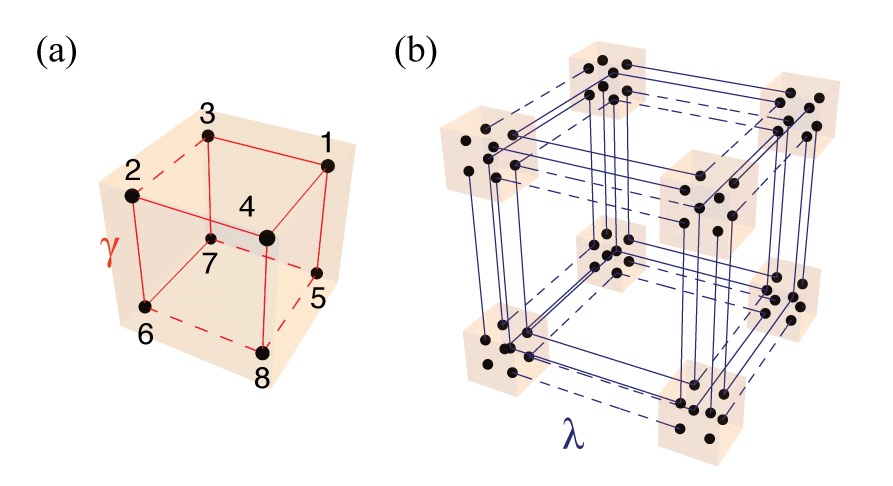}%
\caption{(Color online)  Lattice model of an octupole insulator with Bloch Hamiltonian \eqref{eq:OctupoleHamiltonian}. (a) Degrees of freedom and couplings within the unit cell. (b) Hopping terms in a lattice with eight unit cells. In both (a) and (b) the dashed lines represent a coupling with negative phase factor. As a result of these phase factors a flux of $\pi$ threads each facet.}%
\label{fig:OctupoleLattice}
\end{figure}

In order to have a well-defined octupole moment in the bulk of a 3D insulator, the bulk quadrupole and bulk dipole moments must vanish. Additionally, we require that no Wannier flow exists for Wannier centers along any direction, so as to avoid strong $\mathbb{Z}_2$ insulators and weak topological insulators with layered Chern or $\mathbb{Z}_2$ QSH invariants that would result in metallic boundaries. 
 Using these constraints, we can find  a simple model for an octupole insulator as shown in Fig.~\ref{fig:OctupoleLattice}. It has Bloch Hamiltonian
\begin{align}
h^o_\delta({\bf k})&= \lambda_y \sin(k_y)\Gamma'^1+ [\gamma_y + \lambda_y \cos(k_y)]\Gamma'^2 \nonumber\\
&+\lambda_x \sin(k_x)\Gamma'^3 + [\gamma_x + \lambda_x \cos(k_x)]\Gamma'^4 \nonumber\\
&+\lambda_z \sin(k_z)\Gamma'^5 + [\gamma_z + \lambda_z \cos(k_z)]\Gamma'^6 \nonumber\\
&+\delta \Gamma'^0,
\label{eq:OctupoleHamiltonian}
\end{align}
where $\Gamma'^i=\sigma_3 \otimes \Gamma^i$ for $i=0,1,2,3$, $\Gamma'^4=\sigma_1 \otimes I_{4 \times 4}$, $\Gamma'^5=\sigma_2 \otimes I_{4 \times 4}$, and $\Gamma'^6=i \Gamma'^0 \Gamma'^1 \Gamma'^2 \Gamma'^3 \Gamma'^4 \Gamma'^5$. Here, the internal degrees of freedom follow the numbering in Fig.~\ref{fig:OctupoleLattice}. When $|\lambda_{i}|>|\gamma_{i}|$ for all $i=x, y, z$ this system is an insulator at half filling with four occupied bands and a quantized octupole moment $o_{xyz}=e/2.$
For $\delta=0$, this Hamiltonian has reflection symmetries $M_{x,y,z}$ (up to a gauge transformation, see Section \ref{sec:app_symmetries_up_to_gauge}), with operators
\begin{align}
\hat{M}_x = \tau_0 \otimes \tau_1 \otimes \tau_3\nonumber\\
\hat{M}_x = \tau_0 \otimes \tau_1 \otimes \tau_1\nonumber\\
\hat{M}_x = \tau_1 \otimes \tau_3 \otimes \tau_0
\label{eq:OctupoleReflectionOperators}
\end{align}
which obey $\{\hat{M}_i,\hat{M}_j\} =0$ for $i,j=x,y,z$ and $i \neq j$.
The octupole moment $o_{xyz}$ is odd under each of these symmetries. In the continuum theory, this admits only the solution $o_{xyz}=0$, but the ambiguity in the position of the electrons due to the introduction of the lattice (see Section \ref{sec:PrelimConsiderations}) also allows the solution $o_{xyz}=1/2$ mod $1$. In addition, these symmetries quantize $p_x, p_y, p_z, q_{xy}, q_{xz},$ and $q_{yz},$ all of which must vanish for $o_{xyz}$ to be well-defined.

\begin{figure}[t]%
\centering
\includegraphics[width=\columnwidth]{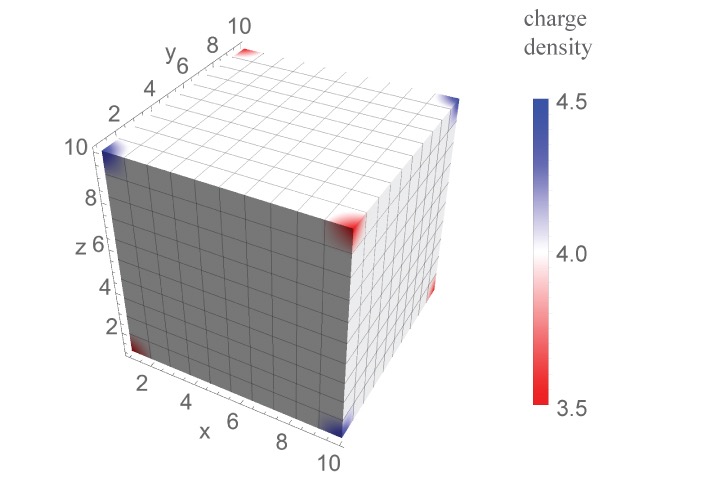}%
\caption{(Color online)  Electronic charge density of the octupole insulator with open boundaries. Corners have a charge of $\pm e/2$ relative to the background charge.}%
\label{fig:OctupoleCharge}
\end{figure}

One signature of the topological octupole moment is the existence of fractional half charges localized on the corners of a cubic sample. Indeed, the non-trivial quantized octupole phase of this model has corner-localized mid-gap modes. We add an infinitesimal $\delta$ in the Hamiltonian that breaks the cubic symmetry of the crystal down to tetrahedral symmetry. This splits the degeneracy of the zero modes, hence fixing the sign of the octupole moment. A plot of the charge density for this crystal is shown in Fig.~\ref{fig:OctupoleCharge}.

The 4-dimensional subspace of occupied energy bands in the Hamiltonian \eqref{eq:OctupoleHamiltonian} has reflection eigenvalues $\{-1,-1,+1,+1\}$ at all high-symmetry points. Consequently, the Wannier centers of the Wilson loop $\W_{z,\bf k}$ come in pairs $\{ \pm \nu^1_z({\bf k_\perp}), \pm \nu^2_z({\bf k_\perp})\}$, where ${\bf k_\perp}=(k_x,k_y)$ (see Table \ref{tab:EigenvaluesRelations}). In the 3D BZ of Hamiltonian \eqref{eq:OctupoleHamiltonian}, the spectrum of the Wilson loop $\W_{z,\bf k}$ yields two, twofold degenerate Wannier bands separated by a Wannier gap, i.e., $\nu^1_z({\bf k_\perp}) = \nu^2_z({\bf k_\perp})$, as seen in Fig.~\ref{fig:octupole_invariant}. Since an octupole is made from two quadrupoles we want to show that each of these two-band Wannier sectors has a topological  quadrupole moment. We now show how to determine this quadrupole moment.

\subsection{Hierarchical topological structure of the Wannier bands}
\label{sec:Oct_HierarchicalTopology}

\begin{figure}
\centering
\includegraphics[width=\columnwidth]{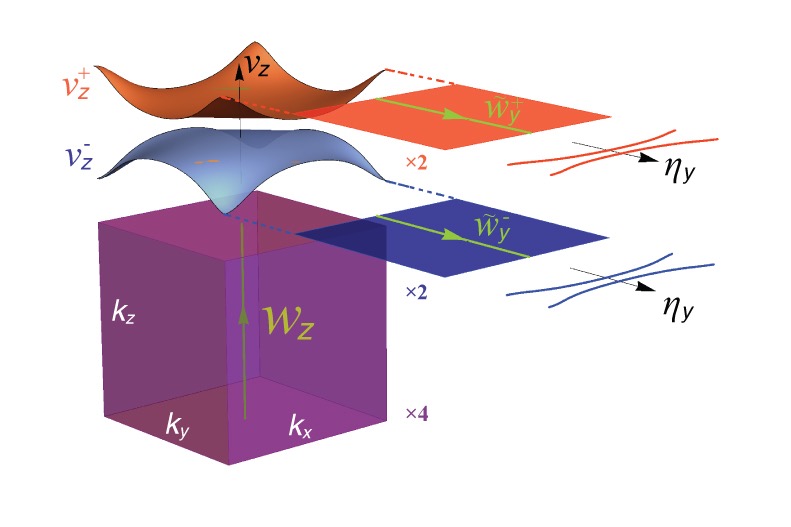}
   \caption{(Color online)  Schematic of the procedure to determine the topology of an octupole moment. A Wilson loop along $z$ over the 3D BZ (purple cube) divides it in two sectors, according to its Wannier value $\nu^\pm_z$ (red and light blue plots over the cube). Each sector has two bands (represented by the red and blue squares) and has quadrupole topology. This can be verified by calculating Wilson loops along $y$ over each sector, which renders two Wannier sectors $\eta^\pm_y$ (red or blue pair of symmetric lines), each of them having a Berry phase of $0$ or $\pi$ in its Wilson loop along $x$ in the $|\lambda_i|<|\gamma_i|$ (for all $i$) or $|\lambda_i|>|\gamma_i|$ (for all $i$) regime, respectively.}
   \label{fig:octupole_invariant}
\end{figure}

Microscopically, a bulk octupole can be thought of as arising from two spatially separated quadrupoles with opposite sign. Thus, since a quadrupole insulator requires two occupied bands, an octupole insulator requires a minimum of four occupied bands. Our model \eqref{eq:OctupoleHamiltonian} is then a minimum model with octupole moment. To reveal its topological structure, we begin the analysis by performing a Wilson loop along the $z$-direction,
\begin{align}
\W_{z,\bf k} \ket{\nu^j_{z,\bf k}}= e^{i 2\pi \nu^j_z({\bf k_\perp})} \ket{\nu^j_{z,\bf k}},
\label{eq:Oct_WilsonLoop1_Diag}
\end{align}
where ${\bf k_\perp}=(k_x,k_y).$ The Wilson loop along $z$ is represented by a $4 \times 4$ matrix, which has eigenstates $\ket{\nu^j_{z,\bf k}}$ for $j=1,2,3,4$. In an octupole phase, the Wilson loop splits the four occupied energy bands into two Wannier sectors $\nu^\pm_z({\bf k_\perp})$, separated by a Wannier gap. The existence of the Wannier gap is protected by the non-commutation of reflections operators \eqref{eq:OctupoleReflectionOperators}. Each of the two sectors, $\nu^\pm_z$, has opposite topological quadrupole moment.  The Wannier bands $\nu^\pm_z$ for the minimal octupole insulator with Hamiltonian in Eq. \ref{eq:OctupoleHamiltonian} are shown in red and light blue in Fig.~\ref{fig:octupole_invariant}. 

In order to determine the quadrupole moment of each of the sectors $\nu^\pm_z$, we proceed similarly to  Section \ref{sec:Quadrupole} for either $\nu^+_z$ or $\nu^-_z$. Concretely, let us first re-write Eq.~\ref{eq:Oct_WilsonLoop1_Diag} as
\begin{align}
\W_{z,\bf k} \ket{\nu^{\pm,j}_{z,\bf k}}= e^{i 2\pi \nu^\pm_z({\bf k_\perp})} \ket{\nu^{\pm,j}_{z,\bf k}},
\label{eq:Oct_WilsonLoop1_Diag2}
\end{align}
for $j=1,2$. Without loss of generality, we choose the sector $\nu^+_z$ and construct the Wannier states
\begin{align}
\ket{w^{+_z,j}_{z,\bf k}} = \sum_{n=1}^{N_{occ}}\ket{u^n_{\bf k}} [\nu^{+,j}_{z,\bf k}]^n,
\end{align}
for $j=1,2$. Here, the superscript $+_z$ is short for the Wannier sector $\nu^+_z$. We use this basis to calculate the nested Wilson loop along $y$,
\begin{align}
[\tilde\W^{+_z}_{y,\bf k}]^{j,j'} =& \braket{w^{+_z,j}_{z,{\bf k}+N_y{\bf \Delta_{k_y}}}}{w^{+_z,r}_{z,{\bf k}+(N_y-1) {\bf \Delta_{k_y}}}} \ldots \nonumber\\
&\bra{w^{+_z,r}_{z,{\bf k}+(N_y-1) {\bf \Delta_{k_y}}}} \dots \ket{w^{+_z,s}_{z,{\bf k}+{\bf \Delta_{k_y}}}}\nonumber\\
&\braket{w^{+_z,s}_{z,{\bf k}+{\bf \Delta_{k_y}}}}{w^{+_z,j'}_{z,\bf k}},
\label{eq:Oct_WilsonLoop2}
\end{align}
where ${\bf \Delta_{k_y}} = (0, 2\pi/N_y,0)$. Notice that, since $j,r,\ldots,s,j' = 1,2$, this nested Wilson loop  is non-Abelian. (This Wilson loop was defined in in Eq.~\ref{eq:WilsonLoop_WannierBasis} for 2D crystals, but we reproduce it here in its obvious extension to 3D). We then diagonalize the nested Wilson loop \eqref{eq:Oct_WilsonLoop2},
\begin{align}
\tilde{\W}^{+_z}_{y,\bf k} \ket{\eta^{+_z,\pm}_{y,\bf k}}= e^{i 2\pi \eta^{\pm}_y(k_x)} \ket{\eta^{+_z,\pm}_{y,\bf k}},
\label{eq:Oct_WilsonLoop2_Diag}
\end{align}
which resolves the Wannier sector $\nu^+_z$ into single Wannier bands $\eta^\pm_y(k_x)$ separated by a Wannier gap (red lines on axes $\eta_y$ in Fig.~\ref{fig:octupole_invariant}). This Wannier gap is also protected by the non-commutation of \eqref{eq:OctupoleReflectionOperators}. The quadrupole topology of the Wannier sector $\nu^+_z$ manifests in that each of the sectors $\eta^\pm_y$ has a quantized dipole moment, indicated by a Berry phase of 0 or 1/2. For example, let us choose the sector $\eta^+_y$,  to define the Wannier basis
\begin{align}
\ket{w^{+_z,+_y}_{y,\bf k}} = \sum_{n=1}^{N_{occ}}\ket{u^n_{\bf k}} [\eta^{+_z,+}_{y,\bf k}]^n
\end{align}
to then calculate a third Wilson loop
\begin{align}
\tilde\W^{+_z,+_y}_{x,\bf k} =& \braket{w^{+_z,+_y}_{y,{\bf k}+N_x{\bf \Delta_{k_x}}}}{w^{+_z,+_y}_{y,{\bf k}+(N_x-1) {\bf \Delta_{k_x}}}} \times \nonumber\\
 &\bra{w^{+_z,+_y}_{y,{\bf k}+(N_x-1) {\bf \Delta_{k_x}}}} \dots \ket{w^{+_z,+_y}_{y,{\bf k}+{\bf \Delta_{k_x}}}} \times \nonumber\\
&\braket{w^{+_z,+_y}_{y,{\bf k}+{\bf \Delta_{k_x}}}}{w^{+}_{y,\bf k}}.
\label{eq:Oct_WilsonLoop3}
\end{align}
This Wilson loop is associated with the Wannier-sector polarization
\begin{align}
p^{+_z,+_y}_x=-\frac{i}{2\pi}\frac{1}{N_yN_z} \sum_{k_y,k_z} \mbox{Log} \left[ \tilde\W^{+_z,+_y}_{x,\bf k} \right] 
\label{eq:HingePolarization}
\end{align}
and which for our model takes the values
\begin{align}
p^{+_z,+_y}_x= \left\{\begin{array}{ll}
1/2 & |\gamma_i| > |\lambda_i|\\
0 & |\gamma_i| < |\lambda_i|
\end{array} \right.,
\label{eq:OctupoleWilsonQuantization}
\end{align}
for all $i.$ From this, it follows that the topology of each original Wannier sector $\nu^\pm_z$ is that of a quadrupole, and the topology of the entire Hamiltonian is that of an octupole.

In this calculation, the order of the nested Wilson loops $\W_z \to \W_y \to \W_x$ was arbitrary. The same results as in \eqref{eq:OctupoleWilsonQuantization} are obtained for any order of Wilson loop nesting in a quantized octupole insulator, provided that the non-commuting quantizing symmetries are present.



\subsection{Boundary signatures}
Classically, the octupole moment manifests at the faces of a 3D material by the existence of surface-bound quadrupole moments (see Section \ref{sec:ClassicalMultipoles}). In this formulation, the connection between the bulk topology and the boundary topology is given by the adiabatic map between the Wilson loops spectrum and the spectrum of the physical boundary Hamiltonians (see Section \ref{sec:WannierBandsEdgeHamiltonian}). Thus, in the formulation derived in Section \ref{sec:Oct_HierarchicalTopology} to characterize the bulk topology of an octupole insulator, we can make the identification
\begin{align}
\W_{z,\bf k}= e^{-i H_{surface}({\bf k})}
\label{eq:Oct_Hsurf}
\end{align}
where $\W_{z,\bf k}$ is the Wilson loop along $z$ of Eq. \ref{eq:Oct_WilsonLoop1_Diag}, and $H_{surface}({\bf k})$ has the same topology of the Hamiltonian at the surface of the insulator in the $xy$ plane (we can similarly assign Wilson loops along $x$ ($y$) to Hamiltonians on the surface $yz$ ($zx$)). Similarly, we can make the identification 
\begin{align}
\tilde{\W}^{+_z}_{y,\bf k} = e^{-i H_{hinge}({\bf k})}
\end{align}
where $\tilde{\W}^{+_z}_{y,\bf k}$ is the nested Wilson loop defined in \eqref{eq:Oct_WilsonLoop2}, and $H_{hinge}({\bf k})$ has the same topological properties as the Hamiltonian at the one-dimensional boundaries of the 2D surface $xy$ of the material (i.e., we are now looking into the boundary of the boundary).
Notice that in all levels of nesting of the Wilson loops, their Wannier bands remain gapped, which was a condition imposed to avoid boundary metallic modes. Since the Wannier Hamiltonians and edge Hamiltonians are adiabatically connected this should imply that when the Wannier Hamiltonians are gapped the corresponding boundary Hamiltonians are energy-gapped.

\subsection{Quadrupole pumping}
\begin{figure}[t]%
\centering
\includegraphics[width=\columnwidth]{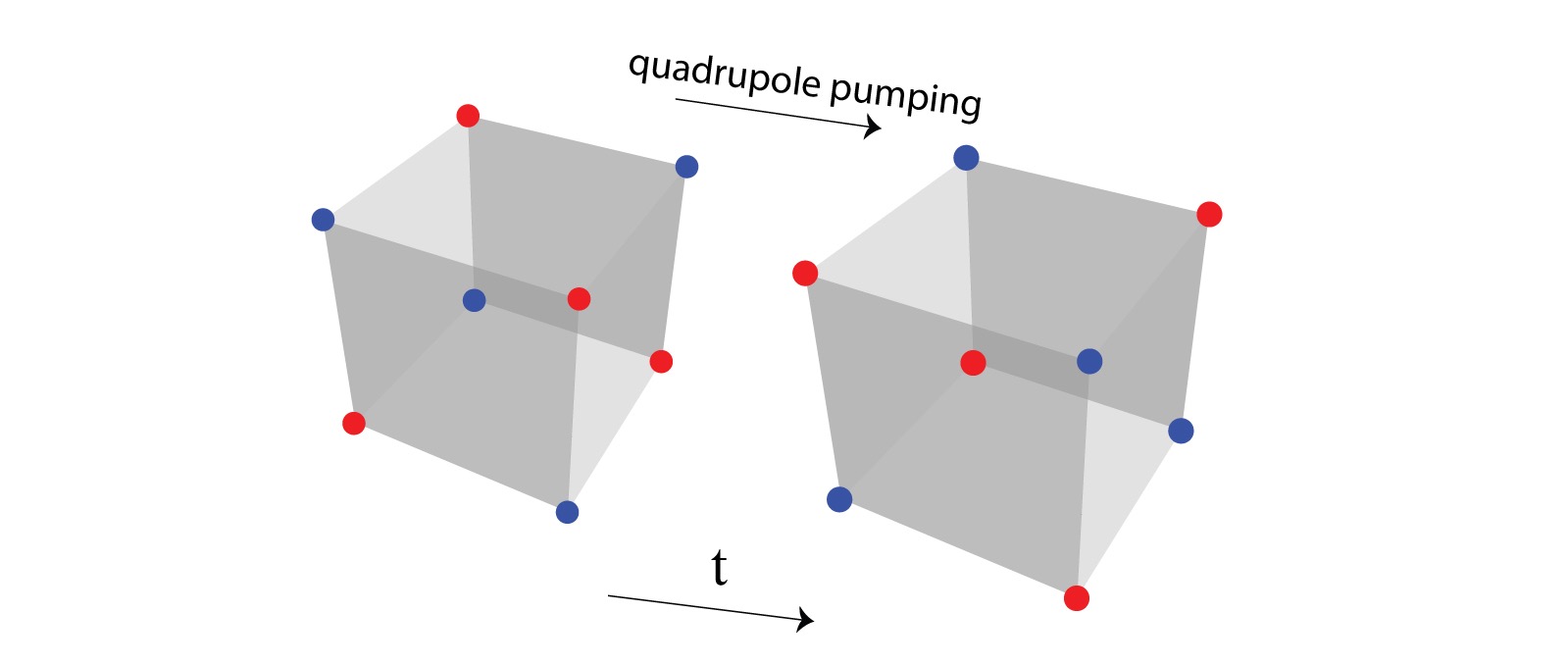}%
\caption{(Color online)  Adiabatic pumping by the Hamiltonian \eqref{eq:OctPumpingTorus}.}%
\label{fig:OctupolePumping}
\end{figure}

Just as a varying dipole generates charge pumping, and a varying quadrupole generates dipole pumping (pumping of charge on the boundary), an adiabatic evolution of the octupole insulator which interpolates between the topological octupole phase and the trivial octupole phase pumps a quantum of quadrupole. This can be achieved by the Hamiltonian
\begin{align}
h^o_{pump}({\bf k},t)&=(-m \cos(t)+1) (\Gamma'^2+\Gamma'^4+\Gamma'^6) \nonumber\\
&- m \sin(t) \Gamma'^0\nonumber\\
&+ (\sin(k_y)\Gamma'^1+ \cos(k_y)\Gamma'^2) \nonumber\\
&+(\sin(k_x)\Gamma'^3 + \cos(k_x)\Gamma'^4) \nonumber\\
&+(\sin(k_z)\Gamma'^5 + \cos(k_z)]\Gamma'^6),
\label{eq:OctPumpingTorus}
\end{align}
for $t \in [0,2\pi)$, where $t$ is the adiabatic parameter. The adiabatic cycle can be characterized by a topological invariant that captures the change in octupole moment:
\begin{align}
\Delta o_{xyz}&= \int_0^{2\pi} d\tau \partial_{\tau} p^{\pm_i \pm_j}_k(\tau)=1
\label{eq:oct_pumping_invariant}
\end{align}
for $i,j,k=x,y,z$ and $i \neq j \neq k$, and where $p^{\pm_i \pm_j}_k(\tau)$ is defined as in Eq. \ref{eq:HingePolarization} for the instantaneous Hamiltonian \eqref{eq:OctPumpingTorus}. This particular pumping process preserves the in-plane $C_2$ symmetries $(x,y,z)\rightarrow(x,-y,-z)$, $(-x,y,-z)\rightarrow(x,y,-z)$, and $(x,y,z)\rightarrow(-x,-y,z)$ at all times $t \in [0,2\pi)$, but breaks the reflection symmetries and the overall inversion symmetry $(x,y,z)\rightarrow (-x,-y,-z)$, except at the SPT phase points at $t=0,\pi$. Breaking the reflection symmetries while preserving the in-plane symmetries allows transport only through the hinges, via surface dipole pumping processes. This occurs at all hinges, so that the octupole configuration is inverted as illustrated in Fig.~\ref{fig:OctupolePumping}. The overall effect amounts to a pumping of a quantum of quadrupole through the 3D bulk. 

\section{Discussion and Conclusion}
\label{sec:Discussion}
In this paper we have systematically addressed the question of whether insulators can give rise to quantized higher electric multipole moments. Starting from the derivation of observables in a classical, continuum electromagnetic setting, we established the physical signatures of these moments and discussed how the definitions could be generalized for an extended quantum mechanical system in a lattice.

\begin{figure}
\centering
\includegraphics[width=\columnwidth]{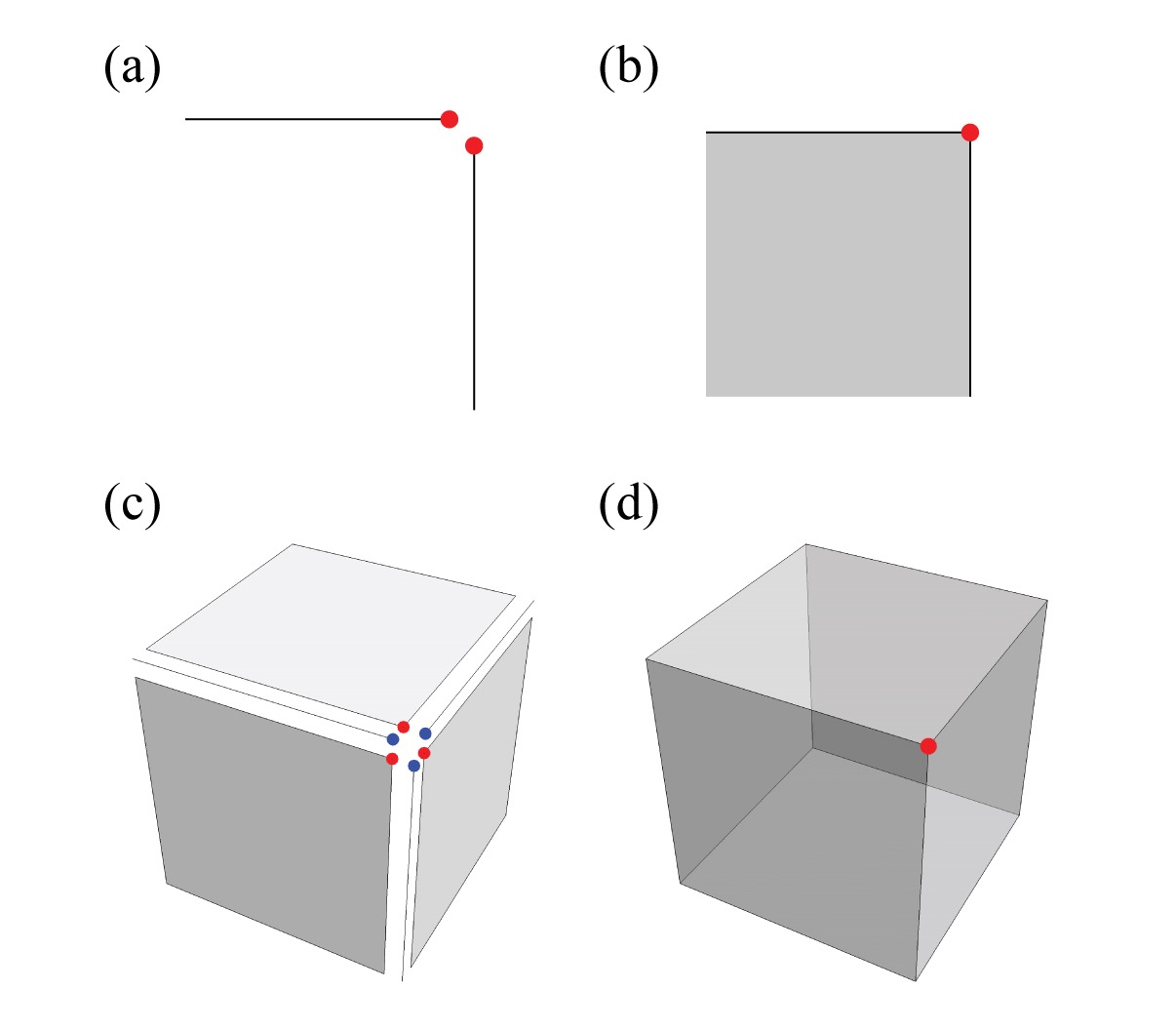}
\caption{(Color online)  Bound states/charges in the quadrupole and octupole SPT phases. (a) Two 1D dipole SPT phases meeting at a corner do not have a corner bound state. (b) A quadrupole SPT phase has edge dipole SPT phases meeting at corners and corner bound state. (c) Three quadrupole and three dipole SPT phases that meet at at a corner do not generate a bound state. (d) An octupole SPT phase, which has three surface quadrupoles and three hinge dipoles as in (c) does harbor a corner bound state.}
\label{fig:anomalousSPT}
\end{figure}

The identification of the higher multipole moments--even in the classical continuum theory--is a subtle matter, especially when a lattice is involved. For example, one signature of a 2D bulk quadrupole moment is a corner charge. However, such a corner charge can arise purely as a surface effect where either free charge is attached to the corner or two edge/surface dipoles converge at a corner. The bulk quadrupole moment exactly captures the failure of the surface dipoles and free charge to account for the corner charge where surfaces intersect (see Fig. \ref{fig:anomalousSPT}).  The octupole moment has similar subtleties connected to the possibility of surface quadrupole moments and hinge polarizations.   If \emph{free} surface quadrupoles and hinge dipoles were attached to the boundaries in an effort to reproduce the same spatial configuration generated by a bulk octupole moment, they would \emph{not} produce the correct value of corner charge associated to a bulk octupole moment  (see Fig.~\ref{fig:anomalousSPT}c,d). Thus, while the quadrupole and octupole moments are bulk properties, their extraction from the associated observable properties, which naturally arise at surfaces and defects, requires care. 

In the crystalline, quantum-mechanical theory, we found that the same macroscopic relations as in the classical continuum theory are maintained. The subtlety that enters at this stage is an inherent ambiguity in the value of the electric moments, where the moments are only well-defined up to a ``quantum." In 1D, 2D, and 3D respectively the dipole, quadrupole, and octupole densities all have units of charge, and each moment is only well-defined up to integer multiples of the electron charge. This ambiguity is countered by the realization that all of the unambiguous observable properties of these moments depend on their changes in time and/or space, and always result in measurements of charges and currents.

One of the most exciting results of our work is that in the presence of symmetries multipole moments can take quantized values. The conventional paradigm to quantize such properties is to enforce a symmetry under which a multipole moment transforms non-trivially.  In certain scenarios the aforementioned ambiguity in the definition of these moments in a lattice system allow for the moment to take a non-zero, quantized value. Insulating phases realizing the quantized value are recognized to have a topological character protected by the enforced symmetries. However, the properties of the quadrupole and octupole topological phases are remarkable in the sense that, instead of exhibiting gapless states on the boundary, these phases have gapped boundaries, which are themselves non-trivial SPT phases of lower spatial dimension. This defies the conventional idea of topological phases of matter as phases with a gapped, featureless bulk that, because of their topological nature, require the existence of gapless states on the boundary. Instead, the picture here is similar to the concept in 3D topological phases that a gapped, symmetry-preserving surface cannot be trivial and must have topological order\cite{vishwanath2013physics,bonderson2013,wang2013gapped,chen2014symmetry,metlitski2015symmetry}. A similar structure follows in the case of the octupole moment; a bulk octupole SPT insulator in 3D generates corner-localized mid-gap bound states, as well as six quadrupole SPT phases on its 2D surfaces, and twelve dipole SPT phases on its 1D hinges, all of which converge at a the 3D corners. Hence, the quadrupole and octupole phases represent a new mechanism for the realization of SPT phases, i.e., surface SPTs. Following this line of reasoning, our work can naturally be extended to the characterization of other 2D or 3D systems exhibiting edges/surfaces that are gapped fermionic/bosonic SPTs or $Z_n$ parafermion chains, with, e.g.,  corners that harbor the corresponding topological bound states such as Majorana fermions.

In this paper we have also shown that the topological structure of these SPT phases is hierarchical in a way that reflects the relationship between a bulk multipole moment and the lower moments realized on the  boundary. For example, the subspace of occupied bands in a quadrupole has two sectors, each having non-trivial dipole topology, while the subspace of occupied bands of an octupole moment has two sectors, each having non-trivial quadrupole topology. Through this hierarchical topological classification, we were able to construct topological invariants that characterize the bulk SPT phases. One can also break the protecting symmetries to generate non-quantized multipole moments, and in such a scenario one can develop protocols where the system is driven in an adiabatic cycle where the multipole moment changes by a quantized amount and topological pumping occurs. 

Such topological pumping processes can also be used to construct topological insulators in one dimension higher where the adiabatic pumping parameter is interpreted as an additional momentum parameter. We provided an example of this in an adiabatic pumping process where the quadrupole moment changes by an integer and gives rise to an associated 3D insulator with chiral states on the hinges of the material. We believe these developments will lead to the discovery of previously unknown topological crystalline phases of matter.

{\bf{Note: }}We have recently become aware of concomitant work on related topics by the groups Refs. \onlinecite{song2017,neupert2017,langbehn2017}.

\begin{acknowledgments}
We thank R. Resta, C. Fang, J. Teo, and A. Soluyanov for useful discussions. WAB and TLH thank the US National Science Foundation under grant DMR 1351895-CAR and the Sloan Foundation for support. BAB acknowledges support from U.S. Department of Energy grant DE-SC0016239, NSF Early-concept Grants for Exploratory Research award DMR-1643312, Simons Investigator award ONR - N00014-14-1-0330, Army Research Office Multidisciplinary University Research Initiative grant W911NF-12-1-0461, NSF-Material Research Science and Engineering Center grant DMR-1420541, and the Packard Foundation and Schmidt Fund for Innovative Research. BAB also wishes to thank Ecole Normale Superieure, UPMC Paris, and Donostia International Physics Center for their generous sabbatical hosting during some of the stages of this work. 
\end{acknowledgments}

\appendix

\section{Details on the definitions of the multipole moments}
\label{sec:app_multipole_definitions}
In this section we provide some intermediate steps used in Section \ref{sec:MultipoleDefinitions} to define the multipole moment densities. Consider the potential \eqref{eq:potential_macroscopic}. 
To simplify the notation, let us define the vector
\begin{align}
\vec{\rr} = \vec{r} - \vec{R} 
\end{align}
which spans from a point in the material $\vec{R}$ to the observation point $\vec{r}.$ The potential is
\begin{align}
\phi(\vec{r})=\frac{1}{4\pi \epsilon} \sum_{\vec{R}} \int_{v(\vec{R})} d^3\vec{r'} \frac{\rho(\vec{r'}+\vec{R})}{|\vec{\rr}-\vec{r'}|}.
\label{eq:potentialVoxels}
\end{align}
Now, let us expand the potential \eqref{eq:potentialVoxels} in powers of $1/|\vec{\rr}|$,
\begin{align}
\phi(\vec{r})= \sum_{l=0}^\infty \phi^l(\vec{r}),
\end{align}
where
\begin{align}
\phi^l(\vec{r}) = \frac{1}{4\pi \epsilon} \sum_{\vec{R}} \int_{v(\vec{R})} d^3\vec{r'} \rho(\vec{r'}+\vec{R}) \frac{|\vec{r'}|^l}{|\vec{\rr}|^{l+1}}P_l\left( \frac{\vec{\rr}}{|\vec{\rr}|} \cdot \frac{\vec{r'}}{|\vec{r'}|} \right),
\end{align}
and $P_l(x)$ are the Legendre polynomials. Here the contributions to the total potential are, up to octupole moment,
\begin{align}
\phi^0(\vec{r})&=\frac{1}{4 \pi \epsilon} \sum_{\vec{R}} Q(\vec{R}) \frac{1}{|\vec{\rr}|}\nonumber \\
\phi^1(\vec{r})&=\frac{1}{4 \pi \epsilon} \sum_{\vec{R}} P_i(\vec{R}) \frac{\rr_i}{|\vec{\rr}|^3}\nonumber  \\
\phi^2(\vec{r})&=\frac{1}{4 \pi \epsilon} \sum_{\vec{R}} Q_{ij}(\vec{R}) \frac{3 \rr_i \rr_j-|\vec{\rr}|^2 \delta_{ij}}{2|\vec{\rr}|^5}\nonumber  \\
\phi^3(\vec{r})&=\frac{1}{4 \pi \epsilon} \sum_{\vec{R}} O_{ijk}(\vec{R}) \frac{5 \rr_i \rr_j \rr_k - 3 |\vec{\rr}|^2 \delta_{ij} \rr_k}{2|\vec{\rr}|^7},
\end{align}
 where
 \begin{align}
 Q(\vec{R})&=\int_{v(\vec{R})} d^3\vec{r'} \rho(\vec{r'}+\vec{R})\nonumber \\
P_i(\vec{R})&=\int_{v(\vec{R})} d^3\vec{r'} \rho(\vec{r'}+\vec{R}) r_i'\nonumber \\
Q_{ij}(\vec{R})&=\int_{v(\vec{R})} d^3\vec{r'} \rho(\vec{r'}+\vec{R}) r_i' r_j'\nonumber \\
O_{ijk}(\vec{R})&=\int_{v(\vec{R})} d^3\vec{r'} \rho(\vec{r'}+\vec{R}) r_i' r_j' r_k'.
\label{eq:multipole_moments}
 \end{align}
are the charge, dipole, quadrupole and octupole moments at the voxel centered at $\vec{R}$. 

If the voxels are very small compared to the material as a whole, we treat $\vec{R}$ as a continuum variable. Now we can define the multipole moment densities
 \begin{align}
 \rho(\vec{R})&=\frac{1}{v(\vec{R})} \int_{v(\vec{R})} d^3\vec{r'} \rho(\vec{r'}+\vec{R})\nonumber \\
p_i(\vec{R})&=\frac{1}{v(\vec{R})} \int_{v(\vec{R})} d^3\vec{r'} \rho(\vec{r'}+\vec{R}) r_i'\nonumber \\
q_{ij}(\vec{R})&=\frac{1}{v(\vec{R})} \int_{v(\vec{R})} d^3\vec{r'} \rho(\vec{r'}+\vec{R}) r_i' r_j'\nonumber \\
o_{ijk}(\vec{R})&=\frac{1}{v(\vec{R})} \int_{v(\vec{R})} d^3\vec{r'} \rho(\vec{r'}+\vec{R}) r_i' r_j' r_k'
\label{eq:app_multipole_moment_densities}
 \end{align}
to write the potentials as
\begin{align}
\phi^0(\vec{r})&=\frac{1}{4 \pi \epsilon} \int_{V}d^3\vec{R}  \left(\rho(\vec{R}) \frac{1}{|\vec{\rr}|} \right)\nonumber \displaybreak[0]\\
\phi^1(\vec{r})&=\frac{1}{4 \pi \epsilon} \int_{V}d^3\vec{R} \left(p_i(\vec{R}) \frac{\rr_i}{|\vec{\rr}|^3}\right)\nonumber \displaybreak[0]\\
\phi^2(\vec{r})&=\frac{1}{4 \pi \epsilon} \int_{V}d^3\vec{R} \left(q_{ij}(\vec{R}) \frac{3 \rr_i \rr_j -|\vec{\rr}|^2 \delta_{ij}}{2|\vec{\rr}|^5}\right)\nonumber \displaybreak[0] \\
\phi^3(\vec{r})&=\frac{1}{4 \pi \epsilon} \int_{V}d^3\vec{R} \left(o_{ijk}(\vec{R}) \frac{5 \rr_i \rr_j \rr_k - 3 |\vec{\rr}|^2 \rr_k \delta_{ij}}{2|\vec{\rr}|^7}\right),
\label{eq:app_potential_multipole_moment_densities}
\end{align}
where $V$ is the total volume of the macroscopic material.

\section{Boundary properties of insulators with multipole moment densities}
\label{sec:boundaries_EM_theory}
In this Appendix we derive the boundary properties due to the existence of uniform electric multipole moments.  We do this for each multipole moment separately, always assuming that all lower moments vanish.
\subsection{Dipole moment}
The potential due to a dipole moment density $p_i(\vec{R})$ is
\begin{align}
\phi^1(\vec{r})&=\frac{1}{4 \pi \epsilon} \int_{V}d^3\vec{R} \left(p_i(\vec{R}) \frac{\rr_i}{\rr^3}\right)
\end{align}
(see Eq. \ref{eq:potential_multipole_moment_densities} or \ref{eq:app_potential_multipole_moment_densities}).
Here  $\vec{\rr} = \vec{r} - \vec{R}$, as defined in the previous section. For convenience, in what follows we refer to the multipole moment densities without their arguments, i.e., we will simply write $p_i$ for $p_i(\vec{R})$, etc. Now, we use
\begin{align}
\frac{\partial}{\partial R_i} \frac{1}{\rr} \equiv \partial_i \frac{1}{\rr}&=\frac{\rr_i}{\rr^3}
\end{align}
to write the potential due to a dipole moment per unit volume $p_i$ as 
\begin{align}
\phi^1(\vec{r}) = \frac{1}{4 \pi \epsilon} \int_V d^3\vec{R}  \left( p_i \partial_i \frac{1}{\rr} \right).
\end{align}
The expression in parentheses can be decomposed as
\begin{align}
 p_i \left(\partial_i\frac{1}{\rr}\right) = \partial_i \left(p_i \frac{1}{\rr} \right) - \left(\partial_i p_i \right) \frac{1}{\rr},
\end{align}
where $\partial_i$ in $\partial_i p_i$ acts on the arguments of $p_i(\vec{R})$; furthermore, since summation is implied, $\partial_i p_i$ is just the divergence of $\vec{p}(\vec{R})$: $\vec{\nabla} \cdot \vec{p}(\vec{R})$. We use this expression to write the potential as
\begin{align}
\phi^1(\vec{r}) = \frac{1}{4 \pi \epsilon} \int_V d^3\vec{R}  \left[ \partial_i \left(p_i \frac{1}{\rr} \right) - \left(\partial_i p_i \right) \frac{1}{\rr} \right].
\end{align}
Using the divergence theorem on the first term we have
\begin{align}
\phi^1(\vec{r}) = \frac{1}{4 \pi \epsilon} \oint_{\partial V} d^2\vec{R}  \left( n_i p_i \frac{1}{\rr} \right) + \frac{1}{4 \pi \epsilon} \int_V d^3\vec{R} \left(-\partial_i p_i \frac{1}{\rr} \right).
\end{align}
where $\partial V$ is the boundary of the material. To aid our understanding, we rewrite this expression in terms of the original variables
\begin{align}
\phi^1(\vec{r}) &= \frac{1}{4 \pi \epsilon} \oint_{\partial V} d^2\vec{R}  \left( n_i p_i \frac{1}{|\vec{r}-\vec{R}|} \right) \nonumber\\
&+ \frac{1}{4 \pi \epsilon} \int_V d^3\vec{R} \left(-\partial_i p_i \frac{1}{|\vec{r}-\vec{R}|} \right).
\end{align}
Since both terms scale as $1/|\vec{r}-\vec{R}|$, where $|\vec{r}-\vec{R}|$ is the distance from a point in the material $\vec{R}$ to the observation point $\vec{r}$, we can define the charge densities
\begin{align}
\rho &= -\partial_i p_i\nonumber \\
\sigma &= n_i p_i.
\label{eq:app_charge_densities_dipole}
\end{align}
The first term is the volume charge density due to a divergence in the polarization, and the second is the areal charge density on the boundary of a polarized material. The consequence of the bulk dipole moment is thus the existence of charge at the boundary, as shown in Fig.~\ref{fig:continuous_dipole}. 

\subsection{Quadrupole moment}
The potential due to a quadrupole moment per unit volume $q_{ij}$ (see Eq. \ref{eq:potential_multipole_moment_densities} or or \ref{eq:app_potential_multipole_moment_densities}) is 
\begin{align}
\phi^2(\vec{r}) = \frac{1}{4 \pi \epsilon} \int_V d^3\vec{R} \left(q_{ij}\frac{3 \rr_i \rr_j - \rr^2 \delta_{ij}}{2\rr^5} \right),
\end{align}
where $\vec{\rr} = \vec{r} - \vec{R}$, as defined in the previous section. We make use of
\begin{align}
\partial_j \partial_i \frac{1}{\rr} &= \frac{3 \rr_i \rr_j - \rr^2\delta_{ij}}{\rr^5}
\end{align}
to write the potential as
\begin{align}
\phi^2(\vec{r}) = \frac{1}{4 \pi \epsilon} \int_V d^3\vec{R}  \left( \frac{1}{2} q_{ij} \partial_i \partial_j \frac{1}{\rr} \right).
\end{align}
Let us rearrange this expression. We use
\begin{align}
q_{ij} \partial_i \partial_j \frac{1}{\rr} &= \partial_i \left( q_{ij} \partial_j \frac{1}{\rr} \right) - \left(\partial_i q_{ij} \right) \partial_i \frac{1}{\rr}\nonumber \\
&= \partial_i \partial_j \left( q_{ij} \frac{1}{\rr} \right) - 2 \partial_i \left[ \left(\partial_j q_{ij} \right) \frac{1}{\rr} \right] + \left( \partial_i \partial_j q_{ij} \right) \frac{1}{\rr}
\end{align}
in the previous expression to find
\begin{align}
\phi^2(\vec{r}) = \frac{1}{4 \pi \epsilon} \int_V d^3\vec{R} &\left[ \frac{1}{2} \partial_i \partial_j \left(  q_{ij} \frac{1}{\rr} \right) \right.\nonumber\\
&- \partial_i \left[ \left(\partial_j q_{ij} \right) \frac{1}{\rr} \right] \nonumber\\
&\left.+ \left( \frac{1}{2} \partial_i \partial_j q_{ij} \right) \frac{1}{\rr}  \right].
\end{align}
Applying the divergence theorem on the first two terms we have
\begin{align}
\phi^2(\vec{r}) &= 
\frac{1}{4 \pi \epsilon} \oint_{\partial V} d^2\vec{R} \left[ \frac{1}{2} n_i \partial_j \left( q_{ij}  \frac{1}{\rr} \right) \right] \nonumber\\
&+\frac{1}{4 \pi \epsilon} \oint_{\partial V} d^2\vec{R} \left(- n_i \partial_j q_{ij} \right)  \frac{1}{\rr}\nonumber\\
&+\frac{1}{4 \pi \epsilon} \int_V d^3\vec{R} \left(\frac{1}{2}\partial_j \partial_i q_{ij}\right) \frac{1}{\rr}.
\label{eq:app_quad_potential_1}
\end{align}

\begin{figure}
\centering
\includegraphics[width=\columnwidth]{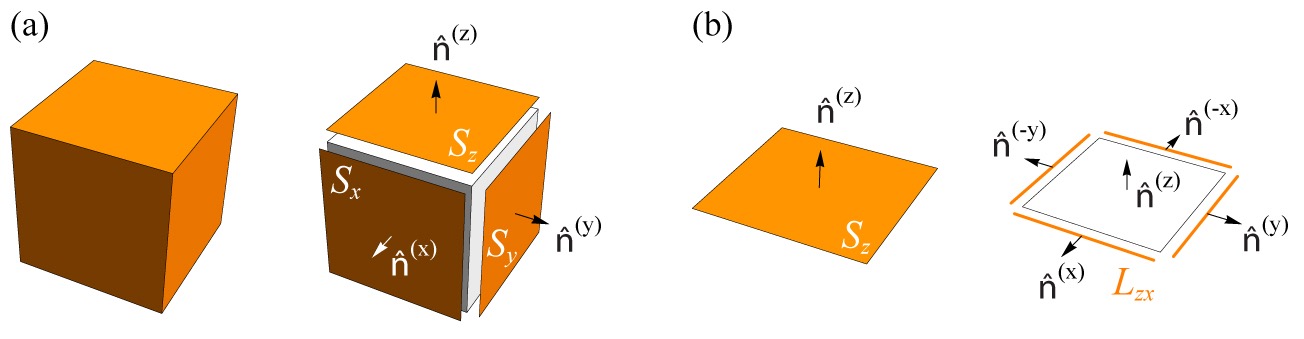}
\caption{(Color online)  Boundary segmentation for the calculation of quadrupole signatures. (a) Separation of 2-dimensional boundary into its flat faces. (b) Separation of a 1D boundary into its straight lines.}
\label{fig:app_boundaries}
\end{figure}

Now let us specialize to a simple geometry.  Consider a cube-shaped material where the boundary consists of flat faces. At the (sharp) intersection of different faces, the normal vector is discontinuous. To avoid this complication we can break the integral over the entire boundary up into a sum over the faces that compose it, as seen in Fig.~\ref{fig:app_boundaries}a:
\begin{align}
\oint\displaylimits_{\partial V} d^2\vec{R}  \left[ \frac{1}{2} n_i \partial_j \left(q_{ji}\frac{1}{\rr} \right) \right] 
&= \sum_a \int\displaylimits_{S_a} d^2\vec{R} \left[ \frac{1}{2} n^{(a)}_i\partial_j \left(q_{ji}\frac{1}{\rr} \right) \right].
\end{align}
For the sake of clarity, we have explicitly written the sum over the flat faces $S_a$ with normal vector $\hat{n}^{(a)}$. Notice that, in this construction, $\hat{n}^{(a)}$ has components $n^{(a)}_i = s_a \delta^{|a|}_i$, where $s_{a=\pm}=\pm 1$ encodes the direction. Now, we apply the divergence theorem over the open surfaces $S_a$. We thus have
\begin{align}
\oint\displaylimits_{\partial V} d^2\vec{R}  \left[ \frac{1}{2} n_i\partial_j \left( q_{ji}\frac{1}{\rr} \right) \right] 
&= \sum_{a,b} \int\displaylimits_{L_{ab}} d\vec{R} \left(\frac{1}{2} n^{(a)}_j n^{(b)}_i q_{ji} \right) \frac{1}{\rr},
\end{align}
where $L_{ab}$ is the one-dimensional boundary of $S_a$ when it meets $S_b$ (see Fig.~\ref{fig:app_boundaries}b). Joining the pieces together, the contributions to the potential from a quadrupole moment are
\begin{align}
\phi^2(\vec{r}) &=
\frac{1}{4 \pi \epsilon} \sum_{a,b} \int_{L_{ab}} d\vec{R} \left( \frac{1}{2} n^{(a)}_i n^{(b)}_j q_{ij}  \right) \frac{1}{\rr}\nonumber\\
&+\frac{1}{4 \pi \epsilon} \sum_a \int_{S_a} d^2\vec{R}  \left( - \partial_j n^{(a)}_i q_{ij}  \right) \frac{1}{\rr} \nonumber\\
&+\frac{1}{4 \pi \epsilon} \int_V d^3\vec{R} \left( \frac{1}{2}\partial_j \partial_i q_{ij}\right) \frac{1}{\rr},
\label{eq:app_quad_potential_3}
\end{align}
Since all the potentials scale as $1/\rr$, where $\vec{\rr}=\vec{r}-\vec{R}$ is the distance from the point in the material to the observation point, all the expressions in parentheses can each be interpreted as charge densities, thus, we define the charge densities
\begin{align}
\rho &= \frac{1}{2}\partial_j \partial_i q_{ij}\nonumber \\
\sigma^{face\;a} &= - \partial_j \left( n^{(a)}_i  q_{ij} \right)\nonumber \\
\lambda^{hinge\;a,b} &= \frac{1}{2} n^{(a)}_i n^{(b)}_j q_{ij}.
\end{align}
The first term is the direct contribution of the quadrupole moment density to the volume charge density in the bulk of the material. The second term is the areal charge density at the boundary surfaces of the material due to a divergence in the quantity $n^{(a)}_i q_{ij}.$ Finally, the third term is the charge density per length at the hinges of the material. For a cube or square with a constant quadrupole moment $q_{xy}$ the charges indicated by the expression for $\lambda^{hinge\;a,b}$ are shown in Fig.~\ref{fig:continuous_quadrupole}. In addition, the expression for the surface charge density $\sigma^{face\;a}$ can be rewritten as
\begin{align}
\sigma^{face\;a} &= - \partial_j p^{face\;a}_j,
\end{align} 
where  $p^{face\;a}_j = n^{(a)}_i  q_{ij}$. This expression resembles the one for the volume charge density $\rho$ in Eq. \ref{eq:app_charge_densities_dipole}. Thus, written this way, we can interpret $p^{face\;a}_j$ as a dipole moment density (per unit area). This polarization exists on the surface of the boundary perpendicular to $n^{(a)}_i$ and is parallel to the surface.  An illustration of this for a cube with constant quadrupole moment is shown in Fig.~\ref{fig:continuous_quadrupole}.

\subsection{Octupole moment}
Making the change of variables $\vec{\rr} = \vec{r} - \vec{R}$, as in the previous sections, the potential due to a octupole moment per unit volume $o_{ijk}$ from Eq. \ref{eq:potential_multipole_moment_densities} or \eqref{eq:app_potential_multipole_moment_densities} is 
\begin{align}
\phi^3(\vec{r}) = \frac{1}{4 \pi \epsilon} \int_V d^3\vec{R} \left(o_{ijk} \frac{5 \rr_i \rr_j \rr_k - 3 \rr^2 \delta_{ij} \rr_k}{2 \rr^7} \right).
\label{eq:app_potential_multipole_densities}
\end{align}
Using the expression
\begin{align}
\partial_k \partial_j \partial_i \frac{1}{\rr} &= 3\frac{5 \rr_i \rr_j \rr_k - 3 \rr^2 \delta_{ij} \rr_k}{\rr^7} 
\end{align}
we write the potential as
\begin{align}
\phi^3(\vec{r}) = \frac{1}{4 \pi \epsilon} \int_V d^3\vec{R} \left( \frac{1}{6} o_{ijk} \partial_i \partial_j \partial_k \frac{1}{\rr} \right).
\end{align}
To find the potential exclusively as arising from charge distributions, we can proceed as before by partial intergation. The result is
\begin{align}
\phi^3(\vec{r}) &= \frac{1}{4 \pi \epsilon} \int_V d^3\vec{R} \left(-\frac{1}{6} \partial_i \partial_j \partial_k o_{ijk} \right) \frac{1}{\rr}\nonumber\\
&+ \frac{1}{4 \pi \epsilon} \sum_a \int_{S_a} d^2\vec{R} \left( \frac{1}{2} n^{(a)}_i \partial_j \partial_k o_{ijk} \right) \frac{1}{\rr}\nonumber\\
&+ \frac{1}{4 \pi \epsilon} \sum_{a,b} \int_{L_{ab}} d\vec{R} \left( -\frac{1}{2} n^{(a)}_i n^{(b)}_j \partial_k o_{ijk} \right) \frac{1}{\rr}\nonumber\\
&+\frac{1}{4 \pi \epsilon} \sum_{a,b,c} \frac{1}{6} n^{(a)}_i n^{(b)}_j n^{(c)}_k o_{ijk} \frac{1}{r},
\end{align}
from which we read off the charge densities
\begin{align}
\rho &= -\frac{1}{6} \partial_i \partial_j \partial_k o_{ijk}\nonumber \\
\sigma^{face\;a} &= \frac{1}{2} n^{(a)}_i \partial_j \partial_k o_{ijk}\nonumber \\
\lambda^{hinge\;a,b} &= -\frac{1}{2} n^{(a)}_i n^{(b)}_j \partial_k o_{ijk}\nonumber \\
\delta^{corner\;a,b,c} &= \frac{1}{6} n^{(a)}_i n^{(b)}_j n^{(c)}_k o_{ijk}
\end{align}\noindent where the new quantity $\delta^{corner\;a,b,c}$ is the corner charge accumulated where three surfaces of the cube intersect. 

Comparing with the expressions for dipole and quadrupole moments  we find we can recast them as
\begin{align}
\rho &= -\frac{1}{6} \partial_i \partial_j \partial_k o_{ijk}\nonumber \\
\sigma^{face\;a} &= \frac{1}{2} \partial_j \partial_k q^{face\;a}_{jk}\nonumber \\
\lambda^{hinge\;a,b} &= -\partial_k  p^{hinge\;a,b}_k\nonumber \\
\delta^{corner\;a,b,c} &= \frac{1}{6} n^{(a)}_i n^{(b)}_j n^{(c)}_k o_{ijk},
\end{align}
where $q^{face\;a}_{jk} = n^{(a)}_i o_{ijk}$, and $p^{hinge\;a,b}_k = \frac{1}{2} n^{(a)}_i n^{(b)}_j o_{ijk}$ are, respectively, the quadrupole moment per unit area on faces perpendicular to $\hat{n}^{(a)},$ and the dipole moment per unit length on hinges perpendicular to both $\hat{n}^{(a)}$ and $\hat{n}^{(b)}$. These boundary properties are illustrated in Fig.~\ref{fig:continuous_octupole} for a cube with uniform octupole moment.

\section{Wilson line in the thermodynamic limit}
\label{sec:app_WilsonLine_ThermodynamicLimit}
Consider the Wilson line element $[G_k]^{mn}= \braket{u^m_{k+\Delta_k}}{u^n_k}$, where $\Delta_k=(k_f - k_i)/N$. For large values of $N$, we expand $\bra{u^m_{k+\Delta_k}} = \bra{u^m_{k}} + \Delta_k \partial_k \bra{u^m_k}+\ldots$, and write the Wilson line element as 
\begin{align}
[G_k]^{mn} = \braket{u^m_k}{u^n_k} + \Delta_k \braket{\partial_k u^m_k}{u^n_k}+\ldots
\end{align}
Now, since $\braket{u^m_k}{u^n_k}=\delta^{mn}$, we have that $\braket{\partial_k u^m_k}{u^n_k} = - \braket{u^m_k}{\partial_k u^n_k}$. Using this in our expansion while keeping only terms linear in $\Delta_k$ we have
\begin{align}
[G_k]^{mn} &= \delta^{mn} - \Delta_k \braket{u^m_k}{\partial_k u^n_k}\nonumber\\
&=\delta^{mn} - i \Delta_k [\A_k]^{mn},
\end{align}
where we have defined the Berry connection 
\begin{align}
[\A_k]^{mn} = -i \matrixel{u^m_k}{\partial_k}{ u^n_k}
\end{align}
which is a purely real quantity. Now, suppose that we evolve the Wilson loop from $k_i$ to $k_f$ in the thermodynamic limit $N \to \infty$. This is achieved by the  (path ordered) matrix multiplication
\begin{align}
\W_{k_f \leftarrow k_i} &= \lim_{N \to \infty} \prod_{n=1}^N \left[ I - i \Delta_k \A_{k+n\Delta_k} \right] \nonumber\\
&= \mbox{exp}\left[{- i \int_{k_i}^{k_f} \A_k dk} \right].
\label{eq:app_WilsonLine_ThermodynamicLimit}
\end{align}

\section{Symmetry constraints on Wilson loops}
\label{sec:WilsonLoopsSymmetry}
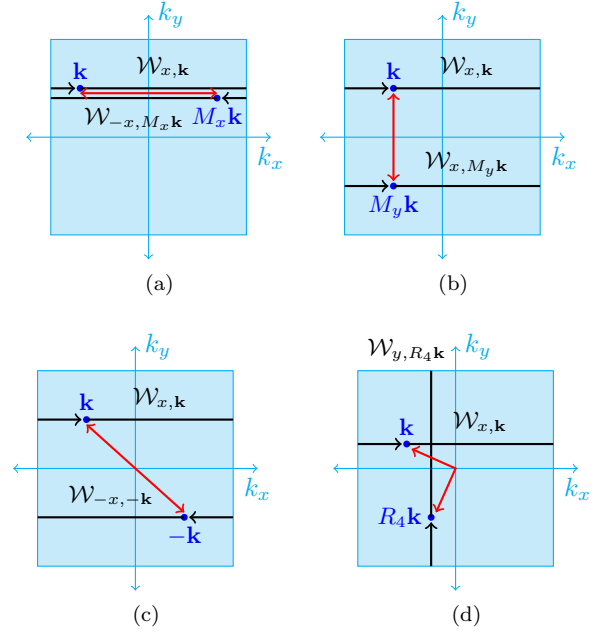
\begin{figure}
\centering
\subfigure[]{
\begin{tikzpicture}[scale=1.3]

\coordinate (pp) at (1,1);
\coordinate (pn) at (1,-1);
\coordinate (np) at (-1,1);
\coordinate (nn) at (-1,-1);

\fill [cyan!20!] (pp)--(pn)--(nn)--(np);
\draw[cyan](pp)--(pn)--(nn)--(np)--(pp);
\draw[<->,cyan](-1.25,0)--(1.25,0) node[below]{$k_x$};
\draw[<->,cyan](0,-1.25)--(0,1.25) node[right]{$k_y$};

\fill [blue] (-0.7,0.5) circle (1pt) node[above] {\small ${\bf k}$};
\draw[->,black,thick](-1,0.5)--(-0.75,0.5);
\draw[black,thick](-0.7,0.5)--node[above]{\small $\W_{x,{\bf k}}$}(1,0.5);

\fill [blue] (0.7,0.4) circle (1pt) node[below] {\small $M_x{\bf k}$};
\draw[<-,black,thick](0.75,0.4)--(1,0.4);
\draw[black,thick](-1,0.4)--node[below]{\small $\W_{-x,M_x{\bf k}}$}(0.7,0.4);

\draw[<->,red, thick](-0.7,0.45)--(0.7,0.45);

\end{tikzpicture}
}
\subfigure[]{
\begin{tikzpicture}[scale=1.3]

\coordinate (pp) at (1,1);
\coordinate (pn) at (1,-1);
\coordinate (np) at (-1,1);
\coordinate (nn) at (-1,-1);

\fill [cyan!20!] (pp)--(pn)--(nn)--(np);
\draw[cyan](pp)--(pn)--(nn)--(np)--(pp);
\draw[<->,cyan](-1.25,0)--(1.25,0) node[below]{$k_x$};
\draw[<->,cyan](0,-1.25)--(0,1.25) node[right]{$k_y$};

\fill [blue] (-0.5,0.5) circle (1pt) node[above] {\small ${\bf k}$};
\draw[->,black,thick](-1,0.5)--(-0.55,0.5);
\draw[black,thick](-0.5,0.5)--node[above]{\small $\W_{x,{\bf k}}$}(1,0.5);

\fill [blue] (-0.5,-0.5) circle (1pt) node[below] {\small $M_y{\bf k}$};
\draw[->,black,thick](-1,-0.5)--(-0.55,-0.5);
\draw[black,thick](-0.5,-0.5)--node[above]{\small $\W_{x, M_y{\bf k}}$}(1,-0.5);

\draw[<->,red, thick](-0.5,0.45)--(-0.5,-0.45);

\end{tikzpicture}
}
\subfigure[]{
\begin{tikzpicture}[scale=1.3]

\coordinate (pp) at (1,1);
\coordinate (pn) at (1,-1);
\coordinate (np) at (-1,1);
\coordinate (nn) at (-1,-1);

\fill [cyan!20!] (pp)--(pn)--(nn)--(np);
\draw[cyan](pp)--(pn)--(nn)--(np)--(pp);
\draw[<->,cyan](-1.25,0)--(1.25,0) node[below]{$k_x$};
\draw[<->,cyan](0,-1.25)--(0,1.25) node[right]{$k_y$};

\fill [blue] (-0.5,0.5) circle (1pt) node[above] {\small ${\bf k}$};
\draw[->,black,thick](-1,0.5)--(-0.55,0.5);
\draw[black,thick](-0.5,0.5)--node[above]{\small $\W_{x,{\bf k}}$}(1,0.5);

\fill [blue] (0.5,-0.5) circle (1pt) node[below] {\small $-{\bf k}$};
\draw[black,thick](-1,-0.5)--node[above]{\small $\W_{-x, -{\bf k}}$}(0.5,-0.5);
\draw[<-,black,thick](0.55,-0.5)--(1,-0.5);

\draw[<->,red, thick](-0.5,0.45)--(0.5,-0.45);

\end{tikzpicture}
}
\quad
\subfigure[]{
\begin{tikzpicture}[scale=1.3]

\coordinate (pp) at (1,1);
\coordinate (pn) at (1,-1);
\coordinate (np) at (-1,1);
\coordinate (nn) at (-1,-1);

\fill [cyan!20!] (pp)--(pn)--(nn)--(np);
\draw[cyan](pp)--(pn)--(nn)--(np)--(pp);
\draw[<->,cyan](-1.25,0)--(1.25,0) node[below]{$k_x$};
\draw[<->,cyan](0,-1.25)--(0,1.25) node[right]{$k_y$};

\fill [blue] (-0.5,0.25) circle (1pt) node[above] {\small ${\bf k}$};
\draw[->,black,thick](-1,0.25)--(-0.55,0.25);
\draw[black,thick](-0.5,0.25)--node[above]{\small $\W_{x,\bf k}$}(1,0.25);

\fill [blue] (-0.25,-0.5) circle (1pt) node[left] {\small $R_4{\bf k}$};
\draw[black,thick](-0.25,-.5)--(-0.25,1)node[above]{\small $\W_{y, R_4{\bf k}}\quad \quad$};
\draw[->,black,thick](-0.25,-1)--(-0.25,-0.55);

\draw[<->,red, thick](-0.45,0.2)--(0,0)--(-0.2,-0.45);

\end{tikzpicture}
}
\caption{(Color online)  Relation between Wilson loops along $x$ at base point ${\bf k}$ after (a) reflection along $x$, (b) reflection along $y$, (c) inversion, or (d) $\pi/2$ rotation.}
\label{fig:app_Wx_transformations}
\end{figure}

In this section we derive the relations between Wilson loops in the presence of reflection, inversion and $C_4$ symmetries. An initial study that determined relations between Wilson loops can be found in Ref. \onlinecite{alexandradinata2014}. In this appendix we expand on that analysis. We will see that some of these relations lead to a quantization of the Wannier centers, the bulk polarization, or the Wannier-sector polarizations. Additionally, we also provide relations that impose constraints on these observables in the presence of time-reversal, charge conjugation and chiral symmetries. 
Insulators with a lattice symmetry obey
\begin{align}
g_{\bf k} h_{\bf k} g_{\bf k}^\dagger = h_{D_g{\bf k}},
\label{eq:app_Hamiltonian_under_symmetry}
\end{align}
where $g_{\bf k}$ is the unitary operator
\begin{align}
g_{\bf k} = e^{-i(D_g {\bf k})\cdot {\bf \delta}}U_g.
\end{align}
$U_g$ is an $N_{orb} \times N_{orb}$ matrix that acts on the internal degrees of freedom of the unit cell, and $D_g$ is an operator in momentum space sending ${\bf k} \rightarrow D_g {\bf k}$. In real space, on the other hand, we have ${\bf r} \rightarrow D_g{\bf r} + {\bf \delta}$, where $\bf \delta=0$ in the case of symmorphic symmetries, or takes a fractional value (in unit-cell units) in the case of non-symmorphic symmetries. The state $g_{\bf k} \ket{u^n_{\bf k}}$ is an eigenstate of $h_{D_g{\bf k}}$ with energy $\epsilon_{n,{\bf k}}$, as can be seen as follows:

\begin{align}
h_{D_g{\bf k}} g_{\bf k} \ket{u^n_{\bf k}} &= g_{\bf k} h_{\bf k} \ket{u^n_{\bf k}} \nonumber\\
&= \epsilon_{n,{\bf k}} g_{\bf k} \ket{u^n_{\bf k}}.
\end{align}
Therefore, one can expand $g_{\bf k} \ket{u^n_{\bf k}}$ in terms of the basis of $h_{D_g {\bf k}}$:
\begin{align}
g_{\bf k} \ket{u^n_{\bf k}} &= \ket{u^m_{D_g{\bf k}}} \matrixel{u^m_{D_g{\bf k}}}{g_{\bf k}}{u^n_{\bf k}}\nonumber\\
&=\ket{u^m_{D_g{\bf k}}}B^{mn}_{g,\bf k},
\label{eq:app_sewing_matrix_expansion}
\end{align}
where, from now on, summation is implied for repeated band indices \text{only} over occupied bands.
\begin{align}
B^{mn}_{g,\bf k}=\matrixel{u^m_{D_g{\bf k}}}{g_{\bf k}}{u^n_{\bf k}}
\end{align}
is the unitary sewing matrix that connects states at ${\bf k}$ with those at $D_g{\bf k}$ which have the same energy. This matrix obeys
\begin{align}
B^{mn}_{g,\bf k+G} = B^{mn}_{g,\bf k}
\end{align}
as can be shown as follows:
\begin{align}
B^{mn}_{g,\bf k+G} &= \matrixel{u^m_{D_g \bf k}}{V(D_g {\bf G})g_{\bf k+G}V(-{\bf G})}{u^n_{\bf k}}\nonumber\\
&= \matrixel{u^m_{D_g \bf k}}{V(D_g {\bf G})V(-D_g {\bf G})e^{i(D_g {\bf G}).\delta}g_{\bf k+G}}{u^n_{\bf k}}\nonumber\\
&= \matrixel{u^m_{D_g \bf k}}{e^{i(D_g {\bf G}).\delta}g_{\bf k+G}}{u^n_{\bf k}}\nonumber\\
&= \matrixel{u^m_{D_g \bf k}}{g_{\bf k}}{u^n_{\bf k}}\nonumber\\
&=B^{mn}_{g,\bf k},\nonumber
\end{align}
where $V({\bf k})$ is defined in Eq.~\ref{eq:Vmatrix}, and we have used Eq.~\ref{eq:UPeriodicGauge} as well as the relation
\begin{align}
g_{\bf k} V({\bf G}) = e^{-i (D_g {\bf k}).\delta} V(D_g {\bf G}) g_{\bf k}.\nonumber
\end{align}
Using the expansion in Eq. \ref{eq:app_sewing_matrix_expansion}, we can write
\begin{align}
\ket{u^n_{\bf k}}=g_{\bf k}^\dagger \ket{u^m_{D_g {\bf k}}}B^{mn}_{g,\bf k} 
\end{align}
So, an element of a Wilson line from ${\bf k_1}$ to ${\bf k_2}$ is equal to
\begin{align}
\W^{mn}_{{\bf k_2} \leftarrow {\bf k_1}} &= \braket{u^m_{\bf k_2}}{u^n_{\bf k_1}} \nonumber \\
&= B^{\dagger mr}_{g,\bf k_2}\matrixel{u^r_{D_g {\bf k_2}}}{g_{\bf k} g_{\bf k}^\dagger}{u^s_{D_g {\bf k_1}}}B^{sn}_{g,\bf k_1} \nonumber \\
&=  B^{\dagger mr}_{g,\bf k_2} \W^{rs}_{D_g {\bf k_2} \leftarrow D_g {\bf k_1}}B^{sn}_{g,\bf k_1}.
\end{align}
Reordering this we have
\begin{align}
B_{g,\bf k_2} \W_{{\bf {\bf k_2}} \leftarrow {\bf k_1}} B^\dagger _{g,\bf k_1} = \W_{D_g{\bf {\bf k_2}} \leftarrow D_g{\bf k_1}}.
\end{align}
In particular, for a Wilson loop at base point ${\bf k}$ we have
\begin{align}
\boxed{B_{g,\bf k} \W_{\C,{\bf k}} B^\dagger_{g,\bf k} = \W_{D_g \C, D_g {\bf k}}}
\label{eq:app_Wilson_loop_under_symmetry}
\end{align}
where, in the Wilson loop $\W_{\C,{\bf k}}$, the first subindex, $\C$, is the contour along which the Wilson loop is performed, and the second subindex, ${\bf k}$, is the starting point, or `base' point, of the Wilson loop. To simplify notation, from now on we will refer to Wilson loops along the contour $\C =  (k_x,k_y) \rightarrow (k_x + 2 \pi, k_y)$ along increasing (decreasing) values of $k_x$ as $\W_{x,\bf k}$ ($\W_{-x,\bf k}$), where ${\bf k} =(k_x,k_y)$ is the base point of the loop. Similarly, for the path $\C =  (k_x,k_y) \rightarrow (k_x, k_y + 2 \pi)$ along increasing (decreasing) values of $k_y$, we will denote the Wilson loops as $\W_{y,\bf k}$ ($\W_{-y,\bf k}$).
Fig.~\ref{fig:app_Wx_transformations} shows how these Wilson loops transform under the four spatial symmetries we will consider here: reflection in $x$, reflection in $y$, inversion, and $C_4$. In what follows, we study the constraints placed by these symmetries on the Wilson loops over the occupied energy bands, as well as on the Wilson loops over the Wannier sectors.

\subsection{Constraints due to reflection symmetry along $x$}
\label{sec:app_Wilson_loops_under_Mx}
We consider the constraints that reflection symmetry $M_x: x \rightarrow -x$ imposes on the Wilson loops $\W_{x,{\bf k}}$, as well as on the nested Wilson loops $\tilde\W^{\nu_x}_{y,{\bf k}}$.

\subsubsection{On the Wilson loop of the occupied energy bands}
Under reflection symmetry along $x$, the eigenvalues of Wilson loops along $x$ are constrained to be $+1$, $-1$, or to come in complex-conjugate pairs $e^{\pm i 2 \pi \nu}$, as we will see in this section. Consider a system with reflection symmetry along $x$
\begin{align}
\hat{M}_x h_{\bf k} \hat{M}_x^\dagger = h_{M_x {\bf k}},
\end{align}
where $M_x {\bf k} = M_x (k_x,k_y)=(-k_x, k_y)$. This symmetry allows us to write the expansion
\begin{align}
\hat{M}_x\ket{u^n_{\bf k}}=\ket{u^m_{M_x {\bf k}}}B^{mn}_{M_x,\bf k},
\end{align}
where 
\begin{align}
B^{mn}_{M_x,\bf k}=\matrixel{u^m_{M_x {\bf k}}}{\hat{M}_x}{u^n_{\bf k}}
\end{align}
is the unitary sewing matrix ($B^\dagger B = B B^\dagger = 1$), which connects states at ${\bf k}$ with states at $M_x {\bf k}$ having the same energy. In particular, $B^{mn}_{M_x,\bf k} \neq 0$ only if $\epsilon_{m, M_x {\bf k}}= \epsilon_{n,{\bf k}}$. 

The relation between Wilson loops in Eq.~\ref{eq:app_Wilson_loop_under_symmetry} for this symmetry is
\begin{align}
B_{M_x,\bf k} \W_{x,{\bf k}} B^\dagger_{M_x,\bf k} = \W_{-x, M_x {\bf k}}=\W^\dagger_{x,M_x{\bf k}}.
\label{eq:app_Wilson_loop_under_Mx}
\end{align}
An illustration of this relation is shown in Fig.~\ref{fig:app_Wx_transformations}A. Thus, the Wilson loop at ${\bf k}$ is equivalent (up to a unitary transformation) to the Hermitian conjugate of the Wilson loop at $M_x {\bf k}$. Since the eigenvalues of the Wilson loop along $x$ are $k_x$ independent, this directly imposes a restriction on the allowed Wannier centers at each $k_y$, namely, the set of Wilson-loop eigenvalues must obey
\begin{align}
\left\{ e^{i 2\pi \nu_x^i(k_y)} \right\} \stackrel{M_x}{=} \left\{ e^{-i 2\pi \nu_x^i(k_y)} \right\},
\end{align}
or
\begin{align}
\left\{\nu_x^i(k_y) \right\} \stackrel{M_x}{=} \left\{ -\nu_x^i(k_y) \right\} \;\;\mbox{mod 1}.
\label{eq:WannierBands_under_reflection_symmetry}
\end{align}
Thus, at each value of $k_y$ the Wannier centers $\nu_x(k_y)$ are forced to be 0 (centered at unit cell), 1/2 (centered in between unit cells), or to come in pairs $(-\nu,\nu)$ (pairs which are equally displaced from the unit cell but at opposite sides of it). From this it follows that, in order to have a gapped Wannier spectrum, we must have an even number $N_{occ}$ of occupied bands, for if we have an odd $N_{occ}$, at least one of the Wannier centers must have the value $\nu=0$ or $\nu=1/2$, equivalent to having at least one mid-gap state in the Wannier spectrum. Eq. \ref{eq:WannierBands_under_reflection_symmetry} also implies that, under reflection symmetry $M_x$ along $x$, the polarization
\begin{align}
p_x(k_y) = \sum^{N_{occ}}_{j=1} \nu^j(k_y) \;\;\mbox{mod 1} \nonumber
\end{align}
obeys
\begin{align}
p_x(k_y) \stackrel{M_x}{=} -p_x(k_y) \;\;\mbox{mod 1}.
\end{align}
i.e., 
\begin{align}
p_x(k_y) \stackrel{M_x}{=} 0 \mbox{ or } 1/2
\end{align}
for $k_y \in [-\pi,\pi)$.
For gapped systems, the Wannier spectrum is not discontinuous. In this case, the above restriction on $p_x(k_y)$ implies that the total polarization along $x$,
\begin{align}
p_x = \frac{1}{N_y}\sum_{k_y} p_x(k_y) \nonumber
\end{align}
obeys
\begin{align}
p_x \stackrel{M_x}{=} -p_x
\end{align}
i.e., the total polarization is also quantized,
\begin{align}
p_x \stackrel{M_x}{=} 0 \mbox{ or } 1/2.
\end{align}

\subsubsection{On the nested Wilson loop over Wannier sectors}
In Section \ref{sec:Quadrupole} we saw that the topological quadrupole is represented by the topology of the Wannier-sector polarizations \eqref{eq:PolarizationWannierSector}. These are subspaces within the subspace of occupied bands that belong to the same subset of Wannier bands. In this section we impose reflection symmetry on the Hamiltonian to see how the Wannier-sector polarizations are affected. 
For that purpose, we first focus on the Wilson loop eigenfunctions
\begin{align}
\W_{x,{\bf k}} \ket{\nu^i_{x,\bf k}}=e^{i2\pi \nu_x^i(k_y)}\ket{\nu^i_{x,\bf k}}.
\end{align}
Using Eq.~\ref{eq:app_Wilson_loop_under_Mx}, we have that
\begin{align}
 \W^\dagger_{x,M_x {\bf k}} B_{M_x,\bf k} \ket{\nu^i_{x,\bf k}} &= B_{M_x,\bf k} \W_{x,{\bf k}}  \ket{\nu^i_{x,\bf k}} \nonumber\\
 &= e^{i2\pi \nu_x^i(k_y)} B_{M_x,\bf k} \ket{\nu^i_{x,\bf k}}.
 \label{eq:app_Wilson_loop_eigenstate_under_Mx}
\end{align}
$B_{M_x,\bf k} \ket{\nu^i_{x,\bf k}} $ is hence an eigenfunction of $ \W_{x,M_x {\bf k}}$ with eigenvalue $e^{-i2\pi \nu_x^i(k_y)}$. We now expand this function as
\begin{align}
B_{M_x,\bf k} \ket{\nu^i_{x,\bf k}} = \ket{\nu^j_{x,M_x{\bf k}}}  \alpha^{ji}_{M_x,\bf k},
\label{eq:app_Wilson_loop_eigenstate_expansion_Mx}
\end{align}
where
\begin{align}
 \alpha^{ji}_{M_x,\bf k} = \matrixel{\nu^j_{x,M_x {\bf k}}}{B_{M_x,\bf k}}{\nu^i_{x,\bf k}}
\end{align}
is a unitary sewing matrix that connects Wilson-loop eigenstates at base points ${\bf k}$ and $M_x {\bf k}$ having \textit{opposite} Wilson-loop eigenvalues. 
If $ \alpha^{ji}_{M_x,\bf k} \neq 0$, we require that $-\nu_x^j(k_y)=\nu_x^i(k_y)$. This implies that $\alpha_{M_x,\bf k}$ is restricted to be block diagonal in the $\nu=0$ and $\nu=1/2$ sectors and off diagonal \textit{between} the sectors $\nu,-\nu$.

Now, let us act with the reflection operator on $\ket{w^j_{x,\bf k}}$, i.e., the states representing the Wannier basis as defined in Eq.~\ref{eq:WannierBasis},
\begin{align}
\hat{M}_x \ket{w^j_{x,\bf k}}&= \hat{M}_x\ket{u^n_{\bf k}}[\nu^j_{x,\bf k}]^n\nonumber\\
&=\ket{u^m_{M_x\bf k}} \matrixel{u^m_{M_x \bf k}}{\hat{M}_x}{u^n_{\bf k}}[\nu^j_{x,\bf k}]^n\nonumber\\
&=\ket{u^m_{M_x\bf k}} B^{mn}_{M_x,\bf k} [\nu^j_{x,\bf k}]^n\nonumber\\
&=\ket{u^m_{M_x\bf k}} [\nu^i_{x,M_x\bf k}]^m \alpha^{ij}_{M_x,\bf k}\nonumber\\
&=\ket{w^i_{x,M_x\bf k}} \alpha^{ij}_{M_x,\bf k}.
\label{eq:app_Wannier_basis_under_Mx}
\end{align}
From this relation we can write
\begin{align}
\ket{w^j_{x,\bf k}} &= M_x^\dagger \ket{w^i_{x,M_x \bf k}} \alpha^{ij}_{M_x,\bf k}\nonumber \\
\bra{w^j_{x,\bf k}} &=  [\alpha^\dagger_{M_x,\bf k}]^{ji}  \bra{w^i_{x,M_x {\bf k}}} M_x,
\end{align}
where $\nu_x^i(k_y)=-\nu_x^j(k_y)$ for nonzero $\alpha^{ji}_{M_x,\bf k}$. The Wilson line elements for the $\ket{w^j_{x,{\bf k}}}$ holonomy are related by
\begin{align}
\left[\tilde{\W}^{\nu_x}_{{\bf k}_2 \leftarrow {\bf k}_1}\right]^{ij} &= \braket{w^i_{x,{\bf k}_2}}{w^j_{x,{\bf k}_1}} \nonumber\\
&= [\alpha^\dagger_{M_x,{\bf k}_2}]^{i i'} \braket{w^{i'}_{x,M_x {\bf k}_2}}{w^{j'}_{x,M_x {\bf k}_1}} \alpha^{j' j}_{M_x,{\bf k}_1}\nonumber\\
&= [\alpha^\dagger_{M_x,{\bf k}_2}]^{i i'} \left[\tilde{\W}^{\nu'_x}_{M_x {\bf k}_2 \leftarrow M_x {\bf k}_1}\right]^{i'j'}  \alpha^{j' j}_{M_x,{\bf k}_1}.
\end{align}
In particular, for the nested Wilson loops along $y$ in the basis $\ket{w^j_{x}}$ we have
\begin{align}
\left[\tilde{\W}^{\nu_x}_{y,{\bf k}} \right]^{ij} &=  \left[\alpha^\dagger_{M_x,\bf k} \right]^{ii'} \left[ \tilde{\W}^{\nu'_x}_{y, M_x {\bf k}} \right]^{i'j'} \left[\alpha^{\phantom{\dagger}}_{M_x,\bf k} \right]^{j'j}.
\label{eq:app_Wilson_of_Wilson_under_Mx}
\end{align}
Eq.~\ref{eq:app_Wilson_of_Wilson_under_Mx} implies two things: first, since $\nu^j_x(k_y)=-\nu^i_x(k_y)$ for nonzero $\alpha^{ji}_{M_x,\bf k}$, this expression tells us that Wilson loops along $y$ at base point ${\bf k},$ over Wannier sectors $\nu_x=0$ or $1/2,$ are equivalent (up to unitary transformations) to Wilson loops along $y$ over the same Wannier sector at base point $M_x {\bf k}$. Second, suppose that we have gapped Wannier bands $\{\nu_x(k_y),-\nu_x(k_y)\}$ across the entire range $k_y \in (-\pi, \pi].$ Then Eq. \ref{eq:app_Wilson_of_Wilson_under_Mx} tells us that if we calculate the Wilson loop along $y$ over the Wannier sector $\nu_x(k_y)$ at base point ${\bf k}$, this Wilson loop is equivalent (up to a unitary transformation) to the Wilson loop along $y$ at base point $M_x {\bf k}$ over the sector $-\nu_x(k_y)$.
Thus, for the eigenvalues $\exp[i 2\pi \nu^{\nu_x,j}_y(k_x)]$ of the Wilson loop $\tilde{\W}^{\nu_x}_{y, {\bf k}}$ over the Wannier sector $\nu_x$, $M_x$ implies that
\begin{align}
\left\{e^{i2\pi \nu^{\nu_x,j}_y(k_x)}\right\} \stackrel{M_x}{=} \left\{e^{i2\pi \nu^{-\nu_x,j}_y(-k_x)}\right\}
\end{align}
or
\begin{align}
\{\nu^{\nu_x,j}_y(k_x)\} \stackrel{M_x}{=}\{\nu^{-\nu_x,j}_y(-k_x)\} \;\; \mbox{mod 1},
\end{align}
where $j \in 1\ldots N_{\nu_x}$ labels the eigenvalue, and $N_{\nu_x}$ is the number of Wannier bands in the sector $\nu_x$. The Wannier-sector polarization can be written as
\begin{align}
p^{\nu_x}_y = \frac{1}{N_x} \sum_{k_x} \sum_{j=1}^{N_{\nu_x}} \nu^{\nu_x,j}_y(k_x) \;\;\mbox{mod 1}.\nonumber
\label{eq:app_Wannier_sector_polarization}
\end{align}
Hence, since $k_x$ is a dummy variable, $M_x$ symmetry implies that
\begin{align}
p^{\nu_x}_y\stackrel{M_x}{=}p^{-\nu_x}_y \;\;\mbox{mod 1},
\end{align}
which is the first expression in \eqref{eq:WannierPolarization_SymmetryConstraints}.

\subsection{Constraints due to reflection symmetry along $y$}
\label{sec:app_Wilson_loops_under_My}
We now derive the constraints that reflection symmetry $M_y: y \rightarrow -y$ imposes on the Wilson loops along $x$, $\W_{x,{\bf k}}$, and on the nested Wilson loops along $y$ over Wannier sector $\nu_x$, $\tilde\W^{\nu_x}_{y,{\bf k}}$.

\subsubsection{On the Wilson loop over energy bands}
Consider a system with reflection symmetry along $y$
\begin{align}
\hat{M}_y h_{\bf k} \hat{M}_y^\dagger = h_{M_y {\bf k}},
\end{align}
where $M_y {\bf k} = M_y (k_x,k_y)=(k_x,-k_y)$. This symmetry allows us to write the expansion
\begin{align}
\hat{M}_y\ket{u^n_{\bf k}}=\ket{u^m_{M_y {\bf k}}}B^{mn}_{M_y,\bf k},
\label{eq:}
\end{align}
where 
\begin{align}
B^{mn}_{M_y,\bf k}=\matrixel{u^m_{M_y {\bf k}}}{\hat{M}_y}{u^n_{\bf k}}
\end{align}
is the unitary sewing matrix, which connects states at ${\bf k}$ with states at $M_y {\bf k}$. In particular, $B^{mn}_{M_y,\bf k} \neq 0$ only if $\epsilon_{m,M_y {\bf k}}= \epsilon_{n,{\bf k}}$. 

Under this symmetry, the Wilson loop along $x$ starting at base point ${\bf k}$ obeys
\begin{align}
B_{M_y,\bf k} \W_{x,{\bf k}} B^\dagger_{M_y,\bf k} = \W_{x,M_y {\bf k}}.
\label{eq:app_Wilson_loop_under_My}
\end{align}
A schematic of this relation is shown in Fig.~\ref{fig:app_Wx_transformations}b. The Wilson loops based at ${\bf k}$ and $M_y {\bf k}$ are equivalent up to a unitary transformation. Hence, the sets of their eigenvalues are the same, namely,
\begin{align}
\left\{ e^{i 2\pi \nu_x^i(k_y)} \right\} \stackrel{M_y}{=} \left\{ e^{i 2\pi \nu_x^i(-k_y)} \right\}
\end{align}
or
\begin{align}
\left\{ \nu_x^i(k_y) \right\} \stackrel{M_y}{=} \left\{ \nu_x^i(-k_y) \right\} \;\;\mbox{mod 1}
\end{align}
which leads to 
\begin{align}
p_x(k_y) \stackrel{M_y}{=} p_x(-k_y)\;\;\mbox{mod 1}.
\end{align}
Notice that the overall polarization along $x$, Eq. \ref{eq:Polarization_2D_discrete}, is not constrained by reflection symmetry along $y$.

\subsubsection{On the nested Wilson loop over Wannier sectors}

Now, we focus on the Wilson loop eigenfunctions
\begin{align}
\W_{x,{\bf k}} \ket{\nu^i_{x,\bf k}}=e^{i2\pi \nu_x^i(k_y)}\ket{\nu^i_{x,\bf k}}.
\end{align}
Rewriting Eq. \ref{eq:app_Wilson_loop_under_My} as $B_{M_y,\bf k} \W_{x,{\bf k}} = \W_{x,M_y {\bf k}} B_{M_y,\bf k}$, we have that
\begin{align}
 \W_{x,M_y {\bf k}} B_{M_y,\bf k} \ket{\nu^i_{x,\bf k}} &= B_{M_y,\bf k} \W_{x,{\bf k}}  \ket{\nu^i_{x,\bf k}} \nonumber\\
 &= e^{i2\pi \nu_x^i(k_y)} B_{M_y,\bf k} \ket{\nu^i_{x,\bf k}}.
 \label{eq:app_Wilson_loop_eigenstate_under_symmetry}
\end{align}
$B_{M_y,\bf k} \ket{\nu^i_{x,\bf k}} $ is an eigenfunction of $ \W_{x,M_y {\bf k}}$ with eigenvalue $e^{i2\pi \nu_x^i(k_y)}$. We now expand this function as
\begin{align}
B_{M_y,\bf k} \ket{\nu^i_{x,\bf k}} = \ket{\nu^j_{x,M_y{\bf k}}}  \alpha^{ji}_{M_y,\bf k},
\label{eq:app_Wilson_loop_eigenstate_expansion}
\end{align}
where
\begin{align}
 \alpha^{ji}_{M_y,\bf k} = \matrixel{\nu^j_{x,M_y {\bf k}}}{B_{M_y,\bf k}}{\nu^i_{x,\bf k}}
\end{align}
is a sewing matrix that connects Wilson-loop eigenstates at base point ${\bf k}=(k_x,k_y)$ and base point $M_y {\bf k}=(k_x,-k_y)$ having the same Wilson-loop eigenvalues; if $ \alpha^{ji}_{M_y,\bf k} \neq 0$, we require that $\nu_x^j(-k_y)=\nu_x^i(k_y)$. 
Following the same procedure as in \eqref{eq:app_Wannier_basis_under_Mx} for the Wannier sectors $\ket{w^j_{x,{\bf k}}}$, we have
\begin{align}
\hat{M}_y \ket{w^j_{x,\bf k}}&=\ket{w^i_{x,M_y\bf k}} \alpha^{ij}_{M_y,\bf k},
\end{align}
from which it follows that
\begin{align}
\ket{w^j_{x,\bf k}} &= \hat{M}_y^\dagger \ket{w^i_{x,M_y \bf k}} \alpha^{ij}_{M_y,\bf k}.
\end{align}
Using these expressions, there is the following relation for a Wilson line element
\begin{align}
\left[\tilde{\W}^{\nu_x}_{{\bf k}_2 \leftarrow {\bf k}_1}\right]^{ij} = \left[\alpha^\dagger_{M_y,{\bf k}_2}\right]^{i i'} \left[\tilde{\W}^{\nu'_x}_{M_y {\bf k}_2 \leftarrow M_y {\bf k}_1}\right]^{i'j'}  \left[\alpha^{\phantom{\dagger}}_{M_y,{\bf k}_1}\right]^{j'j}.
\end{align}
In particular, the nested Wilson loop along $y$ in the basis $\ket{w^j_{x}}$ obeys
\begin{align}
\left[\tilde{\W}^{\nu_x}_{y,{\bf k}} \right]^{ij} &=  \left[\alpha^\dagger_{M_y,\bf k} \right]^{ii'} \left[ \tilde{\W}^{\nu'_x}_{-y, M_y {\bf k}} \right]^{i'j'} \left[\alpha^{\phantom{\dagger}}_{M_y,\bf k} \right]^{j'j},
\end{align}
which looks similar to the one in Section \ref{sec:app_Wilson_loops_under_Mx}, but with the important difference in the structure of $\alpha^{ji}_{M_y,\bf k}$, which connects Wilson-loop eigenstates such that $\nu^j_x(k_y) = \nu^i_x(-k_y)$. Another important difference is the fact that $\hat{M}_y$ reverses the loop contour along $y$ and preserves it along $x$. This expression tells us that the Wilson-loop eigenvalues are related by
\begin{align}
\left\{e^{i 2 \pi \nu^{\nu_x,j}_y(k_x)}\right\} \stackrel{M_y}{=} \left\{e^{-i 2\pi \nu^{\nu_x,j}_y(k_x)}\right\}
\end{align}
or
\begin{align}
\left\{\nu^{\nu_x,j}_y(k_x) \right\} \stackrel{M_y}{=} \left\{ - \nu^{\nu_x,j}_y(k_x)\right\} \;\; \mbox{mod 1},
\end{align}
from which it follows that $\nu^{\nu_x}_y(k_x)$ is either $0$, $1/2$, or comes in pairs $\nu,-\nu$. $M_y$ thus implies that the polarization \eqref{eq:app_Wannier_sector_polarization} over the Wannier sector $\nu_x$ obeys 
\begin{align}
p^{\nu_x}_y \stackrel{M_y}{=} -p^{\nu_x}_y \;\;\mbox{mod 1},
\end{align}
from which it follows that 
\begin{align}
p^{\nu_x}_y \stackrel{M_y}{=} 0 \mbox{ or } 1/2,
\end{align}
which is the second expression in \eqref{eq:WannierPolarization_SymmetryConstraints}.
In particular, values of $\nu^{\nu_x}_y(k_x)$ that come in pairs $\nu,-\nu$ do not contribute to $p^{\nu_x}_y$.

The results in this subsection, and in the previous one, provide the constraints due to both reflection symmetries on the Wilson loops $\W_{y,{\bf k}}$ and $\tilde\W^{\nu_y}_{x,{\bf k}}$. The constraints due to these reflection symmetries on $\W_{x,{\bf k}}$ and $\tilde\W^{\nu_x}_{y,{\bf k}}$ can be obtained simply by replacing the labels $x \leftrightarrow y$ in the results of these two subsections.

\subsection{Constraints due to inversion symmetry}
\label{sec:app_Wilson_loops_under_inversion}
We now derive the constraints that inversion symmetry imposes on the Wilson loops $\W_{x,{\bf k}}$ and on $\tilde\W^{\nu_x}_{y,{\bf k}}$.

\subsubsection{On the Wilson loop over energy bands}
Consider the constrains imposed by inversion symmetry,
\begin{align}
\hat{\I} h_{\bf k} \hat{\I}^\dagger = h_{-{\bf k}},
\end{align}
under which the Wilson loop obeys (see Eq. \ref{eq:app_Wilson_loop_under_symmetry})
\begin{align}
B_{\I,\bf k} \W_{x,{\bf k}} B^\dagger_{\I,\bf k} = \W_{-x,-{\bf k}}=\W^\dagger_{x,-{\bf k}},
\label{eq:app_Wilson_loop_under_Inversion}
\end{align}
where
\begin{align}
B^{mn}_{\I,\bf k}=\matrixel{u^m_{-{\bf k}}}{\hat{\I}}{u^n_{\bf k}}
\end{align}
connects energy eigenstates at ${\bf k}$ and $-{\bf k}$ having the same energy. A schematic of this relation is shown in Fig.~\ref{fig:app_Wx_transformations}c. 
Eq. \ref{eq:app_Wilson_loop_under_Mx} implies that the set of eigenvalues obey
\begin{align}
\left\{ e^{i 2\pi \nu_x^i(k_y)} \right\} \stackrel{\I}{=} \left\{ e^{-i 2\pi \nu_x^i(-k_y)} \right\},
\label{eq:app_Wilson_loop_eigenvalues_inversion_symmetry}
\end{align}
or 
\begin{align}
\{\nu_x^i(k_y)\} \stackrel{\I}{=}\{- \nu_x^i(-k_y)\} \;\;\mbox{mod 1}.
\end{align}
In particular, for values $k^*_y=0,\pi$ of the $y$ coordinate of the Wilson-loop base point, we recover the identical condition as for reflection symmetry along $x$. Thus, at these points the Wilson-loop eigenvalues are either $0$, $1/2$ or come in pairs $\nu, -\nu$. We also have
\begin{align}
p_x(k_y) \stackrel{\I}{=} -p_x(-k_y)\;\;\mbox{mod 1},
\label{eq:app_pxkyUnderInversion}
\end{align}
so that the polarization obeys
\begin{align}
p_x \stackrel{\I}{=} -p_x\;\;\mbox{mod 1},
\end{align}
i.e., it is quantized,
\begin{align}
p_x \stackrel{\I}{=} 0 \mbox{ or } 1/2.
\end{align}
which is the relation given in Eq. \ref{eq:pxUnderInversion}.

\subsubsection{On the nested Wilson loop over Wannier sectors}
For the Wilson-loop eigenstates
\begin{align}
\W_{x,{\bf k}} \ket{\nu^i_{x,\bf k}}=e^{i2\pi \nu_x^i(k_y)}\ket{\nu^i_{x,\bf k}}
\end{align}
one can use Eq. \ref{eq:app_Wilson_loop_under_Inversion}, to show that $B_{\I,\bf k} \ket{\nu^i_{x,\bf k}}$ is an eigenstate of $\W_{x,-{\bf k}}$ with eigenvalue $e^{-i 2\pi \nu_x^i(k_y)}$. Thus, in the expansion
\begin{align}
B_{\I,\bf k} \ket{\nu^i_{x,\bf k}} = \ket{\nu^j_{x,-{\bf k}}}  \alpha^{ji}_{\I,\bf k},
\end{align}
the sewing matrix
\begin{align}
 \alpha^{ji}_{\I,\bf k} = \matrixel{\nu^j_{x,-{\bf k}}}{B_{\I,\bf k}}{\nu^i_{x,\bf k}}
\end{align}
connects Wilson-loop eigenstates at base points ${\bf k}$ and $-{\bf k}$ having opposite Wannier centers; i.e., $ \alpha^{ji}_{\bf k} \neq 0$ only if $\nu_x^i(k_y) = -\nu_x^j(-k_y)$. 
For the Wannier sectors $\ket{w^j_{x,{\bf k}}}$, we have
\begin{align}
\hat{\I} \ket{w^j_{x,\bf k}}&=\ket{w^i_{x,-\bf k}} \alpha^{ij}_{\I,\bf k},
\end{align}
from which it follows that
\begin{align}
\ket{w^j_{x,\bf k}} &= \hat{\I}^\dagger \ket{w^i_{x,- {\bf k}}} \alpha^{ij}_{\I,\bf k}.
\end{align}

Using these expressions, there is the following relation for a Wilson line element
\begin{align}
\left[\tilde{\W}^{\nu_x}_{{\bf k}_2 \leftarrow {\bf k}_1}\right]^{ij} = \left[\alpha^\dagger_{\I,{\bf k}_2}\right]^{i i'} \left[\tilde{\W}^{\nu'_x}_{- {\bf k}_2 \leftarrow - {\bf k}_1}\right]^{i'j'}  \left[\alpha^{\phantom{\dagger}}_{\I,{\bf k}_1}\right]^{j'j}.
\end{align}
In particular the Wilson loop along $y$ obeys
\begin{align}
\left[\tilde{\W}^{\nu_x}_{y,{\bf k}} \right]^{ij} &=  \left[\alpha^\dagger_{\I,\bf k} \right]^{ii'} \left[ \tilde{\W}^{\nu'_x}_{-y, - {\bf k}} \right]^{i'j'} \left[\alpha^{\phantom{\dagger}}_{\I,\bf k} \right]^{j'j}.
\end{align}
Thus, the Wilson-loop eigenvalues are related by
\begin{align}
\left\{e^{i 2 \pi \nu^{\nu_x,j}_y(k_x)}\right\} \stackrel{\I}{=} \left\{e^{-i 2\pi \nu^{-\nu_x,j}_y(-k_x)}\right\}
\end{align}
or
\begin{align}
\left\{\nu^{\nu_x,j}_y(k_x) \right\} \stackrel{\I}{=} \left\{ - \nu^{-\nu_x,j}_y(-k_x)\right\}\;\; \mbox{mod 1}.
\end{align}
Thus, we have that 
\begin{align}
p^{\nu_x}_y(k_x) \stackrel{\I}{=} -p^{-\nu_x}_y(-k_x)\;\;\mbox{mod 1},
\end{align}
and the Wannier-sector polarization \eqref{eq:app_Wannier_sector_polarization} under inversion symmetry obeys
\begin{align}
p^{\nu_x}_y \stackrel{\I}{=} -p^{-\nu_x}_y\;\;\mbox{mod 1}.
\label{eq:app_WannierPolarizationUnderInversion}
\end{align}
which is the third expression in \eqref{eq:WannierPolarization_SymmetryConstraints}.

\subsection{Constraints due to $C_4$ symmetry}
\label{sec:app_Wilson_loops_under_C4}
We now derive the constraints that $C_4$ symmetry imposes on the Wilson loops $\W_{x,{\bf k}}$ and on $\tilde\W^{\nu_x}_{y,{\bf k}}$.

\subsubsection{On the Wilson loop over energy bands}
Now, we consider $C_4$ symmetry
\begin{align}
\hat{r}_4 h_{\bf k} \hat{r}_4^\dagger = h_{R_4 {\bf k}}
\end{align}
under which the Wilson loop obeys
\begin{align}
B_{C_4,\bf k} \W_{x,{\bf k}} B^\dagger_{C_4,\bf k} &= \W_{y,R_4{\bf k}}\nonumber\\
B_{C_4,\bf k} \W_{y,{\bf k}} B^\dagger_{C_4,\bf k} &= \W_{-x,R_4{\bf k}}
\label{eq:app_Wilson_loop_under_C4}
\end{align}
where
\begin{align}
B^{mn}_{C_4,\bf k}=\matrixel{u^m_{R_4{\bf k}}}{\hat{r}_4}{u^n_{\bf k}}
\end{align}
is the sewing matrix with elements $B^{mn}_{C_4,\bf k} \neq 0$ only if $\epsilon_{m,R_4 {\bf k}} = \epsilon_{n,{\bf k}}$. The Wannier values are then related by
\begin{align}
\left\{ \nu^i_x(k_y) \right\} &\stackrel{C_4}{=} \left\{ \nu^i_y(k_x=-k_y)\right\}\;\;\mbox{mod 1} \nonumber\\
\left\{ \nu^i_y(k_x) \right\} &\stackrel{C_4}{=} \left\{ -\nu^i_x(k_y=k_x)\right\}\;\;\mbox{mod 1}.
\end{align}
These in turn lead to 
\begin{align}
p_x(k_y) &\stackrel{C_4}{=} p_y(k_x=-k_y)\;\;\mbox{mod 1} \nonumber\\
p_y(k_x) &\stackrel{C_4}{=} -p_x(k_y=k_x)\;\;\mbox{mod 1}.
\end{align}
Notice that the successive application of one of these relations after the other one leads to \eqref{eq:app_pxkyUnderInversion}, which is nothing but the constraint due to $C_2$ symmetry (in the absence of spin). The constraint over the polarization is then
\begin{align}
p_x \stackrel{C_4}{=} p_y \stackrel{C_4}{=} 0 \mbox{ or } 1/2.
\end{align}

\subsubsection{On the nested Wilson loop over Wannier sectors}
The two relations in \eqref{eq:app_Wilson_loop_under_C4} allow us to write the expansions
\begin{align}
B_{C_4,\bf k} \ket{\nu^i_{x,\bf k}} &= \ket{\nu^j_{y, R_4 {\bf k}}} \alpha^{ji}_{C_4,\bf k}\nonumber\\
B_{C_4,\bf k} \ket{\nu^i_{y,\bf k}} &= \ket{\nu^j_{x,R_4 {\bf k}}} \beta^{ji}_{C_4,\bf k}
\end{align}
where $\alpha_{C_4,\bf k}$ and $\beta_{C_4,\bf k}$ are the sewing matrices
\begin{align}
\alpha^{ji}_{C_4,\bf k} &= \matrixel{\nu^j_{y,R_4 {\bf k}}}{B_{C_4,\bf k}}{{\nu^i_{x,\bf k}}} 
\end{align}
with $\alpha^{ji}_{C_4,\bf k} \neq 0$ only if  $\nu_x^i(k_y) = \nu_y^j(k_x=-k_y)$, and
\begin{align}
\beta^{ji}_{C_4,\bf k} &= \matrixel{\nu^j_{x,R_4 {\bf k}}}{B_{C_4,\bf k}}{\nu^i_{y,\bf k}}
\end{align}
with $\beta^{ji}_{C_4,\bf k} \neq 0$ only if $\nu_y^i(k_x) = -\nu_x^j(k_y=k_x)$.

The Wannier sectors $\ket{w^j_{x,{\bf k}}}$ and $\ket{w^j_{y,{\bf k}}}$ transform as
\begin{align}
\hat{r}_4 \ket{w^j_{x,\bf k}}&=\ket{w^i_{y,R_4\bf k}} \alpha^{ij}_{C_4,\bf k}\nonumber\\
\hat{r}_4 \ket{w^j_{y,\bf k}}&=\ket{w^i_{x,R_4\bf k}} \beta^{ij}_{C_4,\bf k}.
\end{align}
Using these expressions and their hermitian conjugates, we find the following relations for the Wilson line elements
\begin{align}
\left[\tilde{\W}^{\nu_x}_{{\bf k}_2 \leftarrow {\bf k}_1}\right]^{ij} &= \left[\alpha^\dagger_{C_4,{\bf k}_2}\right]^{i i'} \left[\tilde{\W}^{\nu_y}_{R_4 {\bf k}_2 \leftarrow R_4 {\bf k}_1}\right]^{i'j'}  \left[\alpha^{\phantom{\dagger}}_{C_4,{\bf k}_1}\right]^{j'j}\nonumber\\
\left[\tilde{\W}^{\nu_y}_{{\bf k}_2 \leftarrow {\bf k}_1}\right]^{ij} &= \left[\beta^\dagger_{C_4,{\bf k}_2}\right]^{i i'} \left[\tilde{\W}^{\nu_x}_{R_4 {\bf k}_2 \leftarrow R_4 {\bf k}_1}\right]^{i'j'}  \left[\beta^{\phantom{\dagger}}_{C_4,{\bf k}_1}\right]^{j'j}.
\end{align}
In particular, the nested Wilson loops along $x$ and $y$ obey
\begin{align}
\left[\tilde{\W}^{\nu_x}_{y,{\bf k}} \right]^{ij} &=  \left[\alpha^\dagger_{C_4,{\bf k}} \right]^{ii'} \left[ \tilde{\W}^{\nu_y}_{-x, R_4 {\bf k}} \right]^{i'j'} \left[\alpha^{\phantom{\dagger}}_{C_4,{\bf k}} \right]^{j'j}\nonumber\\
\left[\tilde{\W}^{\nu_y}_{x,{\bf k}} \right]^{ij} &=  \left[\beta^\dagger_{C_4,{\bf k}} \right]^{ii'} \left[ \tilde{\W}^{\nu_x}_{y, R_4 {\bf k}} \right]^{i'j'} \left[\beta^{\phantom{\dagger}}_{C_4,{\bf k}} \right]^{j'j}.
\end{align}
These expressions tell us that the Wilson-loop eigenvalues are related by
\begin{align}
\left\{e^{i 2 \pi \nu^{\nu_x}_y(k_x)}\right\} &\stackrel{C_4}{=} \left\{e^{-i 2\pi \nu^{\nu_y}_x(k_x)}\right\}\nonumber\\
\left\{e^{i 2 \pi \nu^{\nu_y}_x(k_y)}\right\} &\stackrel{C_4}{=} \left\{e^{i 2\pi \nu^{-\nu_x}_y(-k_y)}\right\}
\end{align}
or
\begin{align}
\left\{\nu^{\nu_x}_y(k_x) \right\} &\stackrel{C_4}{=} \left\{ - \nu^{\nu_y}_x(k_x)\right\}\;\;\mbox{mod 1}\nonumber\\
\left\{\nu^{\nu_y}_x(k_y) \right\} &\stackrel{C_4}{=} \left\{ \nu^{-\nu_x}_y(-k_y)\right\}\;\;\mbox{mod 1}
\end{align}
or, for the Wannier-sector polarizations:
\begin{align}
p^{\nu^\pm_x}_y&\stackrel{C_4}{=}-p^{\nu^\pm_y}_x\;\;\mbox{mod 1}\nonumber\\
p^{\nu^\pm_y}_x&\stackrel{C_4}{=}p^{\nu^\mp_x}_y\;\;\mbox{mod 1}.
\end{align}
Notice that the sequential application of these two equations, which amounts to a $C_2$ rotation (or inversion), results in \eqref{eq:app_WannierPolarizationUnderInversion}.

\subsection{Constraints due to time-reversal symmetry}
\label{sec:app_Wilson_loops_under_TRS}
In this section we derive the constraints that time reversal symmetry $TR: t \rightarrow -t$ imposes on the Wilson loops $\W_{x,{\bf k}}$ and on the nested Wilson loops $\tilde\W^{\nu_x}_{y,{\bf k}}$.

\subsubsection{On the Wilson loop over energy bands}
The time reversal operator is $\hat{T}=QK$, where $K$ is complex-conjugation and $Q$ is a unitary operator, so that  $Q^{-1}=Q^\dagger$. For spinless systems, $\hat{T}^2=1$. For spinfull systems, $\hat{T}^2=-1$.
Time reversal symmetry (TRS) is stated as
\begin{align}
\hat{T} h_{\bf k}\hat{T}^\dagger=h_{-{\bf k}}.
\end{align}
As before, it is possible to expand
\begin{align}
\hat{T}\ket{u^n_{\bf k}}&=\ket{u^m_{-{\bf k}}}\matrixel{u^m_{-{\bf k}}}{T}{u^n_{\bf k}}\nonumber\\
&=\ket{u^m_{-{\bf k}}} V^{mn}_{\bf k},
\end{align}
where 
\begin{align}
V^{mn}_{\bf k}=\matrixel{u^m_{-{\bf k}}}{T}{u^n_{\bf k}}=\matrixel{u^m_{-{\bf k}}}{Q}{u^{n*}_{\bf k}}
\end{align}
is the sewing matrix, which is unitary. Here, the asterix represents complex-conjugation. The sewing matrix has nonzero elements $V^{mn}_{\bf k} \neq 0$ only if $\epsilon_n({\bf k}) = \epsilon_m(-{\bf k})$. 

For spinfull systems, $\hat{T}^2=-1$ leads to $Q^T=-Q$, which can be seen from joining the two expressions:
\begin{align}
&T^2=QKQK=QQ^*=-1\nonumber\\
&QQ^\dagger =1 \rightarrow Q^*Q^T=1\nonumber,
\end{align}
so that $Q^*Q^T=-Q^*Q$, or $Q^*(Q^T+Q)=0$, from which $Q^T=-Q$. This implies that $V^T_{\bf k}=-V_{-{\bf k}}$, as can be seen as follows:
\begin{align}
V^{mn}_{\bf k}&=\matrixel{u^m_{-{\bf k}}}{Q}{u^{n*}_{\bf k}}\nonumber\\
V^{\dagger mn}_{\bf k}&=[V^{T*}_{\bf k}]^{mn}=\matrixel{u^{m*}_{\bf k}}{Q^{T*}}{u^{n}_{-{\bf k}}}\nonumber\\
[V^T_{\bf k}]^{mn}&=\matrixel{u^m_{\bf k}}{Q^{T}}{u^{n*}_{-{\bf k}}}\nonumber\\
[V^T_{\bf k}]^{mn}&=-\matrixel{u^m_{\bf k}}{Q}{u^{n*}_{-{\bf k}}}\nonumber\\
V^T_{\bf k}&=-V_{-{\bf k}}.
\end{align}
At time reversal invariant momenta, ${{{\bf k}_{*}}} = -{{{\bf k}_{*}}}$, this relation reduces to $V^T_{{\bf k}_*}=-V_{{\bf k}_*}$, which is not possible if $V_{{\bf k}^*}$ is one-dimensional. Since $V^{mn}_{\bf k} \neq 0$ only if $\epsilon_m(-{\bf k}) = \epsilon_n({\bf k})$, this means that at the time-reversal invariant momenta the energy spectrum is at least twofold degenerate.
Reordering terms in the expansion of $T\ket{u^n_{\bf k}}=Q\ket{u^{n*}_{\bf k}}$ above we have:
\begin{align}
\ket{u^n_{-{\bf k}}}&=Q\ket{u^{m*}_{\bf k}}V^{\dagger mn}_{\bf k}\nonumber\\
\bra{u^n_{-{\bf k}}}&=V^{nm}_{\bf k} \bra{u^{m*}_{\bf k}}Q^\dagger.
\end{align}
Now, consider two momenta ${\bf k}_1$, ${\bf k}_2$, which are very close to each other. There is the following relation between Wilson lines:
\begin{align}
\W_{-{\bf k}_2 \leftarrow -{\bf k}_1}^{mn}&=\braket{u^m_{-{\bf k}_2}}{u^n_{-{\bf k}_1}}\nonumber\\
&=V^{mr}_{{\bf k}_2} \matrixel{u^{r*}_{{\bf k}_2}}{Q^\dagger Q}{u^{s*}_{{\bf k}_1}}V^{\dagger sn}_{{\bf k}_1}\nonumber\\
&=V^{mr}_{{\bf k}_2} \W_{{\bf k}_2 \leftarrow {\bf k}_1}^{rs*}V^{\dagger sn}_{{\bf k}_1},
\end{align}
or, more compactly,
\begin{align}
\W_{-{\bf k}_2 \leftarrow -{\bf k}_1}=V_{{\bf k}_2} \W_{{\bf k}_2 \leftarrow {\bf k}_1}^* V^\dagger_{{\bf k}_1}.
\end{align}
For Wilson loops starting at a base point ${\bf k}$ we have
\begin{align}
\W_{{-x},-{\bf k}}=\W^\dagger_{x,-{\bf k}}=V_{\bf k} \W^*_{x,{\bf k}} V^\dagger_{\bf k}
\label{eq:app_Wilson_loop_under_TR}
\end{align}
Thus, as before, there is an equivalence up to a unitary transformation. Thus, the set of eigenvalues must obey
\begin{align}
\left\{e^{-i 2\pi \nu_x^i(-k_y)}\right\} \stackrel{TR}{=} \left\{e^{-i 2\pi \nu_x^i(k_y)}\right\}
\end{align}
or
\begin{align}
\left\{ \nu_x^i(-k_y) \right\} \stackrel{TR}{=} \left\{ \nu_x^i(k_y) \right\},
\end{align}
which implies that
\begin{align}
p_x(k_y) \stackrel{TR}{=} p_x(-k_y).
\end{align}
This does not impose a constraint on the values of polarization. 

\subsubsection{On the nested Wilson loop over Wannier sectors}
Let us calculate how the Wilson loop eigenstates transform under TR. Acting with the Wilson loop $\W^\dagger_{x,-{\bf k}}$ on $V_{\bf k} \ket{\nu^{i*}_{x,{\bf k}}},$ and making use of \eqref{eq:app_Wilson_loop_under_TR}, we have
\begin{align}
\W^\dagger_{x,-{\bf k}} V_{\bf k} \ket{\nu^{i*}_{x,{\bf k}}} &= V_{\bf k} \W^*_{x,{\bf k}} \ket{\nu^{i*}_{x,{\bf k}}}\nonumber\\
&= e^{-i 2\pi \nu_x^i(k_y)} V_{\bf k} \ket{\nu^{i*}_{x,{\bf k}}}
\end{align}
So, $V_{\bf k} \ket{\nu^{i*}_{x,{\bf k}}}$ is an eigenstate of $\W^\dagger_{x,-{\bf k}}$ with eigenvalue $ e^{-i 2\pi \nu^i_x(k_y)}$. Thus, we can write the expansion
\begin{align}
V_{\bf k} \ket{\nu^{i*}_{x,{\bf k}}} = \ket{\nu^j_{x,-{\bf k}}}\alpha^{ji}_{T,\bf k}
\end{align}
where 
\begin{align}
\alpha^{ji}_{T,\bf k} = \matrixel{\nu^j_{x,-{\bf k}}}{V_{\bf k}}{\nu^{i*}_{x,{\bf k}}}
\end{align}
is the sewing matrix connecting $\ket{\nu^{i*}_{x,\bf k}}$ with $\ket{\nu^j_{x,-{\bf k}}}$. In particular, $\alpha^{ji}_{T,\bf k} \neq 0$ only if $\nu_x^i(k_y)=\nu_x^j(-k_y)$. 

Now, we act with the TR operator on the Wannier basis:
\begin{align}
\hat{T} \ket{w^j_{x,\bf k}}&= \hat{T}\ket{u^n_{\bf k}}[\nu^{j*}_{x,\bf k}]^n\nonumber\\
&=\ket{u^m_{-\bf k}} V^{mn}_{\bf k} [\nu^j_{x,\bf k}]^n\nonumber\\
&=\ket{u^m_{-\bf k}} [\nu^i_{x,-\bf k}]^m \alpha^{ij}_{T,\bf k}\nonumber\\
&=\ket{w^i_{x,-\bf k}} \alpha^{ij}_{T,\bf k}.
\label{eq:app_Wannier_basis_under_TR}
\end{align}
At the time reversal invariant momenta we have
\begin{align}
\hat{T} \ket{w^j_{x,\bf k_*}}&=\ket{w^i_{x,\bf k_*}} \alpha^{ij}_{T,\bf k_*},
\end{align}
which implies that, when $\hat{T}^2=-1$, there has to be Kramers degeneracy in the Wannier centers at these invariant momenta. To see this, one can act with the TR operator twice:
\begin{align}
\hat{T}\left( \hat{T} \ket{w^i_{x,\bf k_*}} \right)&= T\left( \ket{w^j_{x,\bf k_*}} \alpha^{ji}_{T,\bf k_*} \right),\nonumber\\
&= \left(T \ket{w^j_{x,\bf k_*}}\right) \alpha^{*ji}_{T,\bf k_*},\nonumber\\
&= \ket{w^k_{x,\bf k_*}} \alpha^{kj}_{T,\bf k_*} \alpha^{*ji}_{T,\bf k_*}\nonumber\\
\end{align}
while, on the other hand,
\begin{align}
\hat{T}\left( \hat{T} \ket{w^i_{x,\bf k_*}} \right)&= T^2 \ket{w^i_{x,\bf k_*}}\nonumber\\
&=-\ket{w^i_{x,\bf k_*}}.
\end{align}
Thus, the sewing matrix needs to obey
\begin{align}
-\delta^{ki}=\alpha^{kj}_{T,\bf k_*} \alpha^{*ji}_{T,\bf k_*},
\end{align}
a restriction that is impossible to meet if $\alpha_{T,\bf k_*}$ is a single number, i.e., if there are no degeneracies. Since $\alpha^{ji}_{T,\bf k_*} \neq 0$ only if $\nu_x^i(k_{*y})=\nu_x^j(k_{*y})$, this means that at the TRIM points of the BZ, the Wannier centers are at least twofold degenerate. When $\hat{T}^2=1$, on the other hand, the Wannier centers are not required to be degenerate.

From \eqref{eq:app_Wannier_basis_under_TR} we have the transformation
\begin{align}
\tilde{\W}^{\nu_x*}_{y,{\bf k}}  &\stackrel{TR}{=} \alpha^\dagger_{T,{\bf k}}  \tilde{\W}^{\nu_x}_{-y, {-\bf k}} \alpha_{T,{\bf k}} \nonumber\\
&\stackrel{TR}{=} \alpha^\dagger_{T,{\bf k}}  \tilde{\W}^{\nu_x \dagger}_{y, {-\bf k}} \alpha_{T,{\bf k}} 
\end{align}
from which it follows that
\begin{align}
\{\nu^{\nu_x,j}_y(k_x)\} \stackrel{TR}{=}\{\nu^{\nu_x,j}_y(-k_x)\} \;\; \mbox{mod 1},
\end{align}
where $j \in 1\ldots N_{\nu_x}$ labels the eigenvalue, and $N_{\nu_x}$ is the number of Wannier bands in the sector $\nu_x$. This implies that
\begin{align}
p^{\nu_x}_y(k_x) = p^{\nu_x}_y(-k_x),
\end{align}
which does not impose further constraints on the Wannier-sector polarization.

\subsection{Constraints due to chiral symmetry}
\label{sec:app_Wilson_loops_under_chiral}
In this section we derive the constraints that chiral symmetry imposes on the Wilson loops $\W_{x,{\bf k}}$ and on the nested Wilson loops $\tilde\W^{\nu_x}_{y,{\bf k}}$.

\subsubsection{On the Wilson loop over energy bands}
Under chiral symmetry, the Bloch Hamiltonian obeys
\begin{align}
\hat{\Pi} h_{\bf k} \hat{\Pi}^\dagger = -h_{\bf k}.
\end{align}
where $\Pi$ is the chiral operator, which is unitary. Chiral symmetry relates the Wilson loops on opposite sides of the energy gap,
\begin{align}
B_{\Pi, \bf k} \W^{occ}_{x,{\bf k}} B^\dagger_{\Pi, \bf k} = \W^{unocc}_{x,\bf k}
\label{eq:app_Wilson_loop_under_Chiral}
\end{align}
where $occ$ ($unocc$) stands for Wilson loops over occupied (unoccupied) bands. The sewing matrix $B^{mn}_{\bf k} = \matrixel{u^m_{\bf k}}{\hat{\Pi}}{u^n_{\bf k}}$ connects states at ${\bf k}$ on opposite sides of the energy gap, i.e., such that $\epsilon_n({\bf k}) = -\epsilon_m(\bf k)$ when the Fermi level is at $\epsilon=0$. Thus, for a non-zero $B^{mn}_{\bf k}$, if $m$ labels a state in the occupied band, $n$ labels a state in the unoccupied band, or vice versa. Let us denote $\ket{\nu}$ as the eigenstates for $\W^{occ}$ and $\ket{\eta}$ as the eigenstates for $\W^{unocc}$. Likewise, we denote the eigenvalues of $\W^{occ}$ ($\W^{unocc}$) as $e^{i 2\pi\nu}$ ($e^{i 2\pi\eta}$).
From \eqref{eq:app_Wilson_loop_under_Chiral}, it follows that chiral symmetry relates the Wannier values of occupied and unoccupied bands according to
\begin{align}
\left\{ \nu^i_x(k_y) \right\} &\stackrel{\Pi}{=} \left\{ \eta^i_x(k_y)\right\}\;\;\mbox{mod 1}.
\end{align}
This in turn leads to 
\begin{align}
p^{occ}_x(k_y) &\stackrel{\Pi}{=} p^{unocc}_x(k_y)\;\;\mbox{mod 1}.
\label{eq:pxkyChiral}
\end{align}
Since the Hilbert space of the Hamiltonian (occupied and unoccupied energy bands included) at each $k_y$ is complete, and thus always has trivial topology, we have that
\begin{align}
p^{occ}_x(k_y)+p^{unocc}_x(k_y) &= 0\;\;\mbox{mod 1},
\end{align}
or 
\begin{align}
p^{occ}_x(k_y) &= -p^{unocc}_x(k_y)\;\;\mbox{mod 1}.
\label{eq:PolarizationHilbertSpace}
\end{align}
Using \eqref{eq:pxkyChiral} and \eqref{eq:PolarizationHilbertSpace} we conclude that
\begin{align}
p^{occ}_x(k_y) \stackrel{\Pi}{=} p^{unocc}_x(k_y) \stackrel{\Pi}{=}  0 \mbox{ or } 1/2,
\label{eq:app_PolarizationQuantizationChiral}
\end{align}
i.e., under chiral symmetry the polarization at each $k_y$ is quantized. This implies that the overall polarization is also quantized.
\begin{align}
p^{occ}_x &\stackrel{\Pi}{=} p^{unocc}_x \stackrel{\Pi}{=}0 \mbox{ or } 1/2.
\end{align}

\subsubsection{On the Wilson loop over Wannier sectors}
From \eqref{eq:app_Wilson_loop_under_Chiral} it follows that $B_{\Pi,\bf k} \ket{\nu^i_{x,\bf k}}$ is an eigenstate of $\W^{unocc}_{x,\bf k}$ with eigenvalue $e^{i 2\pi \nu_x^i(k_y)}$. Thus, in the expansion
\begin{align}
B_{\Pi,\bf k} \ket{\nu^i_{x,\bf k}} = \ket{\eta^j_{x,\bf k}}  \alpha^{ji}_{\Pi,\bf k}
\end{align}
the sewing matrix
\begin{align}
\alpha^{ji}_{\Pi, \bf k} = \matrixel{\eta^j_{x, \bf k}}{B_{\Pi, \bf k}}{\nu^{i}_{x,{\bf k}}}
\end{align}
connects eigenstates of Wilson-loop over occupied and unoccupied energy bands at base points ${\bf k}$ and having the same Wannier centers [$ \alpha^{ji}_{\bf k} \neq 0$ only if $\nu_x^i(k_y) = \eta_x^j(k_y)$]. 
Let us consider the Wannier sectors 
\begin{align}
\ket{w^{occ,j}_{x,{\bf k}}} &= \sum^{N_{occ}}_{n=1}\ket{u^n_{\bf k}} [\nu^j_{x,\bf k}]^n\nonumber\\
\ket{w^{unocc,j}_{x,{\bf k}}} &= \sum^{N}_{n=N_{occ}+1}\ket{u^n_{\bf k}} [\eta^j_{x,\bf k}]^n
\end{align}
where in the first (second) equation $n$ runs over occupied (unoccupied) bands. Under chiral symmetry, the Wannier sectors obey
\begin{align}
\hat{\Pi} \ket{w^{occ,j}_{x,\bf k}}&=\ket{w^{unocc,i}_{x,\bf k}} \alpha^{ij}_{\Pi,\bf k}.
\end{align}
Using this expression one arrives to the following relation for a Wilson line element
\begin{align}
\left[\tilde{\W}^{\nu_x}_{{\bf k}_2 \leftarrow {\bf k}_1}\right]^{ij} = \left[\alpha^\dagger_{\Pi,{\bf k}_2}\right]^{i i'} \left[\tilde{\W}^{\eta_x}_{{\bf k}_2 \leftarrow {\bf k}_1}\right]^{i'j'} \left[\alpha^{\phantom{\dagger}}_{\I,{\bf k}_1}\right]^{j'j},
\end{align}
where the sewing matrices $\alpha_{\Pi, \bf k}$ only Wilson connect lines eigenstates such that $\nu_x = \eta_x$. In particular the nested Wilson loop along $y$ obeys
\begin{align}
\left[\tilde{\W}^{\nu_x}_{y,{\bf k}} \right]^{ij} &=  \left[\alpha^\dagger_{\Pi,\bf k} \right]^{ii'} \left[ \tilde{\W}^{\eta_x}_{y, {\bf k}} \right]^{i'j'} \left[\alpha^{\phantom{\dagger}}_{\Pi,\bf k} \right]^{j'j}.
\end{align}
Thus, the Wilson-loop eigenvalues are related by
\begin{align}
\left\{e^{i 2 \pi \nu^{\nu_x,j}_y(k_x)}\right\} = \left\{e^{i 2\pi \nu^{\eta_x,j}_y(k_x)}\right\}
\end{align}
or
\begin{align}
\left\{\nu^{\nu_x,j}_y(k_x) \right\}=\left\{\nu^{\eta_x,j}_y(k_x)\right\}\;\; \mbox{mod 1},
\end{align}
which implies that
\begin{align}
p^{\nu_x}_y(k_x) \stackrel{\Pi}{=} p^{\eta_x}_y(k_x)\;\; \mbox{mod 1}.
\label{eq:WannierPolarizationChiral}
\end{align}
Hence, the Wannier-sector polarization \eqref{eq:app_Wannier_sector_polarization}, under chiral symmetry obeys
\begin{align}
p^{\nu_x}_y\stackrel{\Pi}{=}p^{\eta_x}_y.
\end{align}
Since the Hilbert space of the Hamiltonian at each $k_x$ is complete, and thus it has trivial topology, we also have
\begin{align}
p^{\nu_x}_y(k_x) + p^{-\nu_x}_y(k_x) + p^{\eta_x}_y(k_x) + p^{-\eta_x}_y(k_x) = 0 \;\;\mbox{mod 1}.
\label{eq:WannierSectorPolarizationHilbertSpaceKx}
\end{align}
which results in the relation for the Wannier-sector polarizations
\begin{align}
p^{\nu_x}_y + p^{-\nu_x}_y + p^{\eta_x}_y + p^{-\eta_x}_y = 0 \;\;\mbox{mod 1}.
\label{eq:WannierSectorPolarizationHilbertSpace}
\end{align}
Notice that \eqref{eq:WannierSectorPolarizationHilbertSpaceKx}, along with \eqref{eq:WannierPolarizationChiral}, is insufficient to quantize the Wannier-sector polarization. At most, we have
\begin{align}
p^{\nu_x}_y(k_x) + p^{-\nu_x}_y(k_x) \stackrel{\Pi}{=}
p^{\eta_x}_y(k_x) + p^{-\eta_x}_y(k_x) \stackrel{\Pi}{=} 0 \mbox{ or } 1/2.
\end{align}
which is compatible with \eqref{eq:app_PolarizationQuantizationChiral}, since $p^{occ}_y(k_x) = p^{\nu_x}_y(k_x) + p^{-\nu_x}_y(k_x)$ and $p^{unocc}_y(k_x) = p^{\eta_x}_y(k_x) + p^{-\eta_x}_y(k_x)$.

\subsection{Constraints due to charge conjugation symmetry}
\label{sec:app_Wilson_loops_under_CC}
Finally, we derive the constraints that charge conjugation symmetry imposes on the Wilson loops $\W_{x,{\bf k}}$ and on the nested Wilson loops $\tilde\W^{\nu_x}_{y,{\bf k}}$.

\subsubsection{On the Wilson loop over energy bands}
Under charge conjugation symmetry the Bloch Hamiltonian obeys
\begin{align}
\hat{C}h_{\bf k}\hat{C}^{-1} = -h_{- {\bf k}}.
\end{align}
Here we will treat $\hat{C}$ as being anti unitary such that $\hat{C}=\hat{Q}K$ is the charge conjugation operator, composed of a unitary matrix $\hat{Q}$ and complex-conjugation $K$. The Wilson loop transforms as
\begin{align}
B_{C,\bf k} \W^{occ*}_{x,{\bf k}} B^\dagger_{C,\bf k} &= \W^{unocc}_{-x,-{\bf k}}= \W^{unocc \dagger}_{x,-{\bf k}}.
\label{eq:app_Wilson_loop_under_ChargeConjugation}
\end{align}
The sewing matrix $B^{mn}_{\bf k} = \matrixel{u^m_{-{\bf k}}}{\hat{C}}{u^n_{\bf k}}= \matrixel{u^m_{-{\bf k}}}{\hat{Q}}{u^{n*}_{\bf k}}$ connects states at ${\bf k}$ with states at $-{\bf k}$ such that $\epsilon_n({\bf k}) = -\epsilon_m(-{\bf k})$. Thus, for a non-zero $B^{mn}_{\bf k}$, if $m$ labels a state in the occupied band, $n$ labels a state in the unoccupied band, or vice versa. 
Let us denote $\ket{\nu}$ as the eigenstates for $\W^{occ}$ and $\ket{\eta}$ as the eigenstates for $\W^{unocc}$. Likewise, let us denote the eigenvalues of $\W^{occ}$ ($\W^{unocc}$) as $e^{i 2\pi\nu}$ ($e^{i 2\pi\eta}$). Eq. \ref{eq:app_Wilson_loop_under_ChargeConjugation} implies that the Wannier centers obey
\begin{align}
\left\{ \nu^i_x(k_y) \right\} \stackrel{CC}{=} \left\{ \eta^i_x(-k_y) \right\} \;\;\mbox{mod 1},
\end{align}
which implies that
\begin{align}
p^{occ}_x(k_y) \stackrel{CC}{=} p^{unocc}_x(-k_y)
\end{align}
and
\begin{align}
p^{occ}_x \stackrel{CC}{=} p^{unocc}_x.
\label{eq:pxCC}
\end{align}
Hence, charge-conjugation symmetry relates the polarization of occupied bands with the polarization of unoccupied bands.
Using \eqref{eq:pxCC} and \eqref{eq:PolarizationHilbertSpace} we conclude that
\begin{align}
p^{occ}_x \stackrel{CC}{=} p^{unocc}_x \stackrel{CC}{=}  0 \mbox{ or } 1/2,
\end{align}
i.e., under charge-conjugation symmetry the polarization is quantized.

\subsubsection{On the Wilson loop over Wannier sectors}
From Eq. \ref{eq:app_Wilson_loop_under_ChargeConjugation} it follows that $B_{C,\bf k} \ket{\nu^{i*}_{x,\bf k}}$ is an eigenstate of $\W^{unocc \dagger}_{x,-\bf k}$ with eigenvalue $e^{-i 2\pi \nu_x^i(k_y)}$. Thus, in the expansion
\begin{align}
B_{C,\bf k} \ket{\nu^{i*}_{x,\bf k}} = \ket{\eta^{j}_{x,-\bf k}}  \alpha^{ji}_{C,\bf k}
\end{align}
the sewing matrix
\begin{align}
\alpha^{ji}_{C, \bf k} = \matrixel{\eta^{j}_{x, -\bf k}}{B_{C, \bf k}}{\nu^{i*}_{x,{\bf k}}}
\end{align}
has $ \alpha^{ji}_{C,\bf k} \neq 0$ only if $\nu_x^i(k_y) = \eta_x^j(-k_y)$. 
Let us consider the Wannier sectors 
\begin{align}
\ket{w^{occ,j}_{x,{\bf k}}} &= \sum^{N_{occ}}_{n=1}\ket{u^n_{\bf k}} [\nu^j_{x,\bf k}]^n\nonumber\\
\ket{w^{unocc,j}_{x,{\bf k}}} &= \sum^{N}_{n=N_{occ}+1}\ket{u^n_{\bf k}} [\eta^j_{x,\bf k}]^n
\end{align}
where in the first (second) equation $n$ runs over occupied (unoccupied) bands. Under charge conjugation symmetry, the Wannier sectors obey
\begin{align}
\hat{C} \ket{w^{occ,i}_{x,\bf k}}&=\ket{w^{unocc,j}_{x,-\bf k}} \alpha^{ji}_{C,\bf k}.
\end{align}
Using this expression one arrives to the following relation for the nested Wilson loop along $y$:
\begin{align}
\left[\tilde{\W}^{\nu_x*}_{y,{\bf k}} \right]^{ij} &=  \left[\alpha^\dagger_{C,\bf k} \right]^{ii'} \left[ \tilde{\W}^{\eta_x}_{-y, -{\bf k}} \right]^{i'j'} \left[\alpha^{\phantom{\dagger}}_{C,\bf k} \right]^{j'j}.
\end{align}
Thus, the Wilson-loop eigenvalues are related by
\begin{align}
\left\{e^{i 2 \pi \nu^{\nu_x,j}_y(k_x)}\right\} \stackrel{CC}{=} \left\{e^{i 2\pi \nu^{\eta_x,j}_y(-k_x)}\right\}
\end{align}
or
\begin{align}
\left\{\nu^{\nu_x,j}_y(k_x) \right\}=\left\{\nu^{\eta_x,j}_y(-k_x)\right\}\;\; \mbox{mod 1}.
\end{align}
This implies that
\begin{align}
p^{\nu_x}_y(k_x) \stackrel{CC}{=} p^{\eta_x}_y(-k_x)\;\;\mbox{mod 1}.
\end{align}
Hence, the Wannier-sector polarization \eqref{eq:app_Wannier_sector_polarization}, under charge conjugation symmetry obeys
\begin{align}
p^{\nu_x}_y\stackrel{CC}{=}p^{\eta_x}_y.
\end{align}
Using this expression along with \eqref{eq:WannierSectorPolarizationHilbertSpace}, we also have the relations
\begin{align}
p^{\nu_x}_y + p^{-\nu_x}_y \stackrel{CC}{=}
p^{\eta_x}_y + p^{-\eta_x}_y \stackrel{CC}{=} 0 \mbox{ or }1/2.
\end{align}

\section{Plaquette flux and its relation to the commutation of reflection operators}
\label{sec:app_symmetries_up_to_gauge}
In this section we study the conditions under which reflection symmetry is compatible with non-zero flux on a plaquette. The existence of reflection symmetry (up to a gauge transformation), and the commutation relations of the $x$ and $y$ reflections depend on the value of the flux. This is important in the model for a quadrupole insulator \eqref{eq:QuadHamiltonian} due to the requirement that reflection operators must not commute in order to have gapped Wannier bands (see Section \ref{sec:GappedWannierBands_conditions}). Furthermore, the cases in which plaquettes have $0$ or $2\pi$ fluxes are gapless at half-filling, and therefore are not useful in the construction of a 2D quadrupole Hamiltonian. On the other hand, plaquettes with $\pi$ flux are gapped at half-filling and obey $[\hat{M}_x,\hat{M}_y] \neq 0.$ Thus, they can be used in the construction of a non-trivial quadrupole model built from arrays of such plaquettes. Indeed, the quadrupole insulator  \eqref{eq:QuadHamiltonian} is built exactly this way using plaquettes with $\pi$ flux.

\begin{figure}[h]
\centering
\includegraphics[width=\columnwidth]{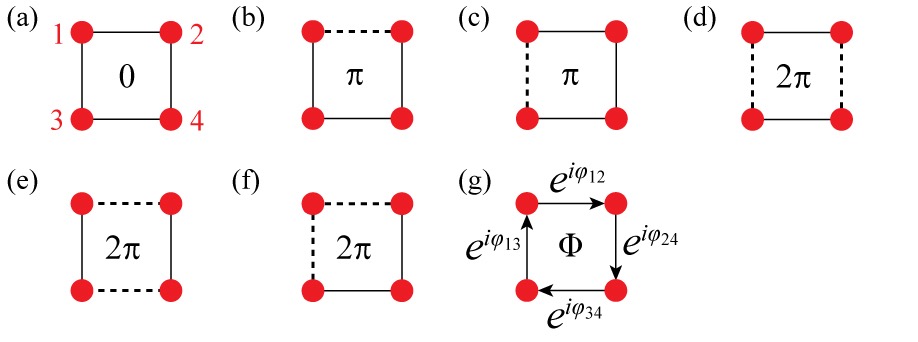}
   \caption{(Color online)  Hopping configurations on a plaquette with four sites. Dotted lines indicate a flipped sign compared to solid lines. (a)-(d) have either $0$ or $2\pi$ flux, while (e),(f) are different configurations with $\pi$ flux. (g) is a generic configuration with flux $\Phi.$}
\label{fig:app_square}
\end{figure}

Let us start with the simple square configuration in Fig.~\ref{fig:app_square}a, which has no flux. Its Hamiltonian is
\begin{equation}
H_0=\left(\begin{matrix} 0 & 1 & 1 &0\\
1 & 0 & 0 &1\\
1& 0 & 0 &1\\
0 & 1 & 1 &0\\\end{matrix}\right),
\end{equation} or, more compactly, $H_0=\mathbb{I}\otimes \sigma^x+\tau^x\otimes\mathbb{I}.$ This plaquette has reflection symmetries that exchange left and right $\hat{M}^{0}_x=\mathbb{I}\otimes\sigma^{x}$ and up and down $\hat{M}^{0}_y=\tau^x\otimes\mathbb{I}.$ These operators multiply to give the inversion operator $\mathcal{I}=\hat{M}^{0}_x \hat{M}^{0}_y=\tau^x\otimes \sigma^x$. In this case we have $[\hat{M}^{0}_x ,\hat{M}^{0}_y]=0.$ Hence $\mathcal{I}^{2}=\hat{M}^{0}_x \hat{M}^{0}_y \hat{M}^{0}_x \hat{M}^{0}_y=(\hat{M}^{0}_x)^2 (\hat{M}^{0}_y)^2 = +1.$ This system has energies $\{-2, 0, 0, +2\}$ and therefore is gapless at half filling. 

Now let us consider configurations with $\pi$ flux. When the flipped bond is between sites 1 and 2 (see Fig.~\ref{fig:app_square}b) we have
\begin{equation}
H_{12}=-\tau^z\otimes \sigma^x+\tau^x\otimes\mathbb{I}.
\end{equation} The energies of $H_{12}$ are $\{-1 ,-1, +1, +1\}$, and hence this system is gapped at half filling. This system has a reflection symmetry in the $x$-direction, but \emph{does not} have an exact reflection symmetry in the $y$-direction, it only has a reflection symmetry times (up to) a gauge transformation. This is because, although the magnetic field is invariant under reflection, the vector potential is not, and we must multiply a pair of the second-quantized operators by a $-1$ in order to recover the symmetry. This $-1$ is the gauge transformation. As such the reflection operator that sends $x \to -x$ does not change, i.e.,  $\hat{M}^{12}_{x}=\mathbb{I}\otimes \sigma^{x}.$ However, the reflection operator in the $y$-direction now has additional signs and we have $\hat{M}^{12}_{y}=\tau^{x}\otimes \sigma^{z}=\hat{M}^{0}_{y}(\mathbb{I}\otimes \sigma^z)$ where $\mathcal{G}=(\mathbb{I}\otimes \sigma^z)$ is one choice for the gauge transformation (another would be $\mathcal{G}=-(\mathbb{I}\otimes \sigma^z)$). This gauge transformation multiplies either $c^{\dagger}_1$ and $c^{\dagger}_3$ or $c^{\dagger}_2$ and $c^{\dagger}_4$ by a minus sign depending on our choice of $\mathcal{G}$, and leaves the other operators unchanged. 
In this case, the commutation relations have now change to $\{\hat{M}^{12}_x, \hat{M}^{12}_y\}=0$. 

Let us consider another $\pi$ flux configuration such that the flipped bond is between sites 1 and 3, as in Fig.~\ref{fig:app_square}c. The Hamiltonian is 
\begin{equation}
H_{13}=\mathbb{I}\otimes \sigma^x- \tau^x\otimes \sigma^z.
\end{equation} This has reflection in $y$, but reflection only up to gauge transformation in the $x$-direction. The gauge transformation in this case is $\mathcal{G}=\tau^z\otimes \mathbb{I}.$ The reflection operators are $\hat{M}^{13}_x=\tau^z\otimes\sigma^x$ and $\hat{M}^{13}_{y}=\tau^{x}\otimes \mathbb{I}.$ These also have a non-vanishing commutator. 

Let us see what happens if we have $2\pi$ flux through the plaquette. If the bonds are arranged as Fig.~\ref{fig:app_square}d,e, the system has reflection symmetries in both the $x$ and $y$ directions with reflection operators $\hat{M}^{0}_x$ and $\hat{M}^{0}_y$ as above, which commute. However, there is another option where the bonds are as in Fig.~\ref{fig:app_square}f. In this case both reflection symmetries are only good up to a gauge transformation and the operators are $\hat{M}^{2\pi}_x=\hat{M}^{13}_x$ and $\hat{M}^{2\pi}_y=\hat{M}^{12}_y$. However, the two operators commute. Hence, only gauge transformations associated with odd numbers of $\pi$ flux lead to non-commuting operators in the spinless case. 

\emph{\underline{General formulation}:}
Let us now consider the general case shown in Fig.~\ref{fig:app_square}g. Let us take the general Hamiltonian for a square with flux $\Phi:$
\begin{equation}
H_\Phi=\left(\begin{matrix} 0 & e^{i\varphi_{12}} & e^{-i\varphi_{13}} &0\\
e^{-i\varphi_{12}} & 0 & 0 &e^{i\varphi_{24}}\\
e^{i\varphi_{13}}& 0 & 0 &e^{-i\varphi_{34}}\\
0 & e^{-i\varphi_{24}} & e^{i\varphi_{34}} &0\\\end{matrix}\right),
\end{equation}\noindent where the total flux through a plaquette is $\Phi=\varphi_{12}+\varphi_{24}+\varphi_{34}+\varphi_{13}.$ Now let us consider the reflection operators $\hat{M}_{x}=\hat{M}_{x}^{0}\mathcal{G}_x$ and $\hat{M}_{y}=\hat{M}_{y}^{0}\mathcal{G}_y$ where $\hat{M}_{x,y}^{0}$ are as above, i.e., the reflection operators with vanishing flux and 
\begin{equation}
\mathcal{G}_{x,y}={\rm{diag}}\left[ e^{i\varphi_{1x,y}}, e^{i\varphi_{2x,y}}, e^{i\varphi_{3x,y}}, e^{i\varphi_{4x,y}}\right]
\end{equation}
are the gauge choices to account for the flux $\Phi$.
By brute force evaluation we can check the conditions under which $[H_{\phi}, \hat{M}_{x,y}]=0.$ The condition, in both cases, reduces to the constraint $1-e^{2i\Phi}=0$, which is solved by $\Phi=n\pi$ for some integer $n$. This makes physical sense since reflection $M_x$ or $M_y$ flips a magnetic field in the $z$ direction (i.e., the flux threading the plaquette), however, flipping a magnetic flux of $0$, $\pi$ is equivalent to $0$, $-\pi$ through a gauge transformation.

Finally, we consider the commutator between the reflection operators. By brute force evaluation of the commutator one can show that if $1-e^{i\Phi}=0$ the commutator vanishes. Otherwise, if $\Phi$ is an odd multiple of $\pi$ we find $[\hat{M}_{x},\hat{M}_{y}]=2\hat{\I}$ where $\hat{\I}$ is the inversion operator. Furthermore, one can show that $\hat{\I}^2=e^{i\alpha}e^{i\Phi}\mathbb{I}$ where $\alpha$ is a global phase that depends on the gauge choice, and $e^{i\Phi}$ is $\pm 1$ for $\Phi$ an even/odd multiple of $\pi.$ We find $\alpha=3\varphi_{12}-\varphi_{13}+\varphi_{24}-\varphi_{34}.$

\section{Conditions for gapped Wannier bands and subsequent quantized Wannier-sector polarization beyond the $N_{occ}=2$ case.}
\label{sec:GappedWannierBands_conditions2}
In Section \ref{sec:GappedWannierBands_conditions} we established that a crystal with $N_{occ}=2$ occupied bands having reflection and inversion symmetries has gapless Wannier bands if the reflection operators commute. Here we generalize this study to the cases in which $N_{occ} = 4$, $N_{occ} = 4n$, and $N_{occ} = 4n+2$. The cases with odd $N_{occ}$ do not need to be considered, because they automatically generate gapless Wannier spectra. 

\subsection{$N_{occ}=4$: Gapped Wannier bands with trivial \\Wannier polarizations}
Unlike the $N_{occ}=2$ case, if four energy bands are occupied, it is possible to meet the conditions of having $\hat{M}_x$ and $\hat{M}_y$ obeying $[\hat{M}_x,\hat{M}_y]=0$, as well as $\I=\hat{M}_x\hat{M}_y$, such that their eigenvalues over the occupied bands come in pairs $(+-)$ at any high-symmetry point. This occurs only for the choice of states $(\ket{++}, \ket{+-}, \ket{-+}, \ket{--})$, where $m_x$ and $m_y$ in $\ket{m_x,m_y}$ are the eigenvalues of the reflection operators $\hat{M}_x$ and $\hat{M}_y$, respectively. In that case, the Wannier bands at the high-symmetry points are gapped. Using this basis, the sewing matrices for $\hat{M}_x$, $\hat{M}_y$, and $\hat{\I}$ at the high-symmetry points ${\bf k}^*$ take the forms
\begin{align}
B_{M_x, {\bf k}^*}&=\tau^z\otimes\mathbb{I}\nonumber\\
B_{M_y, {\bf k}^*}&=\mathbb{I}\otimes\sigma^z\nonumber\\
B_{\I, {\bf k}^*}&=B_{M_x, {\bf k}^*}B_{M_y, {\bf k}^*}=\tau^z\otimes\sigma^z.
\end{align}
These matrices are useful because they represent the symmetries that the Wannier Hamiltonian must have at the high-symmetry points (see Eq.~\ref{eq:app_Wannier_Hamiltonian_symmetries}). For example,  $H_{\W_x}({\bf{k}})$ must satisfy 
\begin{align}
[H_{\W_{x}}({{{\bf k}^*}}),B_{M_y,{{{\bf k}^*}}}]&=\{H_{\W_{x}}({{{\bf k}^*}}),B_{M_x,{{{\bf k}^*}}}\}\nonumber\\
&=\{H_{\W_{x}}({{{\bf k}^*}}),B_{\mathcal{I},{{{\bf k}^*}}}\}=0.
\end{align}
Similarly, $H_{\W_y}({\bf{k}})$ must satisfy
\begin{align}
[H_{\W_{y}}({{{\bf k}^*}}),B_{M_x,{{{\bf k}^*}}}]&=\{H_{\W_{y}}({{{\bf k}^*}}),B_{M_y,{{{\bf k}^*}}}\}\nonumber\\&=\{H_{\W_{y}}({{{\bf k}^*}}),B_{\mathcal{I},{{{\bf k}^*}}}\}=0.
\end{align}
Imposing these symmetries on all sixteen Hermitian matrices $\tau^i \otimes \sigma^j$, for $i,j =0,x,y,z$ (where $\tau$, $\sigma$ are Pauli matrices and $\tau^0 = \sigma^0 = \mathbb{I}$), the most general form for the Wannier Hamiltonians is
\begin{align}
H_{\W_x}({{{\bf k}^*}})&=\delta_1\tau^{x}\otimes\sigma^z+\delta_2\tau^{x}\otimes\mathbb{I}+\delta_3\tau^{y}\otimes\sigma^z+\delta_4\tau^{y}\otimes\mathbb{I}\nonumber \\
H_{\W_y}({{{\bf k}^*}})&=\beta_1\tau^{z}\otimes\sigma^x+\beta_2\mathbb{I}\otimes\sigma^x+\beta_3\tau^{z}\otimes\sigma^y+\beta_4\mathbb{I}\otimes\sigma^y,
\end{align}
where $\delta_i$ and $\beta_i$ for $i=1,2,3,4$, are coefficients which can vary between the different high-symmetry points. The Wannier energies of $H_{\W_x}$ and $H_{\W_y}$ are, respectively, 
\begin{align}
\theta_x&= 2\pi \nu_x = \begin{cases}\pm \sqrt{(\delta_1 - \delta_2)^2 + (\delta_3 - \delta_4)^2}\\
\pm \sqrt{(\delta_1 + \delta_2)^2 + (\delta_3 + \delta_4)^2} \end{cases},\nonumber \\
\theta_y&= 2\pi \nu_y = \begin{cases}\pm \sqrt{(\beta_1 - \beta_2)^2 + (\beta_3 - \beta_4)^2}\\
\pm \sqrt{(\beta_1 + \beta_2)^2 + (\beta_3 + \beta_4)^2}\end{cases},
\end{align}
mod $2\pi$.
By direct computation we find that the eigenstates of the upper (or lower) bands $\nu_x$ of $H_{\W_x}$ have $(+-)$ eigenvalues under $B_{M_y,{{{\bf k}^*}}}$, for any values of the $\delta$ coefficients. Hence the $\alpha_{M_y,{\bf k}^*}$ sewing matrix at each high-symmetry point has $(+-)$  eigenvalues and thus, the eigenvalues of the Wilson loop over Wannier band $\nu_x$ come in pairs $(\nu^{\nu_x}_y(k_x),-\nu^{\nu_x}_y(k_x))$ at $k_x=0,\pi$. Now, since it is not possible to continuously deform the bands $(\nu^{\nu_x}_y(k_x),-\nu^{\nu_x}_y(k_x))$ at $k_x=0,\pi$ to $[0,1/2]$ or $[1/2,0]$ at any other point in $k_x$ without breaking reflection symmetry along $y$, it follows that the eigenvalues of the Wilson loop over Wannier band $\nu_x$  come in pairs $(\nu^{\nu_x}_y(k_x),-\nu^{\nu_x}_y(k_x))$ at all $k_x \in (-\pi,\pi]$, which results in a vanishing Wannier-sector polarization of Eq.~\ref{eq:PolarizationWannierSector}. For $H_{\W_y}$ a similar statement is true. Hence, the quadrupole moment vanishes when the reflection operators commute.

\subsection{$N_{occ}=4n$: Generalizing the $N_{occ}=4$ case}
Now let us generalize the previous argument. Suppose we have $4n$ occupied bands and the $M_x$, $M_y$, and $\mathcal{I}$ eigenvalues all come in $(+-)$ pairs at each high-symmetry point. We can arrange the basis of occupied energy bands such that 
\begin{eqnarray}
B_{M_x,{{{\bf k}^*}}}&=&\tau^z\otimes\mathbb{I}_{2n}\nonumber\\
B_{M_y,{{{\bf k}^*}}}&=&\mathbb{I}_{2n}\otimes\sigma^z\nonumber\\
B_{\mathcal{I},{{{\bf k}^*}}}&=&\mu^z\otimes\mathbb{I}_{n}\otimes\sigma^z.
\label{eq:app_sewing_matrices_4n_case}
\end{eqnarray} 
Crucially, each Wannier Hamiltonian at a high-symmetry point has to commute with one reflection sewing matrix, and anticommute with the other since one reflection preserves the contour and the other flips it. Consider $H_{\W_x}({\bf k}^*).$ It must satisfy $[H_{\W_{x}}({{{\bf k}^*}}),B_{M_y,{{{\bf k}^*}}}]=\{H_{\W_{x}}({{{\bf k}^*}}),B_{M_x,{{{\bf k}^*}}}\}.$ We can label an eigenstate of $H_{\W_x}({{{\bf k}^*}})$ as $\ket{\nu^j_x,b_{m_y}}$, where $\nu^j_x$ is its Wannier eigenvalue, and $b_{m_y}$ is the eigenvalue under $B_{M_y,{{{\bf k}^*}}}$. For each $\ket{\nu^j_x,b_{m_y}}$ we have another state $B_{M_x,{{{\bf k}^*}}} \ket{\nu^j_x,b_{m_y}}$ which has \emph{opposite} Wannier eigenvalue, but the \emph{same} $b_{m_y}$. This is because $M_x$ complex-conjugates the Wannier eigenvalue, but since the Wannier Hamiltonian commutes (by assumption) with $M_y$, it leaves $b_{m_y}$ invariant.

Now we can see from the form of our sewing matrices in Eq.~\ref{eq:app_sewing_matrices_4n_case} that there are an equal, and \emph{even} number of $\pm$ eigenvalues ($4n$ bands means $2n$ each of $\pm$), which is a necessary and direct result of our need for gapped Wannier bands. This means that each of the gapped Wannier sectors has an equal number of $\pm$ reflection sewing eigenvalues. 

Hence, since the reflection-sewing eigenvalues of a Wannier sector determine its polarization as indicated in Table~\ref{tab:EigenvaluesRelations}, we find that the Wannier centers of the projected Wannier sector must come in complex conjugate pairs, and hence its polarization is trivial. 
This result can be applied \emph{mutatis mutandis} for the other Wilson loop Hamiltonian $H_{W_y}.$ Since the nested Wilson loops must be trivial in both directions the quadrupole is trivial. 

\subsection{$N_{occ}=4n+2$: Gapless Wannier bands}
This case mirrors the $N_{occ}=2$ case. In order to have gapped Wannier bands for any set of $4n+2$ occupied energy bands we must choose an array of occupied states such that there are $2n+1$  eigenvalues $+1$ and $2n+1$ eigenvalues $-1$ of both $M_x$, and $M_y$.  After making this choice we can try to arrange them such that the products of the eigenvalues, i.e., the inversion eigenvalues, also come in $\pm$ pairs, so that the Wannier bands are gapped. To achieve that, we also need $2n+1$ inversion eigenvalues $+1$ and $2n+1$ inversion eigenvalues $-1$. No matter what arrangement we choose, the number of inversion eigenvalues $+1$ and the number of inversion eigenvalues $-1$ is always an \emph{even} number and cannot be $2n+1$. Hence, we cannot ever find exactly matched pairs of $\pm$ inversion eigenvalues. An alternative way of stating this is that we can find exactly matched pairs for $4n$ bands, but the remaining eigenvalues reduce to the $2$ band problem that we have already shown is gapless. 

In conclusion we have shown that with commuting reflection operators the Wannier spectrum is either gapless or has trivial topology.

\section{Proof that non-commuting reflection operators protect the energy degeneracy at the high-symmetry points of the BZ}
\label{sec:EnergyDegeneracyProtection}
The Hamiltonian for the quadrupole insulator \eqref{eq:QuadHamiltonian} is symmetric under reflections in $x$ and $y$, where the reflection operators obey $[\hat{M}_x,\hat{M}_y] \neq 0$. At the high-symmetry points of the BZ, ${{\bf k}_*}= \bf \Gamma$, ${\bf X}$, ${\bf Y}$, and ${\bf M}$, the Hamiltonian commutes with both reflection operators,
\begin{align}
[\hat{M}_j, h^q(\bf k_*)] &= 0,
\end{align}
for $j=x,y$. 
Thus, at these points of the BZ there are two natural bases that satisfy
\begin{align}
\hat{M}_x \ket{u^i_{{\bf k}_*}} &= m^i_x \ket{u^i_{{\bf k}_*}}\nonumber\\
\hat{M}_y \ket{v^i_{{\bf k}_*}} &= m^i_y \ket{v^i_{{\bf k}_*}},
\end{align}
where $i=1,2$ labels the energy states.

In the particular case of the reflection operators \eqref{eq:QuadSymmetryOperators} of the quadrupole Hamiltonian \eqref{eq:QuadHamiltonian}, which obey
\begin{align}
\{\hat{M}_x,\hat{M}_y\} &=0
\label{eq:reflection_symmetries}
\end{align}
we can consider labeling the energy bands at the ${\bf{k}}_{*}$  points according to their $\hat{M}_x$ reflection eigenvalues, so that 
\begin{align}
h^q({\bf k}_*) \ket{u^n_{{\bf k}_*}} &= \epsilon^n({\bf k}_*) \ket{u^n_{{\bf k}_*}}\nonumber\\
\hat{M}_x \ket{u^n_{{\bf k}_*}} &= m^n_x({{\bf k}_*}) \ket{u^n_{{\bf k}_*}}.
\end{align}
Picking $\ket{u^n_{{\bf k}_*}}$ to be a simultaneous eigenstate of $h^q({\bf k}_*)$ and $\hat{M}_x$ is possible since $[\hat{M}_x, h^q({\bf k}_*)]=0$.
Then, we have
\begin{align}
\hat{M}_x \hat{M}_y \ket{u^n_{{\bf k}_*}} &= - \hat{M}_y \hat{M}_x \ket{u^n_{{\bf k}_*}}\nonumber\\
&= -m^n_x({\bf k}_*) \hat{M}_y \ket{u^n_{{\bf k}_*}},
\end{align}
so, for every eigenstate $\ket{u^n_{{\bf k}_*}}$ with reflection eigenvalue $m^n_x({\bf k}_*)$ there is another eigenstate $\hat{M}_y \ket{u^n_{{\bf k}_*}}$ with eigenvalue $-m^n_x({\bf k}_*)$. The energy of this eigenstate is
\begin{align}
h^q({\bf k}_*) \hat{M}_y \ket{u^n_{{\bf k}_*}} &= \hat{M}_y h^q({\bf k}_*) \ket{u^n_{{\bf k}_*}}\nonumber\\
&= \epsilon^n({\bf k}_*)\hat{M}_y \ket{u^n_{{\bf k}_*}}
\end{align}
i.e., it is degenerate in energy to $\ket{u^n_{{\bf k}_*}}$.


\begin{figure}[t]%
\centering
\includegraphics[width=\columnwidth]{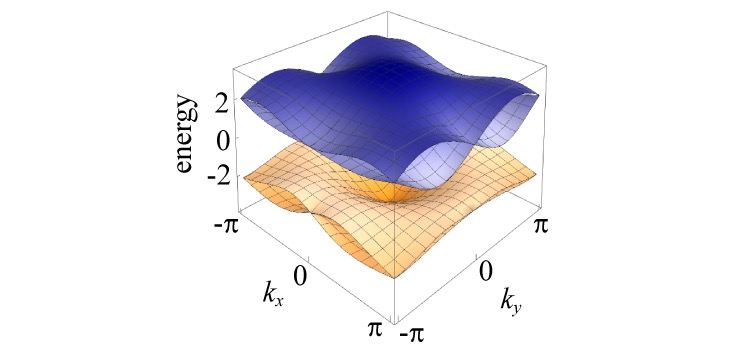}
\caption{(Color online)  Energy bands of Hamiltonian \eqref{eq:quad_symmetry_breaking}, which breaks all symmetries in \eqref{eq:QuadHamiltonian} except the reflection symmetries $M_x$, $M_y$, which have anti-commuting reflection operators. The energies are degenerate at the high symmetry points $\bf k_*=\Gamma, X, Y, M$. In this simulation $\lambda_x=\lambda_y=1$, $\gamma_x=\gamma_y=0.5$, $W=0.75$.}
\label{fig:quad_MxMyProtection}
\end{figure}

\section{Perturbations on the quadrupole Hamiltonian}
\label{sec:app_QuadBreakingSymmetriesSimulation}

\begin{figure}[t]
\centering
\includegraphics[width=\columnwidth]{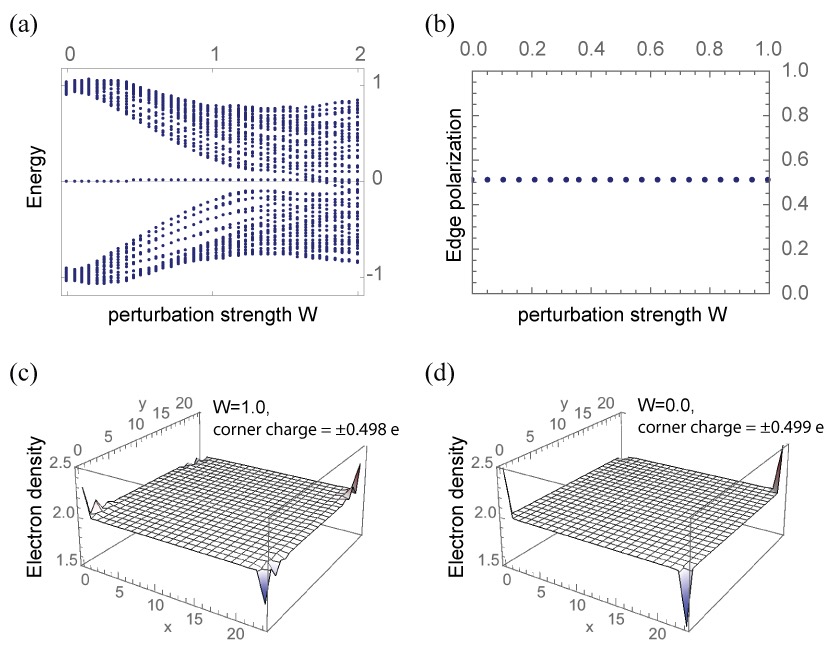}
\caption{(Color online)  Quadrupole with Hamiltonian \eqref{eq:quad_symmetry_breaking}, which breaks charge conjugation symmetry due to the term proportional to $W$. (a) Lowest 100 energies for system with open boundaries as a function of perturbation strength $W$. The lattice has 20 sites per side. (b) Edge polarization as a function of perturbation strength $W$. (c) Electron density of a system with $24 \times 24$ unit cells, at $W=1.0$. The corner charge is $\pm 0.498$. (d) The same simulation as in (c) but with no symmetry breaking perturbation $W=0.0$. The corner charge is $\pm 0.499$. In (b), (c), and (d) a value of $\delta=10^{-3}$ was added to choose the sign of the quadrupole). In these simulations, $\lambda=1$, $\gamma=0.1$.}
\label{fig:quad_PH_breaking}
\end{figure}

In Section \ref{sec:WannierPolarization_quantization}, we mention that the Wannier-sector polarizations \eqref{eq:PolarizationWannierSector}, and consequently the quadrupole invariant \eqref{eq:QuadIndexWannierPolarization2}, are quantized in the presence of reflection symmetries. The analytic proofs of these assertions are in Appendix \ref{sec:WilsonLoopsSymmetry}. In this section we show results of simulations in which all symmetries--other than the noted reflection symmetries--are broken. In particular, we show that breaking charge conjugation symmetry still leaves the  the corner charges and edge polarizations quantized. The Hamiltonian we are considering is
\begin{align}
h&=h^q({\bf k}) \nonumber\\
&+ W \left[ \cos(k_x) R_{e,e} + \sin(k_x) R_{o,e} \right. \nonumber\\
&+\left. \cos(k_y) R_{e,e} + \sin(k_y) R_{e,o} \right]
\label{eq:quad_symmetry_breaking}
\end{align}
where $h^q({\bf k})$ is the quadrupole Hamiltonian defined in \eqref{eq:QuadHamiltonian}, which is in the topological phase for $\delta=0$, and $R_{e,e}$, $R_{e,o}$, and $R_{o,e}$ are random $4 \times 4$ matrices that obey
\begin{align}
[R_{e,i}, \hat{M}_x]=0 \quad 
\{R_{o,e}, \hat{M}_x\}=0 \nonumber\\
[R_{i,e}, \hat{M}_y]=0 \quad 
\{R_{e,o}, \hat{M}_y\}=0.
\end{align}
for $i=e,o$.
Here $\hat{M}_x$ and $\hat{M}_y$ are the reflection operators \eqref{eq:QuadSymmetryOperators}.
These conditions ensure that reflection symmetries  along $x$ and $y$ are preserved while breaking all other symmetries.
In Fig.~\ref{fig:quad_PH_breaking}a we show the energies as a function of the perturbation strength $W$ for a system with open boundaries. 
The lack of charge conjugation symmetry is evident in the asymmetry in the spectrum around zero energy. The in-gap modes, however, remain at an energy close to zero because they are highly localized at the corners, and the perturbation does not include on-site energies, rather, it includes nearest neighbor hopping terms. We also see in Fig.~\ref{fig:quad_PH_breaking}a that the energy gap is maintained at least for values $0<W<1$. Thus, we expect no phase transitions in this range. In Fig.~\ref{fig:quad_PH_breaking}b we show the value of the edge polarization as a function of $W$. The value is sharply quantized at 0.5. In Fig.~\ref{fig:quad_PH_breaking}c we show an electron density plot for $W=1$. The integrated electron density over each quadrant is $\pm 0.498$ relative to the background of 2. In Fig.~\ref{fig:quad_PH_breaking}d we show the simulations when $W=0$ for comparison. The integrated electron density over each quadrant in this case is $\pm 0.499$ relative to the background. In (b), (c), and (d), an additional perturbation of $\delta = 10^{-3}$ (see Eq. \ref{eq:QuadHamiltonian_with_delta}) was added to choose the sign of the quadrupole. This, in addition to finite size effects ($N_x=N_y=24$), explains the (small) departure away from the ideal value of 0.5 in the integrated electron density.
Thus, quantization in our system is maintained in the absence of charge conjugation symmetry, or any other symmetry besides the reflection symmetries $M_x$, $M_y$. In the `clean' Hamiltonian of Eq. \ref{eq:QuadHamiltonian}, the fact that the corner-localized modes are at zero energy is an artifact of the fine-tuned charge conjugation symmetry of our specific model.

\section{Explicit calculation associating reflection eigenvalues to Wannier centers in the quadrupole Hamiltonian in the limit $\gamma_x=\gamma_y=0$}
\label{sec:ReflectionEigenvalues_WannierValues}

The topological properties of Wannier bands can be inferred by inspecting the reflection representation that these bands take at the high-symmetry lines of the BZ shown in red and blue in Fig. \ref{fig:2DBZ}. Since on these lines the eigenstates of well defined position are also the eigenstates of a reflection operator, we can reverse the question: at these high-symmetry points, what is the Wannier value that each $m_x=\pm i$ or $m_y=\pm i$ sector take? In this Appendix we show the explicit calculation that answers this question for the quadrupole model \eqref{eq:QuadHamiltonian} in the limit $\gamma_x=\gamma_y=0$.

To begin, we divide the occupied bands on each of these lines into two sectors according to their reflection representation. We start with the $M_x$ sectors along $(0,k_y)$ and $(\pi,k_y)$ and then with the $M_y$ sectors along $(k_x,0)$ and $(k_x,\pi)$.
The eigenfunctions of the projected reflection operator into the occupied bands $P^{occ}(0, k_y) \hat{M}_x P^{occ}(0, k_y)$ along the line $k_x=0$ are 
\begin{align}
U_{m_x=-i}^{k_x=0}&=\frac{\left\{-\left(1+\sqrt{2}\right) e^{i k_y},1,\left(1+\sqrt{2}\right) e^{i k_y},1\right\}}{2\sqrt{\left(2+\sqrt{2}\right)}}\nonumber\\
U_{m_x=+i}^{k_x=0}&=\frac{\left\{-\left(\sqrt{2}-1\right) e^{i k_y},-1,-\left(\sqrt{2}-1\right) e^{i k_y},1\right\}}{2\sqrt{2- \sqrt{2}}},
\end{align}
where $m_x$ labels the reflection eigenvalue. 
Using each of these states the polarizations are
\begin{align}
\nu_{m_x=-i}^{k_x=0}& = \frac{1}{2\pi} \oint \mathcal{A}_{m_x=-i}^{k_x=0} dk_y = \frac{1}{2}\left(1 +\frac{1}{\sqrt{2}}\right)\nonumber\\
\nu_{m_x=+i}^{k_x=0}& = \frac{1}{2\pi} \oint \mathcal{A}_{m_x=+i}^{k_x=0} dk_y = \frac{1}{2}\left(1 -\frac{1 }{\sqrt{2}}\right).
\label{eq:Py_reflection0}
\end{align}
Similarly, at $k^*_x=\pi$ the states are
\begin{align}
U_{m_x=-i}^{k_x=\pi}&=\frac{\left\{-\left(\sqrt{2}-1\right) e^{i k_y},1,\left(\sqrt{2}-1\right) e^{i k_y},1\right\}}{2\sqrt{2-\sqrt{2}}}\nonumber\\
U_{m_x=+i}^{k_x=\pi}&=\frac{\left\{-\left(1+\sqrt{2}\right) e^{i k_y},-1,-\left(1+\sqrt{2}\right) e^{i k_y},1\right\}}{2\sqrt{\left(2+\sqrt{2}\right)}},
\end{align}
which give the polarizations
\begin{align}
\nu_{m_x=-i}^{k_x=\pi}& = \frac{1}{2\pi} \oint \mathcal{A}_{m_x=-i}^{k_x=\pi} dk_y = \frac{1}{2}\left(1 -\frac{1}{\sqrt{2}}\right)\nonumber\\
\nu_{m_x=+i}^{k_x=\pi}& = \frac{1}{2\pi} \oint \mathcal{A}_{m_x=+i}^{k_x=\pi} dk_y = \frac{1}{2}\left(1 +\frac{1 }{\sqrt{2}}\right).
\label{eq:Py_reflectionPi}
\end{align}
Thus, comparing \eqref{eq:Py_reflection0} and \eqref{eq:Py_reflectionPi}, we conclude that the polarizations along $y$ of the reflection sectors $m_x=\pm i$ at $k_x=0$ and $k_x=\pi$ are opposite (recall $1/2=-1/2\; \mod 1$). Conversely, the reflection representation for each Wannier band is opposite at the $k_x=0$ and $k_x=\pi$ points, which signals a non-trivial topology in the Wannier sectors.

If the projection into sectors is done along $k^*_y=0,\pi$ for the $\hat{M}_y$ operator, similar results are found. The eigenfunctions of $P^{occ}(k_x,0) \hat{M}_y P^{occ}(k_x,0)$ are
\begin{align}
U_{m_y=-i}^{k_y=0}&=\frac{\left\{-1,-\left(\sqrt{2}-1\right) e^{i k_x},\left(\sqrt{2}-1\right) e^{i k_x},1\right\}}{2\sqrt{2-\sqrt{2}}}\nonumber\\
U_{m_y=+i}^{k_y=0}&=\frac{\left\{1,-\left(1+\sqrt{2}\right) e^{i k_x},-\left(1+\sqrt{2}\right) e^{i k_x},1\right\}}{2\sqrt{\left(2+\sqrt{2}\right)}}
\end{align}
which give the polarizations
\begin{align}
\nu_{m_y=-i}^{k_y=0}& = \frac{1}{2}\left(1 -\frac{1}{\sqrt{2}}\right)\nonumber\\
\nu_{m_y=+i}^{k_y=0}& = \frac{1}{2}\left(1 +\frac{1 }{\sqrt{2}}\right),
\label{eq:Px_reflection0}
\end{align}
and the eigenfunctions of $P^{occ}(k_x,\pi) \hat{M}_y P^{occ}(k_x,\pi)$ are
\begin{align}
U_{m_y=-i}^{k_y=\pi}&=\frac{\left\{-1,-\left(1+\sqrt{2}\right) e^{i k_x},\left(1+\sqrt{2}\right) e^{i k_x},1\right\}}{2\sqrt{\left(2+\sqrt{2}\right)}}\nonumber\\
U_{m_y=+i}^{k_y=\pi}&=\frac{\left\{1,-\left(\sqrt{2}-1\right) e^{i k_x},-\left(\sqrt{2}-1\right) e^{i k_x},1\right\}}{2\sqrt{2-\sqrt{2}}},
\end{align}
which give the values
\begin{align}
\nu_{m_y=-i}^{k_y=\pi}& = \frac{1}{2}\left(1 +\frac{1}{\sqrt{2}}\right)\nonumber\\
\nu_{m_y=+i}^{k_y=\pi}& = \frac{1}{2}\left(1 -\frac{1 }{\sqrt{2}}\right).
\label{eq:Px_reflectionPi}
\end{align}
Comparing \eqref{eq:Px_reflection0} and \eqref{eq:Px_reflectionPi}, we conclude that the polarizations along $x$ of the reflection sectors $m_y=\pm i$ at $k_y=0$ and $k_y=\pi$ switch. This implies, just as before, that each Wannier sector has non-trivial topology, and this distinction survives even away from the limit $\gamma_x=\gamma_y=0.$ as long as $\gamma_{x,y}<\lambda_{x,y}$.

By these sets of calculations of Abelian Wilson loops (i.e. Berry phases) we have thus determined that both Wannier-sector polarizations are non-trivial, and thus there is a non-trivial quadrupole moment of $1/2$ in this phase, as per Eq. \ref{eq:QuadIndexWannierPolarization2}.

\section{Derivation of corner zero mode from the low-energy edge Hamiltonian}
\label{sec:DomainWall}
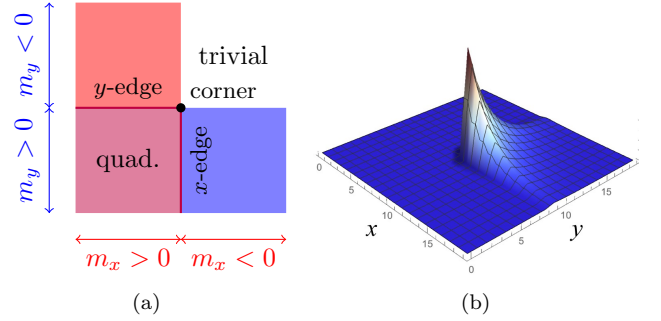
\begin{figure}
	\centering
	\subfigure[]{
		\begin{tikzpicture}[scale=0.7]
		
		\coordinate (pp) at (2,2);
		
		\fill [purple!50] (pp)--(0,2)--(0,0)--(2,0);
		\fill [red!50!] (pp)--(2,4)--(0,4)--(0,2);
		\fill [blue!50!] (pp)--(4,2)--(4,0)--(2,0);
		
		\draw [<->,blue] (-0.5,0)--(-0.5,2);
		\draw [<->,blue] (-0.5,2)--(-0.5,4);
		\node[above,rotate=90,blue] at (-0.5,1) {$m_y>0$};
		\node[above,rotate=90,blue] at (-0.5,3) {$m_y<0$};
		
		\draw [<->,red] (0,-0.5)--(2,-0.5);
		\draw [<->,red] (2,-0.5)--(4,-0.5);
		\node[below,red] at (1,-0.5) {$m_x>0$};
		\node[below,red] at (3,-0.5) {$m_x<0$};
		
		\draw[purple,thick] (0,2)--(pp);
		\node[above] at (1,2){\small $y$-edge};
		
		\draw[purple,thick] (2,0)--(pp);
		\node[below, rotate=90] at (2,1){\small $x$-edge};
		
		\fill [black] (pp) circle (2.5pt) node[above right] {\small corner};
		
		\node[] at (1,1) {quad.};
		\node[] at (3,3) {trivial};
		
		\end{tikzpicture}
	}
	\subfigure[]{
		\includegraphics[width=0.5\columnwidth]{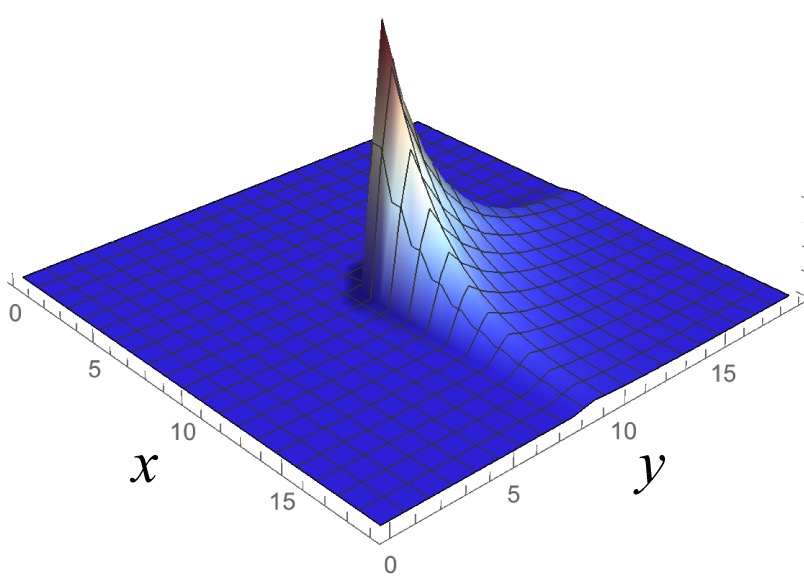}
	}
	\caption{(Color online)  Zero-energy corner-localized mode in the quadrupole phase. (a) Two real space domain walls for $m_x(x)$ and $m_y(y)$ in Hamiltonian \eqref{eq:QuadKP}. The region where the domains $m_x>0$ and $m_y>0$ intersect is in the quadrupole phase (purple region). The zero-energy corner state is localized on the corner shown by the black dot. (b) Probability density function of the zero-energy mode localized at the corner for the configuration of domains shown in (a) for an insulator with $20 \times 20$ sites. The simulation uses Hamiltonian \eqref{eq:QuadHamiltonianL1} with parameters $\gamma_x=-0.01$ for $x \in [0,9]$, $\gamma_x=-1.5$ for $x \in [10,20]$ and $\gamma_y=-0.01$ for $y \in [0,9]$, $\gamma_y=-1.5$ for $y \in [10,20]$.}
	\label{fig:QuadCornerDiagram}
\end{figure}

We claimed that the protected topological corner mode in the quadrupole model is a simultaneous eigenstate of \emph{both} edge Hamiltonians along the $x$ and $y$-edges. To demonstrate this, let us begin with our lattice Hamiltonian \eqref{eq:QuadHamiltonian} with $\lambda=1:$
\begin{align}
H&=\sin k_x \Gamma_3+(\gamma_x+\cos k_x)\Gamma_4\nonumber\\
&+\sin k_y\Gamma_1 +(\gamma_y+\cos k_y)\Gamma_2.
\label{eq:QuadHamiltonianL1}
\end{align}
To simplify the present discussion we will solve a continuum version of the Hamiltonian by assuming that $\gamma_x=-1+m_x$ and $\gamma_y=-1+m_y$ for $m_{x,y}$ small and positive (negative) for the topological (trivial) phase. We can take a continuum limit, or equivalently a $k\cdot p$ expansion about $(k_x, k_y)=0,$ to find the Hamiltonian
\begin{align}
H= k_x \Gamma_3+m_x\Gamma_4 +k_y\Gamma_1+ m_y\Gamma_2.
\label{eq:QuadKP}
\end{align} 
We now use this Hamiltonian to solve for the states localized on the $x$-edge, and then project the Hamiltonian into these states to form the $x$-edge Hamiltonian, from which we can then calculate the corner states. 
We will treat the $x$-edge as a domain wall where $m_x$ steps from positive (inside the topological phase) to negative (outside the topological phase), and the $y$-edge as a domain wall where $m_y$ steps from positive to negative, as shown in Fig.~\ref{fig:QuadCornerDiagram}. 
We use the ansatz $\Psi(x,k_y)=f(x)\Phi_{x}(k_y)$ for the wave function localized at the $x$-edge in the absence of $y$-edges. In this ansatz, $f(x)$ is a scalar function of $x$ and $\Phi_{x}(k_y)$ is a spinor which depends on $k_y$. By inserting this ansatz into the Schrodinger equation with Hamiltonian \eqref{eq:QuadKP} and dividing by $f(x)$ we have
\begin{align}
\left(-i \frac{\partial_x f(x)}{f(x)} \Gamma_3+m_x(x)\Gamma_4\right) \Phi_x(k_y)& \nonumber\\
+ \left( k_y\Gamma_1+ m_y\Gamma_2 \right) \Phi_x(k_y) &= \epsilon \Phi_x(k_y),
\label{eq:quadDomainWall}
\end{align}
where we have replaced $k_x \rightarrow -i\partial_x,$ and $\epsilon$ is the energy. Since the first term in parentheses has all the dependence on $x$, Eq. \ref{eq:quadDomainWall} only has a solution if the first term is a constant. In particular, we choose that constant to be zero (a different value only redefines the zero-point energy of the Hamiltonian),
\begin{align}
(-i\partial_x\Gamma_3+m_x(x)\Gamma_4)f(x)\Phi_x(k_y)=0.
\label{eq:quadDomainWallAux}
\end{align} 
This has the solution $f(x)=C \exp (\int_{0}^{x} m_{x}(x')dx')$, with normalization constant $C$. The matrix equation that results from solving \eqref{eq:quadDomainWallAux} can be simplified to $(\mathbb{I}-\tau^{z}\otimes \sigma^z)\Phi_{x}=0$, from which follows that $\Phi_{x}$ is a positive eigenstate of $\tau^{z}\otimes\sigma^{z},$ i.e., $\Phi_{x1}=(1,0,0,0)$ or $\Phi_{x2}=(0,0,0,1).$ We now project the remaining part of the Hamiltonian into the subspace spanned by these two states to find the low-energy Hamiltonian of the $x$-edge
\begin{align}
H_{edge,\hat{x}}=-k_y \mu^y+m_y\mu^x,
\label{eq:QuadEdgeXHamiltonian}
\end{align} where $\mu^{a}$ are Pauli matrices in the basis $(\Phi_{x1}, \Phi_{x2})$.

Performing an analogous calculation for the $y$-edge we find the matrix equation $(\mathbb{I}-\mathbb{I}\otimes\sigma^z)\Phi_{y}=0$, which has solutions that are positive eigenstates of $\mathbb{I}\otimes\sigma^z,$ i.e., $\Phi_{y1}=(1,0,0,0)$ or $\Phi_{y2}=(0,0,1,0)$. We then project the remaining bulk terms into this basis to find the $y$-edge Hamiltonian
\begin{align}
H_{edge,\hat{y}}=-k_x \gamma^y+m_x\gamma^x
\label{eq:QuadEdgeYHamiltonian}
\end{align} where $\gamma^{a}$ are Pauli matrices in the basis $(\Phi_{y1}, \Phi_{y2}).$

Both of these edge Hamiltonians take the form of massive 1+1d Dirac models, i.e., the natural minimal continuum model for a 1+1d topological insulator (an alternative analysis arriving to this conclusion is found in Ref. \onlinecite{vonOpen2017}). Now the key feature we mentioned earlier, i.e., the simultaneous zero mode can be found by considering a corner, i.e., either the $x$-edge with a $y$-domain wall or the $y$-edge with a $x$-domain wall. 

Let us first look for the zero-energy states localized at a $y$-domain wall on the upper portion of a vertical 1D chain with Hamiltonian \eqref{eq:QuadEdgeXHamiltonian}. The ansatz in this case is of the form $\Phi_x(y) = \exp (\int_{0}^{y} m_{y}(y')dy') \phi_{x,y}$, which, from the Schrodinger equation for Hamiltonian \eqref{eq:QuadEdgeXHamiltonian}, and choosing zero energy, leads to a matrix equation that simplifies to
\begin{align}
(\mathbb{I}-\mu^z)\phi_{x,y}&=0.
\end{align}
Using similar calculations to those above we find the following matrix equation for the $x$-domain wall on the right side of a horizontal 1D insulator with Hamiltonian \eqref{eq:QuadEdgeYHamiltonian}:
\begin{align}
(\mathbb{I}-\gamma^z)\phi_{y,x}&=0.
\end{align}
Hence, the corner mode we find for an $x$-edge with a $y$ domain wall is the positive eigenstate of $\mu^z$ while that for the $y$-edge with an $x$ domain wall is the positive eigenstate of $\gamma^z.$ In both cases the solutions are identical, i.e., they are the first basis elements $\Phi_{x1}=\Phi_{y1}=(1,0,0,0).$ Therefore, the corner zero mode is a simultaneous zero mode of both domain wall Hamiltonians, given by
\begin{align}
\Psi^{corner}(x,y)= e^{\int_0^x m_x(x') dx'}e^{\int_0^y m_y(y') dy'}(1,0,0,0).
\end{align}
Thus, although both edges are 1D topological insulators, they only produce a single zero mode.

\section{Character table of the Quaternion group}
\label{sec:CharacterQuaternion}
The quadrupole insulator with Bloch Hamiltonian \eqref{eq:QuadHamiltonian} has the symmetry of the quaternion group \eqref{eq:quad_quaternion_group}. The character table for this group is shown in Table~\ref{tab:CharacterQuaternion}. There are four one-dimensional representations of this group and one two-dimensional representation. The quadrupole insulator \eqref{eq:QuadHamiltonian} takes the two-dimensional representation at all $\bf k_*$ points of the BZ. 

\begin{table}[h]
\centering
\begin{tabular}{l|rrrrr}
Rep. / class & $\{1\}$ & $\{-1\}$ & $\{\pm \hat{M}_x\}$ & $\{\pm \hat{M}_y\}$ & $\{\pm \hat{I}\}$\\
\hline
$\chi^{triv}$ & 1 & 1 & 1 & 1 & 1\\
$\chi^{M_x}$ & 1 & 1 & 1 & -1 & -1\\
$\chi^{M_y}$ & 1 & 1 & -1 & 1 & -1\\
$\chi^{I}$ & 1 & 1 & -1 & -1 & 1\\
$\chi^{2D}$ & 2 & -2 & 0 & 0 & 0
\end{tabular}
\caption{Character table of the quaternion group}
\label{tab:CharacterQuaternion}
\end{table}

\section{$C_4$-symmetric quadrupole insulator}
\label{sec:QuadC4}
\begin{figure}[t]%
	\centering
	\includegraphics[width=\columnwidth]{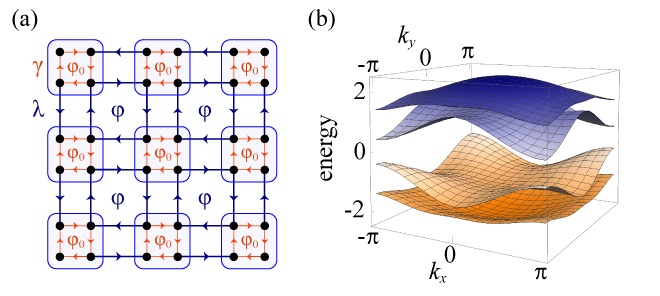}%
	\caption{(Color online)  Quadrupole model that preserves $C_4$ symmetry. (a) $C_4$ symmetric lattice. Red (blue) plaquettes have flux $\varphi_0$ ($\varphi$). Plaquettes sharing red and blue couplings have a flux of $-(\varphi_0 + \varphi)/2$. (b) Energy spectrum when $\varphi_0=0$, $\varphi=\pi$. Despite the fact that the degeneracy of the occupied bands are lifted the quadrupole remains quantized and stable.}
	\label{fig:quad_lattice_c4}
\end{figure}

A schematic representation of a quadrupole with $C_4$ symmetry is shown in Fig.~\ref{fig:quad_lattice_c4}a. It is a variation of the insulator shown in Fig.~\ref{fig:quad_lattice}, but here we set $\gamma_x=\gamma_y=\gamma$ and $\lambda_x=\lambda_y=\lambda$, and allow the fluxes threading each plaquette to be different than $\pi$. The `red' plaquettes, delimited by the intra-cell $\gamma$ couplings have flux $\varphi_0$, while the `blue' plaquettes, delimited by the inter-cell  $\lambda$ hoppings have flux $\varphi$. To simplify the formulation, we take the fluxes into account by replacing 
\begin{align}
\gamma &\rightarrow \gamma e^{i \varphi_0/4}\nonumber\\
\lambda &\rightarrow \lambda e^{i \varphi/4}
\end{align}
in the directions of the arrows in Fig~\ref{fig:quad_lattice_c4}a, or their complex conjugate in the opposite direction.

This implies that plaquettes sharing red couplings and blue hoppings have a flux of $-(\varphi+\varphi_0)/2$. When $\varphi=\varphi_0=0,\pi$ the insulators have reflection symmetries $M_x$, $M_y$. If $\varphi=\varphi_0=0$, we have $[ \hat{M}_x,\hat{M}_y]=0$, and the spectrum is gapless. If $\varphi=\varphi_0=\pi$, on the other hand, we recover the insulator \eqref{eq:QuadHamiltonian}, which has $\{ \hat{M}_x,\hat{M}_y\}=0,$ and realizes a quadrupole SPT phase. In Fig.~\ref{fig:quad_lattice_c4}b we show the energy spectrum for the values $\varphi_0=0$, $\varphi=\pi$. Since the anti-commuting reflection symmetries are lost, so is the protection of the degeneracies at the high-symmetry points of the BZ (cf. Fig.~\ref{fig:quad_MxMyProtection}). The energy and Wannier bands remain gapped during the deformation $\varphi_0=\pi \rightarrow \varphi_0=0$ that connects the quadrupole \eqref{eq:QuadHamiltonian} with the Hamiltonian with the energy spectrum in Fig.~\ref{fig:quad_lattice_c4}b. Thus, the non-trivial topology persists, with the topological signatures shown in Fig.~\ref{fig:quadRotationEigenvalues} and a non-trivial index \eqref{eq:QuadIndexRotationEigenvalues}. Indeed, edge polarization and charge density simulations on this model present the signatures of a quantized quadrupole of Fig.~\ref{fig:quad_polarization}b and Fig.~\ref{fig:quad_charge}b.

\bibliographystyle{ieeetr}
\bibliography{quad_references}

\end{document}